\documentclass[a4paper,10pt,twoside,openleft]{report}

\usepackage[pdftex]{graphicx}

\usepackage{a4wide}
\usepackage{cite}
\usepackage{minitoc}

\usepackage{amssymb,amsmath}

\usepackage{fancyheadings}
\def\GeV{{\ \mbox{GeV}}}
\def\be{\begin{equation}}
\def\ee{\end{equation}}
\newcommand{\thelog}{\log}
\newcommand{\imax}{L}

\usepackage[T1]{fontenc}

\begin{document}




\title{M\'emoire d'habilitation \`a diriger des recherches\\
\vskip 0.5cm
{\bf Quantum chromodynamics at high energy\\
and noisy traveling waves}\\
}


\author{St\'ephane Munier\\
\\
{\it Centre de physique th\'eorique},\\
{\it \'Ecole Polytechnique, CNRS,}\\
{\it Palaiseau, France}}

\maketitle





\maketitle

\centerline{
\bf Abstract}
\vskip 0.5cm

When hadrons scatter at high energies, strong color fields, 
whose dynamics is described
by quantum chromodynamics (QCD), are
generated at the interaction point.
If one represents these fields in terms of partons (quarks and gluons),
the average number densities of the latter
saturate at ultrahigh energies.
At that point, nonlinear effects become predominant
in the dynamical equations.
The hadronic states that one gets in this regime of 
QCD are generically called ``color glass condensates''.

Our understanding of scattering in QCD
has benefited from recent progress in statistical
and mathematical physics. The evolution of hadronic scattering amplitudes 
at fixed impact parameter in the regime 
where nonlinear parton saturation effects 
become sizable was
shown to be similar to the time evolution of a system 
of classical particles undergoing 
reaction-diffusion processes. The dynamics of such a system 
is essentially governed by
equations in the universality class of the stochastic
Fisher-Kolmogorov-Petrovsky-Piscounov equation, 
which is a stochastic nonlinear
partial differential equation. Realizations of that kind of 
equations (that is, ``events'' in a particle physics language)
have the form of noisy traveling waves. 
Universal properties of the latter
can be taken over to scattering amplitudes in QCD.

This review provides an introduction to the basic
methods of statistical physics useful in QCD,
and summarizes the correspondence between
these two fields
and its theoretical and phenomenological implications.\\

\vskip 0.5cm
\hrule
\vskip 1cm

\centerline{
\bf R\'esum\'e}
\vskip 0.5cm

Lors de la diffusion de hadrons \`a haute \'energie,
d'intenses champs de couleur, dont
la dynamique est d\'ecrite par la chromodynamique quantique (QCD),
 sont cr\'e\'es au point d'interaction.
Si on repr\'esente ces champs en termes de partons (quarks et
gluons), la densit\'e de ces derniers sature \`a tr\`es haute \'energie.
Les effets non-lin\'eaires deviennent alors dominants
dans les \'equations dynamiques.
Les \'etats hadroniques que l'on obtient dans ce r\'egime
de la QCD sont g\'en\'eriquement
appel\'es ``condensat de verre de couleur''.

Notre compr\'ehension de la diffusion en QCD a b\'en\'efici\'e
de progr\`es r\'ecents en physique statistique et en physique math\'ematique.
On a montr\'e que 
l'\'evolution des amplitudes de diffusion hadronique \`a param\`etre
d'impact fix\'e dans le r\'egime dans lequel les effets non-lin\'eaires
de saturation des densit\'es de partons deviennent importants
est semblable \`a l'\'evolution temporelle d'un syst\`eme de
particules classiques soumis \`a des processus de type r\'eaction-diffusion.
La dynamique d'un tel syst\`eme est essentiellement gouvern\'ee
par des \'equations dans la classe d'universalit\'e
de l'\'equation de Fisher-Kolmogorov-Petrovsky-Piscounov stochastique,
qui est une \'equation aux d\'eriv\'ees partielles stochastique
et non-lin\'eaire.
Les r\'ealisations de telles \'equations (c'est-\`a-dire les
\'ev\'enements, dans un langage de physique des particules)
ont la forme d'ondes voyageuses bruit\'ees.
Les propri\'et\'es universelles de celles-ci peuvent \^etre
transpos\'ees aux amplitudes d'interactions en QCD.

Ce m\'emoire est une introduction aux m\'ethodes
de physique statistique utiles en QCD, et r\'esume la correspondance
entre ces deux domaines ainsi que ses implications th\'eoriques
et ph\'e\-nom\'e\-nolo\-giques.

\vskip 0.5cm
\hrule

\dominitoc


\chapter*{Preface}
\markboth{PREFACE}{PREFACE}
\addcontentsline{toc}{part}{{Preface}}

The present memoir is based on the review paper of Ref.~\cite{Munier:2009pc}
published as a Physics Report in 2009.
The general structure has not changed, but large parts of the text have been
rewritten in order to propose
different perspectives,
and to incorporate the latest developments.
Most notably, 
recent studies on the density correlations in impact-parameter
space have been added as a new chapter (Chap.~\ref{sec:spatial}),
and a better derivation of the model for the front fluctuations
is given in Chap.~\ref{sec:reviewtraveling}.
Many other more minor modifications have been implemented in the
present version of the review: For example, Sec.~\ref{sec:statisticalmethods}
was simplified.
This thesis was presented on November 30, 2011 to receive the habilitation
degree from the Pierre et Marie Curie University in Paris.\\

The work reported here was completed over a period of about 8 years
starting in 2003.
It was inspired in an essential way by A.H. Mueller and
Bernard Derrida, both by their writings, about the dipole model
for the former and traveling waves for the latter,
and by direct collaboration.
Both of them are leading researchers in their respective fields,
and it has been a honor to be involved in such a collaboration:
My first acknowledgements naturally go to them.
\\

I would like to thank all my collaborators during these years and especially
the once PhD students and postdocs with whom I had the pleasure
to work, the collegues who kindly shared their views
with me, as well as the members of my group, led by Bernard Pire,
and all my labmates at CPHT.\\

I warmly thank the members of the Committee, namely
Jean-Bernard Zuber (chair), Nestor Armesto, 
Fran\c{c}ois G\'elis and
C\'ecile Monthus (referees),
and Yuri Dokshitzer and Lech Szymanowski.\\

Finally, my work from 2006 to 2010
was funded by the Agence Nationale de la Recherche
(France), contract ANR-06-JCJC-0084-02, a grant
received together with Samuel Wallon.\\

\vskip 0.5cm
{
\flushright
{\begin{minipage}{5cm}
\flushright
St\'ephane Munier\\
June 2012
\end{minipage}
}

}

\tableofcontents


\chapter{\label{sec:introduction}Introduction to high energy scattering in QCD}


What is the origin of the mass of ordinary matter?
How are the nucleons ``glued'' together to form stable nuclei?
How can we understand the ``zoo'' of the particles (hadrons) which
are sensitive to the ``strong'' force?

The modern theory of strong interactions, quantum chromodynamics 
(abreviated QCD; For a comprehensive textbook, 
see Ref.~\cite{Muta:1998vi}),
discovered about 40 years ago, seems to have the ability to
help all these problems and many others in a most compact and elegant way.
This theory is parallel in its formulation to the more
well-known theory of electromagnetic interactions, quantum
electrodynamics, and, with some caveats still to be understood, 
to the theory of weak interactions,
consistently with the idea of a (partial)
unification of the elementary forces.
However, QCD poses outstanding mathematical problems,
and it became soon clear that its various
regimes had to be explored by dedicated experiments and specialized
theoretical tools.
While ``simple'' fixed-order perturbation theory has
proved extremely successful to investigate electrodynamics
due to the intrinsic weakness of the force acting 
between charged leptons (the characteristic coupling
is $\alpha_{\mbox{\tiny em}}\simeq \frac{1}{137}$),
chromodynamics has to deal with strong coupling instead (the
coupling $\alpha_s$ is of order 0.1 in the most favorable
cases and up to 1 in general),
and with more subtle nonlinearities which severely limits
the use of fixed-order perturbation theory. 
Perturbative expansions have to be handled with care, and
dedicated tools have to
be invented for QCD.

It has happened that methods were borrowed from
other fields of physics: For example the computation
of low-energy properties of the hadrons (masses, decay rates...)
are investigated using lattice field theory, like in
solid state
physics. More recently,
tools developed by string theorists have proved useful
to address the calculation of specific processes involving
very high-order terms in a perturbative expansion.

This review,
which is a revised version of
the author's publication~\cite{Munier:2009pc}, 
summarizes some recent investigations
in a specific regime of quantum chromodynamics,
the so-called high-energy or, more
technically, ``small-$x$'' regime.
We focus on how this regime is formally
related to some models which appear in
statistical physics and show how this
correspondence may be used.
We shall first go over the recent
history of high-energy QCD in order to better expose
the context of this research.

\paragraph{Short history of the field.}

The study of quantum chromodynamics in the high-energy regime
has undergone a rapid development 
in the last 15 years with the wealth of experimental 
data that have been collected, first at the electron-proton collider DESY-HERA, 
and then at the heavy-ion collider RHIC.
More energy in the collision enables 
the production of objects of higher mass in
the final state, and thus the discovery of new particles.
But higher energies make it also possible to observe more quantum fluctuations
of the incoming objects, that is to say, to study more deeply the structure of the
vacuum.

Analytical approaches to QCD in this regime are based
on a sophisticated handling of
perturbative expansions of observables 
in powers of the strong coupling constant $\alpha_s$ which,
thanks to asymptotic freedom, is justified for carefully
chosen observables in special kinematical regimes.
Some sophistication is needed
because in the evaluation of Feynman graphs,
the coupling constant always comes with ``infrared'' and ``collinear'' logarithms
that are related to the phase space that is available to the reaction, that is to say,
to kinematics,
and may easily push the effective coupling to large values.
Resumming part of these logarithms is mandatory. Resumming all of them
is too difficult.
The question is to carefully select the dominant ones, and this is not at all easy.


An experimental facility able to investigate the high-energy
regime of QCD was
the HERA collider, 
where electrons or positrons scattered off protons at the 
center-of-mass energy $\sqrt{s}$,
exchanging a photon of virtuality $Q$. Through the scattering, one
could probe partonic fluctuations of the proton 
(made of quarks and gluons) of transverse momenta $k\sim Q$, and
longitudinal momentum fractions $x\sim Q^2/(Q^2+s)$.


For a long time, the dominant paradigm had been that the {\em collinear logarithms}
$\ln Q^2$,
that become large when $Q^2$ is large compared to the QCD confinement scale
$\Lambda^2$, were
the most important ones.
As a matter of fact, searches for new particles or for exotic physics require
to scrutinize matter at very small distances, and hence very large $Q^2$ have
to be considered.
Perturbative series of powers of $\alpha_s\ln Q^2$ have to be fully resummed.
The equation that performs this resummation is the celebrated 
Dokshitzer-Gribov-Lipatov-Altarelli-Parisi (DGLAP) equation 
\cite{Gribov:1972ri,Dokshitzer:1977sg,Altarelli:1977zs}.


However, once HERA had revealed its ability to get extremely good statistics
in a regime in which
$Q^2$ is moderate (from $1$ to $100\GeV^2$) and $x$ very small (down to $10^{-5}$)
it became clear that {\em infrared logarithms} ($\ln 1/x$) could show up and even 
dominate the measured observables.
The resummation of the series of infrared logs is performed by the 
Balitsky-Fadin-Kuraev-Lipatov (BFKL) equation
\cite{Lipatov:1976zz,Kuraev:1977fs,Balitsky:1978ic}.
The series $\sum (\alpha_s\ln 1/x)^k$ (with appropriate coefficients)
is the leading order (LO), while the series $\sum \alpha_s(\alpha_s\ln 1/x)^k$
is the next-to-leading order (NLO), which has also been computed 
\cite{Ciafaloni:1998gs,Fadin:1998py}.
The BFKL equation is a linear integro-differential equation.


At ultrahigh energy, the bare BFKL equation seems to violate the Froissart
bound, that states that total hadronic cross-sections cannot rise faster than
$(\ln^2 s)/m_\pi^2$. The latter is a consequence
of the unitarity of the probability of scattering.
 The BFKL equation predicts a power rise with the
energy of the form $s^\varepsilon$, where $\varepsilon$ is positive and
quite large ($0.3$ to $0.5$ according to the effective value of $\alpha_s$ 
that is chosen).
The point at which the BFKL equation breaks down depends on the value of the
typical transverse momentum which characterizes the observable 
(It is the photon virtuality
$Q$ in the case of deep-inelastic scattering).
One may define the energy-dependent {\em saturation scale} $Q_s(x)$ in such a way
that the BFKL equation holds for $Q>Q_s(x)$.
For $Q\sim Q_s(x)$, the probability for scattering to take place
is of order 1, and for $Q<Q_s(x)$, it would be larger than 1 if one trusted
the BFKL equation.
The saturation scale 
is a central observable, which we shall keep discussing in this review:
It signs the point at which the linear (BFKL) formalism has to be corrected
for nonlinear effects. The regime in which nonlinearities manifest themselves
is a regime of strong color fields, sometimes called the {\em color glass
condensate}
(For the etymology of this term,
see e.g. the lectures of Ref.~\cite{McLerran:2001sr};
for a review, see Ref.~\cite{Iancu:2003xm}).

The fact that unitarity is violated
is not only due to the lack of a hadronic scale in the BFKL equation,
which is a perturbative equation; Introducing confinement in the form of
a cutoff would not help this
particular problem:
The violation of unitarity which we are
talking about occur at small distances.
It is just that
still higher orders are needed.
The NLO corrections to the BFKL kernel indeed correct this
behavior in such a way that the description of the HERA data 
in the small-$x$ regime
is possible by the BFKL equation.
However, these corrections are 
not enough to tame the power-like growth of cross-sections as
predicted by the LO BFKL equation.
It seems that a resummation of contributions of arbitrary order would be needed.

New equations were proposed well before the advent of colliders
able to reach this regime.
Gribov-Levin-Ryskin wrote down a model 
for the evolution of the hadronic scattering cross-sections
in the early 80's \cite{Gribov:1981ac,Gribov:1984tu}, and Mueller and Qiu
derived a similar equation from QCD a bit later \cite{Mueller:1985wy}.
These equations are integral evolution equations with a nonlinear term,
which basically takes into account parton saturation effects, that is to say,
recombination or rescattering.
The latter cannot be described in a linear framework such as the BFKL formalism.
Subsequently, more involved QCD evolution equations were derived
from different points of view.
In the 90's, McLerran and Venugopalan 
\cite{McLerran:1993ni,McLerran:1993ka,McLerran:1994vd} proposed a first model,
mainly designed to approach heavy-ion collisions.
Later, Balitsky
\cite{Balitsky:1995ub},
Jalilian-Marian, Iancu, McLerran, Weigert, Leonidov and Kovner (B-JIMWLK)
\cite{JalilianMarian:1997jx,JalilianMarian:1997gr,Iancu:2001ad,Iancu:2000hn,Weigert:2000gi}
worked out QCD corrections to this model, and got equations that reduce
to the BFKL equation in the appropriate limit.
Technically, these equations actually have the form of an infinite hierarchy of
coupled integro-differential equations (in Balitsky's formulation 
\cite{Balitsky:1995ub}), of a functional renormalization group equation,
or alternatively, of a Langevin equation 
(in Weigert's formulation \cite{Weigert:2000gi}).
A much simpler equation was derived in 1996 by Balitsky \cite{Balitsky:1995ub}
and rederived by Kovchegov in 1999
\cite{Kovchegov:1999yj,Kovchegov:1999ua}
in a very elegant way
within a different formalism. The obtained equation is called the Balitsky-Kovchegov
equation (BK).
The latter derivation was based on Mueller's color dipole model \cite{Mueller:1993rr},
which proves particularly suited to represent QCD in the high-energy limit.

The exciting feature of this kinematical regime of
hadronic interaction from a theoretical point of view
is that the color fields are strong, although, at sufficiently high energies,
the QCD coupling is weak, authorizing a perturbative approach,
and thus some of the
analytical calculations outlined above.
In such strong field regime, nonlinear effects become crucial.
But the conditions of applicability of the different
equations that had been found had never been quite clear.
Anyway, these equations are extremely difficult to solve, which had probably been
the main obstacle to more rapid theoretical developments in the field
until recently.

\begin{figure}
\begin{center}
\includegraphics[width=0.8\textwidth]{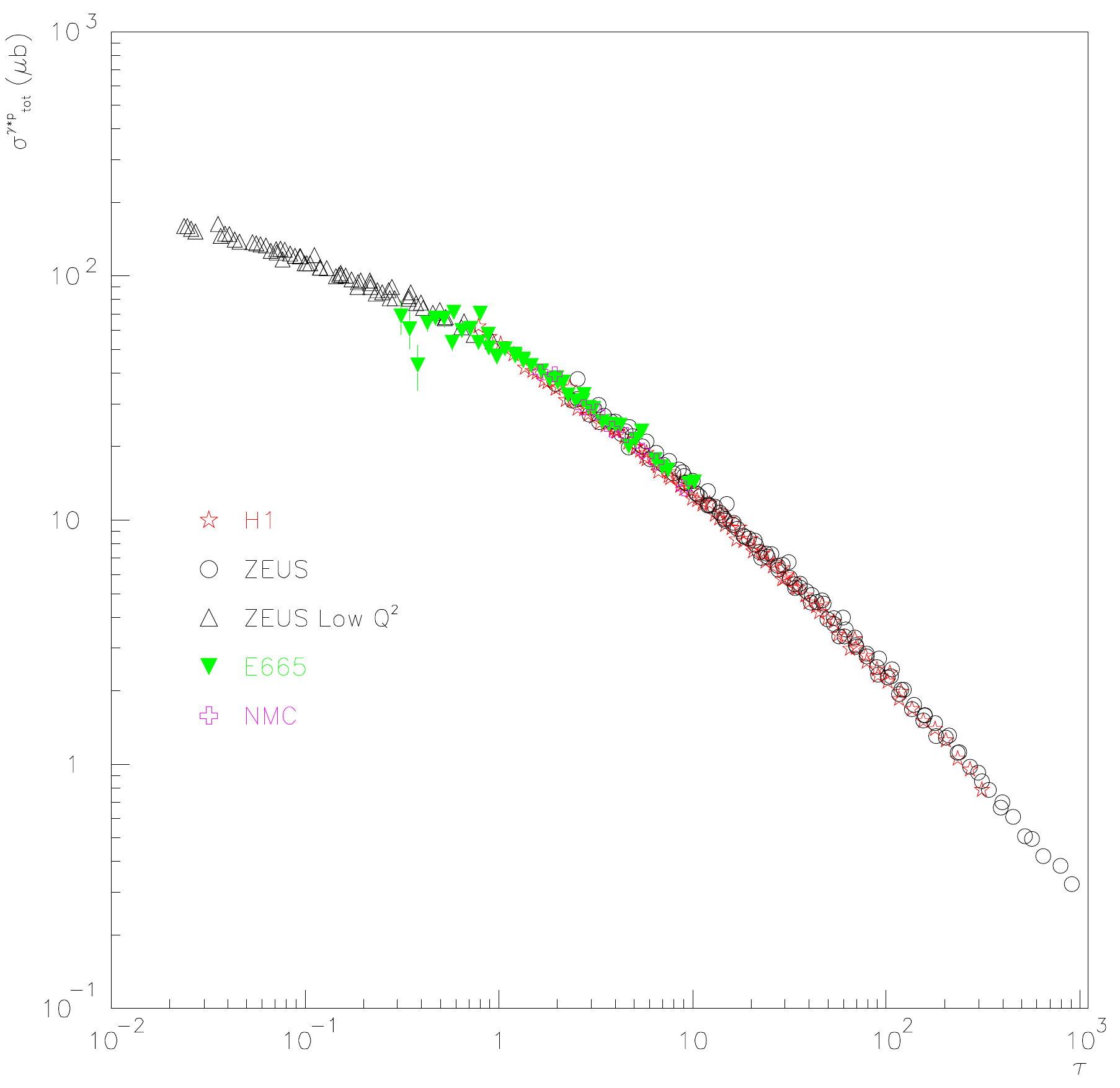}
\end{center}
\caption{\label{fig:geometricscaling}
[From Ref.~\cite{Marquet:2006jb}]
Photon-proton total cross-section from
the most recent set of deep-inelastic scattering data in the low-$x$ regime
plotted as a function of a single scaling-variable 
$\tau=Q^2/Q_s^2(x)$, where $Q$ is the virtuality of the photon and
$Q_s^2(x)\sim\Lambda^2 x^{-0.3}$
is the so-called saturation scale.
Although the cross-section is a priori a function of two variables,
all data fall on the same curve. This phenomenon is called
{\em geometric scaling} and was discovered in Ref.~\cite{Stasto:2000er}.
}
\end{figure}

Furthermore, for a long time, the phenomenological need for
such a sophisticated formalism was not obvious, since linear
evolution equations such as the DGLAP equation were able to
account for almost all available data. 
But Golec-Biernat and W\"usthoff
showed that unitarization effects may have already
been seen at HERA \cite{GolecBiernat:1998js,GolecBiernat:1999qd}. 
Their model predicted, in particular, that
the virtual photon-proton cross-section should
only depend on one single variable $\tau$, made of a combination of
the transverse momentum scale (fixed by the virtuality of the photon $Q$) 
and $x$.
This phenomenon was called ``geometric scaling'' \cite{Stasto:2000er}.
It was found in the HERA data (see Fig.~\ref{fig:geometricscaling}):
This is maybe one of the most
spectacular experimental result from HERA 
in the small-$x$ regime.

This observation has triggered many phenomenological and theoretical works.
Soon after its discovery in the data, 
geometric scaling was shown to be a 
feature of some
solutions of the Balitsky-Kovchegov (BK)
equation, essentially numerically, with some analytical arguments
(see e.g. \cite{Levin:1999mw,Levin:2000mv,Armesto:2001fa,GolecBiernat:2001if}).
The energy dependence of the saturation scale was eventually precisely computed
by Mueller and Triantafyllopoulos \cite{Mueller:2002zm}.
Later, it was shown that the BK equation is actually in the universality
class of the Fisher-Kolmogorov-Petrovsky-Piscounov (FKPP) equation \cite{Fisher,KPP},
and geometric scaling was found to be implied by the fact that the latter equation
admits {\em traveling-wave} solutions \cite{Munier:2003vc}.

A first step beyond the BK equation, in the direction of a full solution
to high energy QCD, was taken by Mueller and Shoshi in 2004 \cite{Mueller:2004sea}.
Actually, they did not solve the B-JIMWLK equations,
which would be the natural candidate for a complete theory.
Instead, they solved the linear BFKL equation with two 
absorptive boundary conditions, which
they argued to be appropriate to represent the expected nonlinearities.
Geometric scaling {\em violations} were found from their calculation, 
which should show up at any energies.

Subsequently, it was shown that high-energy QCD 
at fixed coupling is actually 
in the universality class of {\em reaction-diffusion processes},
studied in statistical physics,
whose dynamics may be encoded in equations similar to the
{\em stochastic} FKPP equation \cite{Iancu:2004es}.
The Mueller-Shoshi solution was shown to be consistent
with solutions to the latter equation.
So high-energy QCD seems to be in correspondence with
disordered systems studied in statistical physics.
This correspondence has provided a new understanding
of QCD in the high-energy regime, and it has
proven very useful to find more features of high-energy
scattering.
%


\paragraph{Outline of this memoir.}

The next chapter is devoted to describing scattering in QCD
from a $s$-channel point of view, relying essentially on the
parton model or, rather, on a realization useful in the
 high-energy limit, the color dipole model.
Once this picture is introduced, it is not difficult to understand 
the correspondence with reaction-diffusion
processes occuring in one spatial dimension, whose dynamics is captured by
equations in the universality 
class of the Fisher-Kolmogorov-Petrovsky-Piscounov (FKPP)
equation.
We then explain how traveling waves appear in this context. 
In Chap.~\ref{sec:zerodimensional}, we study in greater detail a toy model
for which many technics (field theory, statistical methods) may be
worked out completely. This model however ignores spatial dimensions,
and thus, does not account for traveling waves.
We summarize the state-of-the-art research on equations in the universality class
of the FKPP equation in Chap.~\ref{sec:reviewtraveling}.
We then come back to QCD, discussing the issue of the impact
parameter. This will lead us to introduce
new models beyond the simple one-dimensional reaction-diffusion type
models (Chapter~\ref{sec:spatial}).
Finally, we will show how noisy traveling waves may show up
in the actual data.

Over the last few years, several hundreds of papers have appeared related to this
subject, mainly issued from a very active though restricted community.
Obviously, this memoir cannot give a complete account of this abundant literature.
As a matter of fact, some important recent developments had to be left out.
Concerning the correspondence itself, we do not
attempt to establish a definite stochastic nonlinear evolution equation
for QCD amplitudes, for to our judgement, this research line
is not mature enough yet: A better understanding of
the very saturation mechanism at work in QCD is definitely needed before one may
come to this issue. 
Furthermore, it is not clear to us that a stochastic formulation
would be a technical progress, since there are not many known methods to
handle complicated stochastic equations.
We feel that the same is true for
the search for effective actions that would include so-called Pomeron loops.
We also do not address the developments based on the boost-invariance symmetry
that scattering amplitudes should have: This would drive us too far off
the main focus of this review.
As for more phenomenological aspects, 
we only discuss the basic features of total cross-sections without
attempting to address other observables such as diffraction.
We do also not address the issue of next-to-leading effects such as
the running of the QCD coupling. This discussion, though crucial if one
wants to make predictions for actual colliders,
would probably only be technical in its nature: There is no conceptual difference
between the fixed coupling and the running coupling cases.
Here, only basic phenomenological facts brought about
by this new understanding of high-energy QCD are addressed, namely 
geometric scaling and diffusive scaling.


\chapter{\label{sec:schannel}
Hadronic interactions and reaction-diffusion processes
}

{\it We shall introduce here the physical picture
of high-energy scattering in the parton model.
In the first section, the color dipole model \cite{Mueller:1993rr} is described
since it is particularly suited to address high-energy scattering,
especially close to the regime in which nonlinear effects
are expected to play a significant role.
In a second section, we shall argue that high-energy scattering is a peculiar
reaction-diffusion process.}\\

\minitoc

\section{\label{sec:partonmodel}Parton model and dipoles}

\subsection{General picture}

\begin{figure}
\begin{center}
\begin{tabular}{cc}
\includegraphics[width=5cm]{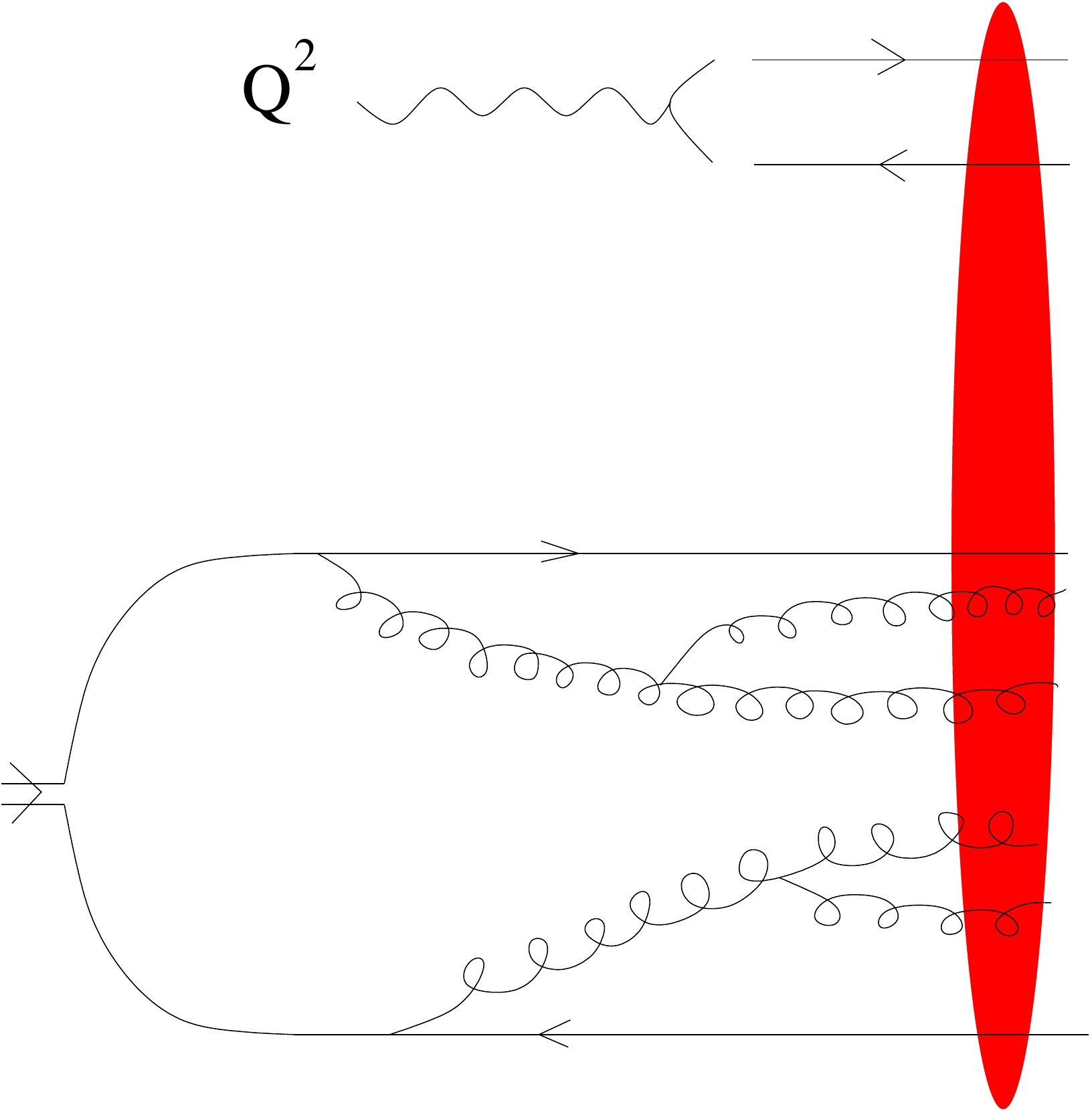}
\phantom{spacespa}
&\includegraphics[width=5cm]{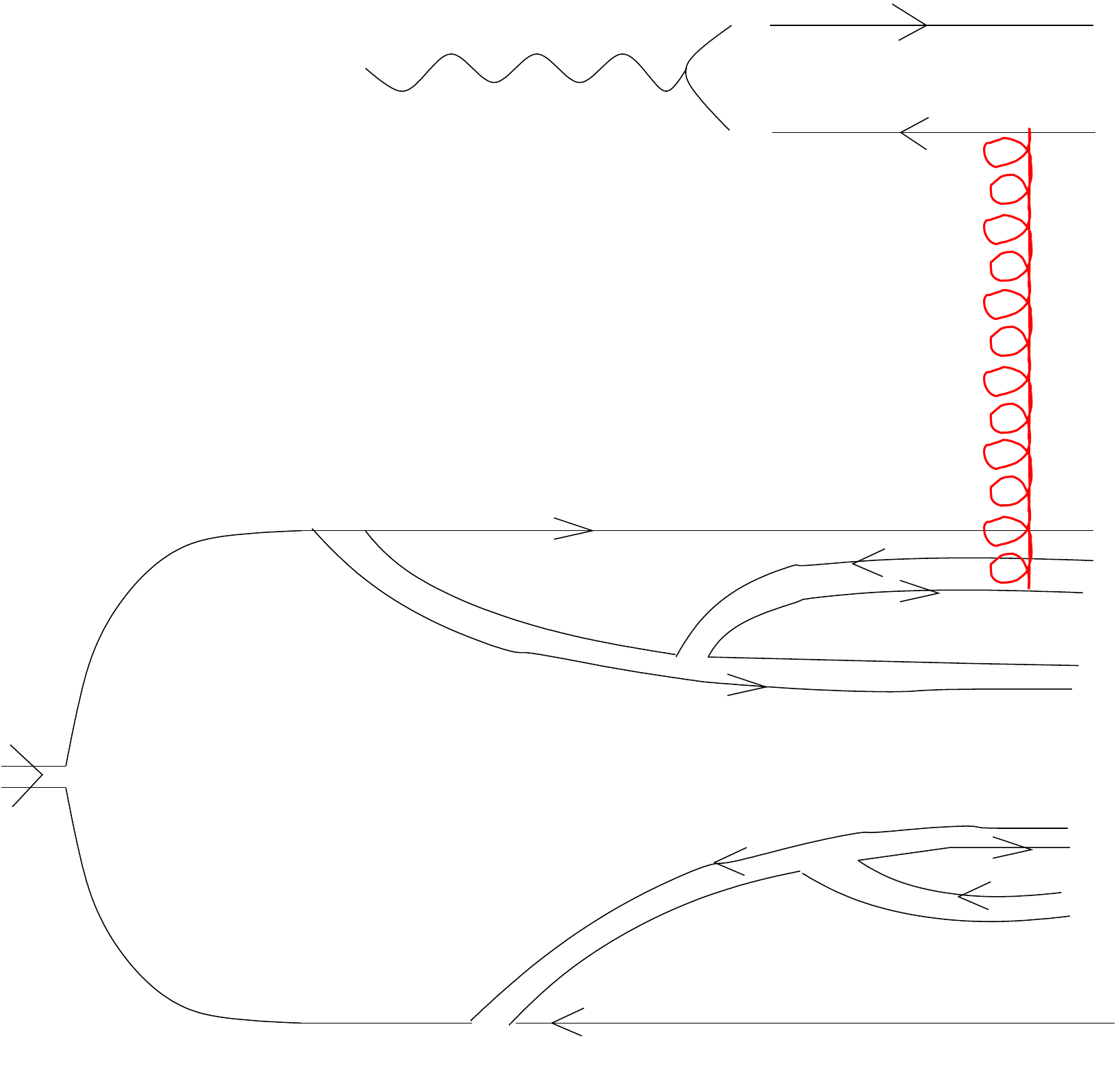}\\
(a) & (b)
\end{tabular}
\end{center}
\caption{\label{fig:basicscat}
{\em (a)}
The scattering of a virtual photon probe off a particular
fluctuation of an evolved target made of a quark and an antiquark
in its bare state.
The photon necessarily goes through a quark-antiquark pair at high
enough energies, when the target is dominated by dense gluon states.
(What is represented in this figure is actually the inelastic amplitude, which
is a cut of the total cross-section or of the forward elastic amplitude).
{\em (b)} In the dipole model, the probe and the 
target may be represented by sets of color dipoles, 
and the interaction proceeds through gluon exchanges.
It is now the elastic amplitude that is represented.
The curly vertical lines represent 2-gluon exchanges between
pairs of dipoles.
}
\end{figure}

For definiteness, let us consider the scattering of a hadronic 
probe off some given target,
in the restframe of the probe and at a fixed impact parameter, that is to
say, at a fixed distance between the probe and the center of the target
in the two-dimensional plane transverse to the collision axis.
In the parton model, the target interacts
through one of its quantum fluctuations, made of a high-occupancy Fock state
if the energy of the reaction is sufficiently high (see Fig.~\ref{fig:basicscat}a).
As will be understood below,
the probe effectively ``counts'' the partons in the Fock state
of the target whose transverse momenta $k$ (or sizes $r\sim 1/k$)
are of the order of the momentum that characterizes the probe: 
The amplitude for the scattering off this particular
partonic configuration is proportional to the number of such partons.

The observable that is maybe the most sensitive to quantum fluctuations
of a hadron is the cross-section for the interaction of a virtual photon
with a hadronic target such as a proton or a nucleus.
The virtual photon is emitted by an electron (or a positron).
What is interesting with this process, called ``deep-inelastic scattering'',
is that the kinematics of the photon is fully controlled
by the measurement of the scattered electron.
The photon can be considered a hadronic object since it interacts through
its fluctuations into a quark-antiquark state. The latter form a
color dipole since although both the quark and the antiquark carry color
charge, the overall object is color neutral due to the color neutrality
of the photon. The probability distribution of these fluctuations may
be computed in quantum electrodynamics (QED). 
Subsequently, the dipole interacts with the target
by exchanging gluons. The dipole-target cross-section factorizes
at high energy.
One typical event is depicted in Fig.~\ref{fig:basicscat}a.

Dipole models \cite{Nikolaev:1990ja,Nikolaev:1991et}
have become more and more popular among phenomenologists
since knowing the dipole cross-section enables one to compute different
kinds of observables. 
Like parton densities, the latter is a universal quantity, that
may be extracted from one process and used to predict other observables.
Different phenomenological models may 
be tried for the dipole cross-section.
QCD evolution equations may even be derived, as we shall explain 
below.
A critical recent study of the foundations of dipole
models may be found in Ref.~\cite{Ewerz:2004vf,Ewerz:2006vd}.

In QCD, the state of a hadronic object, encoded in a set of 
wave functions, is built up
from successive splittings of partons starting from the valence structure.
This is visible in the example of Fig.~\ref{fig:basicscat}a:
The quark and the antiquark that build up, in this example, the target in its
asymptotic state each emit a gluon, which themselves emit, 
later on in the evolution, other gluons.
As one increases the rapidity $y$ by boosting the target,
the opening of the phase space for parton splittings makes the probability
for high occupation numbers larger.
Indeed, the probability to find a gluon that carries a fraction $z$ (up to $dz$)
of the momentum of its parent parton (which may be a quark or a gluon)
is of order $\alpha_s N_c dz/z$ for small $z$ ($N_c$ is the number of colors; $N_c=3$ in
real-life QCD).
As we see, there is a logarithmic singularity in $z$, meaning that 
emissions of very soft gluons (small $z$) are favored if they
are allowed by the kinematics.
The splitting probability is of order 1 when the total rapidity of the scattering 
$y=\ln 1/x$ is increased
by roughly $1/\bar\alpha$  (the convenient notation 
$\bar\alpha=\alpha_s N_c/\pi$ has been introduced).
Only splittings of a quark or of a gluon into a gluon 
exhibit the $1/z$ singularity.
Therefore, at large rapidities, gluons eventually
dominate the partonic content of the hadrons.

The parton model in its basic form, where the fundamental objects of
the theory (quarks and gluons) are directly considered,
is not so easy to handle in the high-energy regime.
One may considerably simplify the problem by going to the limit of
a large number of colors ($N_c\gg 1$), 
in which a gluon may be seen as a zero-size quark-antiquark pair.
Then, color-neutral objects become collections of color dipoles, whose endpoints consist
in ``half gluons'' (see Fig.~\ref{fig:basicscat}b).
There is only one type of objects in the theory, dipoles, which simplifies very much
the picture.
Furthermore, going to {\em transverse coordinate space} 
(instead of momentum space, usually used in the DGLAP formalism) 
by trading the transverse momenta of the gluons for the sizes of the dipoles
(through an appropriate Fourier transform)
brings another considerable simplification.
Indeed, the splittings that contribute to the
amplitudes in the high-energy limit
are the soft ones, for 
which the emitted gluons take only a small fraction of the momentum
of their parent, the latter being very large. Therefore,
the positions of the gluons, and thus of the
edges of the dipoles, in the plane transverse to the collision axis are not
modified by subsequent evolution once the gluons have been created.
Thus, the evolution of each dipole proceeds through completely {\em independent}
splittings to new dipoles.

We will now see how this picture translates into a QCD evolution equation
for scattering amplitudes, first in the regime in which there are no nonlinear
effects. In a second step, 
we will try and understand how to incorporate the latter.

\subsection{BFKL equation from the dipole model}

\begin{figure}
\begin{center}
\includegraphics[width=6cm]{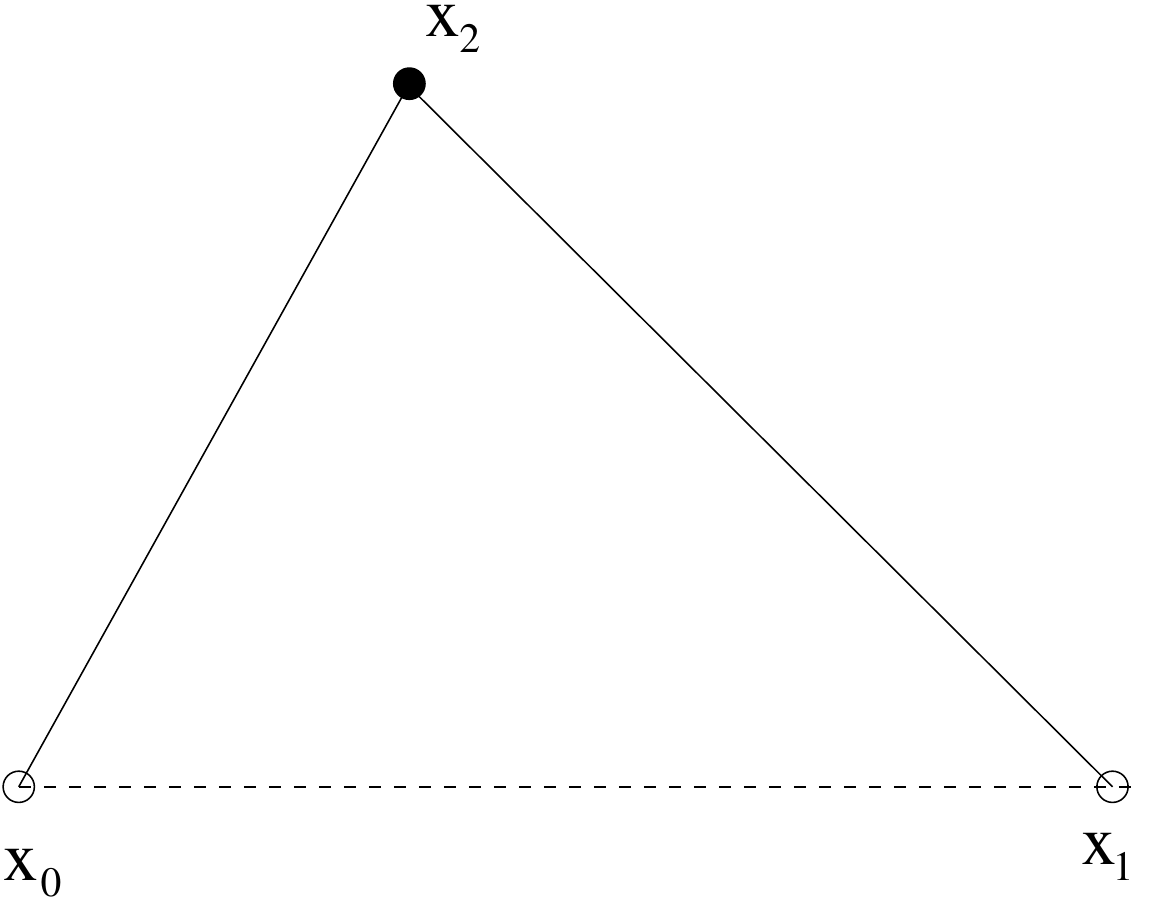}\\
(a)\\
\includegraphics[width=6cm]{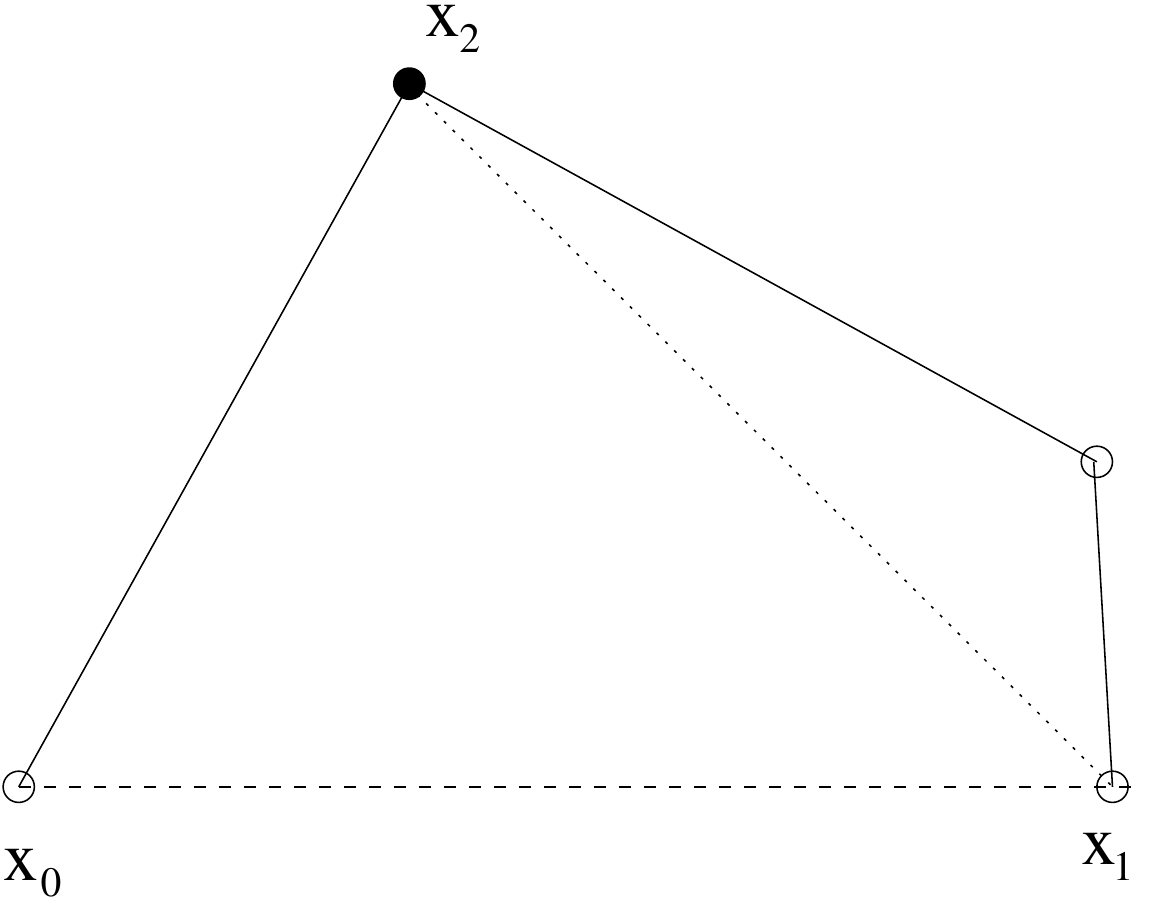}\\
(b)\\
\includegraphics[width=10cm]{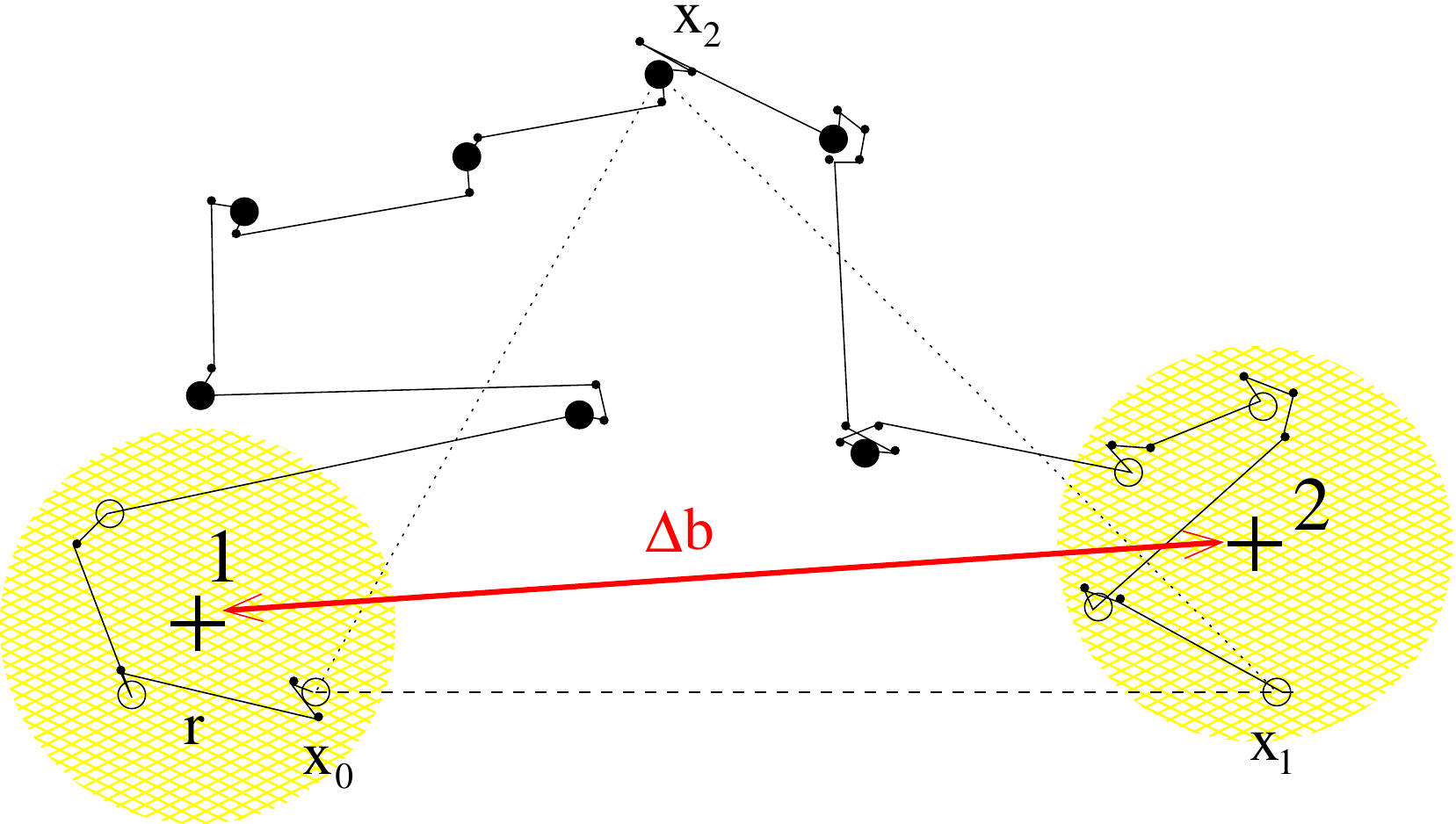}\\
(c)
\end{center}
\caption{\label{fig:scheme}Schematic picture of
a realization of the dipole evolution after the first
two steps of the evolution ((a) and (b)), and after some
larger rapidity evolution (c).
In the first step (a), 
the initial dipole $(x_0,x_1)$ (denoted by a dashed line)
splits to the new dipoles 
$(x_0,x_2)$ and $(x_2,x_1)$ (full lines).
The points represent the edges of each dipole,
that is to say, the position of the gluons.
In the next step (b), the dipole $(x_2,x_1)$ itself splits
in two new dipoles. The splitting process proceeds (c) until
the maximum rapidity is reached.
Many very small dipoles are produced in the vicinity of 
each of these endpoints, due to the infrared singularity visible
in Eq.~(\ref{splitting}) (Only a fraction of them is represented).
The zones 1 and 2 in (c), separated by the transverse 
distance $\Delta b$, would
evolve quasi-independently after the stage
depicted in this figure when saturation effects are included
(See Sec.~\ref{sec:relevance} for the corresponding discussion).
}
\end{figure}

The building up of the states of each hadron is specified by providing the
rate at which a dipole whose endpoints have transverse coordinates
$(x_0,x_1)$ splits
into two dipoles $(x_0,x_2)$ and $(x_2,x_1)$ as the result of a gluon
emission at position $x_2$ when the rapidity of the initial
dipole is increased.
It is computed in perturbative QCD and reads \cite{Mueller:1993rr}
\begin{equation}
\frac{dP}{d(\bar\alpha y)}(x_{01}\rightarrow x_{02},x_{12})
=\frac{|x_{0}-x_1|^2}{|x_{0}-x_2|^2 |x_{1}-x_2|^2}\frac{d^2 x_2}{2\pi}.
\label{splitting}
\end{equation}
Thanks in particular to the large-$N_c$ limit,
dipole splittings are independent. After some rapidity evolution
starting from
a primordial dipole, one gets a chain of dipoles such as the one
depicted in Fig.~\ref{fig:scheme}.

The elementary scattering amplitude for one projectile dipole $(x_0,x_1)$
off a target dipole $(z_0,z_1)$ is independent of the rapidity and 
reads \cite{Mueller:1993rr}
\begin{equation}
T^{\text{el}}((x_0,x_1),(z_0,z_1))=\frac{\pi^2\alpha_s^2}{2}
\ln^2\frac{|x_0-z_1|^2|x_1-z_0|^2}{|x_0-z_0|^2|x_1-z_1|^2}.
\label{eq:Tdipole}
\end{equation}
If the target is an evolved state at rapidity $y$, then it
consists instead in a distribution $n(y,(z_0,z_1))$
of dipoles.
The (forward elastic) scattering amplitude $A(y,(x_0,x_1))$ is just given
by the convolution of $n$ and $T^{\text{el}}$, namely
\begin{equation}
A(y,(x_0,x_1))=
\int \frac{d^2 z_0}{2\pi}\frac{d^2 z_1}{2\pi}
T^{\text{el}}((x_0,x_1),(z_0,z_1))n(y,(z_0,z_1)).
\label{eq:BFKLprojectile}
\end{equation}

Let us examine the properties of $T^{\text{el}}$.
To this aim, it is useful to decompose the coordinates of the
dipoles in their size vector $r_a=x_0-x_1$ (resp. $r_b=z_0-z_1$)
and impact parameter $b_a=\frac{x_0+x_1}{2}$ (resp. $b_b=\frac{z_0+z_1}{2}$).
In the limit in which the relative impact parameters of the dipoles 
$b=b_a-b_b$ is very large compared to their sizes, 
we get the simplified expression
\begin{equation}
T^{\text{el}}(r_a,r_b,b)\underset{|r_a|,|r_b|\ll|b|}{\sim}
 {\alpha_s^2}\frac{r_a^2 r_b^2}{b^4},
\label{eq:Tdipoleapprox1}
\end{equation}
and thus the scattering amplitude decays fast as a function of the relative
impact parameter.
If instead the relative impact parameter is small (of the order of the size of the smallest
dipole), we get for disymmetric sizes
\begin{equation}
T^{\text{el}}(r_a,r_b,b)\underset{|r_a|,|r_b|\sim|b|}{\sim} {\alpha_s^2}
\frac{r_<^2}{r_>^2},
\label{eq:Tdipoleapprox2}
\end{equation}
where $r_<=\min(|r_a|,|r_b|)$, $r_>=\max(|r_a|,|r_b|)$, and
where the integration over the angles has been performed.

Equation~(\ref{eq:Tdipoleapprox1}) means that the dipole interaction 
is local in impact parameter: It vanishes as soon as the relative
distance of the dipoles is a few steps in units of their size.
Eq.~(\ref{eq:Tdipoleapprox2}) shows that only dipoles whose sizes
are of the same order of magnitude interact.
These properties are natural in quantum mechanics.
Thus the amplitude $A$ in Eq.~(\ref{eq:BFKLprojectile}) effectively
``counts'' the dipoles of size of the order of $|x_{01}|$
at the impact parameter $\frac{x_0+x_1}{2}$ (up to $|x_{01}|=|x_0-x_1|$), 
with a weight factor
$\alpha_s^2$.

An evolution equation for the
amplitude $A$ 
with the rapidity of the scattering can be established.
It is enough to know how the dipole density in the target evolves when
rapidity is increased, since all the rapidity dependence is contained
in $n$ in the factorization~(\ref{eq:BFKLprojectile}), and such an equation
may easily be worked out with the help of the splitting
rate distribution~(\ref{splitting}). It reads \cite{Mueller:1993rr}
\begin{equation}
\frac{\partial n(y,(x_0,x_1))}{\partial (\bar\alpha y)}=
\int \frac{d^2 x_2}{2\pi}\frac{|x_{01}|^2}{|x_{02}|^2 |x_{12}|^2}
[
n(y,(x_0,x_2))+n(y,(x_2,x_1))
-n(y,(x_0,x_1))
],
\label{eq:BFKL}
\end{equation}
where $x_{ab}\equiv x_a-x_b$. The very same equation holds for $A$.
The elementary scattering amplitude $T^\text{el}$ only appears
in the initial condition at $y=0$, which is not shown in Eq.~(\ref{eq:BFKL}).

In a nutshell, the integral kernel encodes the branching diffusion of the dipoles.
The total number of dipoles at a given impact parameter grows exponentially,
and their sizes diffuse. The appropriate variable in which diffusion takes
place is $\ln(1/|x_{01}|^2)$. (This is due to the collinear singularities
in Eq.~(\ref{splitting}).)
Equation~(\ref{eq:BFKL}) is nothing but the BFKL equation.
A complete solution to this equation, 
including the impact-parameter dependence, 
is known \cite{Lipatov:1985uk}.

An important property of the amplitude $A$ is that it is boost-invariant.
This property is preserved in the BFKL formulation.
We could have put the evolution in the projectile instead of the
target, or shared it between the projectile and the target: The result
for the scattering amplitude would have been the same.
In a frame in which the target carries $y^\prime$ units of rapidity and
the projectile $y-y^\prime$, the amplitude $A$ reads
\begin{multline}
A(y,(x_0,x_1))=\int \frac{d^2 z_0}{2\pi}\frac{d^2 z_1}{2\pi}
\frac{d^2 z_0^\prime}{2\pi}\frac{d^2 z_1^\prime}{2\pi}
n^\text{projectile}(y-y^\prime,(z_0,z_1)|(x_0,x_1))\\
\times
T^\text{el}((z_0,z_1),(z_0^\prime,z_1^\prime))
n^\text{target}(y^\prime,(z_0^\prime,z_1^\prime)).
\label{eq:BFKLframe}
\end{multline}
$n^\text{projectile}(y-y^\prime,(z_0,z_1)|(x_0,x_1))$
is the density of dipoles $(z_0,z_1)$ found in a dipole of initial
size $(x_0,x_1)$ after evolution over $y-y^\prime$ steps in rapidity.
If $y^\prime=y$, one recovers Eq.~(\ref{eq:BFKLprojectile}).
If $y^\prime=0$, then all the evolution is in the projectile instead.

The amplitude $A$ is related to an interaction probability,
and thus, it must be bounded:
In appropriate normalizations, $A$ has to range between 0 and 1.
But as stated above,
the BFKL equation predicts an exponential rise of $A$
with the rapidity for any dipole size, 
which at large rapidities eventually violates unitarity.
Hence the BFKL equation is not the ultimate representation of
high-energy scattering in QCD.


\subsection{Unitarity and the Balitsky-Kovchegov equation}

It is clear that one important ingredient that has been
left out in the derivation of the BFKL
equation is the possibility of {\em multiple scatterings}
between the probe and the target.
Several among the $n^\text{projectile}$ 
dipoles in Eq.~(\ref{eq:BFKLframe})
may actually interact with the $n^\text{target}$ 
dipoles in the other hadron
simultaneously. The only reason why such interactions
may not take place is that $T^\text{el}\sim\alpha_s^2$ (see Eq.~(\ref{eq:Tdipole})),
and thus the probability for two simultaneous
scatterings is of order $\alpha_s^4$, which
is parametrically
suppressed. But this argument holds only as long as the dipole number densities
are of order $1$. If $n\sim 1/\alpha_s^2$
(which is also the point above which the unitarity of $A$ is no longer
preserved in the BFKL approach), 
then it is clear that multiple scatterings
should occur.

\begin{figure}
\begin{center}
\includegraphics[width=0.4\textwidth]{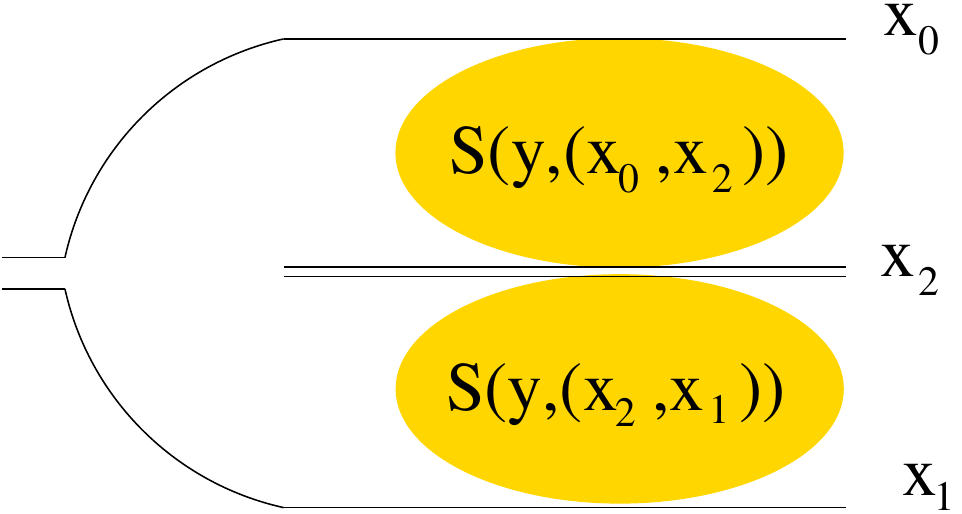}
\end{center}
\caption{\label{fig:BK2}Derivation of the Balitsky-Kovchegov equation.}
\end{figure}

\begin{figure}
\begin{center}
\includegraphics[width=0.35\textwidth]{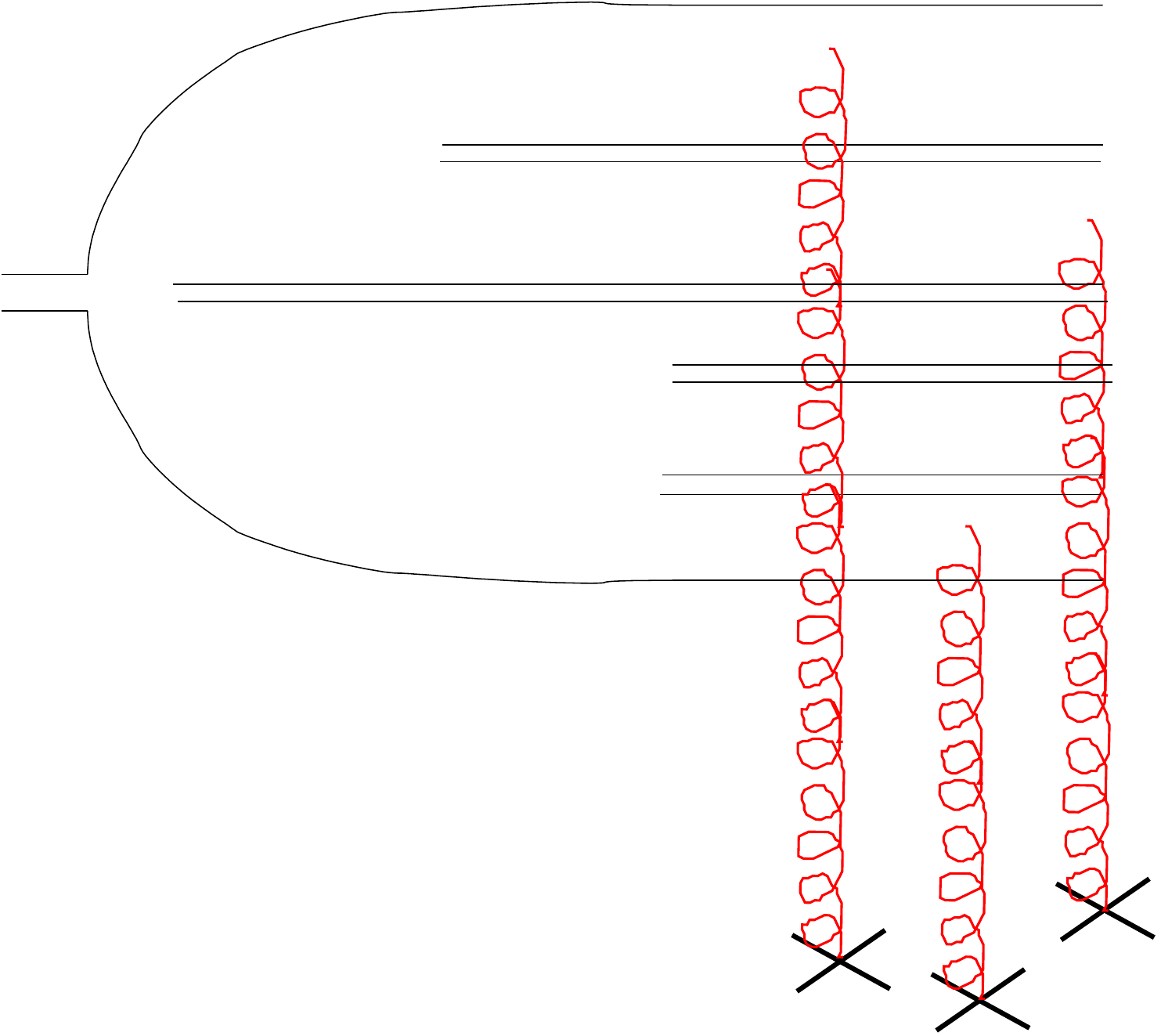}
\end{center}
\caption{\label{fig:bk}
Picture of the BK equation.
All the QCD evolution is put in the probe, which carries the
total rapidity. It develops a high occupancy state of dipoles,
which scatter independently off the target.
}
\end{figure}

In order to try and implement these multiple scatterings, we introduce
the probability that there be no scattering between
a dipole $(x_0,x_1)$ and a given realization of the target 
in a scattering
with total rapidity $y$,
that we shall denote by
$S(y,(x_0,x_1))$.
Let us start with a system in which the evolution is fully contained in the target.
We increase the total rapidity by boosting the {\em projectile} (initially at rest)
by a small amount $dy$.
Then there are two cases to distinguish, depending
on whether the dipole $(x_0,x_1)$ splits in the rapidity interval $dy$.
In case it splits into two dipoles $(x_0,x_2)$ and $(x_2,x_1)$, the probability that 
the projectile does not interact is just the product of the probabilities
that each of these new dipoles do not interact.
This is because once created, dipoles are assumed to be independent.
In summary:
\begin{equation}
S(y+dy,(x_0,x_1))=\begin{cases}
S(y,(x_0,x_1)) \ \text{with proba}\ 1-\bar\alpha dy
\int_{x_2} \frac{dP}{d(\bar\alpha y)}(x_{01}\rightarrow x_{02},x_{12})\\
S(y,(x_0,x_2))S(y,(x_2,x_1))\ \text{with proba}\
\bar\alpha dy \frac{dP}{d(\bar\alpha y)}(x_{01}\rightarrow x_{02},x_{12})
\end{cases}
\label{eq:bk01}
\end{equation}
Taking the average over the realizations of the
target and the limit $dy\rightarrow 0$, we get
\begin{equation}
\frac{\partial}{\partial y}\langle S(y,(x_0,x_1))\rangle=\bar\alpha
\int \frac{d^2 x_2}{2\pi}\frac{x_{01}^2}{x_{02}^2 x_{21}^2}
[
\langle S(y,(x_0,x_2))S(y,(x_2,x_1)) \rangle
 -\langle S(y,(x_0,x_1))\rangle
]
\label{eq:bk02}
\end{equation}
(See Fig.~\ref{fig:BK2} for a graphical representation.)
We see that this equation is not closed:
An evolution equation for the correlator 
$\langle S(y,(x_0,x_2))S(y,(x_2,x_1))\rangle$
is required.
However, we may assume that such correlators factorize
in the following sense:
\begin{equation}
\langle S(y,(x_0,x_2))S(y,(x_2,x_1))\rangle =
\langle S(y,(x_0,x_2))\rangle
\langle S(y,(x_2,x_1))\rangle.
\label{eq:factocorr}
\end{equation}
This assumption is justified if the dipoles scatter off 
uncorrelated targets,
for example, off different nucleons of a very large nucleus.
Writing $A=1-\langle S\rangle$, we get the following closed equation for $A$:
\begin{multline}
\frac{\partial}{\partial y} A(y,(x_0,x_1))=\bar\alpha
\int \frac{d^2 x_2}{2\pi}\frac{x_{01}^2}{x_{02}^2 x_{21}^2}
[
A(y,(x_0,x_2))+A(y,(x_2,x_1))\\
-A(y,(x_0,x_1))-A(y,(x_0,x_2))A(y,(x_2,x_1))
],
\label{eq:BKA}
\end{multline}
which is the Balitsky-Kovchegov (BK) equation \cite{Kovchegov:1999yj,Kovchegov:1999ua}.
Note that if one neglects the nonlinear term, 
one gets back the BFKL equation~(\ref{eq:BFKL}) (written for $A$ instead of $n$).
A graphical representation of this equation is given in Fig.~\ref{fig:bk}.

It is not difficult to see analytically that the BK equation preserves the 
unitarity of $A$: When $A$ becomes of the order of 1, then the nonlinear term
gets comparable to the linear terms in magnitude, and slows down
the evolution of $A$ with $y$, which otherwise would be exponential.
Hence the solution of the BK equation will exhibit essentially
two regimes: A BFKL regime of low density in which $A\ll 1$ and
in which the evolution proceeds linearly, and a high-density
regime $A\sim 1$. At fixed $y$, the transition between these two regimes
is controlled by the so-called saturation scale $Q_s(y)$, which is
the inverse size of the dipoles which scatter with an amplitude $A$ 
equal to some fixed number of order 1, for example $\frac12$.
($Q_s$ also depends a priori on the impact parameter, as will
be discussed in Chap.~\ref{sec:spatial}).

\begin{figure}
\begin{center}
\begin{minipage}[c]{0.4\textwidth}
\centerline{\includegraphics[width=0.8\textwidth]{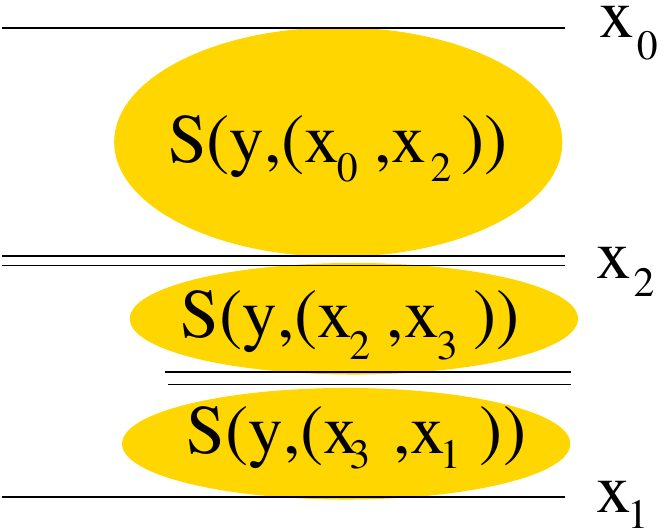}}
\centerline{(a)}
\end{minipage}
\hskip 0.5cm
\begin{minipage}[c]{0.4\textwidth}
\centerline{\includegraphics[width=0.8\textwidth]{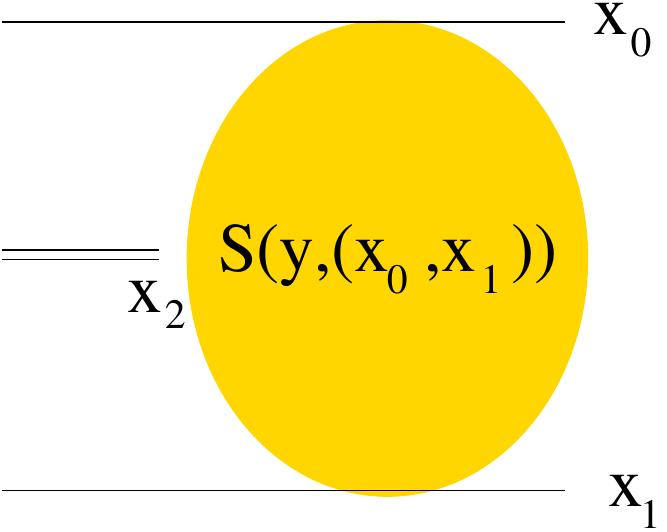}}
\centerline{(b)}
\end{minipage}
\end{center}
\caption{\label{fig:JIMWLK2}(a) Contribution to the B-JIMWLK equation for the 2-point correlator 
restricted to dipoles ($x_2^\prime$
is taken equal to $x_2$ in this figure). (b) A graph that would
also contribute to the 2-point correlator and
that is missing in the B-JIMWLK formalism.}
\end{figure}

Let us go back to Eqs.~(\ref{eq:bk01}),(\ref{eq:bk02}) and instead of assuming the
factorization of the correlators~(\ref{eq:factocorr}), work out
an equation for the two-point correlator $\langle SS\rangle$. From
the same calculation as before, we get
\begin{multline}
\frac{\partial}{\partial y}\langle S_{02}S_{2^\prime 1}\rangle=
\bar\alpha\int \frac{d^2x_3}{2\pi}\frac{x_{02}^2}{x_{03}^2x_{32}^2}
\left(\langle S_{03}S_{32}S_{2^\prime 1}\rangle-\langle S_{02}S_{2^\prime 1}\rangle\right)\\
+\bar\alpha\int \frac{d^2x_3}{2\pi}\frac{x_{12^\prime}^2}{x_{13}^2x_{32^\prime}^2}
\left(\langle S_{2^\prime3}S_{31}S_{02}\rangle-\langle S_{02}S_{2^\prime 1}\rangle\right),
\label{eq:balitskydipoles}
\end{multline}
where we have introduced the notation $S_{ab}\equiv S(y,(x_a,x_b))$.
(See Fig.~\ref{fig:JIMWLK2}a for the corresponding graphical representation.)

This equation calls for a new equation for the 3-point correlators, and so on.
The obtained hierarchy is nothing but the Balitsky hierarchy \cite{Balitsky:1995ub}
(see also Ref.~\cite{Balitsky:1998kc,Balitsky:1998ya,Levin:2004yd})
restricted to dipoles.

\subsection{The B-JIMWLK formalism}

For completeness, let us briefly mention the 
Balitsky-Jalilian Marian-Iancu-McLerran-Weigert-Leonidov-Kovner (B-JIMWLK) formalism.
It is a systematic approach beyond the large-$N_c$ limit,
initially supposed to completely describe high-energy scattering
under appropriate well-defined approximations.
(Nowadays, the status of this approach is less clear, as we
shall discuss later on).
Instead of evolving dipole or parton number densities, 
B-JIMWLK evolves observables
constructed with the help of Wilson operators
\begin{equation}
U(x)=P \exp\left(ig_s\int dx^\mu A_\mu^a(x)t^a\right)
\label{eq:wilson}
\end{equation}
where $A_\mu^a$ is the color 4-potential over which one
eventually averages to arrive at observables, $t^a$
the standard color matrics (fundamental representation),
and $g_s=\sqrt{4\pi\alpha_s}$ the coupling of the
line to the color field.
The integration in the exponent goes along
the trajectory of the quark whose interactions
are described by $U$, which is essentially a straight lightlike line
at very high energies.

There are several equivalent ways to present this physics.
Let us exhibit the so-called ``Balitsky hierarchy'', which
is an infinite system of coupled integro-differential equations
for the correlators of $U$. In terms of the $U$'s, 
the $S$-matrix for dipole scattering off a particular
field configuration reads
\begin{equation}
S_{ab}=\frac{1}{N_c}\mbox{Tr} U(x_a)U^\dagger(x_b).
\end{equation}
The observables are the associated correlators: For example
the dipole amplitude is obtained from $S$ by averaging over
the color fields.

The first equation of the Balitsky hierarchy 
for the observable $\langle S_{01}\rangle$
is identical 
to Eq.~(\ref{eq:bk02}). The second equation,
needed to solve the first one, reads
\begin{multline}
\frac{\partial}{\partial y}
\langle
S_{02}S_{21}
\rangle
=\bar\alpha\int
\frac{d^2x_3}{2\pi}\bigg\{
\bigg[
\frac{x_{02}^2}{x_{03}^2x_{32}^2}
\left\langle
(S_{03}S_{32}-S_{02})S_{21}
\right\rangle
+\frac{x_{01}^2}{x_{03}^2x_{31}^2}
\left\langle
S_{02}(S_{03}S_{31}-S_{01})
\right\rangle
\bigg]\\
+\frac{1}{N_c^2}
\left[
-\frac{(x_0-x_3)\cdot(x_1-x_3)}{(x_0-x_3)^2(x_1-x_3)^2}
-\frac{1}{(x_2-x_3)^2}
+\frac{(x_0-x_3)\cdot(x_2-x_3)}{(x_0-x_3)^2(x_2-x_3)^2}
+\frac{(x_2-x_3)\cdot(x_1-x_3)}{(x_2-x_3)^2(x_1-x_3)^2}
\right]\\
\times\bigg[
 \frac{1}{N_c}
 \left\langle
 \mbox{Tr}\left(U_{x_0}U^\dagger_{x_2}U_{x_3}U^\dagger_{x_1}U_{x_2}U^\dagger_{x_3}\right)
 \right\rangle
+\frac{1}{N_c}
 \left\langle
 \mbox{Tr}\left(U_{x_0}U^\dagger_{x_3}U_{x_2}U^\dagger_{x_1}U_{x_3}U^\dagger_{x_2}\right)
 \right\rangle
-2S_{01}
\bigg]
\bigg\}
\end{multline}
The first line is the same as Eq.~(\ref{eq:balitskydipoles}).
The other terms involve sextupoles and are suppressed at large-$N_c$.
Hence in the dipole approximation, we recover Eq.~(\ref{eq:balitskydipoles}).

The factorized correlators~(\ref{eq:factocorr}) is a 
solution of the whole dipole hierarchy, and turn actually out
to be a good approximation
to the solution of the full B-JIMWLK equations.
This statement was first made after the numerical solution 
to a version of the B-JIMWLK formalism was worked out
in Ref.~\cite{Rummukainen:2003ns}.
We note however that the latter simulations did not cover
a very large range in rapidity, and therefore, they may have missed
physical effects that would differentiate the full B-JIMWLK
equation from its approximate forms.

We may wonder why there are no terms involving one-point functions 
in the right-hand side of the previous
equation. Actually, such terms would correspond to graphs like the one 
of Fig.~\ref{fig:JIMWLK2}b, in which, for example, two dipoles merge.
They are expected to occur if saturation is properly taken into account.
While the restriction of the Balitsky equation to dipoles does a priori
not drastically change the solution for the scattering amplitudes,
such terms would instead have a large effect, as we shall discover in 
the next section.
To simplify the discussion, we will stick to the dipole
approximation, which leads to the evolution 
equations~(\ref{eq:bk02}),(\ref{eq:balitskydipoles}).


\subsection{Saturation}

The BK equation may be well-suited for the ideal case
in which the target is a nucleus made of an infinity of independent nucleons.
But it is not quite relevant to describe the scattering
of more elementary objects such as two dipoles (or two virtual photons,
to be more physical).

\begin{figure}
\begin{center}
\begin{tabular}{cc}
\includegraphics[width=0.3\textwidth]{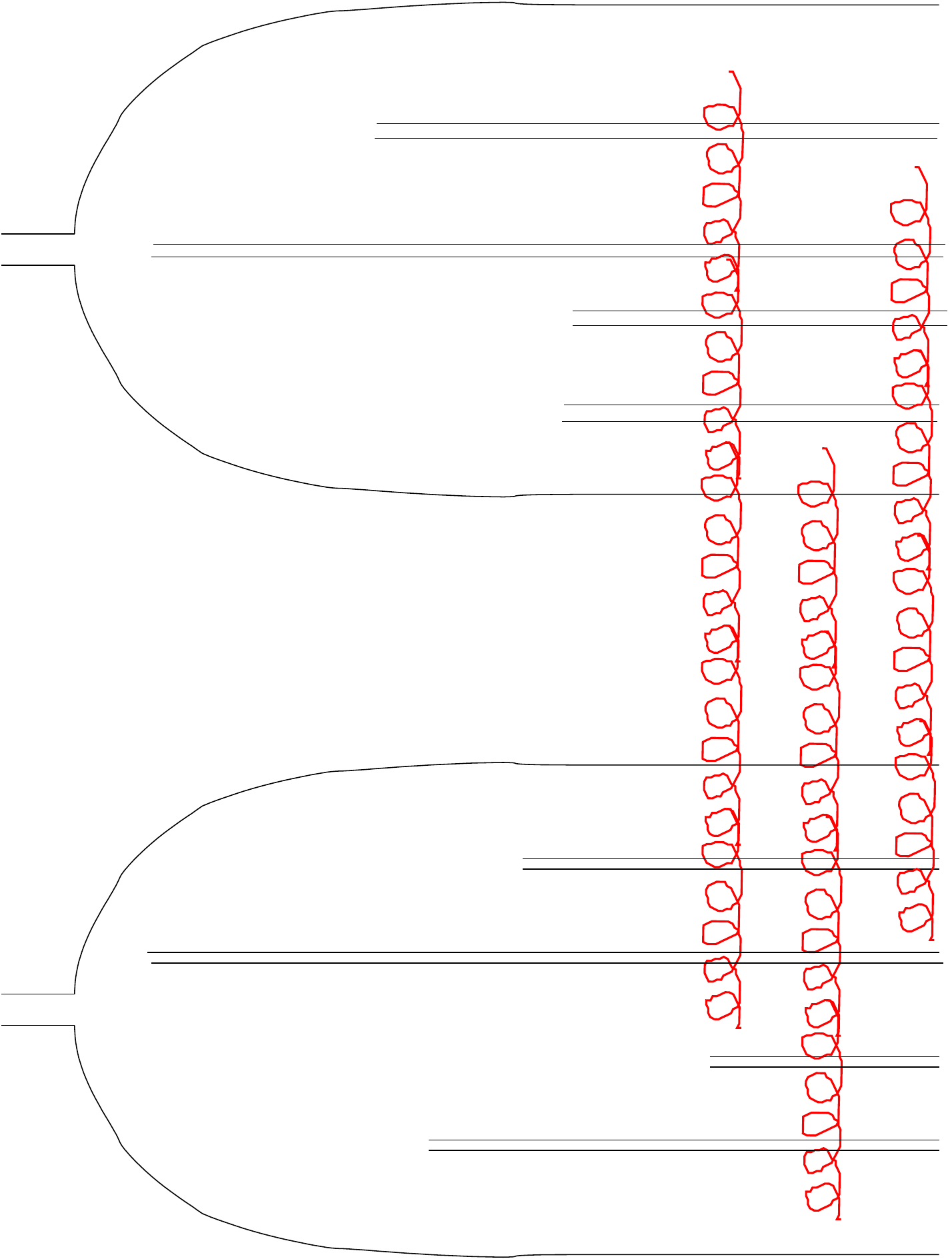}
&
\hskip 0.1\textwidth
\includegraphics[width=0.3\textwidth]{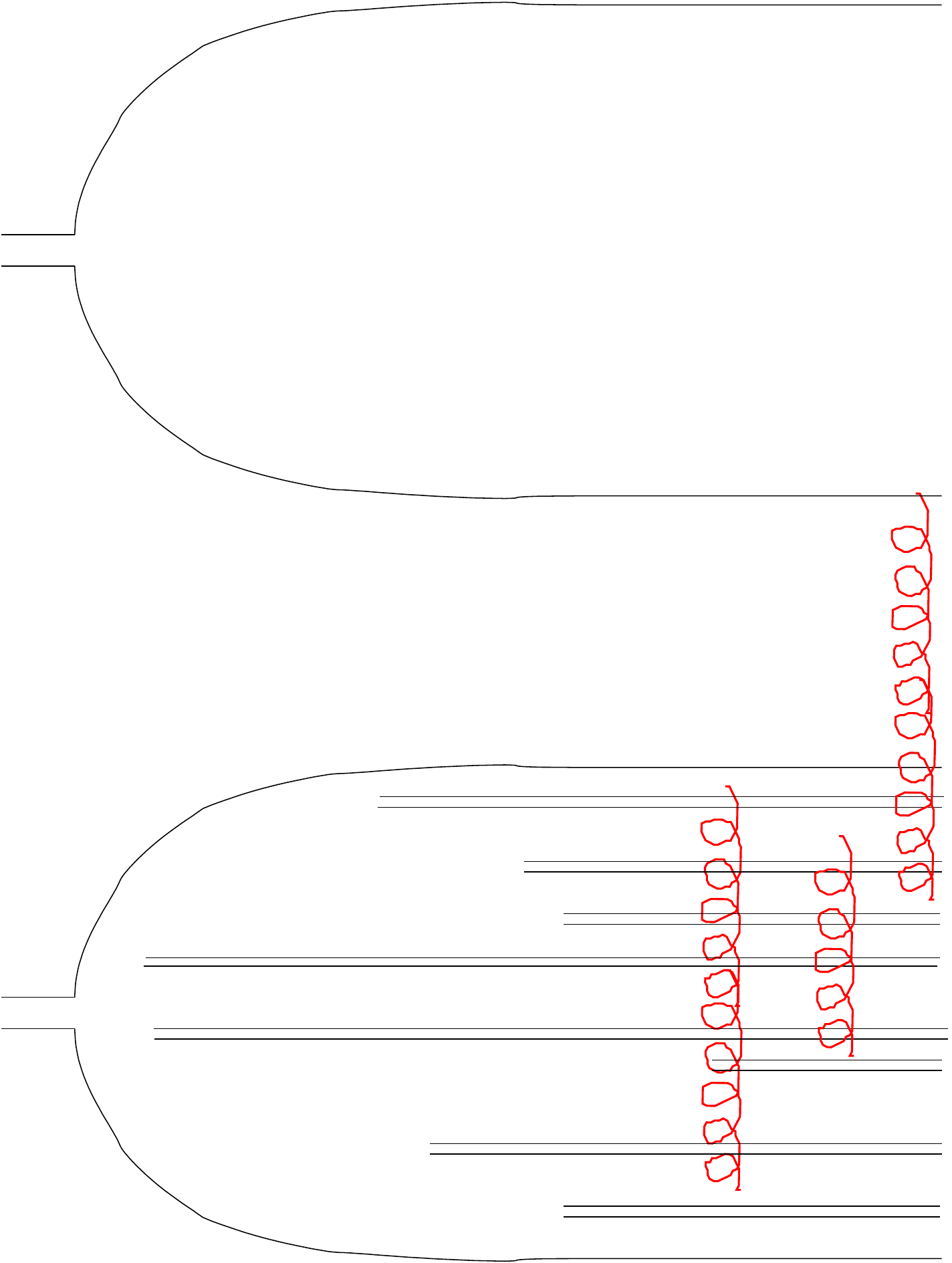}\\
(a) &
\hskip 0.1\textwidth
(b)
\end{tabular}
\end{center}
\caption{\label{fig:com}
(a) Scattering in the dipole model in the center-of-mass frame.
The evolution is shared between the target and the probe.
The amplitude is unitarized through the multiple scatterings
occurring between the two evolved wave-functions.
(b) Boost of the previous graph to the restframe of the projectile.
There is now twice as much evolution in the target
and the nonlinear effects should occur inside its wave-function, in the course
of the evolution.
They may take the form of ``internal'' rescatterings (as depicted), or
dipole merging...
}
\end{figure}

Indeed, 
following Chen and Mueller~\cite{Chen:1995pa} (see also Ref.~\cite{Kovchegov:1997dm}),
let us consider dipole-dipole scattering in the center-of-mass frame,
where the rapidity evolution is equally shared between the projectile and the target
(see Fig.~\ref{fig:com}a).
Then at the time of the interaction, the targets are dipoles that stem from the branching
of a unique primordial dipole. 
Obviously, the assumption of statistical independence of the diffusion
centers,
which was needed for the factorization~(\ref{eq:factocorr}) to hold, is no longer
justified.

So far, we have seen that nonlinear effects which go beyond the factorization
formula~(\ref{eq:BFKLframe}) are necessary to
preserve unitarity
as soon as $n\sim 1/\alpha_s^2$. 
This came out of an analysis of Eq.~(\ref{eq:BFKLframe})
in the restframe of the target. The rapidity $y_\text{BFKL}$ at which the system
reaches this number of dipoles and hence at which the BFKL approach breaks
down may be found from the form
of the typical growth of $n$ with $y$, namely $n(y)\sim e^{\bar\alpha y}$.
Parametrically,
\be
y_\text{BFKL}\sim\frac{1}{\bar\alpha}\ln\frac{1}{\alpha_s^2}.
\ee
Now we may go to the center-of-mass frame, where Eq.~(\ref{eq:BFKLframe})
with $y^\prime=y/2$ would describe the scattering
amplitude in the absence of
nonlinear effects.
There, the typical number of dipoles in the projectile and in the
target are well below $1/\alpha_s^2$:
$n(y_\text{BFKL}/2)\sim 1/\alpha_s$.
We actually see that the evolution of the dipoles 
in each of theses systems
remains linear until $y=2 y_\text{BFKL}$.
In that rapidity interval, nonlinear effects consist in the
simultaneous scatterings of several dipoles from the
target and the projectile but the evolution of $n$ still obeys
the BFKL equation.
Now, performing a boost to the projectile restframe, 
the evolution goes into the target. 
Formula~(\ref{eq:BFKLprojectile}) should then apply for the amplitude $A$.
But if the evolution of the target were kept linear,
then the amplitude would break unitarity because the number of dipoles would
be larger than $1/\alpha_s^2$.
Hence, through some nonlinear mechanism, 
which was represented by multiple scatterings between linearly evolving
objets in the center-of-mass frame,
the dipole number density 
has to be kept effectively lower than $1/\alpha_s^2$ in order to preserve unitarity.
This is called {\em parton saturation}.
The precise saturation mechanism has not been formulated in QCD.
It could be dipole recombinations due to gluon fusion, 
multiple scatterings inside
the target which slow down the production of new dipoles \cite{Mueller:1996te}, 
``dipole swing''
as was proposed more recently \cite{Avsar:2005iz,Avsar:2006jy}, or any
other mechanism.
Some of these mechanisms may be implemented in simplified
toy models; see Chap.~\ref{sec:zerodimensional}.

Hence, unitarity of the scattering amplitudes together
with boost-invariance seem to require some sort of saturation of the
density of partons.
It is not clear whether such a mechanism is included in the B-JIMWLK
formalism, since the latter is not obviously boost-invariant.
What is clear is that some saturation mechanism lacks in the dipole
model.

A pedagogical review of saturation 
and the discussion of the relationship between saturation and unitarity
may be found in Ref.~\cite{Mueller:2001fv}.
Original papers include Refs.~\cite{Mueller:1989st,Mueller:1999wm}.


\subsubsection{Visualizing saturation: Evolution in different models}

We now wish to illustrate
how the different schemes of unitarization 
(BK unitarization, multiple scatterings in the center-of-mass frame,
explicit parton number saturation)
affect the evolution of scattering amplitudes.

\begin{figure}
\begin{center}
\includegraphics[width=0.8\textwidth]{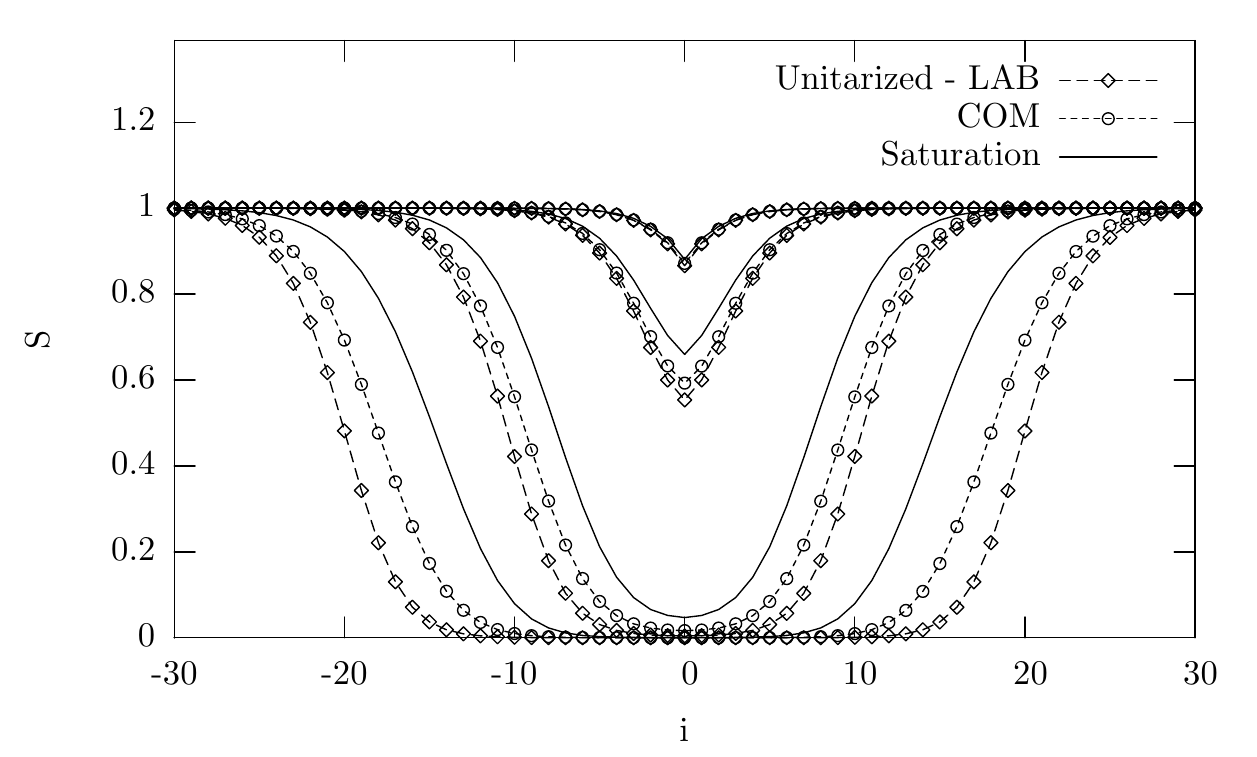}\\
\end{center}
\caption{\label{fig:visualization}
[From Ref.~\cite{Munier:2008rh}]
$S$-matrix element as a function of the logarithm of the dipole size
for $y = 5, 10, 20, 30$ (from the center of the figure towards
the outskirts). $1/\alpha^2$ = 20. For low rapidities ($y=5$ and $y=10$), 
the evolution is linear (of BFKL type).
At $y = 20$, the unitarity limit has been reached in all calculations ($S$ becomes zero in some region of sizes). 
Later ($y = 30$), the region in which $S=0$ propagates outwards. 
Different unitarization mechanisms are tested: simple BK unitarization
(``Unitarized-LAB''), center-of-mass unitarization (``COM'') and
intrinsic saturation.
}
\end{figure}

In Fig.~\ref{fig:visualization}, we plot the $S$-matrix element at different
rapidity and as a function of the (logarithmic) dipole size
resulting from the evolution of toy models with dynamics
similar to QCD
(The use of such models will be justified later, when we will establish
the correspondence of high-energy QCD with more general processes).
We see that $S$ goes to zero in a region of sizes that
extends with rapidity. This phenomenon is slower when there are saturation
effects explicitly included in the evolution, as discussed above in
this section.

\subsection{The Pomeron language}

So far, we have presented in detail a $s$-channel picture of hadronic interactions, 
and it is in this
formalism that we will understand most easily the link with reaction-diffusion
processes. In the $s$-channel formulation, 
all the QCD evolution happens in the form
of quantum fluctuations of the interacting hadrons. However, a picture
maybe more familiar to the reader 
belonging to the ``traditional'' QCD community
is a $t$-channel picture,
where the rapidity evolution is put in the $t$-channel, while the projectile
and target stay in their bare states.
This picture directly stems from the usual Lorentz-invariant
formulation of quantum field theory, while the dipole picture
(or the parton model) is derived in the framework of time-ordered perturbation
theory.

Both pictures have their respective advantages and
drawbacks. The covariant formulation seems to be more
suited for higher-order systematic calculations, since
for a given observable the number of diagrams is smaller 
than in the time-ordered ($s$-channel) formalism.
The time-ordered formalism seems unpractical
beyond the tree-level approximation.
On the other hand, the latter gives maybe a more intuitive picture
of scattering processes
and seems to be particularly useful to study the approach
to unitarity.

\begin{figure}
\begin{center}
\includegraphics[width=0.4\textwidth]{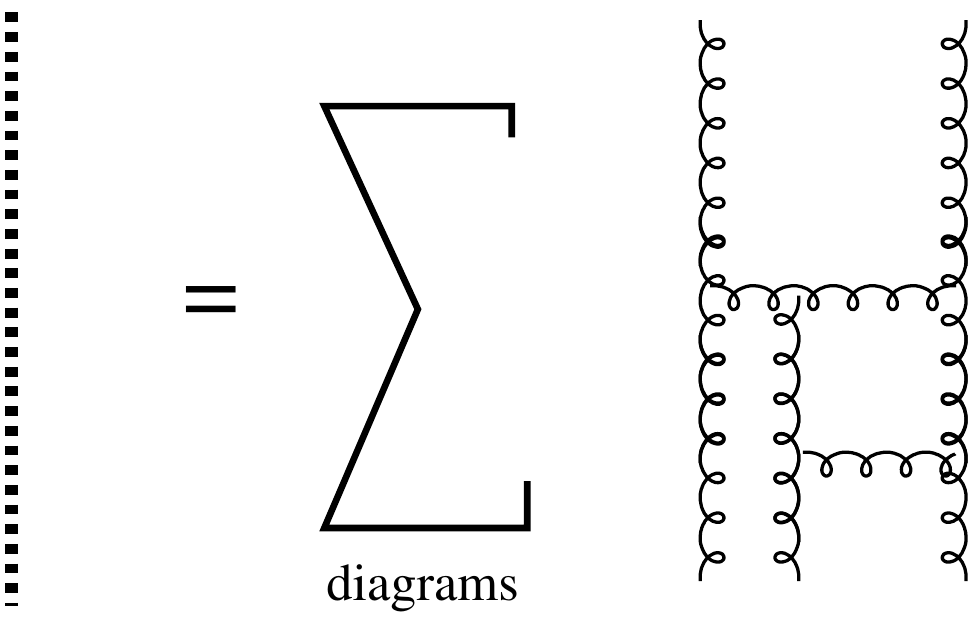}
\end{center}
\caption{\label{fig:pomeron}The BFKL Pomeron is a sum of $t$-channel
gluon Feynman diagrams.}
\end{figure}

In the $t$-channel picture, classes of Feynam diagrams can be grouped into ``Pomerons'' 
(or Reggeized gluons, see Fig.~\ref{fig:pomeron}),
in terms of which scattering processes may be analyzed. (A pedagogical
review on how to derive the BFKL equation in such a formalism is available
from Ref.~\cite{Forshaw:1997dc}).
\begin{figure}
\begin{center}
   \begin{minipage}[c]{.45\textwidth}
\begin{center}
\includegraphics[width=0.5\textwidth]{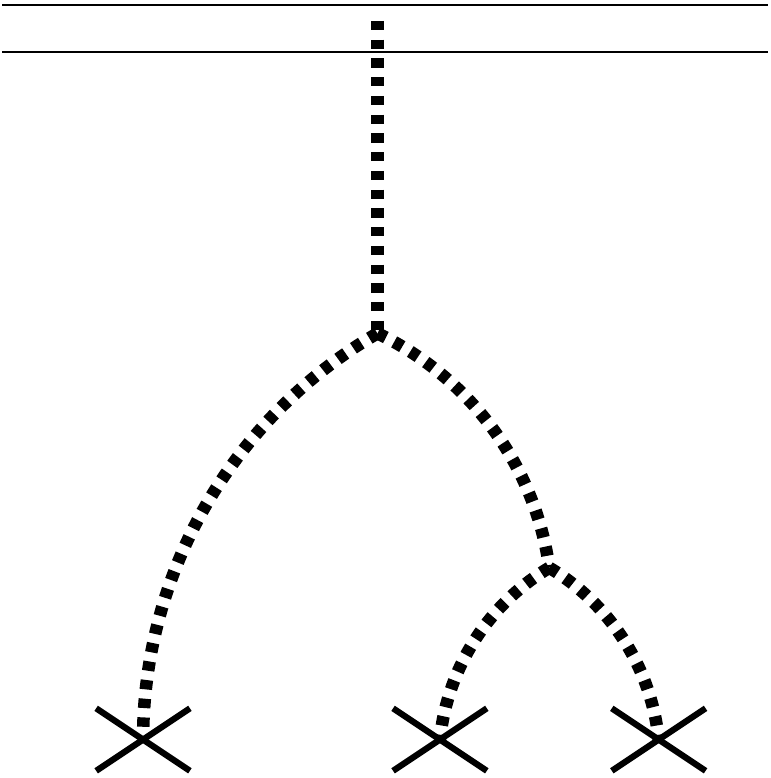}
\end{center}
   \end{minipage}
   \begin{minipage}[c]{.45\textwidth}
\begin{center}
\includegraphics[width=0.5\textwidth]{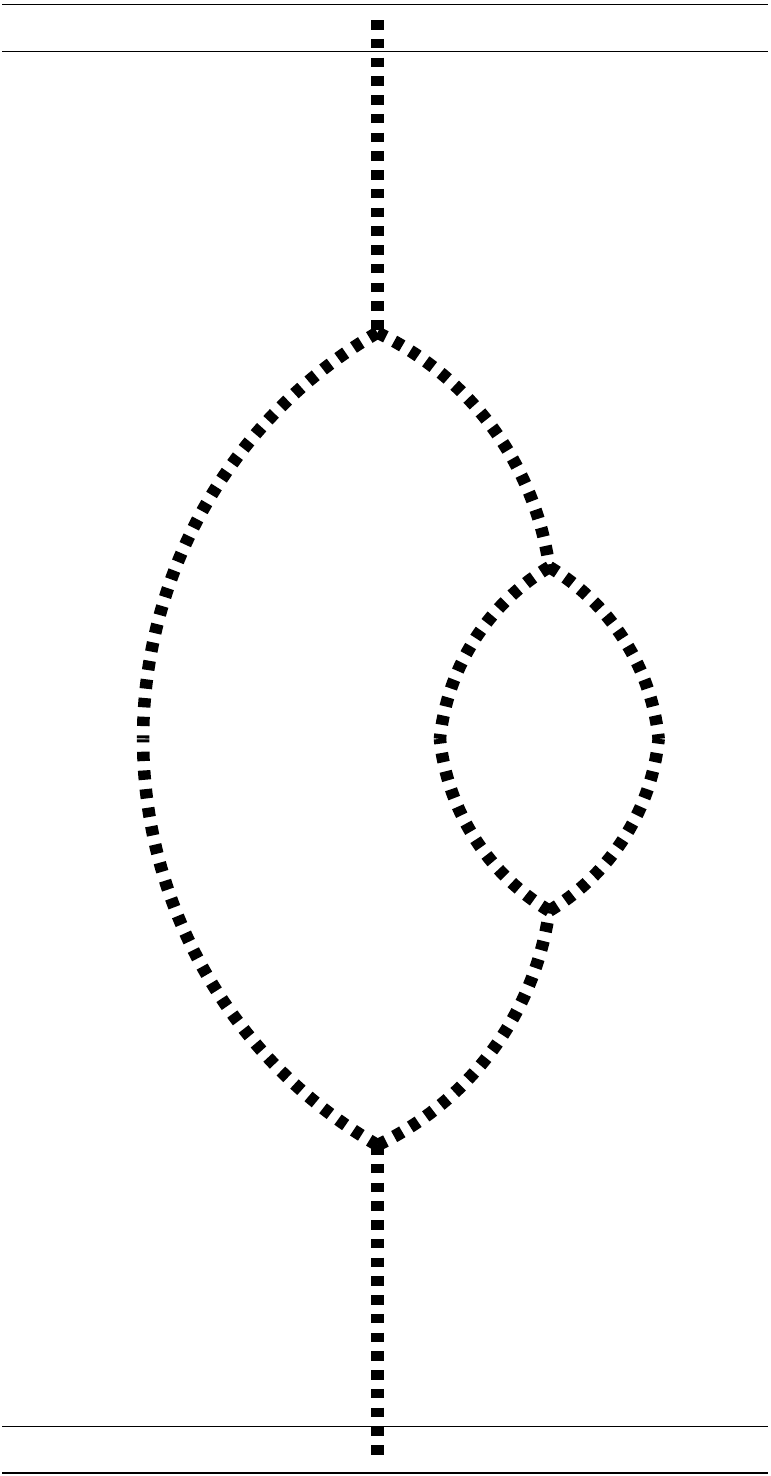}
\end{center}
   \end{minipage}
\vskip 0.4cm
   \begin{minipage}[b]{.45\textwidth}
\begin{center}
(a)
\end{center}
   \end{minipage}
   \begin{minipage}[b]{.45\textwidth}
\begin{center}
(b)
\end{center}
   \end{minipage}
\end{center}
\caption{\label{fig:bkmppomeron}(a) Example of a diagram contributing to the BK
equation in the $t$-channel representation (see Fig.~\ref{fig:bk}).
The dashed lines represent Pomerons. The rapidity is proportional to the length
of the Pomeron lines in the $t$-channel.
(b) Pomeron representation of
a class of diagrams to which Fig.~\ref{fig:com}a belongs.
}
\end{figure}%
An effective action containing Pomeron 
fields and Pomeron vertices may be constructed.
In these terms, the $s$-channel diagrams of 
Fig.~\ref{fig:bk} and~\ref{fig:com}a may
be translated in terms of the diagrams of Fig.~\ref{fig:bkmppomeron}.
The effective action formalism was initially developped in 
Refs.~\cite{Kirschner:1994gd,Kirschner:1994xi,Lipatov:1995pn}.
More recently, there has been some progress 
in the definition of the effective action
\cite{Antonov:2004hh},
some of it with the help of the correspondence with
statistical physics processes \cite{Blaizot:2005vf,Hatta:2005rn}.

We will not expand on this formulation in the present review,
because it is difficult to see the analogy with statistical
physics in this framework. A $s$-channel picture is much more natural.
However, a full solution of high-energy QCD may require to go back to that kind
of calculation and compute accurately the $1\rightarrow n$ Pomeron vertices.
This program was formulated some time ago 
\cite{Bartels:1992ym,Bartels:1993ih}, and there is continuing
progress in this direction (see e.g. \cite{Bartels:1999aw,Ewerz:2003an}).


\section{Analogy with reaction-diffusion processes}

We are now in position to draw the relationship between high-energy QCD
and reaction-diffusion processes.
In the first section below, we will show that the BK equation
is, in some limit, an equation that also appears in the context
of statistical physics.
Second, we will exhibit a simple reaction-diffusion model,
and show in the final section how this model
is related 
to scattering in QCD, even beyond the
approximations implied in the BK equation.

\subsection{The BK equation and the FKPP equation}

Let us first show at the technical level 
that under some well-controlled approximations, the BK equation~(\ref{eq:BKA})
may be mapped exactly to a parabolic nonlinear partial differential equation.
This observation was first made in Ref.~\cite{Munier:2003vc}.

To simplify, we will look for impact-parameter independent solutions:
$A(y,(x_0,x_1))$ is assumed to depend on $y$ and $x_{01}$ only,
not on $\frac{x_0+x_1}{2}$.
We switch to momentum space through the Fourier transformation
\begin{equation}
A(y,k)=\int \frac{d^2 x_{01}}{2\pi x_{01}^2}e^{i \vec k \vec x_{01}} A(y,x_{01}).
\end{equation}
This transformation greatly simplifies the BK equation 
\cite{Kovchegov:1999yj,Kovchegov:1999ua}.
It now reads
\begin{equation}
\partial_{\bar\alpha y} A(y,k)=\chi(-\partial_{\ln k^2})A(y,k)-A^2(y,k).
\end{equation}
The first term on the right-hand side, 
which is a linear term, is actually an 
integral transform whose kernel, 
obtained by Fourier transformation of the BFKL kernel
(first three terms on the right-hand side of Eq.~(\ref{eq:BKA})).
It is most easily expressed in Mellin space
since the powers $k^{-2\gamma}$ 
are its eigenfunctions, with the corresponding eigenvalues
\begin{equation}
\chi(\gamma)=2\psi(1)-\psi(\gamma)-\psi(1-\gamma).
\end{equation}
This kernel may be expanded around some real $\gamma=\gamma_0$, fixed 
between 0 and 1. Keeping the terms up to $\mathcal{O}((\gamma-\gamma_0)^2)$ is the
well-known diffusive approximation, which is a good approximation at
large rapidities.
Introducing the notations $\chi_0\equiv\chi(\gamma_0)$, 
$\chi^\prime_0\equiv\chi^\prime(\gamma_0)$
and $\chi^{\prime\prime}_0\equiv\chi^{\prime\prime}(\gamma_0)$,
the BK equation reads
\be
\partial_{\bar\alpha y}A
={\scriptstyle\frac{\chi_0^{\prime\prime}}{2}}\partial_{\ln k^2}^2 A
+(\gamma_0 \chi_0^{\prime\prime}-\chi^\prime_0)\partial_{\ln k^2} A
+(\chi_0-\gamma_0 \chi_0^{\prime}+{\scriptstyle\frac{\gamma_0^2\chi_0^{\prime\prime}}{2}})A
-A^2.
\ee
Through the linear change of variable $(\bar\alpha y,\ln k^2)\rightarrow (t,x)$,
\be
\begin{split}
\bar\alpha y&= 
\frac{t}{\chi_0-\gamma_0 \chi_0^{\prime}+{\scriptstyle\frac{\gamma_0^2\chi_0^{\prime\prime}}{2}}}\\
\ln k^2    &=
\sqrt{\frac{\chi_0^{\prime\prime}}{2(\chi_0-\gamma_0 \chi_0^{\prime})+{{\gamma_0^2\chi_0^{\prime\prime}}}}}x
-\frac{\gamma_0 \chi_0^{\prime\prime}-\chi^\prime_0}
{\chi_0-\gamma_0 \chi_0^{\prime}+{\scriptstyle\frac{\gamma_0^2\chi_0^{\prime\prime}}{2}}}t,
\end{split}
\ee
one may get rid of the first-order partial derivative in the right handside.
We then find that the rescaled function
\begin{equation}
u(t,x)=\frac{A(y(t),\ln k^2(t,x))}
{\chi_0-\gamma_0 \chi_0^{\prime}+{\scriptstyle\frac{\gamma_0^2\chi_0^{\prime\prime}}{2}}}
\end{equation}
obeys the equation
\begin{equation}
\frac{\partial u(t,x)}{\partial t}=
\frac{\partial^2 u(t,x)}{\partial x^2}
+u(t,x)-u^2(t,x),
\label{eq:BFKLtoFKPP}
\end{equation}
which is the Fisher \cite{Fisher} and 
Kolmogorov-Petrovsky-Piscounov \cite{KPP} 
(FKPP) equation.
This equation was first written down as a model for
gene propagation in a population in the limit of large
number of individuals.
But it turns out to apply directly or indirectly
to many different physical situations, such as
reaction-diffusion processes, but also directed percolation, 
and even mean-field spin glasses \cite{DerridaSpohn}.
A recent comprehensive review on the known mathematics and the
phenomenological implications of the FKPP equation can be found
in Ref.~\cite{vansaarloos-2003-386}.

As a side remark, we note that if $\gamma_0$ is chosen such 
that the equation
$\chi(\gamma_0)=\gamma_0\chi^\prime(\gamma_0)$
is verified, then the mapping
drastically simplifies. Actually, this choice has a physical
meaning, as we will discover in Chap.~\ref{sec:reviewtraveling}
when we try and solve the BK equation.

Beyond the exact mapping~(\ref{eq:BFKLtoFKPP}) 
between an approximate form
of the BK equation and the FKPP equation,
the full BK equation is said to be in the {\em universality
class} of the FKPP equation. All equations in this universality
class share some common properties, as will be understood below.
The exact form of the equation is unessential.
As a matter of fact, recently, 
it has been checked explicitly that the BFKL equation with next-to-leading
order contributions to the linear evolution kernel 
(but keeping the QCD coupling fixed)
is also in the same universality
class. A mapping to a partial differential equation (which involves
higher-order derivatives in the rapidity variable) was
exhibited \cite{Enberg:2006aq}.
What defines physically the universality class of the FKPP equation is
a branching diffusion process with some saturation mechanism. The details
seem unimportant.

We must however keep in mind that there
is for the time being
no theorem that would clearly state
the necessary and sufficient conditions for
a model to belong (or not) to the universality class of the FKPP
equation: Our statements are nothing but conjectures, supported
by arguments and checked against numerical simulations.

In the next section, we shall give a concrete 
example of a reaction-diffusion process:
We will see how the FKPP equation appears as a fluctuationless 
(or ``mean-field'') limit of some stochastic reaction-diffusion process.
In Ref.~\cite{Munier:2003vc}, it had not been realized that 
the analogy of QCD with such processes is in fact much
deeper than the formal mapping between the BK equation and the FKPP equation
that we have just outlined.
But this is actually the case, as we shall shortly argue.

\subsection{\label{sec:reactiondiffusionexample}
Reaction-diffusion processes: an example}

We consider the reaction-diffusion model which was
introduced in Ref.~\cite{Enberg:2005cb}.
It consists in a set of particles
which are evolving in discrete time 
on a one-dimensional lattice. 
The following rules define the dynamics of the system:
At each timestep, a particle may jump
to the nearest position on the left or on the right with respective
probabilities $p_l$ and $p_r$, and may split into two particles with
probability $\lambda$. We also allow each of the $n(t,x)$ particles
on site $x$ at time $t$ to die with probability
$\lambda n(t,x)/N$.

\begin{figure}
\begin{center}
\includegraphics[width=0.8\textwidth]{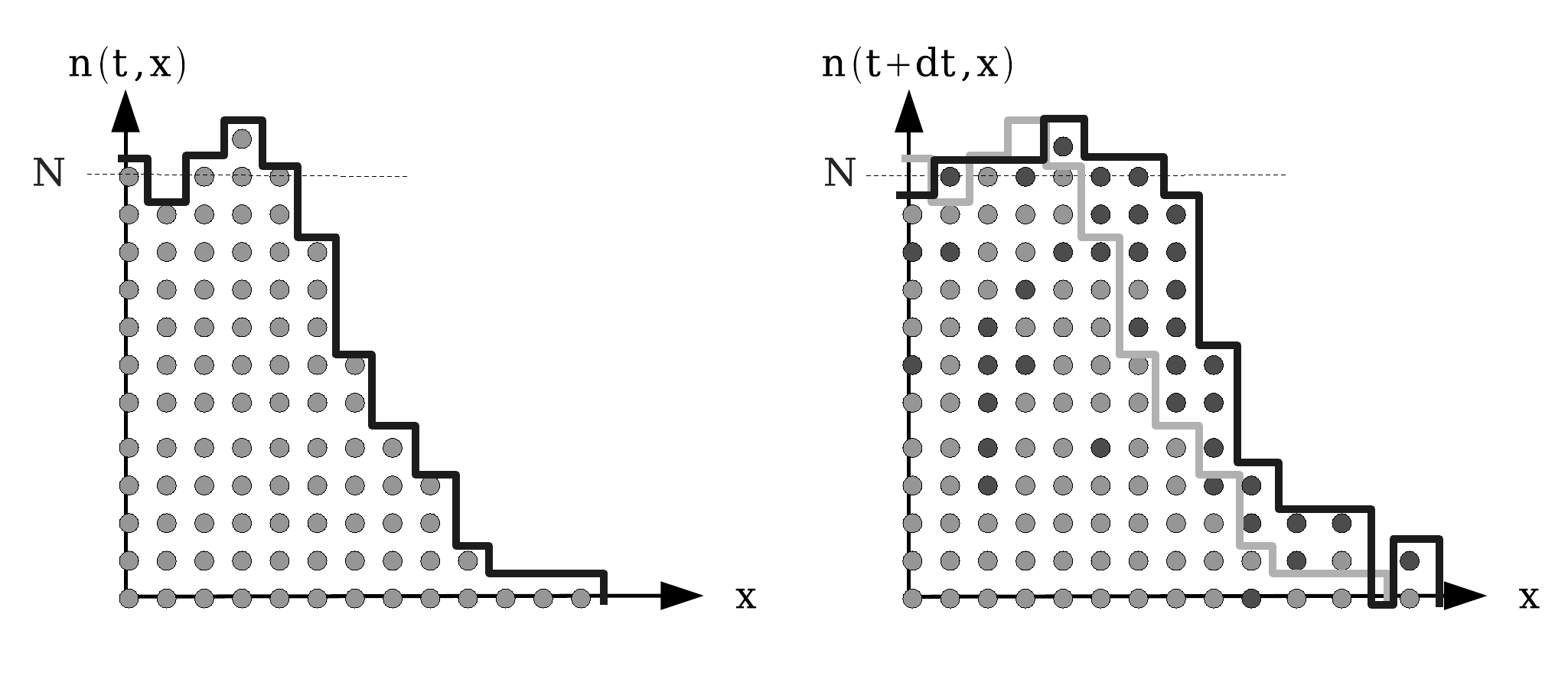}
\end{center}
\caption{\label{fig:model1d}
Picture of a realization of the system of particles at two successive times.
In the bins in which the number of particles is of order $N$, 
some particles disappear, others are created by splittings, but overall the
number of particles is conserved up to
fluctuations of order $\sqrt{N}$.
In the bins in which $n$ is small compared to $N$, the dynamics
is driven by branching diffusion. As a result, $n(t,x)$ looks like a noisy
wave front moving to the right.
}
\end{figure}

We can guess what a realization of this evolution may
look like at large times.
The particles branch and diffuse (they undergo an evolution
which can be represented by a linear finite difference equation)
until their number $n$ becomes of the order of $N$, at which point
the probability that they ``die'' starts to be sizable,
in such a way that their number never 
exceeds $N$ by a large amount, on any site.
If the initial condition is spread on a finite number of lattice sites,
the linear branching-diffusion process may always proceed towards 
larger values of $|x|$, where there were no particles in the beginning of
the evolution.
Hence 
after some lapse of time (typically larger than $\ln N$)
a realization will look like a double front connecting an ensemble
of lattice sites where a quasi-stationary
state in which the number of particles is $N$ (up to fluctuations)
has been reached,
to an ensemble of empty sites. One
front will move towards $x\rightarrow+\infty$, the
other one towards $x\rightarrow-\infty$ 
as the branching diffusion process proceeds.
Let us focus on the front traveling to the right. 
The position of the front $X(t)$ may be defined in different ways,
leading asymptotically to equivalent determinations, up to a constant.
For example, one may define $X(t)$ 
as the rightmost bin in which there are more 
than $N/2$ particles, or, alternatively, as the total number of particles
in the realization whose positions are greater than 0, scaled by $1/N$.
A realization and its time evolution is sketched in Fig.~\ref{fig:model1d}.

Between times $t$ and $t+\Delta t$, $n_l(t,x)$ particles 
out of $n(t,x)$ move to the left and
$n_r(t,x)$ of them move to the right. 
Furthermore, $n_+(t,x)$ particles are replaced by their
two offspring at $x$, and $n_-(t,x)$ particles disappear.
Hence the total variation in the number of particles on site $x$ reads
\begin{subequations}\label{stocha}
\begin{multline}
n(t+\Delta t,x)-n(t,x)
=-n_l(t,x)-n_r(t,x)-n_-(t,x)\\
+n_+(t,x)+n_l(t,x+\Delta x)+n_r(t,x-\Delta x).
\end{multline}
The numbers describing a timestep at position $x$ have 
a multinomial distribution:
\begin{equation}
P(\{n_l,n_r,n_+,n_-\})=
\frac{n!}{n_l!n_r!n_+!n_-!\Delta n!}
p_l^{n_l}p_r^{n_r}
\lambda^{n_+}(\lambda n/N)^{n_-}
(1\!-\!p_l\!-\!p_r\!-\!\lambda \!-\!\lambda n/N)^{\Delta n},
\end{equation}
\end{subequations}
where $\Delta n=n-n_l-n_r-n_+-n_-$, and all quantities 
in the previous equation
are understood at site $x$ and time $t$.
The evolution of $u\equiv n/N$ is obviously stochastic. One could
write the following equation:
\be
u(t+\Delta t,x)=\langle u(t\!+\!\Delta t,x)\rangle
+\sqrt{\langle u^2(t\!+\!\Delta t,x)\rangle
-\langle u(t\!+\!\Delta t,x)\rangle^2}
\,\nu(t+\Delta t,x)
\label{eq:stochamodel}
\ee
where the averages are performed
over the time step that takes
the system from $t$ to $t+\Delta t$. They are conditioned to
the value of $u$ at time $t$.
$\nu$ is a noise, i.e. a random function.
The equation was written in such a way that it has zero mean and
unit variance.
Note that the noise is updated at time $t+\Delta t$ in this equation.

One can compute the mean evolution of $u\equiv n/N$ in one step of time
which appears in the right-hand side of Eq.~(\ref{eq:stochamodel})
from Eq.~(\ref{stocha}).
It reads
\begin{multline}
\langle u(t\!+\!\Delta t,x)|\{u(t,x)\}\rangle\!=\!u(t,x)\!+\!
p_l[u(t,x\!+\!\Delta x)\!-\!u(t,x)]\\
+\!p_r[u(t,x\!-\!\Delta x)\!-\!u(t,x)]\!+\!\lambda u(t,x)[1\!-\!u(t,x)].
\label{mod_mf}
\end{multline}
The mean evolution of the variance of $u$ that appears in 
Eq.~(\ref{eq:stochamodel}) may also be computed. The precise form of the
result is more complicated, but roughly speaking, the variance 
of $u$ after evolution over a unit of time is
of the order of $u/N$ for small $u\sim 1/N$. 
This is related to the fact
that the noise has a statistical origin: Having
$n$ particles on the average
in a system means that each realization typically
consists in $n\pm\sqrt{n}$ particles.

When $N$ is infinitely large, one can replace the $u$'s in Eq.~(\ref{mod_mf})
by their averages: This would be a mean-field approximation.
Obviously, the noise term drops out, and the equation becomes
deterministic.
Note that if we appropriately take
the limits $\Delta x\rightarrow 0$ and $\Delta t\rightarrow 0$,
setting
\begin{equation}
\lambda=\Delta t,\ \ p_R=p_L=\frac{\Delta t}{(\Delta x)^2},
\end{equation}
the obtained mean-field equation is nothing but the
FKPP equation~(\ref{eq:BFKLtoFKPP}).
For the numerical simulations of this model that we will perform in 
Sec.~\ref{sec:reviewtraveling},
we will keep $\Delta t$ and $\Delta x$ fixed,
which is more convenient for computer implementation.

Thus we have seen that the evolution of
reaction-diffusion systems
is governed by a stochastic equation~(\ref{eq:stochamodel}) 
whose continuous limit ($\Delta t\rightarrow 0,\ \Delta x\rightarrow 0$) and
mean-field limit ($N\gg 1$) is a partial differential equation
of the form of (exactly actually, in our simple case study) the FKPP equation.
We shall now argue that partons in high-energy QCD
form a similar system.

\subsection{Universality class of high-energy QCD}

Let us come back to the QCD dipole model discussed in Sec.~\ref{sec:partonmodel}.
We have seen that evolution proceeds through a branching diffusion
process of dipoles. Let us denote by $T(y,r)$ the scattering amplitude
of the probe dipole off one particular realization of the target at rapidity $y$
and at a given fixed impact parameter.
This means that we imagine for a while that we may freeze the target
in one particular realization after the rapidity evolution $y$, 
and probe the latter with
projectiles of all possible sizes.
Of course, this is not doable in an actual experiment, not even in principle.
But it is very important for the statistical picture to 
decompose the physical observables with the help of such
a ``gedanken observable''.
The amplitude $A$, which is related to the measurable total cross-section,
is nothing but the average of $T$ over all possible
realizations of the fluctuations of the target, namely
\begin{equation}
A(y,r)=\langle T(y,r)\rangle.
\end{equation}

\begin{figure}
\begin{center}
\includegraphics[width=0.7\textwidth]{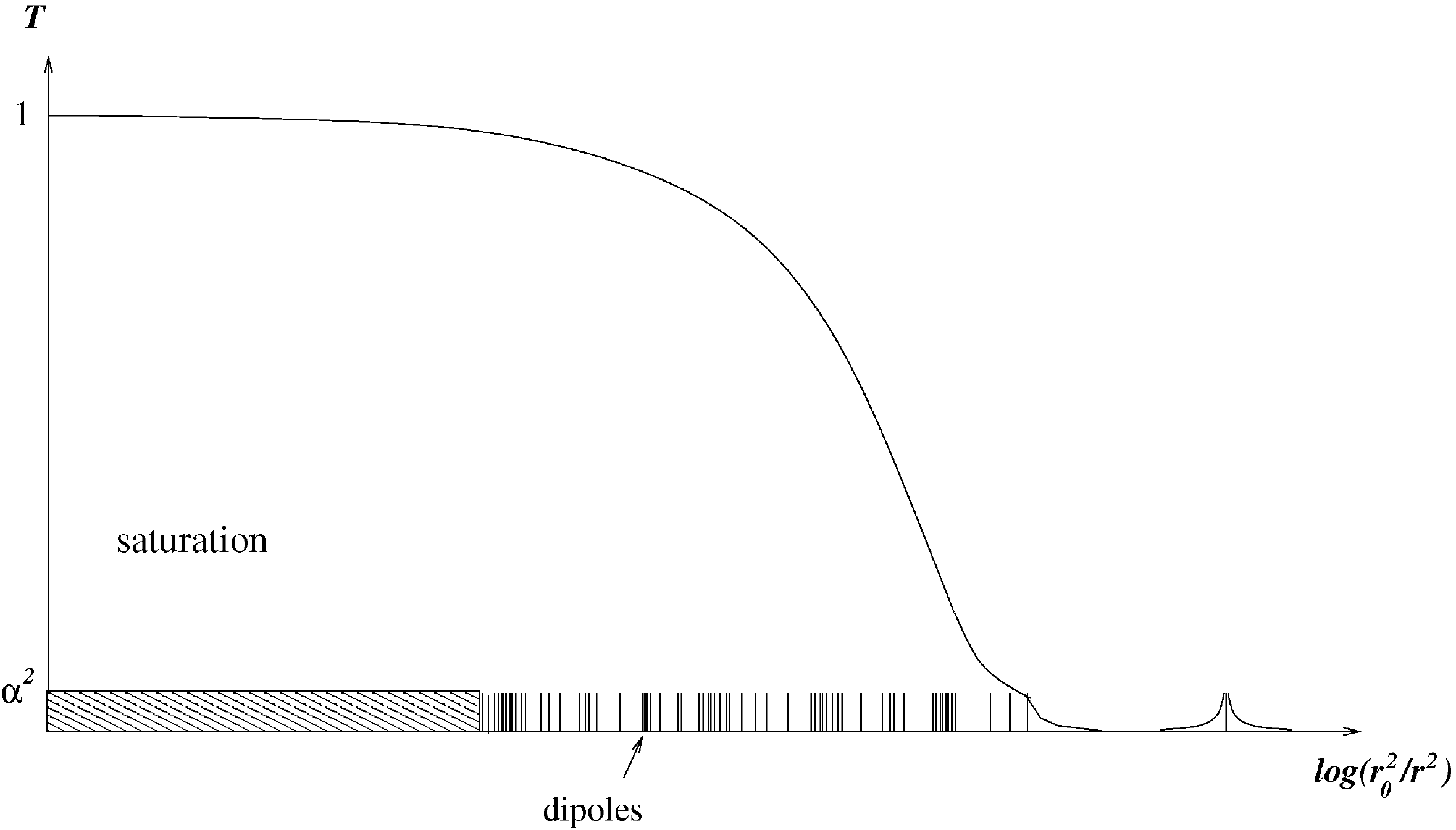}
\end{center}
\caption{\label{fig:T}Sketch of the scattering amplitude $T$ 
of a dipole of size $r$ off a frozen
partonic configuration. The small lines on the axis denote the
dipoles ordered by their logarithmic sizes.
Up to fluctuations, $T$ looks like a wave front.
}
\end{figure}

The branching diffusion of the dipoles essentially occurs in the $\ln(1/r^2)$ variable.
The scattering amplitude is roughly equal to the number of dipoles
in a given bin of (logarithmic) dipole size, multiplied by $\alpha_s^2$.
From unitarity arguments and consistency with boost-invariance, 
we have seen that the branching diffusion process
should (at least) slow down in a given bin as soon as the number of
objects in that very bin is of the order of $N=1/\alpha_s^2$, in such
a way that effectively, the number of dipoles in each bin is limited to $N$.
A typical realization of $T$ is sketched in Fig.~\ref{fig:T}.
As in the case of the reaction-diffusion process, 
from similar arguments,
it necessarily looks like a front.
The position of the front, defined to be the value $r_s$ of $r$ for which
$T$ is equal to some fixed number, say $\frac12$,
is related to the saturation scale defined
in the Introduction: $r_s=1/Q_s(y)$.

We now see that there is a very close analogy between what we are describing
for QCD here and the model that we were introducing in the previous section.
So in particular, one might be able to formulate interaction processes in QCD 
with the help of a stochastic
nonlinear evolution equation for the ``gedanken'' amplitude $T$.
We already know the equation that one should get 
in the mean-field limit in which $N$ is very large:
It is the BK equation, as was rigorously proven above.
Thus we know the equivalent of the term $\langle u(t+\Delta t,x)\rangle$
in Eq.~(\ref{eq:stochamodel}).
The noise term is not known, but since it is of statistical origin,
it must be of the order of the square root of the number of dipoles normalized to $N$, 
that is to say, of order $\sqrt{T/N}$.
We may write an equation of the form
\be
\partial_{\bar\alpha y} T(y,k)=\chi(-\partial_{\ln k^2})T(y,k)
-T^2(y,k)+\alpha_s\sqrt{2T(y,k)}\,\nu(y,k),
\label{eq:stochaqcd}
\ee
where $\nu$ is a noise, uncorrelated in rapidity and transverse momentum,
with zero mean and unit variance.
This equation is to be compared to the following one:
\be
\partial_t u(t,x)=\partial_x^2 u(t,x)+u(t,x)-u^2(t,x)
+\sqrt{\frac{2u(t,x)}{N}}\,\nu(t,x),
\label{eq:RFT}
\ee
which is the so-called ``Reggeon field theory'' equation
when the noise $\nu$ is exactly a normal Gaussian white noise, that is
to say, whose non-vanishing cumulants read
\be
\begin{split}
\langle \nu(t,x)\rangle&=0\\
\langle \nu(t,x)\nu(t^\prime,x^\prime)\rangle&=
\delta(t-t^\prime)\delta(x-x^\prime).
\end{split}
\ee
It is a stochastic extension of Eq.~(\ref{eq:BFKLtoFKPP}).
If the noise term were of the form 
\be
\sqrt{\frac{2u(t,x)(1-u(t,x))}{N}}\,\nu(t,x)
\ee 
instead,
then this equation would be what is usually referred to as the
{\em stochastic Fisher-Kolmogorov-Petrovsky-Piscounov equation}.
The sFKPP equation and the physics that it represents is
reviewed in Ref.~\cite{panja-2004-393}.

Taking averages over events converts this equation into
a hierarchy of coupled equations, which has a lot in common in its structure
with the (modified) Balitsky hierarchy~(\ref{eq:bk02},\ref{eq:balitskydipoles}).
A detailed study may be found in Ref.~\cite{Iancu:2004iy}.
We will perform explicit calculations in this
spirit within simpler models in Chap.~\ref{sec:zerodimensional} below.

Based on these considerations, 
we may establish a dictionary between QCD and reaction-diffusion processes.
The correspondence is summarized in Tab.~\ref{tab:dictionary}.

\begin{table}
\begin{center}
\begin{tabular}{l  l}
\hline\hline
     Reaction-diffusion & QCD \\
\hline
     Occupation fraction $u(t,x)$  & 
\begin{minipage}[t]{7cm}
Scattering amplitude for the probe
off a frozen realization of the target $T(y,k)$
\end{minipage}
\\
\\
\begin{minipage}[t]{7cm}
     Average occupation fraction $\langle u(t,x)\rangle$ \end{minipage}
&\begin{minipage}[t]{7cm} Physical scattering amplitude $A=\langle T\rangle$
\end{minipage}\\
\\
 \begin{minipage}[t]{7cm} Space variable $x$ \end{minipage}
& \begin{minipage}[t]{7cm}$\ln(k^2/\Lambda^2)$ or $\ln(1/x^2\Lambda^2)$\end{minipage}\\
\\
 \begin{minipage}[t]{7cm}    Time variable $t$   \end{minipage}
& \begin{minipage}[t]{7cm}$\bar\alpha y$\end{minipage}\\
\\
\begin{minipage}[t]{7cm}     Average maximum density of particles $N$ \end{minipage}
& \begin{minipage}[t]{7cm}$1/\alpha_s^2$\end{minipage} \\
\\
\begin{minipage}[t]{7cm}     Position of the front $X(t)$ \end{minipage}
&\begin{minipage}[t]{7cm}
Saturation scale $\ln(Q_s^2(y)/\Lambda^2)$\end{minipage}\\
\\
\begin{minipage}[t]{7cm}
Branching-diffusion kernel $\omega(-\partial_x)$\\
($\omega(-\partial_x)=\partial_x^2+1$ in the
FKPP case)
\end{minipage}&
\begin{minipage}[t]{7cm}
BFKL kernel $\chi(-\partial_{\ln k^2})$\\
or its equivalent in coordinate space
\end{minipage}\\
\hline\hline\\
\end{tabular}
\end{center}
\caption{\label{tab:dictionary}Dictionary between QCD and the reaction-diffusion
model for the main physical quantities.
$\Lambda$ is a typical hadronic scale.}
\end{table}

The mechanism for saturation of the parton densities 
(i.e. of the dipole number) is
not known for sure in QCD. 
There is even some evidence that dipole degrees of freedom
are no longer sufficient to describe scattering beyond some
rapidity, as is understood from the appearance of 
sextupoles in the second equation of the full Balitsky hierarchy.
There are also important differences between
the reaction-diffusion model introduced above and QCD
that lie in the ``counting rule'' of the particles (provided by
the form of $T^\text{el}$ in the QCD case).
But from the general analysis of processes described by equations in the
universality class of the stochastic FKPP equation and the underlying evolution
mechanisms presented in Chap.~\ref{sec:reviewtraveling}, 
we will understand that most of the observables have universal properties in 
appropriate limits,
which do not depend on the details of the mechanism at work.
We draw the reader's attention to Refs.~\cite{Mueller:2005ut,Iancu:2005nj}, 
where a precise stochastic
equation was searched for in QCD. Some of the problems one may face
with the use and the very interpretation of
such equations were studied in Ref.~\cite{Iancu:2005dx}.

The way in which we view high energy QCD in this review 
is actually not particularly original: It is nothing but the QCD dipole model,
which was implemented numerically in the form of a Monte Carlo
event generator by Salam \cite{Salam:1995zd,Salam:1995uy,Salam:1996nb} 
(see also \cite{Avsar:2005iz} for another more recent implementation).
He also devised and implemented a ad hoc
saturation mechanism \cite{Mueller:1996te}
that went beyond the original dipole model pictured in Fig.~\ref{fig:com}a,
but which is necessary, as we argued before.

Before discussing more deeply the physical content of equations of the
form of Eq.~(\ref{eq:stochaqcd}),
we shall first study a model in which spatial dimensions are left out,
that we will be able to formulate in different ways.


\chapter{\label{sec:zerodimensional}The simplest saturation model}

{\it
In the previous chapter, we have understood that scattering at high energy
in QCD may be viewed as a branching-diffusion process supplemented
by a saturation mechanism. We have exhibited a simple toy model
with these characteristics, whose dynamics is represented by an equation
of the type~(\ref{eq:RFT}).

Unfortunately, even that toy model is too difficult to solve analytically.
We shall study a still simplified model, where there is no diffusion 
mechanism: Realizations are completely specified by the number of particles
that the system contains at a given time. 
Of course, in this case, a saturation scale cannot be defined, which limits
the relevance of this model for QCD. However, we will be able to
formulate this model in many different ways, and to draw parallels
with QCD.
 
We start by defining precisely the model. Then, two approaches to the
computation of the moments of the number of particles are presented.
The first set of methods relies on field theory (Sec.~\ref{sec:fieldtheory}).
The second method relies on a statistical approach (Sec.~\ref{sec:statisticalmethods})
and will be extended in a phenomenological way to models with a spatial dimension
in Chap.~\ref{sec:reviewtraveling}.
We shall then draw the relation to a scattering-like formulation 
(Sec.~\ref{sec:relscattering}).
Finally (Sec.~\ref{sec:alternative}), some variants of the basic model are reviewed.
}\\

\minitoc

\section{Definition}

Let us consider a simple model in which 
at a given time $t$,
the system is fully characterized
by the number $n_t$ of particles. The evolution rules are
the following.
Between times $t$ and $t+dt$, each particle has a probability $dt$ to split
in two particles. For each pair of particles, there is a probability $dt/N$
that they merge into one.
We may summarize these rules by the following set of equations:
\begin{equation}
n_{t+dt}=\left\{
\begin{aligned}
n_t\!+\!1 & \text{  proba  } n_tdt\\
n_t\!-\!1 & \text{  proba  } \frac{n_t (n_t\!-\!1)dt}{N}\\
n_t   & \text{  proba  } 1\!-\!n_tdt -\frac{n_t(n_t\!-\!1)dt}{N}.
\end{aligned}
\right.
\label{rule0}
\end{equation}
From this, one can easily derive an equation for the 
time evolution of the
probability $P(n,t)$
of observing
$n$ particles in the system at time $t$:
\begin{equation}
\frac{\partial P}{\partial t}(n,t)=
(n-1)P(n-1,t)+\frac{n(n+1)}{N}P(n+1,t)-\left(n+\frac{n(n-1)}{N}\right)P(n,t).
\label{master0d}
\end{equation}
This is the {\em master equation} for the Markovian process under consideration.
The two first terms with a positive sign represent the process of going
from one state containing $n$ particles 
to an adjacent one containing $n+1$ or $n-1$ particles respectively,
while the last term
simply corrects the probability to keep it unitary.

By multiplying both sides of this equation by $n$ and summing over $n$,
we get an evolution equation for the average number of 
particles $\langle n_t\rangle$:
\begin{equation}
\frac{d\langle n_t\rangle}{dt}=\langle n_t\rangle
-\frac{1}{N}\langle n_t(n_t-1)\rangle.
\label{evolutionnnaive}
\end{equation}
Obviously, this equation is not closed, and one would have to establish an
equation for $\langle n_t(n_t-1) \rangle$, which would involve 3-point correlators
of $n_t$, and so on, ending up with an infinite hierarchy of equations, exactly
like in Chap.~\ref{sec:schannel} for QCD 
(see Eqs.~(\ref{eq:bk02}) and~(\ref{eq:balitskydipoles})).

This illustrates the difficulties one has to face
before one can get an analytical expression for $\langle n_t\rangle$,
even in such a simple model.

\section{\label{sec:fieldtheory}``Field theory'' approach}

In the next subsections, 
we will follow different routes to get analytical results
on the moments of the number of particles in the system at a given time $t$. 
The first one will be similar
to the $s$-channel picture of QCD (see Sec.~\ref{sec:schannel}), 
since it will consist
in computing the time (equivalent to the
rapidity in QCD) evolution of realizations of the system.
The second one will be closer to the $t$-channel picture of QCD. 
We will see how ``Pomerons'' may appear in these simple systems.
We will then examine a formulation in terms of a stochastic nonlinear
partial differential equation, which is nothing but the sFKPP equation
in which the space variable ($x$) has been discarded.

\subsection{\label{sec:0dfockstate}Particle Fock states and their weights}

Statistical problems were first formulated as field theories
by Doi \cite{Doi} and Peliti \cite{Peliti}. 
Different authors have used these methods (see Ref.~\cite{tauber-2007-716} for a review).
We shall start by following the presentation given in
Ref.~\cite{PhysRevE.59.3893}.

We would like to interpret the master equation~(\ref{master0d})
 as a quasi-Hamiltonian evolution
equation of the type of the ones that appear in quantum mechanics.
To this aim, we need to introduce the basis of states $|n\rangle$ of
fixed number $n$ of particles.
We define the ladder operators $a$ and $a^\dagger$ 
by their action on these states:
\begin{equation}
a|n\rangle=n|n-1\rangle,\ a^\dagger|n\rangle=|n+1\rangle
\end{equation}
and which obey the commutation relation
\begin{equation}
[a,a^\dagger]=1.
\end{equation}
The $n$-particle state may be constructed from the vacuum 
(zero-particle) state
by repeated application of the ladder operator:
\begin{equation}
|n\rangle=\left(a^\dagger\right)^n|0\rangle.
\end{equation}
The normalization is not standard with respect to what is usually
 taken in quantum mechanics.
In particular, the orthogonal basis $|n\rangle$ 
is defined in such a way that
$\langle m|n\rangle=n!\delta_{m,n}$.
This implies that the completeness relation reads
\be
\sum_n \frac{1}{n!}|n\rangle\langle n|=1.
\ee
We also introduce the state vector of the system at a time $t$
as a sum over all possible Fock states weighted by their probabilities:
\begin{equation}
|\phi(t)\rangle=\sum_n P(n,t)|n\rangle.
\end{equation}
It is straightforward to see that the master equation~(\ref{master0d}) is
then mapped to the Schr\"odinger-type equation
\begin{equation}
\frac{\partial}{\partial t}|\phi(t)\rangle=-{\cal H}|\phi(t)\rangle,
\end{equation}
where $\mathcal{H}$ is the ``Hamiltonian'' operator
\begin{equation}
{\cal H}=(1-a^\dagger)a^\dagger a-\frac{1}{N}(1-a^\dagger)a^\dagger a^2.
\label{hamiltonian0d}
\end{equation}
The first term represents the splitting of particles, while the second one,
proportional to $1/N$,
represents the recombination.
We may rewrite ${\cal H}$ as
\begin{equation}
{\cal H}={\cal H}_0 +{\cal H}_1,
\end{equation}
where
\be
{\cal H}_0 =a^\dagger a
\ee
is the ``free'' Hamiltonian whose eigenstates are the Fock states.
We now go to the interaction picture by introducing the time-dependent
Hamiltonian
\begin{equation}
{\cal H}_I(t)=e^{{\cal H}_0 t} {\cal H}_1 e^{-{\cal H}_0 t}
\end{equation}
and the states $|\phi\rangle_I=e^{{\cal H}_0 t}|\phi\rangle$.
The solution of the evolution reads
\begin{equation}
\begin{split}
|\phi\rangle_I
&=T \exp\left(-\int_0^t dt^\prime {\cal H}_I(t^\prime)\right)|\phi_0\rangle_I\\
&=|\phi_0\rangle_I-\int_0^t dt^\prime {\cal H}_I(t^\prime)|\phi_0\rangle_I
+\int_0^t dt^\prime \int_0^{t^\prime} dt^{\prime\prime} 
{\cal H}_I(t^\prime){\cal H}_I(t^{\prime\prime})|\phi_0\rangle_I+\cdots
\end{split}
\label{Hevolsol}
\end{equation}
We may then compute the weights of the successive Fock states
by applying this formula.
Let us show how it works in detail by computing the state of a single
particle evolved from time 0 to time $t$, in the limit $N=\infty$ in which
there are no recombinations.
We follow the usual method to deal with
such problems in field theory. 
We 
repeatedly insert
complete bases of eigenstates of ${\cal H}_0$ into Eq.~(\ref{Hevolsol}), 
namely
\begin{equation}
|\phi\rangle_I=|1\rangle
-\int_0^t dt^\prime \sum_{n_1}
\frac{1}{n_1!}
|n_1\rangle\langle n_1|
{\cal H}_I(t^\prime)
|1\rangle+\cdots
\end{equation}
(We have kept the first two terms in Eq.~(\ref{Hevolsol}) explicitely).
Using the expression for ${\cal H}_I(t)$ 
as a function of ${\cal H}_0$ and ${\cal H}_1$,
together with the knowledge
that the Fock states are eigenstates of ${\cal H}_0$, we get
\begin{equation}
|\phi\rangle=e^{-t}|1\rangle-\sum_{n_1} e^{-n_1 t}
\int_0^t dt^\prime e^{n_1 t^\prime-t^\prime}
\frac{1}{n_1!}
|n_1\rangle\langle n_1|{\cal H}_1|1\rangle+\cdots
\end{equation}
Inserting the expression
for ${\cal H}_1$, one sees that
in the infinite-$N$ limit, there is only one possible elementary
transition, namely the splitting.
Performing the integration over $t^\prime$
and computing in the same manner the higher orders,
one finally gets the expansion
\begin{equation}
|\phi\rangle=e^{-t}|1\rangle+e^{-t}(1-e^{-t})|2\rangle+\cdots
+e^{-t}(1-e^{-t})^{n-1}|n\rangle+\cdots
\end{equation}
from which one can read the probabilities of the successive Fock states.
This expansion is similar to the expansion in dipole Fock states
introduced in Sec.~\ref{sec:schannel}: The $n$-particle states
correspond to $n$-dipole states in QCD,
and their weights are computed by applying successive splittings to the system,
whose rates are given by Eq.~(\ref{splitting}). (They are just unity in the case of
the zero-dimensional model.)

We see that this method is well-suited to compute the probabilities of the lowest-lying
Fock-states, and their successive corrections
at finite $N$. 
But in general we are rather interested in averages such as $\langle n^k\rangle$,
for which the weights of all Fock states are needed.
We will develop a slightly different (but equivalent) formalism below, 
that will enable us
to get these averages in a much more 
straightforward way.

\subsection{\label{sec:pomeronfieldtheory} ``Pomeron'' field theory}

Let us introduce the generating function of the factorial
moments of the distribution of the number of particles
\begin{equation}
Z(z,t)=\sum_n (1+z)^n P(n,t).
\end{equation}
The evolution equation obeyed by $Z$ can easily be derived
from the master equation~(\ref{master0d}):
\begin{equation}
\frac{\partial Z}{\partial t}=z(1+z)\left(\frac{\partial Z}{\partial z}-
\frac{1}{N}\frac{\partial^2 Z}{\partial z^2}\right).
\end{equation}
We may represent this equation in a second-quantized formalism
by introducing the operators
\begin{equation}
b^\dagger=z,\ b=\frac{\partial}{\partial z}=\bar z
\end{equation}
acting on the set of states $|Z\rangle$ consisting in the analytic
functions of $z$.
Then we may write
\begin{equation}
\frac{\partial Z}{\partial t}=-{\cal H}^{\mathbb{P}}Z,
\end{equation}
where
\begin{equation}
{\cal H}^{\mathbb{P}}={\cal H}_0^{\mathbb{P}}+{\cal H}_1^{\mathbb{P}},\ \text{with}\
{\cal H}_0^{\mathbb{P}}=-b^\dagger b,\ {\cal H}_1^{\mathbb{P}}
=-b^\dagger b^\dagger b+\frac{1}{N}
b^\dagger(1+b^\dagger)b^2.
\label{eq:hamiltonianpft}
\end{equation}
A basis for the states is
\begin{equation}
|k\rangle=z^k,\ \langle k|={\bar z^k}
\end{equation}
which is orthogonal with respect to the scalar product
\begin{equation}
\langle Z_1|Z_2\rangle=\int \frac{dzd\bar z}{2i\pi}e^{-|z|^2} 
\bar Z_1(z,\bar z) Z_2(z,\bar z),
\end{equation}
and obeys the normalization condition $\langle k|l\rangle=k! \delta_{k,l}$.
We shall call these states ``$k$-Pomeron'' states, by analogy with high-energy QCD.
We may apply exactly the same formalism as before, since the operators $b$, $b^\dagger$
have the same properties as the $a$, $a^\dagger$.

From the definition of the scalar product,
it is not difficult to see that the $k$-th factorial moment of $n$ may be obtained
by a mere contraction of the state vector $|Z\rangle$, computed by
solving the Hamiltonian evolution,
with a $k$-Pomeron state. The following identity holds:
\begin{equation}
\langle k|Z\rangle=
\left\langle\frac{n!}{(n-k)!}\right\rangle,
\end{equation}
where the average in the right-hand side goes over the realizations of 
the system.
As for the initial condition, starting the evolution with one particle
means taking as an initial condition the superposition $|0\rangle+|1\rangle$
of zero- and one-Pomeron states respectively. 
The zero-Pomeron state does not contribute to the
evolution, hence effectively a one-Pomeron state is like a one-particle state.

\begin{figure}
\begin{center}
\includegraphics[width=8cm]{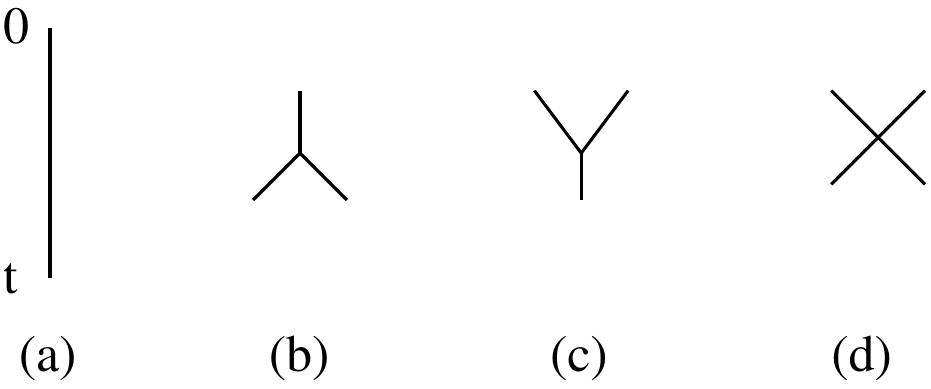}
\end{center}
\caption{\label{fig:feynman0d}
Propagator and vertices for the Pomeron field theory.
Time flows from the top to the bottom.
}
\end{figure}

In order to simplify the systematic computation
of these moments, we may 
use a diagrammatic method and establish Feynman rules.
To this aim, we write the contribution of the graphs with $l$-vertices
(corresponding to the term of order $l$ in the expansion of 
Eq.~(\ref{Hevolsol})), starting with a one-Pomeron state:
\be
\langle k|Z\rangle\supset (-1)^l
\int_0^t dt_1 \int_0^{t_1}dt_2\cdots\int_0^{t_{l-1}}dt_l
\sum_{n_1,\cdots,n_l}
\langle k|n_l\rangle
\frac{1}{n_l!}
\langle n_l |{\cal H}_I^{\mathbb{P}}|n_{l-1}\rangle
\cdots
\frac{1}{n_1!}
\langle n_1 |{\cal H}_I^{\mathbb{P}}|1\rangle.
\ee
Each matrix element that appears in this equation is associated
to a vertex, and propagators connect these vertices.
We read on the expression for the Hamiltonian~(\ref{eq:hamiltonianpft})
that there is one propagator and three vertices in the theory:
one splitting vertex ($1\rightarrow 2$), one recombination
($2\rightarrow 1$) and a $2\rightarrow 2$ elastic diffusion vertex.

The method to compute the 1 to $k$ Pomeron transition amplitude is standard.
First, one draws all possible diagrams for this transition that contain
$l$ vertices, including all possible permutations. (Note that a splitting
may occur in $k$ different ways, if $k$ is the number of
Pomerons before the splitting; A recombination instead
may occur in $k(k-1)/2$ ways).
Then, the propagators (Fig.~\ref{fig:feynman0d}a) are replaced by 
\be
\langle 1|e^{-t {\cal H}_0^\mathbb{P}}|1\rangle = e^t,
\label{eq:PFTpropagator}
\ee
(where $t$ is the time
interval they span).
The $n$-Pomeron state propagates as
$\langle n|e^{-t {\cal H}_0^\mathbb{P}}|n\rangle=e^{nt}$. 
Intermediate times are eventually integrated over.
As for the vertices (Figs.~\ref{fig:feynman0d}b-d), 
the following factors have to be applied:
\begin{equation}
(1\rightarrow 2):\ -1;\ \
(2\rightarrow 1):\ \frac{2}{N};\ \
(2\rightarrow 2):\ \frac{2}{N}.
\end{equation}
In addition, there is a $(-1)^{\#\text{vertices}}$ factor.
Finally, an overall $k!$ factor leads to the expression for the factorial 
moment $\langle n(n-1)\cdots(n-k+1)\rangle$.

The lowest-order diagram for the average particle number, 
consisting in a simple propagator, reads $\langle n\rangle=e^t$.
We now understand that this method leads to a more straightfoward computation
of the moments of the number of particles 
than the one based on the
computation of the probabilities of successive Fock states, for
a single Pomeron already resums an infinity of particle Fock states.
The Pomeron in this case is exactly like the BFKL Pomeron introduced 
in Sec.~\ref{sec:schannel}, which leads to
an exponential increase of the scattering amplitudes with the rapidity 
(Eq.~(\ref{eq:PFTpropagator})).

\begin{figure}
\begin{center}
\includegraphics[width=4.5cm]{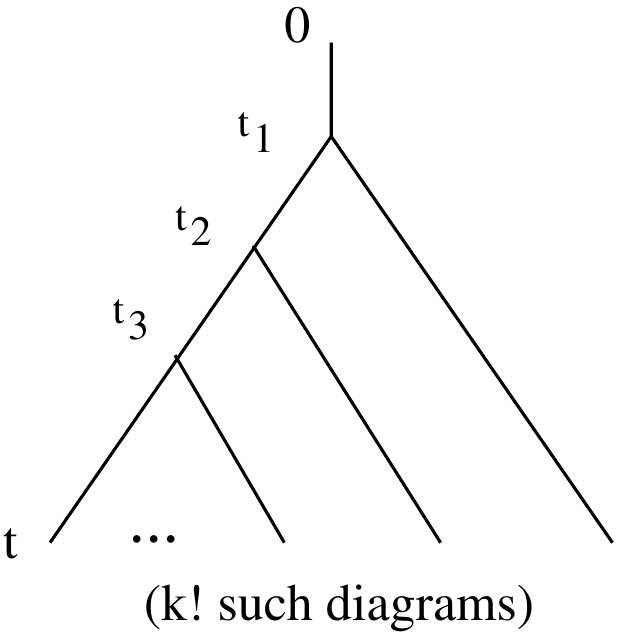}
\end{center}
\caption{\label{fig:diag0dtree}Diagrams contributing to
the one Pomeron $\rightarrow$ $k$-Pomeron transition, which gives the moments
$\langle n(n-1)\cdots(n-k+1)\rangle$ at leading order in a $1/N$ expansion.
}
\end{figure}

We now move on to the computation of higher-order diagrams
in which recombinations are absent.
First, let us recover simple results by taking the infinite-$N$ limit.
We consider the diagrams in Fig.~\ref{fig:diag0dtree}, which are 
the only ones that survive at infinite $N$ in the evaluation of the
moment $\langle n(n-1)\cdots(n-k+1)\rangle$.
Using the Feynman rules, we get for each individual diagram
\begin{equation}
(-1)^k\times(-1)^k\times 
e^{kt}\int_0^t dt_1e^{-t_1}\int_{t_1}^{t}dt_2 e^{-t_2}\cdots\int_{t_{k-1}}^t dt_k e^{-t_k}
=\frac{1}{k!}e^{kt}\left(1-e^{-t}\right)^{k-1}.
\end{equation}
There are $k!$ such diagrams (corresponding to all possible permutations
of the Pomerons), and there is an extra overall $k!$ factor 
to be added in order to get
the relevant factorial moment:
\begin{equation}
\langle n(n-1)\cdots (n-k+1)\rangle=k! e^{kt}\left(1-e^{-t}\right)^{k-1}.
\end{equation}

\begin{figure}
\begin{center}
\includegraphics[width=0.6\textwidth]{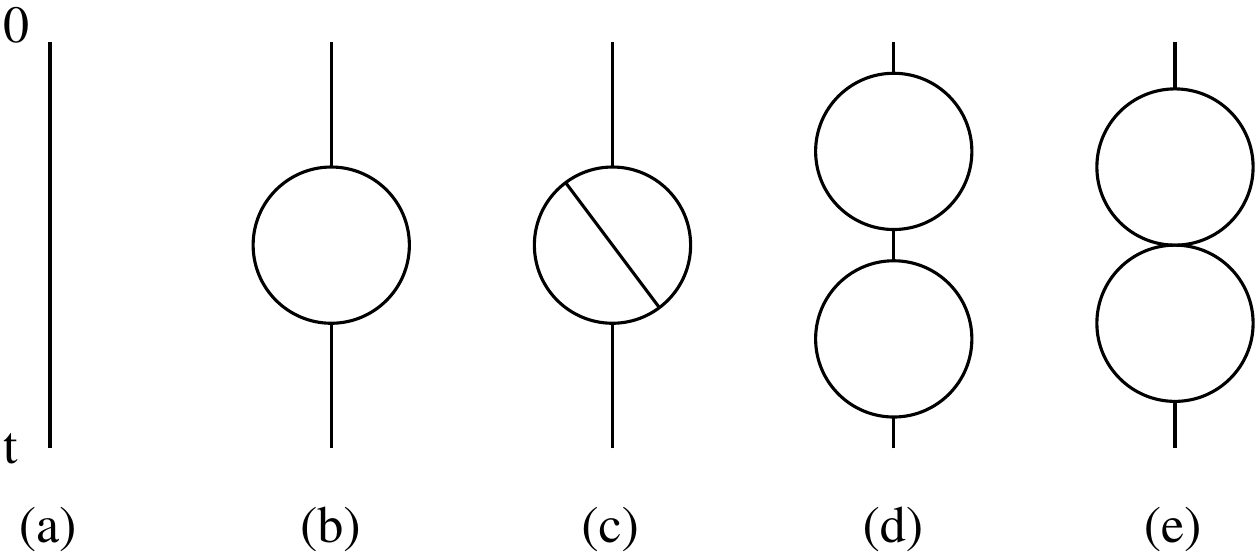}
\end{center}
\caption{\label{fig:diag0dloops}Diagrams up to order $1/N^2$ contributing to
the average of the number of particles in the system after an evolution over the
time interval $t$.}
\end{figure}

Next, we would like to perform the computation of the one-Pomeron $\rightarrow$ 
one-Pomeron transition (which provides the value of $\langle n\rangle$)
within the full theory, including the recombinations.
Some of the lowest-order diagrams are shown in Fig.~\ref{fig:diag0dloops}.
A straightforward application of the Feynman rules 
edicted above
leads to the following results
for the graphs 
that are depicted in Fig.~\ref{fig:diag0dloops}:
\begin{equation}
\begin{split}
\left.\langle n\rangle\right|_\text{tree, Fig.~\ref{fig:diag0dloops}a}&=e^{t}\\
\left.\langle n\rangle\right|_\text{1 loop, Fig.~\ref{fig:diag0dloops}b}
&=-2!\frac{e^{2t}}{N}\left(1-e^{-t}(1+t)\right)\\
\left.\langle n\rangle\right|_\text{2 loops, Fig.~\ref{fig:diag0dloops}c}
&=3!\frac{e^{3t}}{N^2}\left(1+4e^{-t}(1-t)-e^{-2t}(2t+5)\right)\\
\left.\langle n\rangle\right|_\text{2 loops, Fig.~\ref{fig:diag0dloops}d}&
=4 \frac{e^{2t}}{N^2}\left(t-3+e^{-t}\left({\scriptstyle\frac{t^2}{2}}+2t+3\right)\right)\\
\left.\langle n\rangle\right|_\text{2 loops, Fig.~\ref{fig:diag0dloops}e}&
=4 \frac{e^{2t}}{N^2}\left(t-2+e^{-t}\left(t+2\right)\right)
\end{split}
\end{equation}
We may classify these different contributions according to their order
in $e^t/N$: 
We see on the explicit expressions that
the leading terms for large $t$ and $e^t/N\sim 1$
are always of the form $N(e^t/N)^{1+\#\text{loops}}$.
It turns out that we may compute easily these dominant terms at any number
of loops. They stem from the graphs in which all splittings
occur before all recombinations
(such as \ref{fig:diag0dloops}b and \ref{fig:diag0dloops}c). 
These terms build up a series that reads
\begin{equation}
\langle n\rangle=\sum_{k=1}^{\infty}(-1)^{k-1}k!\frac{e^{kt}}{N^{k-1}}.
\label{eq:avn}
\end{equation}
This series is factorially divergent, but is easy to resum
with the help of the Borel transformation.
Indeed, using the identity
\begin{equation}
k!=\int_0^{+\infty}db\, b^k e^{-b},
\end{equation}
replacing it in Eq.~(\ref{eq:avn}),
then exchanging the integration over $b$ and the sum over the number of Pomerons $k$,
one gets
\begin{equation}
\langle n\rangle=N^2 e^{-t}\int_0^{+\infty}db\frac{1}{1+\frac{1}{b}}e^{-Ne^{-t}b}
=N\left(1-N e^{Ne^{-t}}\Gamma(0,N{e^{-t}})\right),
\label{eq:finalntpomeron}
\end{equation}
where $\Gamma$ is the incomplete Gamma function.

This result was obtained for the first time using a diagrammatic method
in Ref.~\cite{Shoshi:2005pf}. The authors of the latter paper
also computed the
next-to-leading order, that is to say, 
the terms of relative order $1/N$ after
the resummation has been performed.
The equivalent of the diffractive processes known in QCD
were also investigated by these very authors
in Ref.~\cite{Shoshi:2006eb}.
More results were obtained on that kind of models
by another group in Ref.~\cite{Levin:2007yv,Kozlov:2006cu}, 
using different techniques, which go beyond the perturbative approach.
Remarkably, the latter calculations can be applied to some extent
to QCD \cite{Kozlov:2007xc,Levin:2007wc}.

\subsection{Stochastic evolution equations}

The model may also be formulated in the form of a stochastic evolution
equation for the number of particles $n_t$ it contains at each time $t$.
The most straightforward way of doing this would be to
first compute the mean and variance of $n_{t+dt}$ given
$n_t$, with the help of the master equation~(\ref{master0d}).
This would enable one to write the time evolution
of $n_t$ in terms of a drift and of a noise of zero mean
and normalized variance, namely:
\begin{equation}
\frac{dn_t}{dt}=n_t-\frac{n_t(n_t-1)}{N}+
\sqrt{n_t+\frac{n_t(n_t-1)}{N}}\nu_{t+dt},
\label{evoln}
\end{equation}
where $\nu$ is such that $\langle \nu_t\rangle=0$ and
$\langle\nu_t\nu_{t^\prime}\rangle=\delta(t-t^\prime)$.
This equation is similar to Eqs.~(\ref{eq:stochaqcd}) and~(\ref{eq:RFT}), except for
it does not have a spatial dimension where some diffusion could take place.
The noise term is of order $\sqrt{n}$, as it should according to
the argumentation of Chap.~\ref{sec:schannel}.
Note that the distribution of $\nu$
depends on $n_t$ and
is not a Gaussian.
This last point is easy to understand: The evolution
of $\nu_t$ is intrinsically discontinuous, since
it stems from a rescaling of
$n_t$, which is an integer at all times.
A Brownian evolution (i.e. with a Gaussian noise) would
necessarily be continuous.
For completeness, we write the statistics of  $\nu_{t+dt}$, which is
deduced from the evolution of $n$:
\begin{equation}
\nu_{t+dt}=
\begin{cases}
\phantom{-}\frac{1}{\sigma\, dt}-\frac{\Delta}{\sigma} & \text{proba}\
{n_t\, dt}\\
\phantom{-\frac{1}{\sigma\, dt}}-\frac{\Delta}{\sigma} & \text{proba}\
1-n_t\, dt-\frac{n_t(n_t-1)}{N}dt
\\
-\frac{1}{\sigma\,dt}-\frac{\Delta}{\sigma} & \text{proba}\
\frac{n_t(n_t-1)}{N}dt,
\end{cases}
\end{equation}
where
$\Delta=n_t-\frac{n_t(n_t-1)}{N}$ and
$\sigma=\sqrt{n_t+\frac{n_t(n_t-1)}{N}}$.
There are jumps, represented 
by the large terms proportional to $1/dt$.

This formulation is not of great interest, neither for analytical
calculations nor for numerical simulations, since it is much
easier to just implement the rules that define the model in 
the first place (Eq.~(\ref{rule0})) 
in the form of a Monte Carlo event generator.

There is a better way to arrive at a stochastic evolution equation for
this model, although it is a bit more abstract. (It is actually equivalent
to the Pomeron field theory formulated before.)
Instead of following states with a definite number of particles like
above, we may introduce coherent states
\begin{equation}
|z\rangle=e^{-z+z a^\dagger}|0\rangle,
\end{equation}
where $z$ is a complex number.
For real positive values of $z$, the state $|z\rangle$ 
is nothing but a Poissonian
state, which is a superposition of $|k\rangle$-particle states, where
the weight of each term follows the 
Poisson law
of parameter $z$.
For the simplicity of the argument, let us restrict ourselves
to Poissonian states.
By applying the Hamiltonian ${\cal H}$ 
(defined in Eq.~(\ref{hamiltonian0d}))
to a Poissonian state $|z_t\rangle$,
one gets a new state $|\phi_{t+dt}\rangle$:
\begin{equation}
|\phi_{t+dt}\rangle=|z_t\rangle-dt\,{\cal H}|z_t\rangle.
\label{evolcoherent0d}
\end{equation}
Of course, that new state is not itself
a Poissonian state in general, but
may be written as a superposition of such states.
One writes
\begin{equation}
|\phi_{t+dt}\rangle=\int dz\,f(z)|z\rangle=\int dz\,f(z)\sum_n e^{-z}\frac{z^n}{n!}|n\rangle.
\label{decompcoherent0d}
\end{equation}
The idea is to interpret the weight function $f(z)$ as the probability to observe
a given Poissonian state $|z\rangle$.
Hence the evolution is viewed as a stochastic path
\begin{equation}
\cdots\rightarrow z_{t-dt}
\rightarrow z_t\rightarrow z_{t+dt}\rightarrow z_{t+2dt}\rightarrow\cdots
\end{equation}
with well-defined transition rates from one Poissonian state to the next one.
Inserting the explicit expression for the Hamiltonian~(\ref{hamiltonian0d}) and the
decomposition~(\ref{decompcoherent0d}) in
Eq.~(\ref{evolcoherent0d}), one gets for each Fock state $|n\rangle$
\begin{equation}
\int dz\,e^{-z}f(z)\frac{z^n}{n!}=e^{-z_t}\frac{z_t^{n}}{n!}
-dt\,e^{-z_t}
\left[
\frac{z_t^n}{(n-1)!}-\frac{z_t^{n-1}}{(n-2)!}-\frac{1}{N}
\left(
\frac{z_t^{n+1}}{(n-1)!}-\frac{z_t^n}{(n-2)!}
\right)
\right].
\end{equation}
Finally, this equation can be inverted for $f(z)$ by 
a weighted
integration over $n$,
$\int \frac{dn}{2i\pi}z_{t+dt}^{-n-1}$, along
an appropriate contour in the complex plane.
After some straightforward algebra, we get
\begin{equation}
f(z_{t+dt})=\delta(z_{t+dt}-z_t)+dt\left(z_{t}-\frac{z_t^2}{N}\right)\delta^\prime(z_{t+dt}-z_t)
+\frac12\left[
2dt\left(z_t-\frac{z_t^2}{N}\right)\delta^{\prime\prime}(z_{t+dt}-z_t)
\right].
\end{equation}
This is a representation for the
Gaussian law centered at $z_t+dt(z_t-\frac{z_t^2}{N})$
of variance $2dt(z_t-\frac{z_t^2}{N})$.
Introducing a normal Gaussian noise $\nu_t$ which satisfies
\begin{equation}
\langle \nu_t\rangle=0\ \ \text{and}\ \ 
\langle \nu_t\nu_{t^\prime}\rangle=\delta(t-t^\prime),
\end{equation}
we may write
\begin{equation}
{
\frac{dz_t}{dt}=z_{t}-\frac{z_t^2}{N}
+\sqrt{2\left(z_{t}-\frac{z_t^2}{N}\right)}\nu_{t+dt}
}
\label{eq:stochastic0d}
\end{equation}
where one must be careful to take
the noise at time $t+dt$, and hence, 
this equation is to be interpreted in the Ito sense.
If $z_{t=0}$ is a real number between 0 and $N$, then
the equation keeps it in this range.
But of course the equation is valid for 
more general coherent states,
with complex $z_t$.

This equation is suitable for numerical simulations:
One may discretize the time in small steps $\Delta t\ll 1$
in which case $\nu_t$ is distributed as
\be
p(\nu_t)=\frac{1}{\sqrt{2\pi \Delta t}}\exp\left(
-\frac{\nu_t^2}{2\Delta t}
\right).
\ee
(In many cases, one has to use more sophisticated methods, see e.g. 
Ref.~\cite{gardiner};
A more rigorous and general derivation
of this stochastic formulation may be obtained from
a path integral formalism
starting 
from the Hamiltonian~(\ref{hamiltonian0d}), see Ref.~\cite{tauber-2007-716}).
Analytical manipulations of this equation
using Ito's calculus are also quite easy. We are going to give
an example of such a calculation below, avoiding unnecessary
formalism. (We refer again the reader to \cite{gardiner} for a textbook
on a more mathematical handling of stochastic equations).

We may transform the stochastic 
equation~(\ref{eq:stochastic0d}) 
to a hierarchy of equations for the factorial moments
of the number of particles, using the relation
\begin{equation}
\langle z_t^k\rangle=\langle n(n-1)\cdots(n-k+1)\rangle\equiv n^{(k)}.
\label{correspfactorials}
\end{equation}
First, let us write Eq.~(\ref{eq:stochastic0d}) in a discretized form:
\begin{equation}
{z_{t+dt}}=z_t+dt\left(z_{t}-\frac{z_t^2}{N}\right)
+dt\sqrt{2\left(z_{t}-\frac{z_t^2}{N}\right)}\nu_{t+dt}.
\end{equation}
We then take the $k$-th power of the left and the right-hand side, and
we average the result over realizations. 
Expanding in powers of $dt$ for small $dt$, we get
\begin{multline}
\left\langle z_{t+dt}^k\right\rangle=\langle z_t^k\rangle
+dt\,k
\left\langle z_{t}^k-\frac{z_t^{k+1}}{N}\right\rangle
+dt\,k \left\langle z_t^{k-1}\sqrt{2\left(z_{t}-\frac{z_t^2}{N}\right)}
\right\rangle
\langle \nu_{t+dt}\rangle\\
+dt^2\,\frac{k(k-1)}{2} \left\langle
2\left(z_{t}^{k-1}-\frac{z_t^{k}}{N}\right)
\right\rangle
\langle\nu_{t+dt}^2\rangle
+\cdots
\end{multline}
We have factorized the averages over the time intervals $[t,t+dt]$ and
$[0,t]$, since the noise $\nu$ is uncorrelated in time.
The term proportional to $dt$ vanishes thanks to the fact that $\nu_{t+dt}$ averages
to zero. One may think that the next term could be neglected for it
is apparently proportional to $dt^2$.
Actually, it
gives a contribution of order $dt$,
because for discretized $t$,
$\langle \nu_{t+dt}^2\rangle=1/dt$.
The dots stand for terms of order $dt^2$ at least. 
Using Eq.~(\ref{correspfactorials}) to identify the
factorial moments of $n$, we eventually get
\begin{equation}
\frac{dn^{(k)}}{dt}=k\left(n^{(k)}-\frac{n^{(k+1)}}{N}\right)+k(k-1)
\left(n^{(k-1)}-\frac{n^{(k)}}{N}\right)
\end{equation}
This equation is similar to the (modified) Balitsky hierarchy in high-energy QCD.
Indeed, let us write explicitly the equations for the first two moments:
\be
\begin{split}
\frac{d\langle n\rangle}{dt}&=\langle n\rangle-\frac{1}{N}\langle n(n-1)\rangle,\\
\frac{d\langle n(n-1)\rangle}{dt}&=2\left(1-\frac{1}{N}\right)\langle n(n-1)\rangle
-\frac{2}{N}{\langle n(n-1)(n-2)\rangle}+2\langle n\rangle.
\end{split}
\ee
We note the similarity in structure with 
Eqs.~(\ref{eq:bk02}),~(\ref{eq:balitskydipoles}), except for
the term $2\langle n\rangle$ in the right-hand side
of the second equation. This term stems precisely
from the particle recombinations, and was absent in the B-JIMWLK/BK formalism.


\section{\label{sec:statisticalmethods}Statistical methods}

\begin{figure}
\begin{center}
\includegraphics[width=0.7\textwidth]{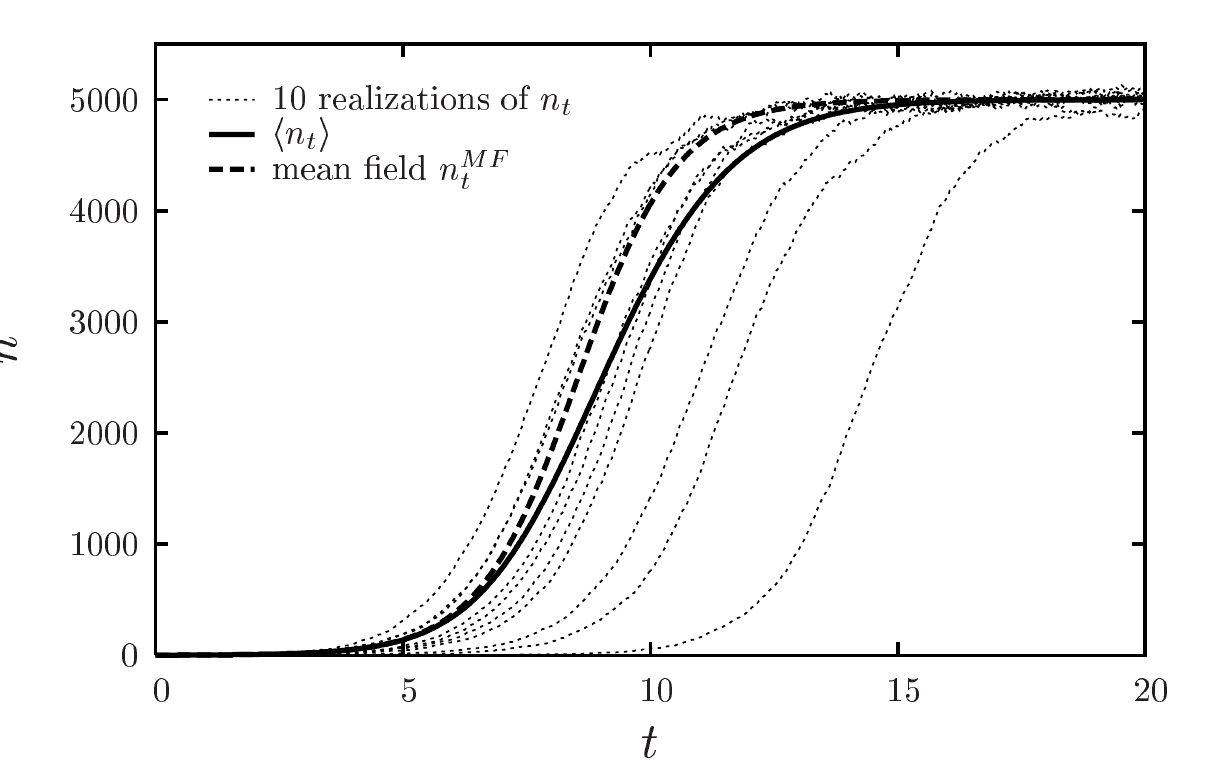}
\end{center}
\caption{\label{fig:realizations0d}
[From Ref.~\cite{Munier:2006um}]
Ten different realizations of the stochastic evolution of the 
zero-dimensional model (dotted lines; $N=5\times 10^3$).
All realizations look the same, up to a shift in time. They are all parallel to
the solution to the mean-field equation~(\ref{eq:MFequation}) (dashed line).
Note the significant difference between the latter and the average of the particle
number over the realizations (full line).
}
\end{figure}

The field theory methods presented above provide a systematics to solve the evolution
of the system to arbitrary orders in $1/N$, at least theoretically.
(In practice, identifying and resumming the relevant diagrams becomes
increasingly difficult).
However, it would look quite unreasonable
to get into such an involved formalism
if one were only interested in computing
the lowest order in a large-$N$ expansion.
Indeed, as we shall demonstrate it below,
in the case of this simple model, an intuitive and economical
calculation leads to the right answer \cite{Munier:2006um}.
We work it out here because this line of reasoning is at
the basis of the solution of more complicated models, 
closer to QCD, that we shall
address in the next chapter (Chap.~\ref{sec:reviewtraveling}).

As before, we denote by $n_t$ the value of the number
of particles in a given realization of the system.
We further introduce
$p_{\bar n}(\bar t)$ the distribution of the times
at which the number of particles in the system reaches some given
value $\bar n$ for the first time,
and $\langle n_t|n_{\bar t}\rangle$ the conditional average number of
particles at time $t$ given that there were $n_{\bar t}$ particles in the system
at time $\bar t$.
One may write the following factorization formula:
\begin{equation}
\langle n_t\rangle=\int_0^\infty d\bar t p_{\bar n}(\bar t)\langle n_t|n_{\bar t}\rangle.
\label{facto0d}
\end{equation}
This formula holds exactly for any value of $\bar n$. In particular, if $N$ is large enough,
one may choose $\bar n$ such that $1\ll \bar n\ll N$.

Observing at a few realizations generated numerically
(Fig.~\ref{fig:realizations0d}), 
one sees that the curves that represent $n_t$ look like the solution
to the mean-field equation obtained by neglecting the noise term in Eq.~(\ref{evoln}),
up to a translation of the origin of times by some random $t_0$.
(The curves look also slightly noisy around the average trend, but the noise
would still be much weaker for larger values of $N$.)
This suggests that once there are enough particles in the system 
(for $n_t>\bar n\gg 1$), 
the evolution
becomes essentially deterministic and in that stage of the evolution, the noise
can safely be discarded.
Thus stochasticity only manifests itself in the initial stages of the evolution,
but in a crucial way.
Indeed, as one can see in Fig.~\ref{fig:realizations0d}, after averaging, $\langle n_t\rangle$
differs significantly from the mean-field result, and this difference stems from
rare realizations in which the particle number stays low for a long time.
Therefore, in individual realizations, stochasticity should be accurately
taken into account as long as $n_t<\bar n$. Fortunately, when the number
of particles in the system is small compared to the parameter
$N$ that fixes the typical maximum number of particles
in a realization, the stochastic evolution is essentially governed
by a linear equation.

Thanks to this discussion,
we may assume that the evolution is linear as long as there are 
less than $\bar n$ particles
in the system and deterministic when $n_t>\bar n$.
It is then enough to compute $p_{\bar n}(\bar t)$ for an evolution without recombinations,
and $\langle n_t|n_{\bar t}\rangle$ for an evolution without noise.
The second quantity is most easily computed by replacing the averages of
powers of $n_t$ in Eq.~(\ref{evolutionnnaive}) by $n_t^\text{MF}$ and
discarding the term of order $1/N$.
One gets a closed equation for $n_t^\text{MF}$ in the form
\begin{equation}
\frac{dn_t^\text{MF}}{dt}=n_t^\text{MF}-\frac{(n_t^\text{MF})^2}{N}
\label{eq:MFequation}
\end{equation}
which is solved by
\begin{equation}
n_{t-\bar t|\bar n}^\text{MF}=\frac{N}{1+\frac{N}{\bar n}e^{-(t-\bar t)}}
\label{eq:MFsolution}
\end{equation}
where the initial condition has been chosen in such a way that
$n_{0}=\bar n$.

As for the distribution $p_{\bar n}(\bar t)$ for the 
waiting times $\bar t$ to observe
$\bar n$ particles in the system, its derivation is a bit more subtle.

Let us introduce $R(n,t)$ the probability 
distribution of the first passage time at the given
population size $\bar n$, starting with $n$ individuals
at time 0.
The probability $p_{\bar n}(\bar t)$ we are looking
for is nothing but $R(1,\bar t)$.

We now establish an evolution equation for $R$.
Recall that the evolution equation for $P$ was obtained by
considering the variation in the number of particles in the system
between times $t$ and $t+dt$.
Here we consider the beginning of the time evolution, between times $0$ and $dt$.
The probability that the system has $\bar n$ particles
for the first time at $t+dt$ 
starting with $n$ particles at time 0, 
$R(n,t+dt)$, is the
probability $ndt$ that the system gains 
a particle between times $0$ and $dt$ multiplied by
$R(n+1,t)$, minus a unitarity-preserving term.
In this way, after having taken the limit $dt\rightarrow 0$,
we get
\begin{equation}
\frac{\partial R(n,t)}{\partial t}=
n\left(R(n+1,t)-R(n,t)\right),
\mbox{ with the condition } R(\bar n,t)=\delta(t).
\end{equation}
This equation is valid when we neglect recombination
processes, which is the relevant approximation here
since we stick to the dilute regime.
In order to find a solution, 
we introduce the generating function 
for the moments of $n$:
\begin{equation}
G(u,t)=\sum_{n=0}^\infty u^n R(n,t)
\end{equation}
and the Laplace transform
\begin{equation}
\tilde G(u,s)=\int_0^{+\infty}dt\,e^{-st}G(u,t).
\end{equation}
The evolution of $R$ implies the following equation for $\tilde G$:
\begin{equation}
(1-u)\frac{d\tilde G}{du}=\left(s+\frac{1}{u}\right)\tilde G.
\label{eq:dGdt}
\end{equation}
This equation is straightforward to integrate.
Its solution reads
\begin{equation}
\tilde G(u,s)=C u(1-u)^{-1-s}.
\end{equation}
The constant $C$ must be determined from the initial condition,
namely from the equation $R(\bar n,t)=\delta(t)$,
which after Laplace transform reads $\tilde R(\bar n,s)=1$.
The latter means that the $\bar n$-th order
in the expansion of $\tilde G$ in powers of $u$
should be set to one, which writes
\begin{equation}
C\frac{\Gamma(s+\bar n)}{\Gamma(s+1)\Gamma(\bar n)}=1.
\end{equation}
For large $\bar n$, the Stirling formula
enables one to cast the equation in the simplified form
$C=\Gamma(s+1)\bar n^{-s}$.
We then see that $\tilde R(1,s)=\Gamma(1+s)\bar n^{-s}$.
The inverse Laplace transform of this function is just the
Gumbel distribution:
\begin{equation}
R(1,\bar t)=p_{\bar n}(\bar t)=\bar n e^{-\bar t -\bar n e^{-\bar t}}.
\label{eq:gumbel}
\end{equation}

Plugging Eqs.~(\ref{eq:gumbel}) and~(\ref{eq:MFsolution}) into Eq.~(\ref{facto0d}), 
we get for the average number of particles after $t$ time units 
of evolution:
\begin{equation}
\langle n_t\rangle=N\int_0^\infty d\bar t\frac{\bar n e^{-\bar t-\bar n e^{-\bar t}}}
{1+\frac{N}{\bar n}e^{-(t-\bar t)}}.
\end{equation}
Because the Gumbel distribution is strongly damped for $\bar t<0$, the lower
integration
boundary may safely be extended to $-\infty$. Indeed, it is easy to see that
a conservative upper bound for the contribution of the domain $]-\infty,0]$ to the
integral is $e^{-\bar n}$, which is very small in the limit $\bar n\gg 1$.
Finally, we perform the change of variable $b=\bar n e^{t-\bar t}/N$
to arrive at the form
\begin{equation}
\langle n_t\rangle=N^2 e^{-t}\int_0^\infty db\frac{1}{1+\frac{1}{b}}e^{-Ne^{-t}b}.
\label{eq:finalntstat}
\end{equation}
It can be checked that it is exactly the form found through
the diagrammatic approach to Pomeron field theory (compare Eq.~(\ref{eq:finalntstat}) 
to Eq.~(\ref{eq:finalntpomeron})).

The factorization in Eq.~(\ref{facto0d})
and the convenient approximations that it subsequently allows
are actually very important.
Indeed, we realized that we may write the average number of particles
at time $t$, whose expression would {\em a priori} be given
by the solution of a {\em nonlinear stochastic differential equation},
by solving two much simpler problems.
The key observation was the following. When the number of particles
in the system is low compared to the maximum average number of particles
$N$ allowed by the reaction process, then the nonlinearity is
not important, but the noise term is instead crucial.
On the other hand, when the number of particles is large compared to 1,
then the noise may be discarded, but
the nonlinearity of the evolution equation, which corresponds
to recombinations of particles, must be treated accurately.
From this method, one gets an expression for $\langle n_t\rangle$ 
for any time up to relative
corrections of order $1/N$.

When we address the problem of reaction-diffusion with one spatial dimension, we will
rely on the very same observation.
It is essentially the latter which will enable us to find analytical results
also in that case.


\section{\label{sec:relscattering}
Relation to high energy scattering and the parton model approach}

So far, we have focussed on the factorial moments of the number $n$ of
particles in the system. We have seen how they may be computed from ``Pomeron'' 
diagrams, which are quite similar to the diagrams that appear in effective
formulations of high-energy QCD.
However, the relation to scattering amplitudes, which are the observables
in QCD, may not be clear to the reader at this stage.
In particular, we do not understand yet what would correspond to boost invariance
of the QCD amplitudes.
The aim of this section is to try and clarify these points.

Let us consider a realization of the system of particles, evolved up to time $t$,
that we may call the projectile.
A convenient formalism to compute the weights of
Fock states was presented in Sec.~\ref{sec:0dfockstate}.
At time $t$, the system
of particles scatters off a target consisting of a single particle, and
can have at most one exchange with the target, which ``costs''  a factor $1/N$.
All the particles in the system have an equal probability to scatter.
Hence the probability that the system scatters reads $T=n_t/N$.

This way of viewing the evolution of the system makes it obviously 
very similar to the
QCD dipole model introduced in Chap.~\ref{sec:schannel}, 
provided one identifies the number of particles to the number
of dipoles and the time to the rapidity variable. 
The average of $T$ over
realizations is the elastic scattering amplitude.

From this analogy, there is a property similar 
to boost invariance that should hold.
Instead of putting all the evolution in the projectile, we may share it between
the projectile and the target. Let us call $n_{t^\prime}$ the number of particles
in the projectile at the time of the interaction, and $m_{t-t^\prime}$ the number
of particles in the target. The total evolution time is the same as before.
To establish the expression for $T$ in this frame, it is easier to
work with the probability $S=1-T$ that there is no interaction.
If any number of interactions were allowed between each pair of particles
from the projectile and the target, then one would simply write 
$S=\exp(-n_{t^\prime}m_{t-t^\prime}/N)$.
But since the number of scatterings should be limited to one per particle,
one has to decrease $n$ and $m$ for each new power of $1/N$, i.e. for each
additional rescattering:
\begin{multline}
S=1-\frac{1}{N}nm+\frac1{2!}\frac{1}{N^2}[n(n-1)][m(m-1)]
-\frac{1}{3!}\frac{1}{N^3}
[n(n-1)(n-2)][m(m-1)(m-2)]\cdots
\end{multline}
where the time dependences are understood, in order to help the reading.
This is like a ``normal ordering'' of the expression 
to which we would arrive by assuming
any number of exchanges.
Note that $S$ is not necessarily positive in a given event,
and hence looses its probabilistic interpretation
once one has performed the normal ordering.

Taking the average over realizations, one gets
\begin{equation}
\langle S\rangle =\sum_{k=0}^\infty 
\left\langle \frac{n!}{(n-k)!}\right\rangle_{t^\prime} 
\left\langle \frac{m!}{(m-k)!}\right\rangle_{t-t^\prime}
\frac{(-1)^k}{k! N^k}.
\end{equation}
Let us analyze this expression.

\begin{figure}
\begin{center}
\includegraphics[width=0.4\textwidth]{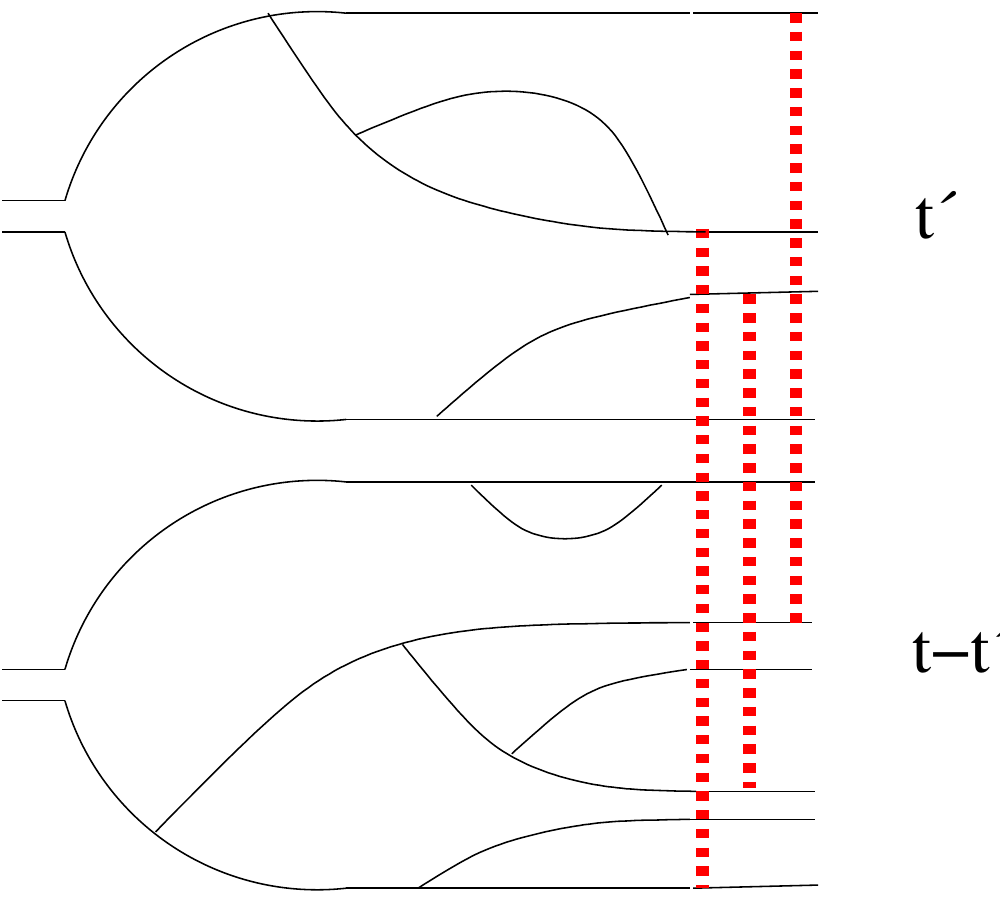}
\end{center}
\caption{\label{fig:scat0d}Representation of the scattering of two
systems of particles. The systems evolve in time from the
left to the right.
The horizontal lines represent the particles, and the vertical
dashed lines the interactions between the systems. Each of
the elementary scatterings comes with a power of $1/N$.
Note the strong similarity with the QCD diagram in Fig.~\ref{fig:com}a, except
that in the present case, 
recombinations are included in the evolution 
of each of the systems.}
\end{figure}

First, if $t^\prime=t-t^\prime$ (``center-of-mass frame''), the first two factors 
in each term of the series are of course identical after averaging.
The sum runs over the number of actual exchanges between the probe and the target.
A realization of the evolution, which would correspond to an event in QCD,
is represented in Fig.~\ref{fig:scat0d}.
Note that the figure is very similar to Fig.~\ref{fig:com}a, except
that particle mergings are allowed, while they have not been properly
formulated in QCD yet.

Second, this expression should be independent of $t^\prime$.
It is not difficult to check that this is indeed true by taking
the derivative of $\langle S\rangle$ with respect to $t^\prime$.
Expressing the averages of the factorial moments of the number
of particles with the help of the probability distributions
$P(n,t^\prime)$ and $P(m,t-t^\prime)$ respectively,
each term of the sum over $k$ and $m,n$ reads
\begin{equation}
\left.\frac{d\langle S\rangle}{dt^\prime}\right|_{n,m,k\ \text{fixed}}
=(\dot P_n P_m-P_n \dot P_m)\frac{n!}{(n-k)!}
\frac{m!}{(m-k)!}\frac{(-1)^k}{k!N^k}.
\end{equation}
The time dependence is again implicit,
and we introduced the notation $P_n=P(n,\cdot)$ and $\dot P_n=\partial_t P(n,\cdot)$ 
to get a more
compact expression. The time variabe that should be used for each factor
is unambiguous since it is in one-to-one correspondence with 
the particle number index.
We may use the master equation~(\ref{master0d})
 to express the time derivatives:
\begin{equation}
\dot P_n P_m-P_n \dot P_m=\left[(n-1)P_{n-1}+\frac{n(n+1)}{N}P_{n+1}-
\left(n+\frac{n(n-1)}{N}\right)P_n\right]P_m 
- [n\leftrightarrow m].
\end{equation}
Recalling that there are sums over $m$, $n$ and $k$ which go from 0 to $\infty$,
one may shift first the indices $m$ and $n$ in order to factorize
$P_nP_m$ in each term. The factors $1/N$ may then be absorbed by shifting
$k$ for the relevant terms. Then cancellations occur between the
terms of both squared brackets in such a way that 
once the summations over $n$, $m$ and $k$ have been performed,
the global result is 0.
This proves the independence of $\langle S\rangle$ 
upon $t^\prime$, that is, ``boost invariance'' in a quantum field theory
language.

We have seen that we may formulate scattering amplitudes
in the zero-dimensional toy model, exactly in the same way as in QCD.
We have seen in particular how crucial it is to include
particle mergings consistently with the form of the interaction
between the states of the projectile and of the target at the time of the interaction, 
in order to get a boost-invariant amplitude.


\section{\label{sec:alternative}Alternative models in zero dimension}

For the sake of completeness, we shall now construct some variants of
the zero-dimensional model introduced above,
since the latter were also discussed in the literature.
We review two of the most popular models.

\subsection{Allowing for multiple scatterings between pairs of particles}

Instead of assuming that there is at most one single exchange between each pairs
of partons, one may allow for any number of exchanges. Then the definition
of $S$ is modified as follows:
\begin{equation}
\langle S\rangle=\left\langle e^{-\frac{nm}{N}}\right\rangle
=\sum_{n,m\geq 1}P(n,t^\prime) P(m,t-t^\prime) e^{-\frac{mn}{N}}.
\end{equation}
One sees immediately that if the probabilities $P$
satisfy the master equation~(\ref{master0d}), then this expression
cannot be boost-invariant (i.e. independent of $t^\prime$).
Indeed, if Eq.~(\ref{master0d}) holds, then
\begin{equation}
P(n,t\rightarrow\infty)=\delta_{n,N}\ \text{and}\ 
P(n,t=0)=\delta_{n,1}.
\end{equation}
It follows that in the frame in which
the projectile is at rest,
\begin{equation}
\langle S\rangle_{t^\prime=0,t\rightarrow\infty}=e^{-1}
\end{equation}
while in the center-of-mass frame (if the projectile and the target
share an equal fraction of the evolution),
\begin{equation}
\langle S\rangle_{t^\prime=\frac{t}{2},t\rightarrow\infty}=e^{-{N}}
\end{equation}
which is very different.
Actually, in this model, the average 
number of particles cannot saturate at a fixed
value $N$.
It would not be compatible with boost invariance.

In order to preserve boost-invariance, one has to modify the master equation.
We may write the following general equation for the evolution
of the probability:
\begin{equation}
\dot P_n=\sum_{k\ne 0}(\alpha^k_{n-k} P_{n-k}-\alpha^k_{n}P_n).
\end{equation}
The coefficients $\alpha^k_n$ are the transition rates from
a $(n-k)$-particle state to a $n$-particle state.
We determine the $\alpha^k_n$ from the
boost-invariance requirement. 
Actually, only the coefficient $\alpha^{k=1}_n$
is needed in the case of this model.

Using the same method as the one employed for checking the boost invariance
in the previous model, we write
\begin{equation}
\frac{d\langle S\rangle}{dt^\prime}
=\sum_{n,m}(\dot P_n P_m-P_n \dot P_m)\left\langle e^{-\frac{mn}{N}}\right\rangle,
\end{equation}
and express $\dot P_n, \dot P_m$ with the help of the master equation.
Requiring that the sum over $n$ and $m$
in the right-hand side cancels
leads to the rates
\begin{equation}
\alpha^1_{n}=N\left(1-e^{-n/N}\right),
\end{equation}
where the overall constant is determined from the rate
in the unsaturated version of the model, which should hold for small
values of $n\ll N$.
This model was first proposed by Mueller and Salam \cite{Mueller:1996te}.

We see that the saturation mechanism is quite different than in the previous model.
Indeed, the average number of particles in the system keeps growing, but at a rate
that slows down and depends on the number of particles in the system itself.
Unitarity of the scattering probability $T$ is ensured first 
by multiple scatterings rather than 
by the saturation of the number of particles to a constant number $N$ 
(up to fluctuations).

This model was studied in detail in Ref.~\cite{Blaizot:2006wp}. The conclusions
drawn in there is that the saturation mechanism implied by the above model
is likely to be quite close to the one at work in QCD.
We could get analytical results for this model using one of the methods
presented above.
In particular, the statistical method outlined in Sec.~\ref{sec:statisticalmethods}
would apply and lead in a straightforward way to the expression for
$\langle n\rangle$, up to corrections of relative order $1/N$.

\subsection{Reggeon field theory}

Starting from the field theory formulation 
in Sec.~\ref{sec:pomeronfieldtheory}, 
we may discard the 4-Pomeron vertex (term $(b^\dagger)^2b^2/N$ 
in Eq.~(\ref{eq:hamiltonianpft})).
The new Hamiltonian then reads
\begin{equation}
{\cal H}^{RFT}=-b^\dagger b-(b^\dagger)^2 b+\frac{1}{N}b^\dagger b^2.
\label{hamiltonianRFT}
\end{equation}
The stochastic formulation reads
\begin{equation}
\frac{dz}{dt}=z-\frac{z^2}{N}+\sqrt{2z}\,\nu_{t+dt}
\end{equation}
(Compare to Eq.~(\ref{eq:stochastic0d})).
This is the zero-dimensional version of the stochastic equation
defining the so-called Reggeon field theory, which was intensely 
studied in the 70's as a pre-QCD model for hadronic interactions.

This model has peculiar properties if one insists on interpreting it
as a particle model. Indeed,
the Hamiltonian~(\ref{hamiltonianRFT}) corresponds to 
a generating function for the factorial moments of the number
$n$ of particles in the system at a given time $t$
that satisfies
\be
\frac{\partial Z(z,t)}{\partial t}=z(1+z)\frac{\partial Z(z,t)}{\partial z}
-\frac{z}{N}\frac{\partial^2 Z(z,t)}{\partial z^2}
\ee
and the corresponding master equation, obeyed by
the probability $P(n,t)$ to find
$n$ particles in the system at time $t$, writes
\begin{multline}
\frac{\partial P(n,t)}{\partial t}=-n P(n,t)+(n-1)P(n-1,t)\\
+\frac{1}{N}(n+1)(n+2)P(n+2,t)
-\frac{1}{N}n(n+1)P(n+1,t).
\end{multline}
One can read off this equation the rates for particle creation/disappearance.
One has a $1\rightarrow 2$ splitting, with rate $dt$; 
a $2\rightarrow 0$ annihilation
with rate
$dt/N$; and a $2\rightarrow 1$ recombination with rate $-dt/N$.
This is a negative number, and of course, it is unacceptable for a physical 
probability not to take its values between $0$ and $1$.
But we should not reject {\em a priori} negative probabilities as a formal calculation
tool \cite{Feynman:1984ie}, as long as the physical probabilities
are well-defined.
However, a Monte-Carlo code based on these negative rates turns out
to be extremely unstable, and thus of no practical use.\footnote{
We thank Al Mueller and Bo-Wen Xiao for interesting discussions
on this topic, and Krzysztof Golec-Biernat for having brought
Ref.~\cite{Feynman:1984ie} to our attention.
}

Note that the statistical approach teaches us that
in the large $N\gg 1$ limit, the moments of the number of particles
in the system should not be 
very different than for the model with 3 and 4-Pomeron vertices, 
since it is essentially the form of the fluctuations
in the dilute regime which determines the moments at all times.

A detailed study of the special properties of this model as well
as a comparison with reaction-diffusion-like models may be found 
in Ref.~\cite{Bondarenko:2006rh}.


\chapter{
\label{sec:reviewtraveling}
General results on stochastic traveling-wave equations}

{\it In chapter~\ref{sec:schannel}, we have shown the relevance of the stochastic
FKPP equation for high-energy QCD. The latter represents (classical)
particle models that undergo a branching-diffusion process in one dimension,
supplemented by a saturation mechanism.
Chapter~\ref{sec:zerodimensional} was dedicated to a detailed study,
from different points of view, of simplified models obtained from the
former ones by switching off diffusion.
We now go back to the study of one-dimensional models.
We proceed by steps: First, we shall address the deterministic FKPP equation
(which is equivalent to the BK equation in QCD)
(Sec.~\ref{sec:deterministicFKPP}). 
Second, we shall introduce fluctuations
to get solutions for equations 
in the universality class of the sFKPP equation
(Sec.~\ref{sec:combining} and~\ref{sec:beyond}).
}\\

\minitoc

\section{\label{sec:deterministicFKPP}Deterministic case: the FKPP equation}

We address the simplest reaction-diffusion equation, namely
the FKPP equation
\begin{equation}
\partial_t u=\partial_x^2 u+u-u^2.
\label{eq:FKPP}
\end{equation}
This equation 
was found to describe scattering in QCD under some assumptions,
see Chap.~\ref{sec:schannel}.

It is a mathematical theorem \cite{Bramson} that this equation admits 
{\em traveling waves} as solutions, that is to say, solitonic-like solutions such
that
\begin{equation}
u(t,x)=u(x-vt)
\label{eq:travelsolution}
\end{equation}
where $v$ is the velocity of the wave.
$u$ is a front that smoothly connects 1 (for $x\rightarrow -\infty$)
to 0 (for $x\rightarrow +\infty$).
The velocities of the traveling waves and their shapes for large $x$
are also known mathematically. 
Starting with some given initial condition which itself
is not necessarily a traveling wave such as Eq.~(\ref{eq:travelsolution})
(but which satisfies some conditions, see below), 
the FKPP equation turns it
into a stationary wave front
at large times, namely a function which may be written
in the form~(\ref{eq:travelsolution}).
The front velocity
during this phase may also be predicted asymptotically.
We informally review these results in this section.

\subsection{\label{sec:generalanalysis}General analysis and wave velocity}

The FKPP equation~(\ref{eq:FKPP}) encodes 
a diffusion in space (through the
term $\partial_x^2 u$ in the right-hand side), 
a growth (term $u$), and a saturation of this growth
(term $-u^2$).
It admits two fixed-points:
the constant functions $u(t,x)=0$ and $u(t,x)=1$.
A linear stability analysis shows that $0$ is unstable, while $1$ is stable.
Indeed, thanks to the growth term $u$ in the right-hand side, a small perturbation
$u(t,x)=\varepsilon\ll 1$
grows exponentially with time. On the other hand, a perturbation near 1 
of the form $u(t,x)=1-\varepsilon$ goes back to the fixed point $1$ through evolution.
Hence the FKPP equation describes the transition from an
unstable to a stable state.
Therefore, we expect that the linear part of the equation drives the
motion of the traveling wave, since the role of the nonlinear
term is just to stabilize the fixed point $u=1$.

We shall cast the linear part of the equation into a more general
form:
\begin{equation}
\partial_t u(t,x)=\omega(-\partial_x)u(t,x),
\label{eq:linear}
\end{equation}
where $\omega(-\partial_x)$ is a branching diffusion kernel.
It may be an integral or differential operator.
An appropriate kernel is, in practice, an operator
such that the ``phase velocity'' 
$v_\phi(\gamma)=\omega(\gamma)/\gamma$ (see below)
has a minimum in its domain of analyticity.
The FKPP equation is obtained from the choice 
$\omega(-\partial_x)=\partial_x^2+1$.

Let us follow the wave front in the vicinity of a specific value of $u$.
To this aim, we define a new coordinate $x_{\text{WF}}$
such that
\begin{equation}
x=x_{\text{WF}}+vt.
\end{equation}
The solution of the linearized equation~(\ref{eq:linear})
writes most generally
\begin{equation}
u(t,x)=\int_{\mathcal{C}} \frac{d\gamma}{2i\pi}u_0(\gamma)
\exp\left[-\gamma(x_{\text{WF}}+vt)+\omega(\gamma)t
\right],
\label{eq:sollinear}
\end{equation}
where $\omega(\gamma)$ is the Mellin transform of the linear kernel
$\omega(-\partial_x)$ 
(and thus $\gamma$ corresponds to $-\partial_x$),
and defines the dispersion relation of the
linearized equation. $u_0(\gamma)$ is the Mellin transform of the initial
condition $u(t=0,x)$.
Let us assume for definiteness
that the initial condition is a function
smoothly connecting 1 at $x=-\infty$ to 0 at $x=+\infty$, with
asymptotic decay of the form $u(t=0,x)\sim e^{-\gamma_0 x}$.
Then $u_0(\gamma)$ has singularities on the real negative axis, and on
the positive axis starting from $\gamma=\gamma_0$ and extending towards $+\infty$.
Let us take a concrete example: If $u(0,x\leq 0)=1$ and $u(0,x>0)=e^{-\gamma_0 x}$,
then $u_0(\gamma)=1/\gamma+1/(\gamma_0-\gamma)$.
The integration contour $\mathcal{C}$ should go parallel to the imaginary
axis in the complex $\gamma$-plane and cross
the interval $[0,\gamma_0]$.

Each partial wave of wave number $\gamma$ has a phase velocity
\be
v_\phi(\gamma)=\frac{\omega(\gamma)}{\gamma},
\ee
whose expression is found by imposing that the exponential 
factor in the integrand of Eq.~(\ref{eq:sollinear})
be time-independent when the velocity $v$ of the frame
is set to $v=v_\phi(\gamma)$.

\begin{figure}
\begin{center}
\includegraphics[width=0.5\textwidth]{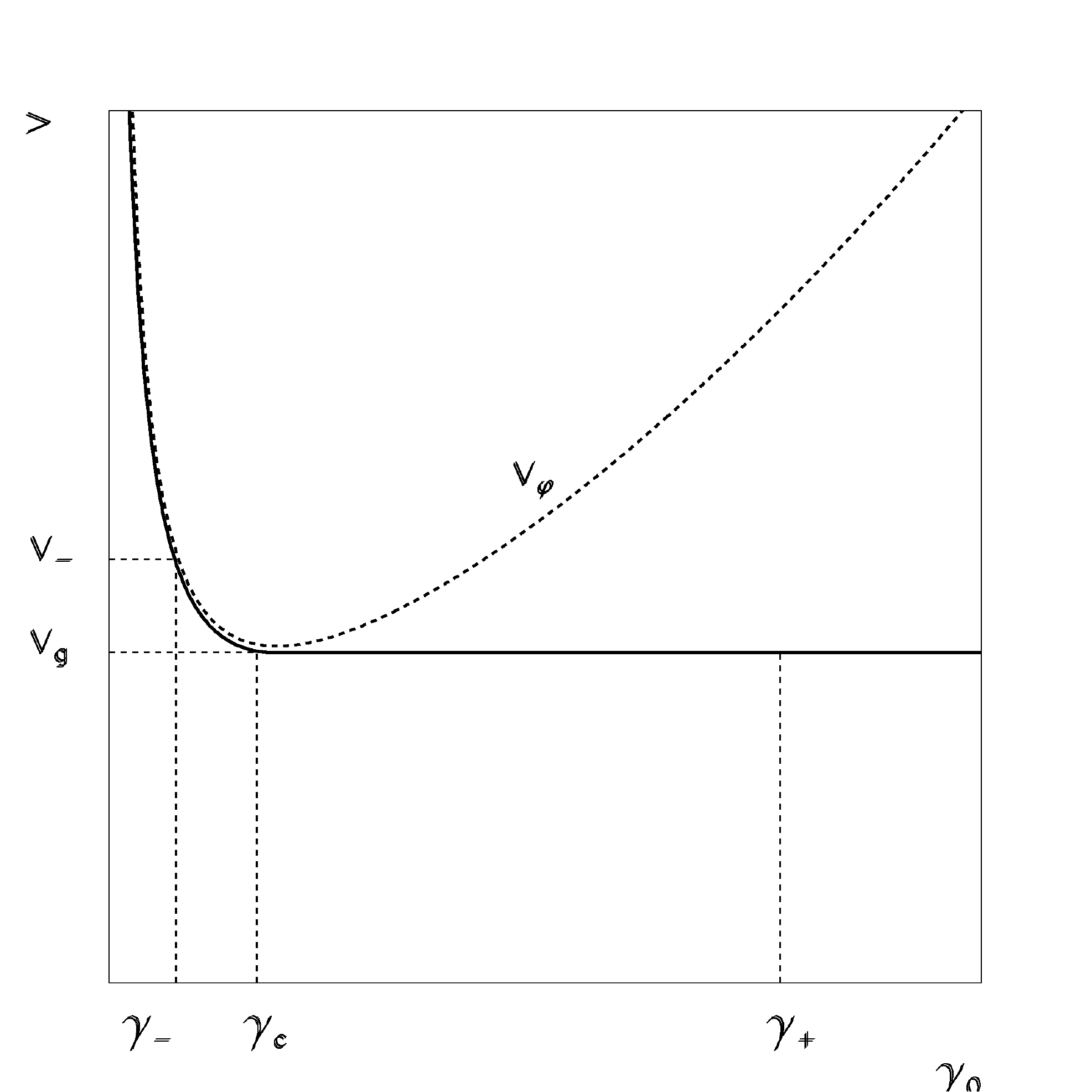}
\end{center}
\caption{\label{fig:velocitytw.eps}
Front velocity as a function of its asymptotic
decay rate $\gamma_0$ (dashed curve). 
It has a minimum at $\gamma=\gamma_c$. 
The full line represents the actual velocity 
that would be selected starting with
an initial condition decaying as $e^{-\gamma_0 x}$ for large $x$.
If $\gamma_0=\gamma_-<\gamma_c$ 
(initial condition less steep than $\gamma_c$), then
the asymptotic velocity is the phase velocity of a front 
which has the same asymptotics as the initial condition.
For any $\gamma_0=\gamma_+>\gamma_c$, the velocity of the front is the minimum 
of the phase velocity $v_\phi(\gamma)$.
}
\end{figure}

We are interested in the large-time behavior of $u(t,x)$.
The integrand in Eq.~(\ref{eq:sollinear}) admits a saddle point
at a value $\gamma_c$ of the integration variable
such that
\begin{equation}
\omega^\prime(\gamma_c)=v,
\end{equation}
that is to say, when $v$ coincides with the group velocity of the wave packet.
But the large-time solution is not necessarily given by the saddle point:
This depends on the initial condition $u_0(\gamma)$.
In order to understand this point, let us
 work out in detail the simple example of initial condition
quoted above.
The integral has two contributions for large $t$:
\begin{equation}
u(t,x)=e^{-\gamma_0(x_{\text{WF}}+vt)+\omega(\gamma_0)t}+
\kappa
e^{-\gamma_c(x_{\text{WF}}+vt)+\omega(\gamma_c)t},
\label{eq:solsaddle}
\end{equation}
up to a relative $\mathcal{O}(1)$ factor $\kappa$.
The time invariance of $u(t,x)$ in the frame of the wave
may only be achieved by tuning $v$
to one of the following two values:
\be
\mathit{(i)}\ \  v_0=\frac{\omega(\gamma_0)}{\gamma_0},\ \ \ 
\mathit{(ii)}\ \  v_c=\frac{\omega(\gamma_c)}{\gamma_c}=\omega^\prime(\gamma_c).
\ee
In the second case, $v$ coincides with 
the minimum of the phase velocity $\omega(\gamma)/\gamma$ and in particular,
$v_c\leq v_0$.
The relevant value of $v$ depends on the shape of the initial condition:
\begin{itemize}
\item If $\gamma_0<\gamma_c$, i.e. the decay of the initial
condition is less steep than the decay of the wave from the saddle-point,
then one has to pick the first choice {\it (i)} for the velocity.
Indeed, this is the only one for which the first term in Eq.~(\ref{eq:solsaddle})
is time-independent, and the second term vanishes at large time.
Due to the fact that $v_c<v_0$, choice {\it (ii)} would make the first term
in Eq.~(\ref{eq:solsaddle}) blow up exponentially, $u\sim e^{\gamma_0 (v_0-v_c)t}$.

\item If instead $\gamma_0>\gamma_c$, then it is the second
choice {\it (ii)} that has to be made. The saddle point dominates,
and the wave velocity at large time is independent of the initial condition.

\end{itemize}
Figure~\ref{fig:velocitytw.eps} summarizes these two cases.

The limiting case $\gamma_0=\gamma_c$ requires a special treatment.
Since it is not relevant for the physics of QCD traveling waves,
we refer the interested reader to the review paper of 
Ref.~\cite{vansaarloos-2003-386}
for a complete treatment also of that case.

There exists a rigorous mathematical proof of these solutions
in the case of the straight FKPP equation \cite{Bramson}.
These results are largely confirmed in numerical simulations
for various other branching diffusion
kernels, including the ones of interest for QCD 
(see e.g.~\cite{Albacete:2005ef,Enberg:2005cb}, 
and Ref.~\cite{Levin:2001et,Armesto:2001fa,Albacete:2003iq} for
earlier simulations of the BK equation).

Actually, in QCD as well as in many problems in statistical physics,
the initial condition is localized or has a finite support, and hence,
its large-$x$ decay is always very steep.
Thus for the physical processes of interest in this review, the asymptotic
front velocity, that we shall denote by $V_\infty$ for reasons that
will become clear later, reads
\be
V_\infty=v_c=\frac{\omega(\gamma_c)}{\gamma_c}=\omega^\prime(\gamma_c),
\ee
where the last equality defines $\gamma_c$ as the value of
$\gamma$ for which $v_\phi(\gamma)=\omega(\gamma)/\gamma$ is minimum.
Note that in the context of particle physics, this result was already known from 
the work of Gribov, Levin, Ryskin~\cite{Gribov:1984tu}, and was rederived later
in the framework of the BK equation 
\cite{GolecBiernat:2001if,Iancu:2002tr,Mueller:2002zm}.

\begin{figure}
\begin{center}
\includegraphics[width=0.7\textwidth]{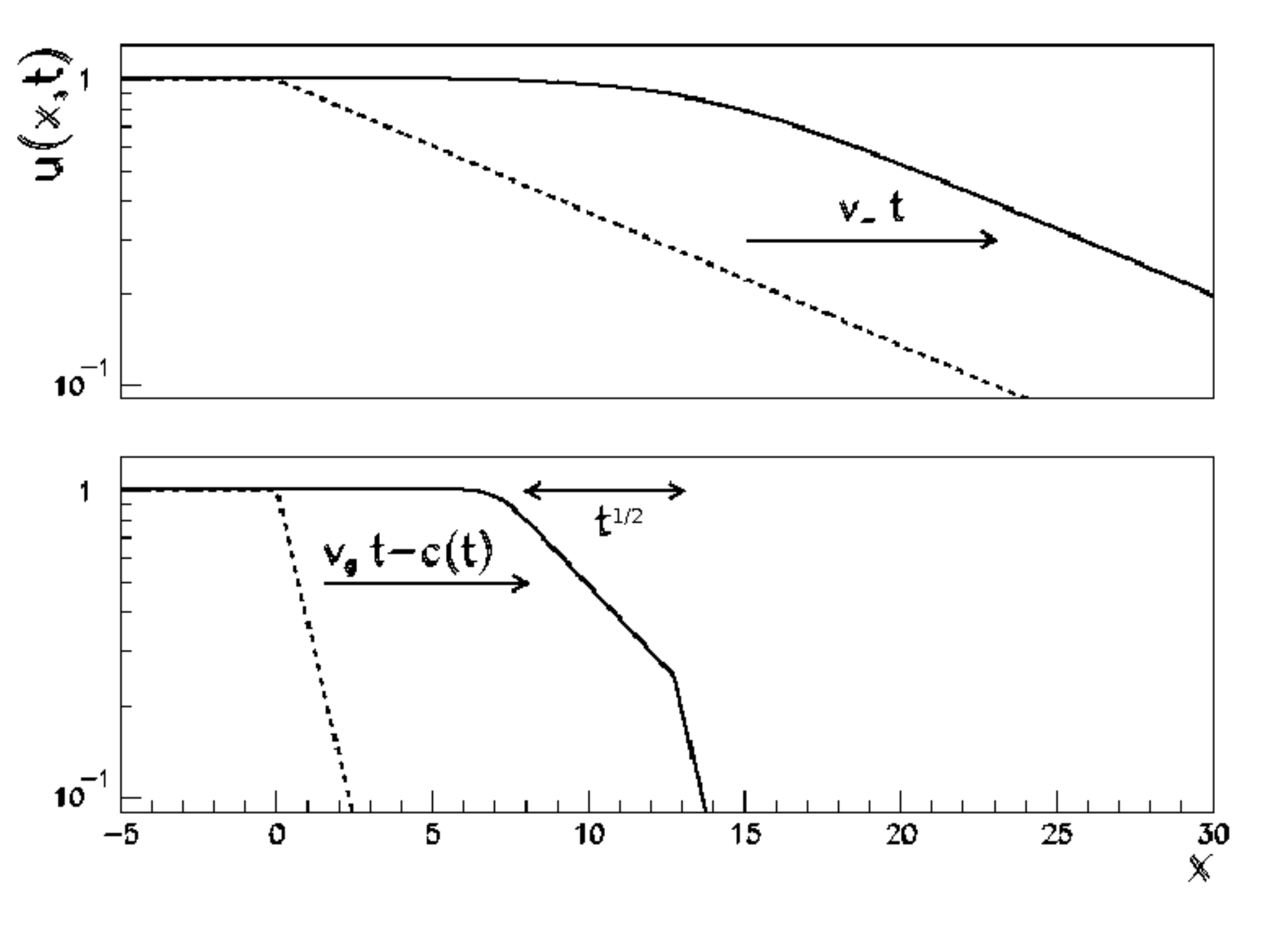}
\end{center}
\caption{\label{fig:plot2.eps}
Sketch of the shape of the front according to the large-$x$
behavior of the
initial condition
$u(t=0,x)\sim e^{-\gamma_0 x}$.
{\em Top:} $\gamma_0<\gamma_c$. 
The asymptotic
shape of the initial condition is conserved. The relaxation of the front
is fast.
{\em Bottom:} $\gamma_0>\gamma_c$. The asymptotic shape of the front
is $e^{-\gamma_c x}$, and the velocity for $t=\infty$ is $v_c=\omega(\gamma_c)/\gamma_c$.
The asymptotic shape is reached over a distance $\sqrt{t}$ ahead of the front,
and the velocity at finite time is less than the asymptotic velocity
by $\frac{3}{2\gamma_c t}$.
}
\end{figure}

So far, we have discussed the asymptotic velocity of the solutions
to the FKPP equation as a function of the initial condition.
When the initial condition is steep enough, then the asymptotic front velocity
takes a fixed value which is the minimum of $\omega(\gamma)/\gamma$. 
In the opposite case, the shape of the initial condition is retained
(see Fig.~\ref{fig:plot2.eps}).
We wish to know more detailed properties of the wave front,
such as its shape and the way its velocity approaches
the asymptotic velocity.
There are several methods to arrive at this result
(which is known from rigorous mathematics, see \cite{Bramson}).
At the level of principle, they all rely on a matching between a solution
near the fixed point $u=1$, and a solution of the linearized equation
which holds in the tail $u\ll 1$.

\subsection{Diffusion equation with a boundary and 
the approach to the asymptotic traveling wave}

We now come back to the original FKPP equation~(\ref{eq:FKPP}).
We have seen that the nonlinearity $-u^2$ has the effect of taming
the growth induced by the linear term $u$, when $u$ gets close to 1.
But nonlinear partial differential equations are
very difficult to address mathematically.
It may be much simpler to address the linear equation 
\begin{equation}
\partial_t u=\partial_x^2u+u
\label{eq:FKPPlinear}
\end{equation}
supplemented with an absorptive
(moving with time) boundary condition that ensures that $u(t,x)$ has
a maximum value of $1$ at any time.
We need to work out the solution of Eq.~(\ref{eq:FKPPlinear})
with this kind of boundary condition.
Here, we reformulate the approach proposed in the QCD context by Mueller
and Triantafyllopoulos \cite{Mueller:2002zm}
(see also Ref.~\cite{Triantafyllopoulos:2002nz} for an account of the next-to-leading order
BFKL kernel).

A solution to Eq.~(\ref{eq:FKPPlinear}) with the initial condition
$u(t=0,x)=\delta(x-x_0)$ is given, for positive times, by
\begin{equation}
u(t,x)=\frac{1}{\sqrt{4\pi t}}\exp\left(t-\frac{(x-x_0)^2}{4t}\right).
\label{eq:u0}
\end{equation}
This solution holds if the boundary condition is at spatial infinity.

We note that the lines $x$ of constant $u(t,x)=C$ are given by
\be
x=x_0+2t-\frac12\ln t-\ln(C\sqrt{4\pi})
+\text{terms vanishing for $t\rightarrow\infty$}.
\label{eq:cstugaussian}
\end{equation}
(We have selected the rightmost front $x>x_0$).
This would be the correct expression of the position of the front
if it were enough to solve the linearized FKPP equation,
to stay around some line of constant amplitude closing an eye
on the exponential growth behind the latter line
(which would be tamed by the nonlinearity).
The asymptotic velocity is 2, which coincides with the critical velocity
$v_c$ of the FKPP equation discussed above.
It is corrected by a logarithmic term. We will see that
the actual solution has the same logarithm except for the
coefficient. We will be able to get the solution
by setting an appropriate absorptive boundary which will be
time dependent.
We will proceed by steps, implementing first some
fixed boundary condition in order to gain intuition
on the form of the solution.

\begin{figure}
\begin{center}
\includegraphics[width=0.6\textwidth]{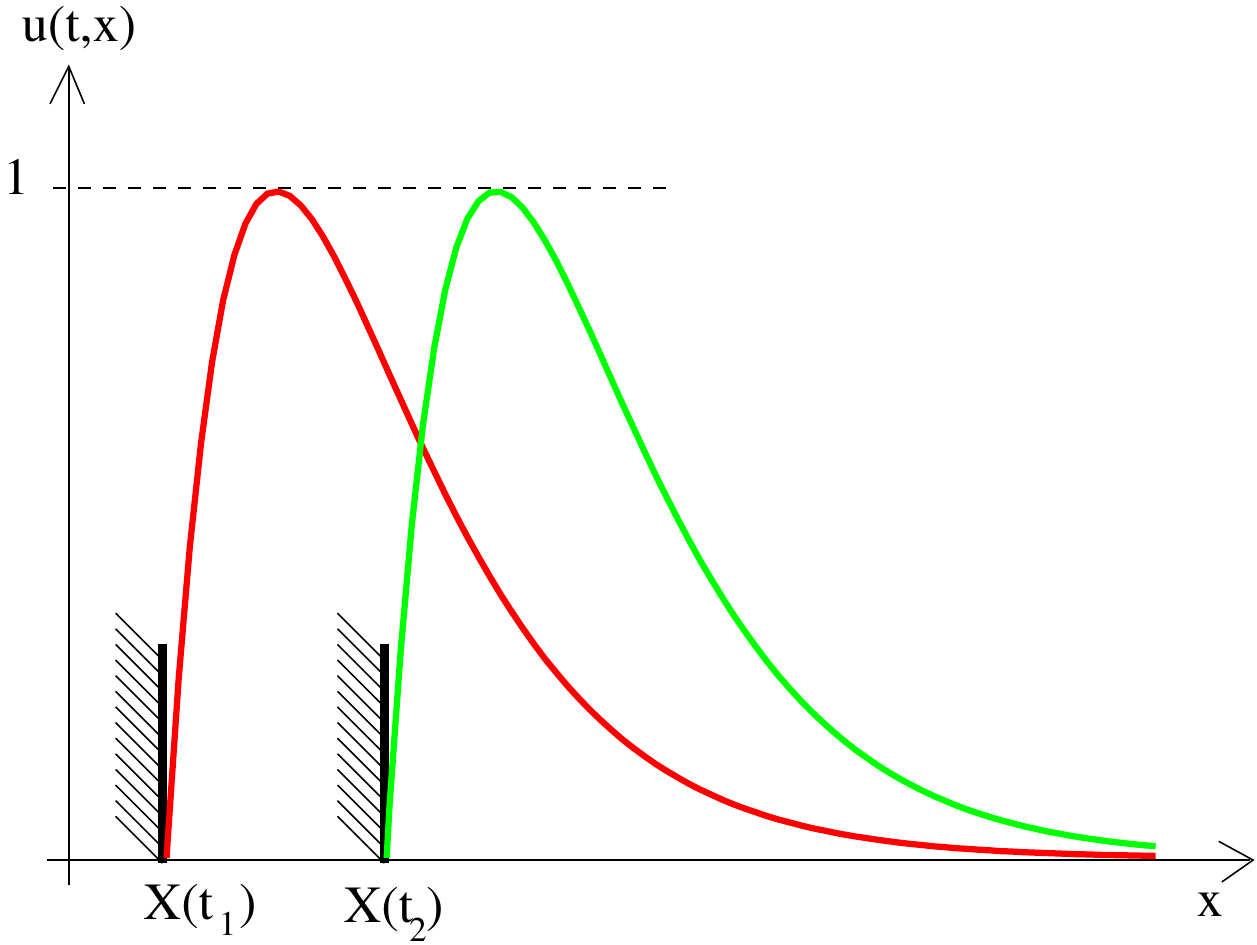}
\end{center}
\caption{\label{fig:cutoff2}Shape of the solution of 
the branching diffusion equation~(\ref{eq:FKPPlinear}) with
a moving cutoff, whose position is adjusted in such a way that the 
maximum of $u(t,x)$ be 1 at all times. The solution is represented at 
two different times $t_1$ and $t_2$, 
showing the soliton-like behavior of the solution.}
\end{figure}

So if instead of the boundary condition at infinity 
there is an absorptive barrier
at say $x=X$, i.e. if $u(t,x=X)=0$ for any $t$, then a solution
may be found through a linear combination of the latter solution
with different initial conditions, in such a way as the sum vanishes at $x=X$.
This is known as the method of images.
It is based on the elementary 
observation that any linear combination
of Eq.~(\ref{eq:u0}) also solves Eq.~(\ref{eq:FKPPlinear}).
From the solution with initial condition $\delta(x-x_0)$,
we subtract the solution of the same equation but 
with initial condition $\delta(x-(2X-x_0))$, in such a way that 
this linear combination
vanishes for $x=X$, at any time.
We get
\begin{equation}
u_X(t,x)=\frac{e^t}{\sqrt{4\pi t}}\left(
e^{-\frac{(x-x_0)^2}{4t}}-e^{-\frac{(x-2X+x_0)^2}{4t}}
\right).
\end{equation}
At this point, let us already comment that
we do not expect the solution to this problem to represent accurately
the solution to the full FKPP equation near the boundary $x\sim X$
since in that region, the details of the nonlinearity must matter.
So the region of interest will be significantly
ahead of the boundary, while the
starting point $x_0$ of the evolution 
is at some finite distance of the boundary:
\begin{equation}
x-X\gg 1\ \ \text{and}\ \ x_0-X\sim 1.
\end{equation}
One may then expand the two Gaussian terms:
\begin{equation}
u_X(t,x)=\frac{x_0-X}{\sqrt{4\pi}}\frac{x-X}{t^{3/2}}
\exp\left(t-\frac{(x-X)^2}{4t}\right).
\label{eq:abs}
\end{equation}
But in this equation, $X$ does not yet depend on time.
We cannot implement in a straightforward way
a time-dependent absorptive boundary. We may
get to such a solution by successive iterations: The main trick
is to go to a frame in which the solution of the branching diffusion
with a boundary is stationary for large times.
We went through the steps of this procedure in Ref.~\cite{Munier:2009pc}.
Here we wish to simply argue the form of the solution
from the elements we have learned so far.

As we see in Eq.~(\ref{eq:abs}), the presence of the boundary at
$X$ requires $u$ to vanish linearly at $x\sim X$. We expect this 
property to be preserved when we promote $X$ to a function of time.
Moreover, we know from our earlier investigations
that the large-$x$ asymptotic shape is $e^{-(x-X(t))}$.
From Eq.~(\ref{eq:abs}), we see that this shape is reached diffusively;
Hence there must be a factor of the form $e^{-(x-X(t))^2/(4t)}$.

Putting everything together, we are lead to the ansatz
\begin{equation}
u(t,x)=Ce^{-X} (x-X(t))e^{-(x-X(t))}\exp\left(-\frac{(x-X(t))^2}{4t}\right),
\label{eq:front0}
\end{equation}
where 
$C$ and $X$ are constants.

We know that the front velocity at large time is $X^\prime(t)\sim v_c=2$
and we expect a logarithmic correction $c(t)\sim \ln t$, so we write
\begin{equation}
X(t)=2t-c(t).
\label{eq:ansatzX}
\end{equation}
Inserting Eq.~(\ref{eq:front0}) and~(\ref{eq:ansatzX})
into Eq.~(\ref{eq:FKPPlinear}) and setting $x=X(t)+a$ ($a$
is a constant), we arrive at the equation
\begin{equation}
c^\prime(t)\left[2t(a-1)+a^2\right]=3a
\end{equation}
which means that for large $a$ and $t$, $c(t)\sim \frac32\ln t$.
Hence
\begin{equation}
X(t)\equiv x-x_2+X=2t-\frac32\ln t+\mathcal{O}(1).
\label{eq:posfront0}
\end{equation}
The latter quantity is
the position of the absorptive boundary for large times, 
and thus also the position of the front. The constant 
$X$ is the position of the front in the moving frame
(while $X(t)$ is its position in the initial
reference frame).
Setting $X=-1$ and $C=1$, the maximum of $u$ is reached at $x=X(t)+1$, and 
is indeed equal to 1.

For large $t$ or in the region $x-X(t)\leq \sqrt{t}$ 
which expands with time, 
the Gaussian factor goes to 1, and we see that
$u(t,x)$ only depends on one single variable $x-X(t)$.
This was expected: It is precisely the defining property of traveling waves. 
But in addition
to these asymptotic solutions, we get from this calculation 
the first finite-$t$ correction to the front shape and front velocity.

Actually, the speed of the front is intimately related
to its shape.
At time $t$, the front has reached its asymptotic shape
over the distance $\sqrt{t}$
from the saturation point.
This remark will be important in the following.

We have derived the solution of a problem
that was not exactly the initial one,
however, we believe that the shape of the front
in its forward part ($u\ll 1$, namely for $x-X(t)\gg 1$)
as well as its velocity are quite universal.
Heuristically, these properties are completely 
derived from the linear part of the equation.
For this reason, the front is said to be ``pulled''
by its tail.
The nonlinearity only tames the growth of $u$ near $u\sim 1$, 
and so its precise
form should not influence the front position itself, at least
at large enough times.
Thus we expect these solutions to have a broad validity,
only depending on the diffusion kernel, and so,
may be obtainable from our calculation up to
the replacement of the relevant parameters.

For the more general branching diffusion kernel
in Eq.~(\ref{eq:linear}), the velocity of the front would read
\begin{equation}
{
\frac{dX(t)}{dt}=\frac{\omega(\gamma_c)}{\gamma_c}-\frac{3}{2\gamma_ct}+\cdots
}
\label{eq:velocitygeneral0}
\end{equation}
where $\gamma_c$ solves $\omega(\gamma_c)=\gamma_c\omega^\prime(\gamma_c)$,
as was explained in Sec.~\ref{sec:generalanalysis}.
The front shape in its forward part $x-X(t)\gg 1$
is represented by the equation
\be
{
u(t,x)\propto(x-X(t))e^{-\gamma_c(x-X(t))}
\exp\left(-\frac{(x-X(t))^2}{2\omega^{\prime\prime}(\gamma_c) t}\right),
}
\label{eq:frontgeneral0}
\ee
up to an overall constant.
Fig.~\ref{fig:cutoff2} represents a sketch of the solution 
at two different times.
The large-time shape is an exponential decay,
\begin{equation}
u(t,x)\sim e^{-\gamma_c(x-X(t))}
\label{eq:asymptoticdecay}
\end{equation}
up to a linear growth,
and from Eq.~(\ref{eq:frontgeneral0}), this shape extends over a range
\be
L=x-X(t)\sim \sqrt{2\omega^{\prime\prime}(\gamma_c)t}.
\ee
In other words, the time needed for the front to reach its asymptotic
shape over a range $L$ reads
\be
t\sim\frac{L^2}{2\omega^{\prime\prime}(\gamma_c)}.
\label{eq:asymptotictime}
\ee

Through our simple arguments and calculation,
we got the lowest order in an expansion of the front shape
and position at large times.
The next corrections to $X(t)$
would be of order 1 (this constant depends
on the way we define the position of the front), 
followed by an algebraic series in $t$ whose
terms all vanish at large $t$.
The first next-to-leading term in the series has been computed 
(see Ref.~\cite{ebert-2000-146}): It turns out to be of
order $1/\sqrt{t}$.
We will not reproduce the calculations that lead to it
because they are rather technical and
there is already a comprehensive review paper available on
the topic \cite{vansaarloos-2003-386}.
But let us write the result for the position and the shape of the front
at that level of accuracy, for the more general branching diffusion
kernel given by Eq.~(\ref{eq:linear}).
To that accuracy, the front position reads \cite{ebert-2000-146,Munier:2004xu}
\begin{equation}
X(t)
=\frac{\omega(\gamma_c)}{\gamma_c} t-\frac{3}{2\gamma_c}\ln t
-\frac{3}{\gamma_c^2}
\sqrt{\frac{2\pi}{\omega^{\prime\prime}(\gamma_c)}}\frac{1}{\sqrt{t}}
+{\cal O}(1/t).
\label{eq:fullvelocity}
\end{equation}
For the simple FKPP case, we recall that $\omega(-\partial_x)=\partial_x^2+1$,
then $\gamma_c=1$ and $\omega(\gamma_c)=2$.
The first two terms in the last equations match the ones found in 
Eq.~(\ref{eq:velocitygeneral0}).
The shape of the front in its forward part has the following form
\cite{ebert-2000-146,Munier:2004xu}:
\begin{multline}
u(t,x)=C_1 e^{-\gamma_c(x-X(t))}\exp\left({-z^2}\right)
\times\\
\Bigg\{\gamma_c(x-X(t))]+C_2
+\left(3-2C_2+\frac{\gamma_c\omega^{(3)}(\gamma_c)}{\omega^{\prime\prime}(\gamma_c)}
\right)
z^2\\
-\left({\frac23}\frac{\gamma_c\omega^{(3)}(\gamma_c)}{\omega^{\prime\prime}(\gamma_c
)}
+
\frac13{}_2\!F_2\left[1,1;{\scriptstyle\frac52},3;
z^2\right]\right)z^4\\
+6\sqrt{\pi}
\left(1-{}_1\!F_1\left[-{\scriptstyle \frac12},{\scriptstyle \frac32};
z^2
\right]\right)z+{\cal O}(1/\sqrt{t})
\Bigg\},
\label{eq:front1}
\end{multline}
where
\be
z=\frac{x-X(t)}{\sqrt{2 \omega^{\prime\prime}(\gamma_c) t}}.
\ee
$C_1$ and $C_2$ are constants,
and ${}_2F_2$, ${}_1F_1$ are generalized hypergeometric functions.
The terms in the first and second lines match with the
result of our calculation (Eq.~(\ref{eq:frontgeneral0})).
These expressions should apply also to QCD, up to the relevant replacements
given in Tab.~\ref{tab:dictionary}.

So far, we have considered equations of the type of Eq.~(\ref{eq:FKPP})
as saturation equations, in the sense that they
describe the diffusive growth of a continuous function $u$ until
it is tamed for $u\sim 1$.
We will see below
that these equations 
may actually be given a different physical interpretation.


\subsubsection{Relevance of this formalism to the BK equation}

\begin{figure}
\begin{center}
\includegraphics[width=12cm]{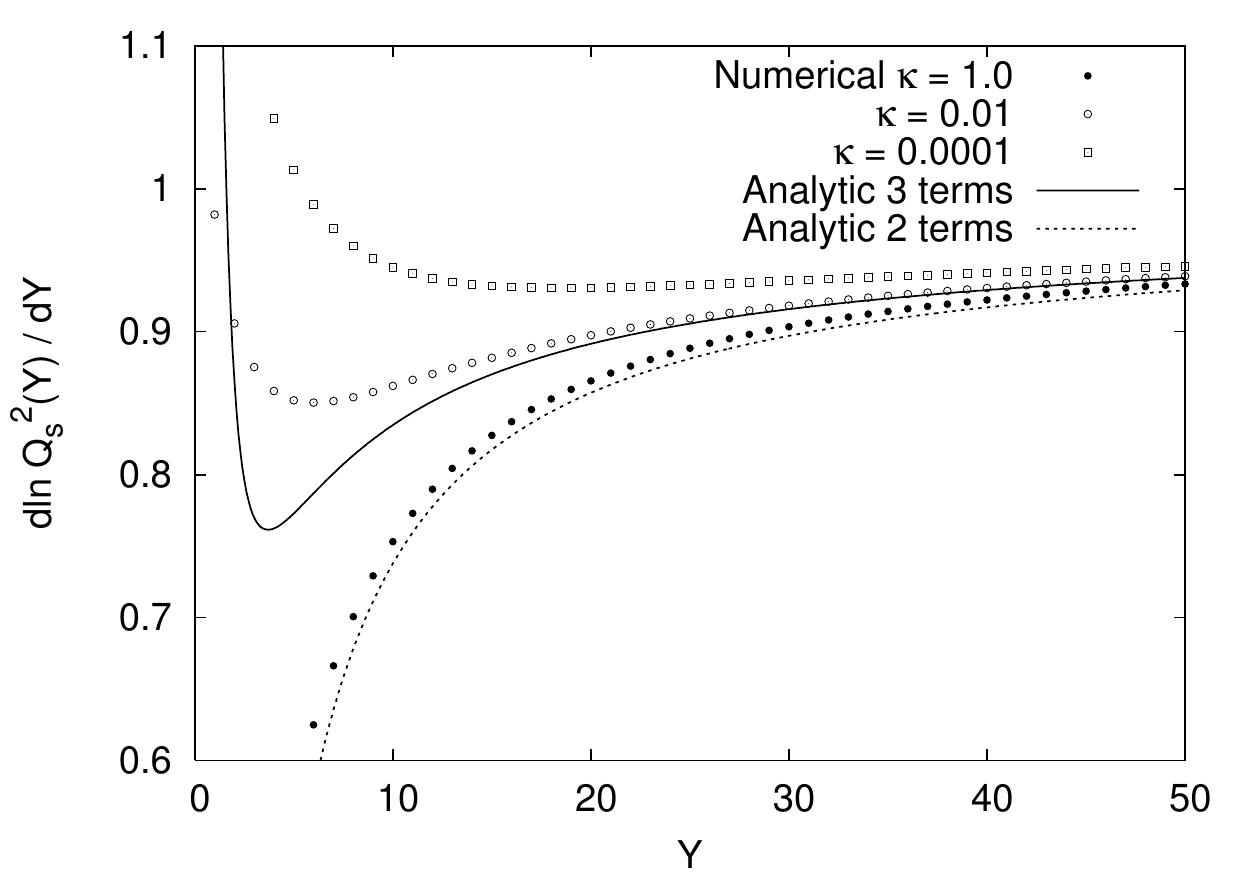}
\end{center}
\caption{\label{fig:enberg}
Velocity of the QCD traveling wave as a function of the rapidity.
The different curves correspond to the numerical simulation,
the saturation scale being defined in various ways 
with the help of the parameter $\kappa$
(see the text for the definition of the latter),
and to the analytical formula of Eq.~(\ref{eq:fullvelocity}),
either truncated after the second term (``Analytic 2 terms''), or
complete (``Analytic 3 terms'') (see the dictonary in  Tab.~\ref{tab:dictionary}
for the notations).
}
\end{figure}

In order to check that the formalism used to arrive at a solution
to FKPP-like equation applies to the BK equation in QCD, we performed
in Ref.~\cite{Enberg:2005cb} a numerical simulation of the BK equation.
We compared the velocity of the obtained traveling wave with 
Eq.~(\ref{eq:fullvelocity}), using either the full expression with three
terms or a truncation of it keeping only the two dominant terms.
We defined the saturation scale by the equation $A(y,Q_s(y))=\kappa$,
and we chose different values of $\kappa$.
We see in the plot of Fig.~\ref{fig:enberg}
that the numerical result is consistent with the analytical expectations 
(using the dictionary in Tab.~\ref{tab:dictionary}),
although the complete discussion is quite subtle.
All details of the numerics and the discussion of the results
were published in Ref.~\cite{Enberg:2005cb}.

\subsection{\label{sec:discretebranchingdiffusion}Discrete branching diffusion}

We have investigated the solutions of the FKPP equation
in a mathematical way, without discussing the physics that may lead to such an equation.
The absorptive boundary that we have put
replaces the nonlinear term in the FKPP equation, whose role is to
make sure that $u$ never exceeds the limit $u=1$.
Hence we have thought of this boundary as a way to enforce the {\em saturation}
of some density of particles.
Actually, the FKPP equation~(\ref{eq:FKPP})
may stem from a branching diffusion 
process in which the number of particles is unlimited, and
thus, for which there is no saturation at all.
As a matter of fact, this is 
what the BK equation describes in QCD: An exponentially growing 
number of dipoles,
stemming from the rapidity evolution of a hadronic probe, scatters off some
target. The overall interaction probability is unitary because
multiple scatterings are allowed (the interaction probability of $n$ dipoles
is actually of the form $1-e^{-\alpha_s^2 n}$), but not because
there is a saturation of the number of dipoles in the wavefunction
of the probe. We refer the reader back
to Fig.~\ref{fig:bk} for a picture
of the process.

\begin{figure}
\begin{center}
\includegraphics[width=10cm]{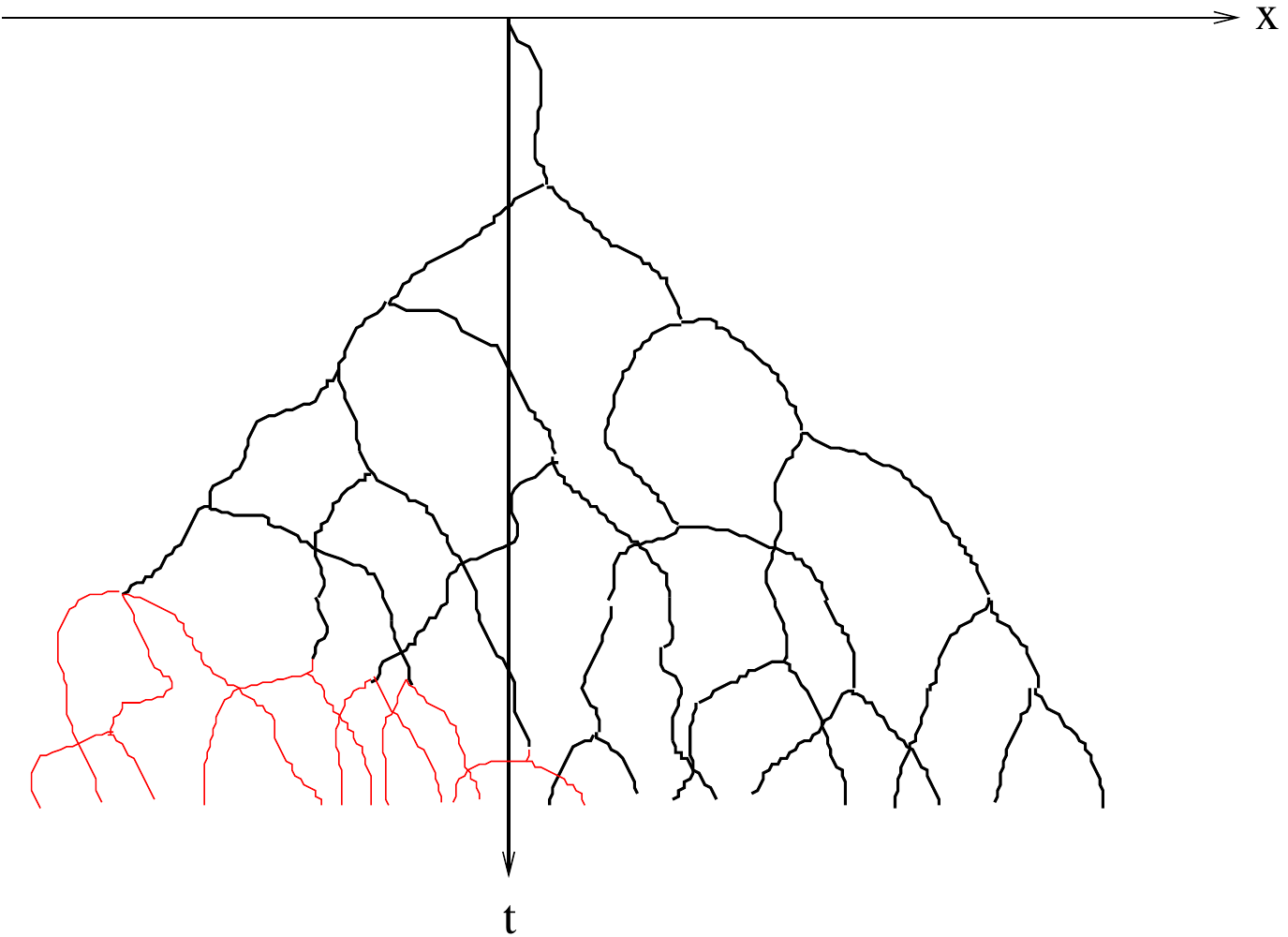}
\end{center}
\caption{\label{fig:branchingdiff}Example of branching diffusion process
on a line (see the text for a mathematical description of the evolution rules).
If the number of individuals is limited by a selection process which,
at each new branching, eliminates the individual  
sitting at the smallest $x$ as soon
as the total number of individuals reaches say $N$ ($N=10$ in this figure),
then only the branches drawn in thick line survive.
}
\end{figure}

To illustrate how the FKPP equation
arises in such a simple model of branching diffusion,
let us consider a set of particles
on a line, each of them being indexed by a continuous variable $x$.
(Such a model was considered for instance in Ref.~\cite{DerridaSpohn}).
We let the system evolve according to the following rules.
During the time interval $dt$, each particle has a probability $dt$
to split in 2 particles. Unless it splits, it moves
of the small random amount $\delta x$, which is a Gaussian
variable distributed like
\begin{equation}
p(\delta x)=\frac{1}{\sqrt{4\pi dt}}
\exp\left(-\frac{(\delta x)^2}{4 dt}\right).
\label{eq:pgauss}
\end{equation}
Let us consider the number of particles $n(t,x)$ contained in an interval
of given size $\Delta x$ centered around the coordinate $x$. At time $t=0$, the
system is supposed to consist in a single particle sitting at the origin $x=0$.
A sketch of a realization of this model is shown in Fig.~\ref{fig:branchingdiff}.
From the evolution rules, 
we easily get an equation for the average number of particles $\langle n\rangle$:
\begin{equation}
\langle n(t+dt,x)\rangle=dt\,2\langle n\rangle
+(1-dt)\int d(\delta x)p(\delta x)\langle n(t,x-\delta x)\rangle
\end{equation}
which reads, after replacing $p$ by Eq.~(\ref{eq:pgauss}) and after
the limit $dt\rightarrow 0$ has been taken,
\begin{equation}
\frac{\partial\langle n\rangle}{\partial t}=\langle n\rangle
+\frac{\partial^2 \langle n\rangle}{\partial x^2}.
\end{equation}
All the dependence on the size $\Delta x$ 
of the ``bin''
is contained in the initial condition.
It is clear that for large enough times,
the solution to this equation is given by Eq.~(\ref{eq:u0}).

Let us now define
\begin{equation}
S(t,x)=e^{-n(t,x)/N}
\end{equation}
where $N$ is some (large) constant.
This definition is reminiscent of the $S$-function, related
to the scattering amplitude, introduced in the discussion
of the BK equation in Chap.~\ref{sec:schannel}.
At a fixed time and for large enough $x$, $n(t,x)\ll N$ and thus 
$1-S(t,x)\simeq n(t,x)/N\rightarrow 0$.
For any $x$, 
the exponential makes sure that $S$ ranges between $0$ and $1$.
Thus $S$ (or $1-S$) has the shape of a traveling wave.
Its position $X(t)$ is the value of $x$ for
which $n(t,x)$ is some given constant say of the order of $N$.
In the mean-field limit in which $n$ is replaced by its
average $\langle n\rangle$, it is very easy to compute $X(t)$
from the form of the solution~(\ref{eq:u0}). We get
(see Eq.~(\ref{eq:cstugaussian}))
\be
X(t)=2t-\frac12\ln t
\ee
up to a constant.

On the other hand however, the average of $S$ over events, namely $A=1-\langle S\rangle$
obeys the FKPP equation.
Indeed
\begin{equation}
\langle S(t+dt,x)\rangle=
dt\langle S(t,x)\rangle^2
+(1-dt)\int d(\delta x)p(\delta x)\langle S(t,x-\delta x)\rangle.
\end{equation}
In the limit $dt\rightarrow 0$ and rewriting the equation 
with the help of $A$, we get
\begin{equation}
\frac{\partial A}{\partial t}=\frac{\partial^2 A}{\partial x^2}+A-A^2.
\end{equation}
Hence $A$ is a traveling wave at large times, and its position $X(t)$
is given by Eq.~(\ref{eq:posfront0}).
It is obviously behind by a term $\ln t$ with respect to
the value of $x$ for which the average number of particles
has a given constant value
(compare Eq.~(\ref{eq:posfront0}) and Eq.~(\ref{eq:cstugaussian})).
Furthermore, 
the probability distribution of the
position of the rightmost particle (or of the $k$-th rightmost particle
for any given $k$)
may also be derived from the FKPP equation.
(Brunet and Derrida have recently given an advanced discussion
of the statistics of the position of these particles,
see Ref.~\cite{BD2009,BD2011}).
It turns out that in any event, the average $x$ for which
$n(t,x)$ has a given value, say $n_0$, 
moves with the FKPP velocity
which can be read off from Eq.~(\ref{eq:posfront0}). This is much slower
than the rate of change
of $X(t)$ when the latter is defined as the implicit solution of the equation
$\langle n(t,X(t))\rangle=n_0$.

All this may seem a bit paradoxical.
But actually, it is just related to the fact that 
$\langle e^{-n/N}\rangle$ cannot be
approximated by $e^{-\langle n\rangle/N}$.
We may understand it in the following way.
By taking the average of $n$, we have somewhat forgotten a fundamental
property of $n$: its {\em discreteness}. Indeed, it only takes integer
values, and in particular, the distribution of $n$ in a realization
has a finite support: At any time, there is a value of $x$
to the right of which there are no particles at all. $n$ obeys a stochastic equation.
This is not the case for $\langle n\rangle$, which just obeys
an ordinary branching diffusion equation.

\begin{figure}
\begin{center}
\includegraphics[width=8cm]{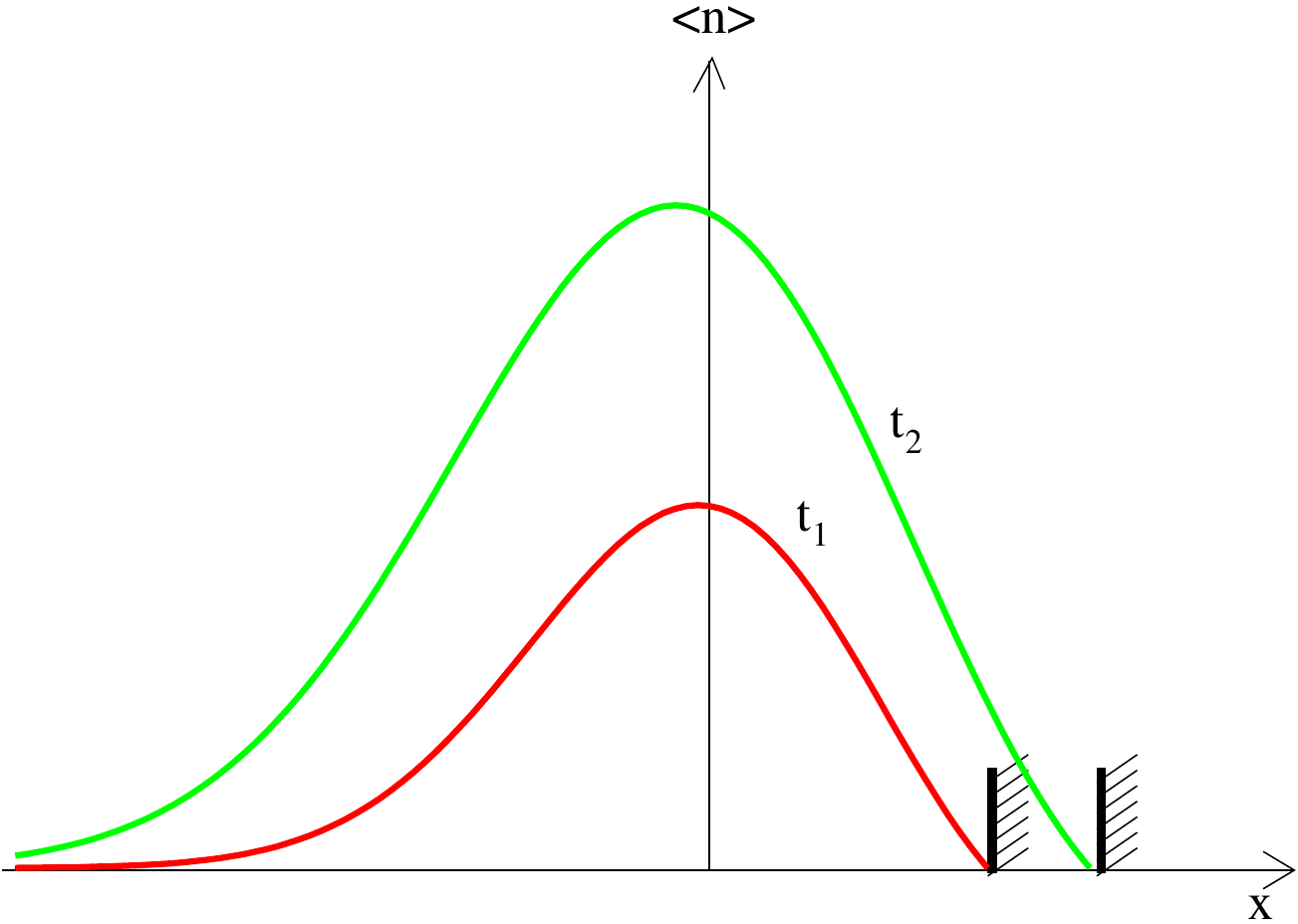}
\end{center}
\caption{\label{fig:cutoff1}Solution of the branching 
diffusion equation~(\ref{eq:absorpboundary})
with a moving  absorptive boundary that forces $\langle n\rangle$ to vanish at 
the point $X(t)$ such that $\langle n(t,X(t)-1)\rangle=1$.
Two different times are represented.}
\end{figure}

In order to recover the effect of
the discreteness of $n$ and compute the velocity,
we may again use the absorptive boundary trick.
Let us solve the linear equation
\begin{equation}
\partial_t\langle n\rangle=\partial_x^2\langle n\rangle+\langle n\rangle
\label{eq:absorpboundary}
\end{equation}
with an absorptive boundary.
The latter will be placed in such a way that at a distance of order one to
its left (we will focus on the right-moving wave), $\langle n\rangle=1$
(see Fig.~\ref{fig:cutoff1}).
There is no difference in principle with the boundary calculation that we have performed
before, except that the absorptive boundary is now placed to the right of the front (i.e.
$x_0<X$ in the notations used above).
Thus we find without any further calculation
that the realizations of $n$ move, on the average,
with the FKPP velocity~(\ref{eq:velocitygeneral0}).

\section{\label{sec:combining}Combining saturation and discreteness}

We have seen that physically, 
the FKPP equation (or the BK equation in QCD) may be interpreted either
as an equation for the growth, diffusion and
saturation of a continuous function, or as
the evolution equation for the average of a bounded function
of a discrete (thus stochastic) branching diffusion process.
For each of these interpretations,
we may find the main features of the solutions by imposing one absorptive
boundary on the linear partial differential equation encoding branching diffusion.
In one case, the boundary is a cutoff that prevents $u$ to be larger than 1:
It represents saturation, i.e. the explicit nonlinearity present in the
FKPP equation. In the other case, the boundary forces 
the function $n$ that represents the number of particles
to vanish quickly when $n$ becomes
less than 1. Formally, it actually models the stochasticity due to the intrinsic
discreteness of the number $n$ of particles, and avoids to address a stochastic
equation directly.

\begin{figure}
\begin{center}
\includegraphics[width=0.8\textwidth]{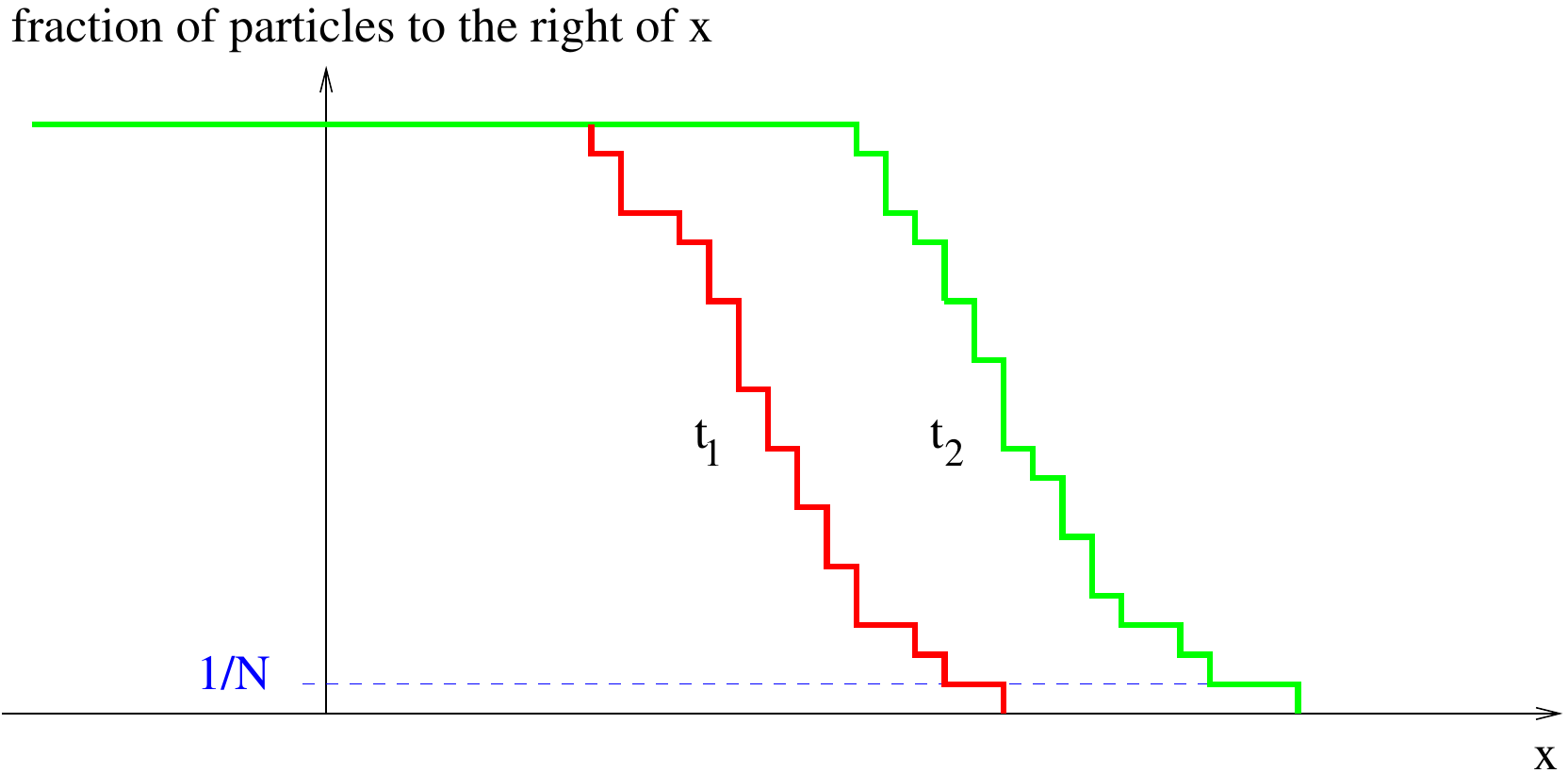}
\end{center}
\caption{\label{fig:frontbranching}Branching diffusion model of 
Sec.~\ref{sec:discretebranchingdiffusion} with
selection that limits the total number of particles to $N$.
The function ${\cal U}(t,x)$, which is the number of particles
to the right of $x$ normalized to the maximum number of particles $N$,
is represented.
One sees that the fraction of particles to the right of $x$ looks like
a traveling wave front.
}
\end{figure}

In physical cases such as reaction-diffusion processes for finite $N$,
we define $u(t,x)$ as the number of particles per site 
(or per bin) in $x$ normalized to $N$.
Hence it takes discrete values: $1/N$, $2/N$ etc...
While for large $N$ discreteness is unlikely to play a role in the region $u\sim 1$,
it is expected to be crucial when $u\sim 1/N$.
It is thus natural to impose the two boundaries: one representing
saturation of the particle number, the other one 
taking care of the
discreteness of the same
quantity.
A model that these two cutoffs may represent is for example, the branching diffusion
model in Sec.~\ref{sec:discretebranchingdiffusion}, but in which the total
number of particles is limited to $N$ by keeping only the $N$ rightmost ones
at each new branching.
It is clear that the function ${\cal U}(t,x)$ defined to be the number of particles to
the right of some position $x$ normalized to the maximum number $N$
is, for large enough times, a front connecting 1 (for $x\rightarrow -\infty$) to 0 
(for $x\rightarrow +\infty$) (see Fig.~\ref{fig:frontbranching}).

Reaction-diffusion problems 
(described by nonlinear stochastic partial differential equations)
were interpreted  
as branching diffusion problems taking place between two absorptive boundaries
for the first time
by Brunet and Derrida in Ref.~\cite{brunet-1997-57} and later, independently,
by Mueller and Shoshi in the case of QCD in Ref.~\cite{Mueller:2004sea}.
Note however that in the context of the QCD parton model,
the present interpretation of the cutoffs was only found
in Ref.~\cite{Iancu:2004es}.
Mueller and Shoshi introduced the both cutoffs for reasons related to the
boost-invariance of the QCD amplitude.
(The discreteness cutoff was thought as the symmetric
of the saturation cutoff under some boost).
The duality of the two boundaries, that is to say of the dense and dilute 
regimes of the traveling wave, was studied more deeply in 
Refs.~\cite{Kovner:2005jc,Kovner:2005en,Kovner:2005uw,Kovner:2005aq,Kovner:2007zu}.

Before moving on to the technical derivation of the shape and position
of the front in this case, let us
figure out what we expect to find.

Starting from the initial condition which we assume
to be one or a few particles, the front builds up
and its velocity increases with $t$ (see Eq.~(\ref{eq:velocitygeneral0}))
until it reaches its asymptotic shape, which is a decreasing
exponential $e^{-\gamma_c(x-X(t))}$
that holds for all $x-X(t)\gg 1$. ($X(t)$ is here the position
of the bulk of the front, say for example of the leftmost
surviving particle).
But if the front is made of discrete particles, then
it has a finite support, and the exponential shape 
may not extend to infinity to the right, since 
$u(t,x)$ has to be either larger than $1/N$, or zero.
It cannot take values that would be a fraction of $1/N$ in realizations,
and thus, we cannot accommodate the shape $e^{-\gamma_c(x-X(t))}$
for arbitrarily large values of $x$, since
it would mean authorizing arbitrarily small positive
values of $u(t,x)$.
From Eq.~(\ref{eq:asymptotictime}) and from the shape of
the asymptotic front~(\ref{eq:asymptoticdecay}), 
the exponential shape sets down to $u=1/N$ at time
\begin{equation}
t_\text{relax}=\frac{c}{2\omega^{\prime\prime}(\gamma_c)}\left(\frac{\ln N}{\gamma_c}\right)^2.
\label{eq:relaxtime}
\end{equation}
Beyond, the front cannot develop any longer, and thus, its shape and
velocity remain fixed.
$t_\text{relax}$ is the time that is needed 
for the front to relax from any perturbation,
which is why we have put the subscript ``relax''.

From Eq.~(\ref{eq:velocitygeneral0}) evaluated at $t=t_\text{relax}$, we get 
the new asymptotic velocity, 
which takes into account the effects of discreteness,
in the form
\be
\frac{dX(t)}{dt}=\frac{\omega(\gamma_c)}{\gamma_c}
-\frac{3}{c}\frac{\gamma_c\omega^{\prime\prime}(\gamma_c)}{\ln^2 N}.
\ee
The calculation of the constant 
$c$ requires a proper account of the exact shape of the front.
We shall turn to this calculation now.

As announced, we are now going to solve the 
linear branching 
diffusion equation with two absorptive boundaries: one representing
saturation, the other one discreteness.
\begin{figure}
\begin{center}
\includegraphics[width=0.7\textwidth]{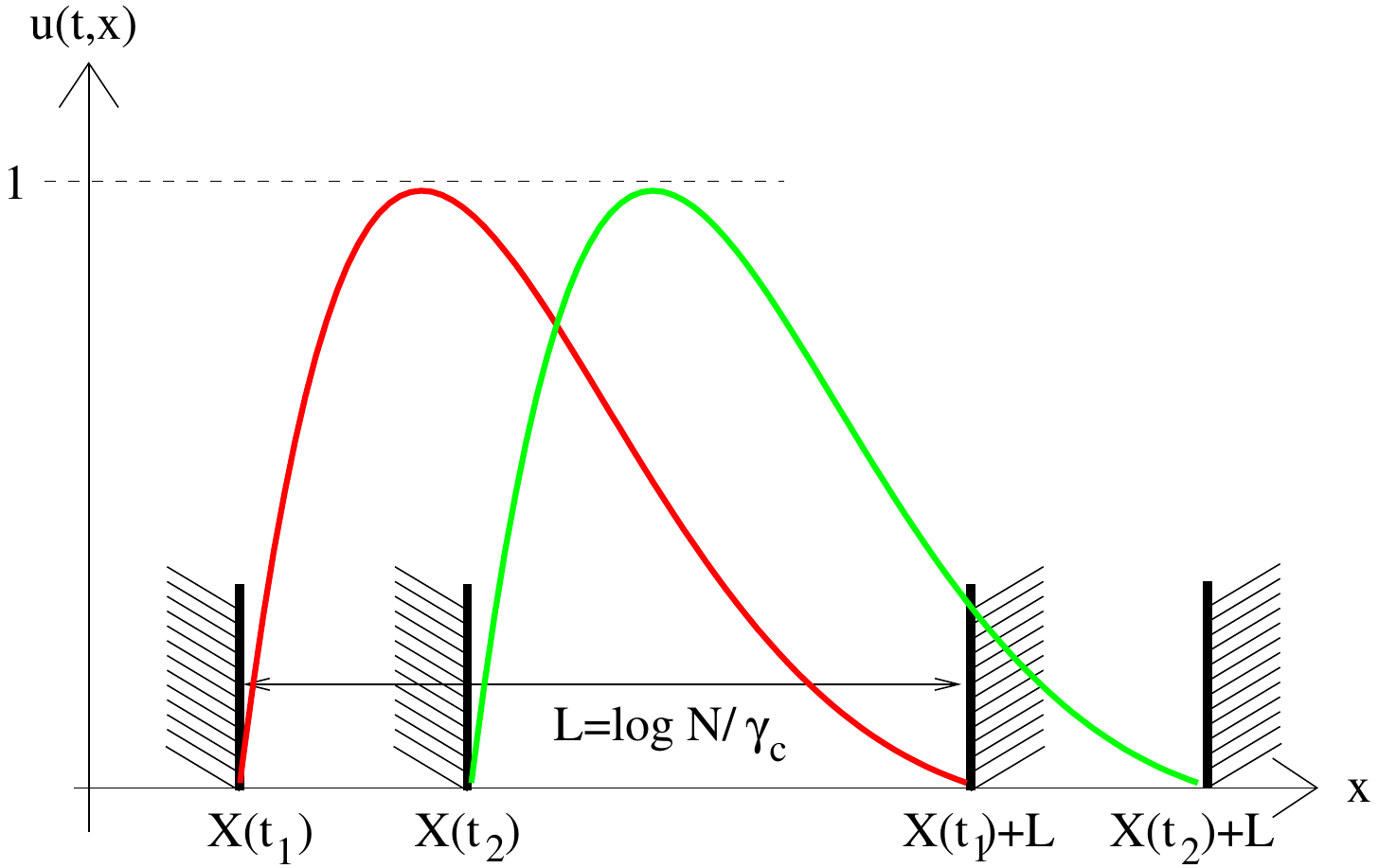}
\end{center}
\caption{\label{fig:cutoff3}
Sketch of the
solution to the branching diffusion equation with two
boundaries.}
\end{figure}
Using the intuition gained from the study of the
deterministic FKPP equation,
we write the ansatz
\be
u(t,x)=e^{-\gamma_c(x-X(t))}
L\psi\left(\frac{2\omega^{\prime\prime}(\gamma_c)t}{L^2},
\frac{x-X(t)}{L}\right).
\label{eq:ansatz}
\ee
$L$ is a constant which will represent 
the size of the front, which is essentially equal to
$L_0=\ln N/\gamma_c$
for large $N$.
When $\omega$ is expanded to second order around
the eigenvalue $\gamma_c$,
then $\psi$ obeys the partial differential equation
\be
\partial_y\psi=\frac{1}{4}\partial_\rho^2\psi
+\frac{\gamma_c L^2}{2\omega^{\prime\prime}(\gamma_c)}
(\omega^\prime(\gamma_c)-X^\prime(t))\psi,
\label{eq:diffusion0}
\ee
where we have defined
\be
y=\frac{2\omega^{\prime\prime}(\gamma_c)t}{L^2}\ \ 
\text{and}\ \ 
\rho=\frac{x-X(t)}{L}.
\label{eq:changevar}
\ee
We have only kept the dominant terms
for large $L$.
We see that $\omega^\prime(\gamma_c)-X^\prime(t)$ has to scale
like $1/L^2$ for all terms of this equation to be relevant,
as was already guessed heuristically.
The coefficient of $1/L^2$ must be chosen in such a way that
in the large-$y$ limit, there is a nontrivial stationary solution.
We will check that the correct ansatz is
\be
X^\prime(t)=\omega^\prime(\gamma_c)-\frac{\pi^2\omega^{\prime\prime}(\gamma_c)}
{2\gamma_c L^2}+o(1/L^2)
\label{eq:Xprime}
\ee
Equation~(\ref{eq:diffusion0}) then becomes
\be
\partial_y\psi=\frac{1}{4}\partial_\rho^2\psi
+\frac{\pi^2}{4}
\psi,
\label{eq:diffusion}
\ee
up to higher-order terms when $L$ is large.

We now implement the 
absorptive boundaries at
$\rho=0$ and another one at $\rho=1$ (which corresponds to a distance
$L$ between the boundaries in $x$-coordinates, i.e. to the natural
size of the stationary front).
The boundary conditions formally read
\be
\psi(y,\rho=0)=0\ \ \mbox{and}\ \ \psi(y,\rho=1)=0.
\label{eq:bc}
\ee

As for the initial condition,
for reasons that will become clear later,
put a localized mass close to the rightmost boundary, namely we write
\be
\psi(y=0,\rho)=\delta(\rho-1+\bar a)\frac{e^{\gamma_c\delta}}{L^2},
\label{eq:ic}
\ee
where $\bar a$ is a constant of order $1/L$, and therefore $\bar a\ll 1$.
The value of the
weight $e^{\gamma_c\delta}$ actually corresponds to putting one single
particle
at a distance $\delta$ to the {\it right} of the rightmost
boundary. As we will see, such a weight corresponds
to a fluctuation added to a stationary solution.
But in this section, we shall only focus on the large-time behavior:
The initial condition will be forgotten through the time evolution.

The solution of Eq.~(\ref{eq:diffusion}) with the conditions~(\ref{eq:bc})
and~(\ref{eq:ic}) reads
\be
\psi_\delta(y,\rho)=
\frac{2e^{\gamma_c\delta}}{L^2}
\sum_{n=1}^\infty
(-1)^{n+1}
\sin\pi n \bar a\,
\sin\pi n \rho\,
e^{-\frac{\pi^2 (n^2-1) y}{4}}.
\label{eq:psi}
\ee
While the full solution with all harmonics
will be of interest later, we shall discuss here
only the stationary solution.
We see that for large $y$, 
the higher harmonics are suppressed
exponentially with respect to the fundamental mode $n=1$,
which gives the following contribution:
\be
\psi_{\delta_0}(y,\rho)=
\frac{2e^{\gamma_c\delta_0}}{L^2}
\sin \pi \bar a
\sin \pi \rho
\underset{\bar a\ll 1}{\simeq}
\frac{2\pi\bar a e^{\gamma_c\delta_0}}{L^2}\sin\pi\rho.
\label{eq:psi0}
\ee
Thanks to the choice~(\ref{eq:Xprime}) for $X'(t)$,
this solution has no $y$ dependence, and leads
to a stationary $u$ in the frame of the front.
The expression~(\ref{eq:psi0}) is independent of the initial
condition except for the overall normalization.
The value of $\delta_0$,
which characterizes the initial condition,
will be adjusted later.
Undoing the changes of variables which trade
$u$ for $\psi$, $x$ for $\rho$ and $t$ for $y$
(Eq.~(\ref{eq:ansatz})),
the stationary solution $u_{\delta_0}$ reads
\be
u_{\delta_0}(t,x)=
e^{-\gamma_c(x-X(t))}\frac{2\pi \bar a e^{\gamma_c\delta_0}}{L^2}
\left[
L\sin\frac{\pi(x-X(t))}{L}
\right].
\label{eq:u0bis}
\ee
We further require that $u_0(t,x)\sim 1$ for $x=X(t)+aL$,
where $aL$ is a constant of order~1.
This condition is satisfied if we set 
$\delta_0\sim 3\ln L/\gamma_c$.
Indeed, with this choice,
\be
u_{\delta_0}(t,X(t)+aL)\simeq 2\pi^2\bar a a L^2 e^{-\gamma_c a L}.
\label{eq:pointnorm}
\ee
Since $\bar a L$ and $a L$ are constants,
the right-hand side is just a number of order 1.

All in all, 
the final solution reads
\be
{
u(t,x)\propto\kappa\, e^{-\gamma_c(x-X(t))}L\sin\frac{\pi(x-X(t))}{L}
}
\ee
(see Fig.~\ref{fig:cutoff3})
where the size of the front is
\be
L=\frac{\ln N}{\gamma_c}
\ee
and its velocity reads, from Eq.~(\ref{eq:Xprime}),
\begin{equation}
{
v_{\text{BD}}\equiv\frac{dX(t)}{dt}=
V_{\infty}
-\frac{\pi^2\omega^{\prime\prime}(\gamma_c)}
{2\gamma_c L^2}
=\frac{\omega(\gamma_c)}{\gamma_c}
-\frac{\pi^2\gamma_c\omega^{\prime\prime}(\gamma_c)}{2\ln^2 N}
}.
\label{eq:velocityBD}
\end{equation}
The subscript BD stands for ``Brunet-Derrida'' after the
first authors who wrote down such an expression.
In the FKPP case, namely for $\omega(\gamma)=\gamma^2+1$, $\gamma_c=1$
and $\omega(\gamma_c)=\omega^{\prime\prime}(\gamma_c)=2$.

\section{\label{sec:beyond}
Beyond the deterministic equations: Effect of the fluctuations}

So far, we have actually solved deterministic equations although we were
addressing a model with a discrete number of particles, that therefore
had necessarily fluctuations. Our procedure gave the leading effects.
We shall now incorporate more fluctuation effects,
in a phenomenological way.
(We shall essentially review Ref.~\cite{Brunet:2005bz}).

\subsection{Phenomenological model and analytical results}

The two-boundary procedure has led to the following result: 
The front propagates at a velocity $v_\text{BD}$ in Eq.~(\ref{eq:velocityBD})
lower than the velocity predicted by the mean-field equation~(\ref{eq:velocitygeneral0}), 
and its shape is the decreasing exponential $e^{-\gamma_c(x-X(t))}$
down to the position 
\be
x_\text{tip}(t)=v_\text{BD}t
+\frac{\ln N}{\gamma_c},
\ee
at which it
is sharply cut off by an absorptive boundary.
This boundary was meant to make the front vanish
typically
over one unit in $x$, hence to implement discreteness
on a deterministic equation.
\begin{figure}
\begin{center}
\includegraphics[width=0.7\textwidth]{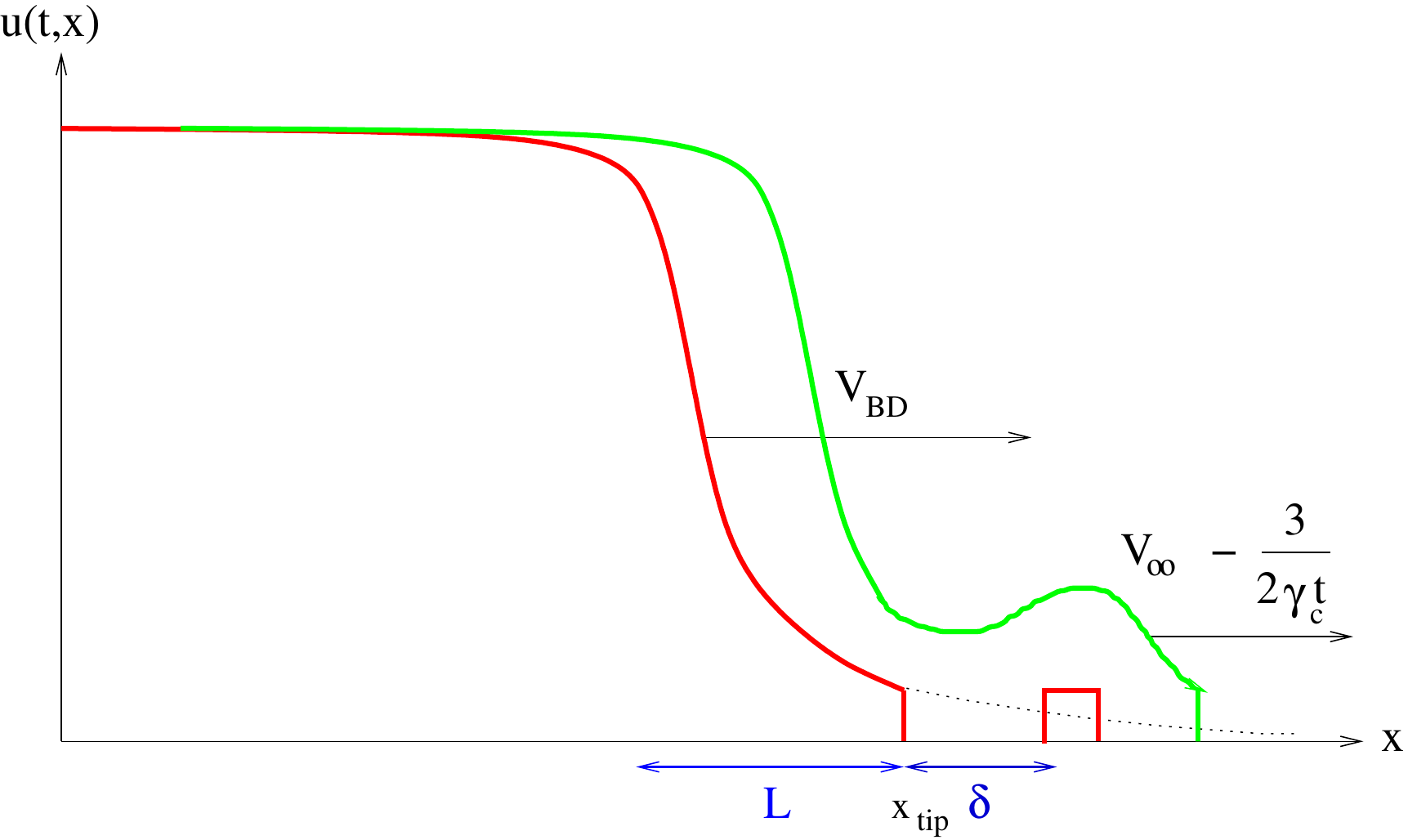}
\end{center}
\caption{\label{fig:fluct}
Evolution of the front 
with a forward fluctuation.
At time $t_0$, the primary front extends over a size 
$L$ and is a solution of the branching diffusion equation
with two appropriate boundaries.
An extra particle is stochastically generated at
a distance $\delta$ with respect to the tip of the
primary front.
At a later time, the latter grows deterministically into
a secondary front that is a bit slower, and
that will add up to the primary one.
The overall effect, after relaxation, is a shift to
the right of the distance $R(\delta)$
with respect to the position of the front if a
fluctuation had not occured.
}
\end{figure}

But since the evolution is not deterministic,
it may happen that a few extra 
particles are sent stochastically
ahead of the tip of the front (See Fig.~\ref{fig:fluct}).
Their evolution would pull the front forward.
To model this effect, we assume that the probability 
per unit time
that there be a particle
sent at a distance $\delta$ ahead of the tip simply 
continues the asymptotic
shape of the front, that is to say, the distribution of $\delta$ is
\begin{equation}
p(\delta)=C_1 e^{-\gamma_c\delta},
\label{proba}
\end{equation}
where $C_1$ is a constant.
Heuristic arguments to support this assumption were presented
in Ref.~\cite{Brunet:2005bz}.
Note that while the exponential shape is quite natural since it is the continuation
of the deterministic solution~(\ref{eq:frontgeneral0}) in the linear regime, 
the fact that $C_1$ need to be strictly
constant (and cannot be a slowly varying function of $\delta$)
is a priori more difficult to argue.

Once 
a particle has been produced at position $x_\text{tip}+\delta$, 
say at time $t_0$,
it starts to multiply (see Fig.~\ref{fig:fluct}) and
it eventually develops its own front (after a time 
$t_\text{relax}$
of the order of $L^2$), 
that will add up to
the deterministic primary front made of
the evolution of the bulk of the particles.

Note that the philosophy of our phenomenological approach to
the treatment of the fluctuations is identical to the spirit of the 
statistical
approach in Sec.~\ref{sec:statisticalmethods}
developped for the zero-dimensional model. Whenever the number of particles
is larger than $\bar n$ ($\bar n=1$ here), we apply a deterministic
nonlinear evolution. Fluctuations instead are produced with a probability
which stems from a linear equation.
The difficulty here is that we are
unable to solve either of these equations
for arbitrary initial conditions and thus we have to make
conjectures on the form of their solutions.

Let us estimate the shift in the position of the front
induced by these extra forward particles.
The solution
of the diffusion equation
is the superposition of the large-time stationary solution
$u_{\delta_0}$ given by Eq.~(\ref{eq:u0bis})
with~$\delta_0=3\ln L/\gamma_c$,
and of the solution $u_\delta$ of the diffusion equation
with the generic initial condition
characterized by $\delta$
(see Eq.~(\ref{eq:psi})), up to a multiplicative
constant $C_2$ of order 1 that we do not control
in this calculation, since it certainly 
depends on the detailed shape
of the fluctuations. 
We write
\be
u(t,x)=u_{\delta_0}(t,x)+C_2 u_\delta(t,x)
=e^{-\gamma_c(x-X(t))} L \left[\psi_{\delta_0}(y,\rho)+C_2 \psi_\delta(y,\rho)
\right]
\label{eq:u}
\ee
up to the replacement of the variables
by their expressions~(\ref{eq:changevar}).
\begin{figure}
\begin{center}
\includegraphics[width=0.8\textwidth]{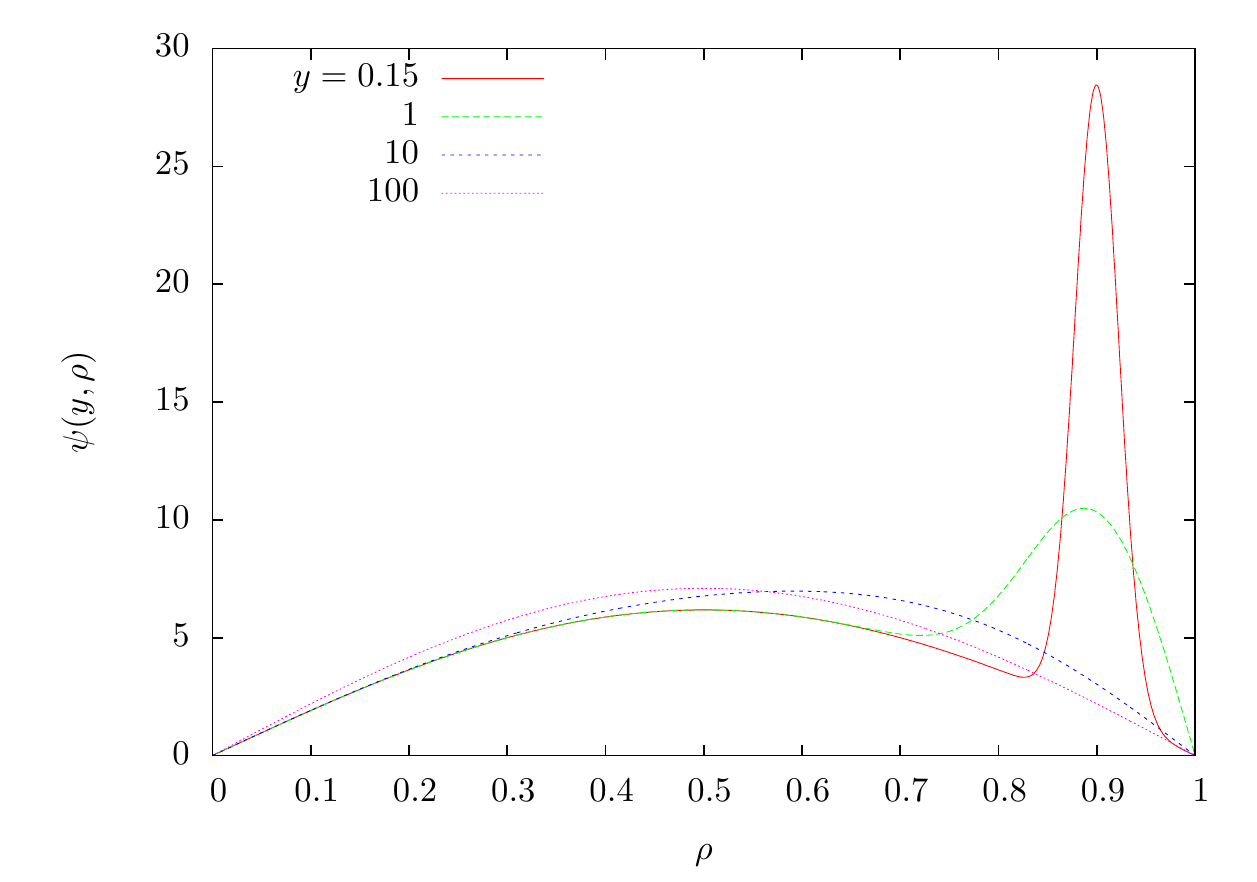}
\end{center}
\caption{\label{fig:psi}$\psi_{\delta_0}(y,\rho)+\psi_\delta(y,\rho)$
[see Eqs.~(\ref{eq:psi0}), (\ref{eq:psi})]
for different values of 
the reduced time variable $y$
after a fluctuation of size $\delta=5$ has occurred at $y=0$.
In this plot,
the size of the front is $L=10$, and $\bar a=0.1$.
We see how the fluctuation, initially localized at the tip of the front,
gets smeared uniformly over the width of the front as $y$ gets large.
Eventually, a small forward shift $X\rightarrow X+R$ 
would be needed in order to
absorb it and recover the stationary front.
}
\end{figure}
The presence of the
second term alters the shape of the front
(the front eventually relaxes back to the sine shape
in Eq.~(\ref{eq:u0bis})),
see Fig.~\ref{fig:psi}. %
But of course,
we want to keep the normalization condition
for $u$, namely for some appropriate value of $x$,
$u$ is required to equate Eq.~(\ref{eq:pointnorm})
at all $t$. This is possible by shifting
the value of $x$ at which we 
enforce the normalization condition 
from $x=X(t)+aL$ to say $x=X(t)+aL+R(t,\delta)$.
This is equivalent to shifting the
position of the front
$X(t)\rightarrow X(t)+R(t,\delta)$. 
Equation~(\ref{eq:u}) then leads to
\be
u(t,X(t)+aL+R(t,\delta))
=2\pi^2 \bar a a L^2
e^{-\gamma_c a L-\gamma_c R(t,\delta)} \left[1+
C_2\frac{\psi_\delta(y,a)}{2\pi^2 \bar a a\, L}
\right].
\label{eq:ufin}
\ee
Equating the right-hand sides of
Eq.~(\ref{eq:ufin}) and Eq.~(\ref{eq:pointnorm}),
we get
\be
R(t,\delta)=\frac{1}{\gamma_c}
\ln
\left[
1+{C_2} 
\frac{\psi_\delta\left(\frac{2\omega^{\prime\prime}(\gamma_c)t}{L^2},
a\right)}
{2\pi^2\bar a a\, L}
\right],
\label{eq:R}
\ee
where only the lowest orders in $\bar a$, $a$ 
in the expansion of $\psi_\delta$
must be kept.
With the help of Eq.~(\ref{eq:psi}), it is then
straightforward to arrive at an explicit 
expression of $R$.

In the large time limit in which only the
fundamental mode survives in the expression of $\psi$,
we get the shift 
\begin{equation}
R(\delta)=\frac{1}{\gamma_c}\ln\left(1+C_2\frac{e^{\gamma_c \delta}}{L^3}\right).
\label{R}
\end{equation}
The probability distribution~(\ref{proba}) 
and the front shift~(\ref{R}) 
due to a forward fluctuation
define
an effective theory for the evolution of the position of the front $X(t)$:
\begin{equation}
X(t+dt)=\begin{cases}
X(t)+v_\text{BD} dt & \text{proba.}
\ \ 1-dt \int_0^\infty d\delta p(\delta)\\
X(t)+v_\text{BD} dt+R(\delta)& \text{proba.}
\ \ p(\delta)d\delta dt.
\end{cases}
\label{eq:effectivetheory}
\end{equation} 
From these rules,
we may compute all cumulants of $X(t)$, by writing the evolution
of their generating function, deduced from the effective theory~(\ref{eq:effectivetheory}):
\be
\frac{\partial}{\partial t}\ln\left\langle
e^{\lambda X(t)}\right\rangle=
\lambda v_\text{BD}+\int d\delta\, p(\delta)
\left(e^{\lambda R(\delta)}-1\right).
\ee
The left hand-side is a power series in $\lambda$
whose coefficients are
the time derivatives of the cumulants of $X(t)$.
Identifying the powers of $\lambda$ in the left and 
right-hand sides, we get
\begin{equation}
\begin{split}
V-v_\text{BD}&=\int d\delta p(\delta)R(\delta)
 =\frac{C_1C_2}{\gamma_c}\frac{3\ln L}{\gamma_c L^3}\\
\frac{[\text{$n$-th cumulant}]}{t}&=\int d\delta p(\delta)[R(\delta)]^n
 =\frac{C_1C_2}{\gamma_c}\frac{n!\zeta(n)}{\gamma_c^n L^3}.
\end{split}
\label{eq:cumulants}
\end{equation}
We see that the statistics of the position of the front still depend 
on the product $C_1 C_2$ of
the undetermined constants $C_1$ and $C_2$. We
need a further assumption to fix
its value.

We go back to the expression for the correction to
the mean-field front velocity, given in
Eq.~(\ref{eq:velocityBD}).
From the expressions of $R(\delta)$ (Eq.~(\ref{R})) and of
$p(\delta)$ (Eq.~(\ref{proba})), we see that
the integrand defining $V-v_\text{BD}$ in Eq.~(\ref{eq:cumulants})
is almost a constant function of $\delta$ for 
$\delta<\delta_0=3\ln L/\gamma_c$, and is decaying exponentially for
$\delta>\delta_0$. Furthermore, $R(\delta_0)$ is
of order 1, which means that when a fluctuation is sent out at
a distance $\delta\sim\delta_0$ ahead of the tip of the front,
it evolves into a front that matches in position the
deterministic primary front.
We also notice that when a fluctuation has $\delta<\delta_0$,
its evolution is completely linear until it is incorporated
to the primary front, whereas fluctuations with $\delta>\delta_0$
evolve nonlinearly but at the same time have a very 
suppressed probability.
We are thus led to the natural conjecture that the average 
front velocity is given by $v_\text{BD}$ in
Eq.~(\ref{eq:velocityBD}), with the replacement 
\be
L\rightarrow L_\text{eff}=\frac{\ln N}{\gamma_c}+\delta_0
=\frac{\ln N}{\gamma_c}+3\frac{\ln\ln N}{\gamma_c},
\ee
namely
\be
V=\frac{\omega(\gamma_c)}{\gamma_c}-\frac{\pi^2\omega^{\prime\prime}(\gamma_c)}
{2\gamma_c\left(
\frac{\ln N}{\gamma_c}+\frac{3\ln\ln N}{\gamma_c}
\right)^2}.
\ee
The large-$N$ expansion of the new expression of the velocity
yields a correction  of the order of
$\ln\ln N/\ln^3 N$
to the Brunet-Derrida result,
more precisely
\be
{
V=\frac{\omega(\gamma_c)}{\gamma_c}
-\frac{\pi^2\gamma_c\omega^{\prime\prime}(\gamma_c)}{2\ln^2 N}
+\pi^2\gamma_c\omega^{\prime\prime}(\gamma_c)
\frac{3\ln\ln N}{\ln^3 N}.
}
\label{eq:vcorr}
\ee
Eqs.~(\ref{eq:cumulants}) and~(\ref{eq:vcorr}) match
for the choice 
\be
C_1 C_2=\pi^2\omega^{\prime\prime}(\gamma_c).
\label{eq:C1C2}
\ee 
From this determination of $C_1 C_2$, 
we also get the full expression of the
cumulants of the position of the front:
\begin{equation}
{
\frac{[\text{$n$-th cumulant}]}{t}=
\pi^2\gamma_c^2\omega^{\prime\prime}(\gamma_c)
\frac{n!\zeta(n)}{\gamma_c^n\ln^3 N}.
}
\label{eq:cumcorr}
\end{equation}
We note that all cumulants are of order 
unity for $t\sim \ln^3 N$, which 
is the sign that the distribution of the front position 
is far from being a trivial 
Gaussian. This makes it particularly interesting.
On the other hand, the cumulants 
are proportional to $\kappa=t/\ln^3 N$, 
which is the sign that
the position of the front is the result of the sum of $\kappa$ 
independent random variables,
and as such, becomes Gaussian when $\kappa$ is very large.
The properties of the statistics of the front position
were investigated in some more details in Ref.~\cite{Marquet:2006xm}.

Thanks to our discussion in Chap.~\ref{sec:schannel}, we see that
these results should apply to QCD with the relevant substitution
of the kernel $\omega$ and of the parameter $N$ 
according to Tab.~\ref{tab:dictionary}.

\subsection{Numerical simulations}

The results obtained so far rely on a number of 
conjectures that no-one has been able to prove
so far. 
In order to check our results,
let us consider again the model
introduced in Sec.~\ref{sec:reactiondiffusionexample}.
The first step to take before being able to apply our results 
to this particular model
is to extract from the linear part of Eq.~(\ref{mod_mf})
the corresponding function $\omega(\gamma)$, and then to compute $\gamma_c$.
Setting $\Delta x=\Delta t=1$, we get
\begin{equation}
\omega(\gamma)=\ln\left[1+\lambda+p_l(e^{-\gamma}-1)+p_r(e^{\gamma}-1)\right],
\label{eq:omegaEGBM}
\end{equation}
and $\gamma_c$ is defined by $\omega(\gamma_c)=\gamma_c\omega^\prime(\gamma_c)$.

\begin{figure}
\begin{center}
\includegraphics[width=0.8\textwidth]{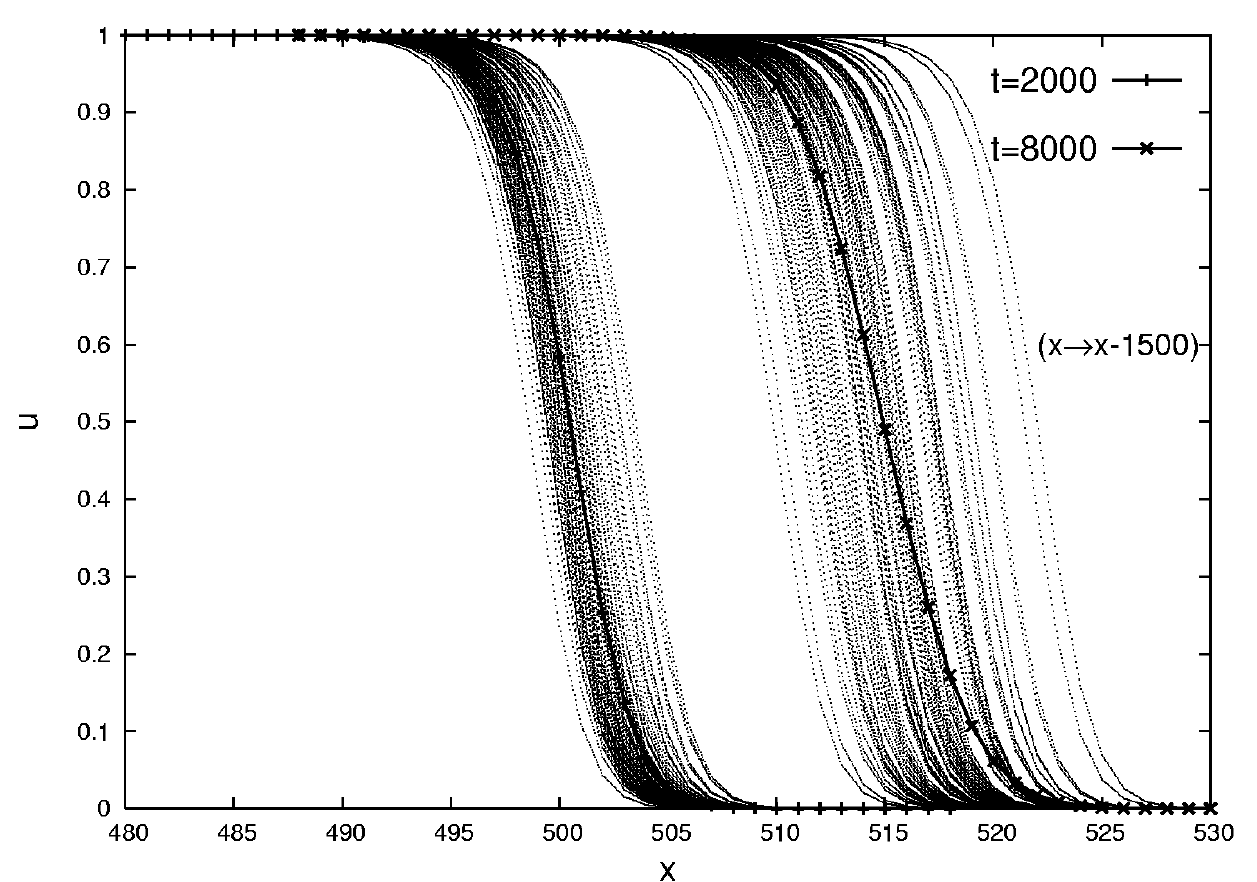}
\end{center}
\caption{\label{fig:diffscaling}
1000 realizations of the model introduced in Sec.~\ref{sec:reactiondiffusionexample}
at two different times (dotted lines), and the average of $u$ over the realizations
(full line). One clearly sees that 
$\langle u\rangle$ does not keep its shape upon time evolution,
which shows that the traveling wave property of the FKPP equation
is lost due to the stochasticity.
This point is addressed in some detail in Chap.~\ref{sec:application}.
}
\end{figure}

For the purpose of our numerical study, we set
\begin{equation}
p_l=p_r=0.1\text{\quad and\quad}\lambda=0.2\,.
\end{equation}
Simulated realizations for this set of parameters are shown 
in Fig.~\ref{fig:diffscaling}.

From (\ref{eq:omegaEGBM}), this choice leads to
\begin{equation}
\begin{split}
&\gamma_c=1.352\cdots\ ,\ \ \omega^\prime(\gamma_c)=0.2553\cdots,\\
&\omega^{\prime\prime}(\gamma_c)=0.2267\cdots.
\end{split}
\end{equation}
Predictions for all cumulants 
of the position of the front 
are obtained
by replacing the values of these parameters in 
Eqs.~(\ref{eq:vcorr}),(\ref{eq:cumcorr}).

Technically,
in order to be able to go to very large values of $N$, we replace the 
full stochastic model by its deterministic
mean field approximation $u\rightarrow \langle u\rangle$, where 
the evolution of $\langle u\rangle$
is given by Eq.~(\ref{mod_mf}), in all bins in which the number of
particles is larger than $10^3$ (that is, in the bulk of the front).
Whenever the number of particles is smaller, we use the full 
stochastic evolution~(\ref{stocha}).
We add an appropriate boundary condition 
on the interface between the
bins described by the deterministic equation and the bins described by
the stochastic equation so that the flux of particles is
conserved~\cite{moro-2004-70}.
This version of the model will be called ``model~I''.
Eventually, we shall use the 
mean field approximation
everywhere except in the rightmost bin (model~II): 
at each time step, a new bin is filled immediately
on the right of the rightmost nonempty site with a number 
of particles given by a Poisson law
of average
$\theta=N\langle u(x,t\!+\!1)|\{u(x,t)\}\rangle.$
We checked numerically that
this last approximation gives indistinguishable results from
those obtained within model~I as far as the statistics of the position of the front
is concerned.

We define the position of the front at time $t$ by
\begin{equation}
X_t=\sum_{x=0}^\infty u(x,t).
\end{equation}
We start at time $t=0$ from the initial condition $u(x,0)=1$ for $x\leq 0$ and
$u(x,0)=0$ for $x>0$. We evolve it up to time $t=\ln^2 N$ to get rid
of subasymptotic effects related to the building of the asymptotic shape of the 
front, and we measure the mean velocity between times 
$\ln^2 N$ and $16\times \ln^2 N$. 
For model I (many stochastic bins),  we average the results over $10^4$ such realizations.
For model II (only one stochastic bin), we generate  $10^5$ such realizations
for $N\leq 10^{50}$ and $10^4$ realizations for $N>10^{50}$.
In all our simulations, models~I and~II give
numerically indistinguishable results for the values of $N$ where both
models were simulated, as can be seen on the figures (results for model~I are
represented by a circle and for model~II by a cross).

\begin{figure}
\begin{center}
\includegraphics[width=0.7\textwidth]{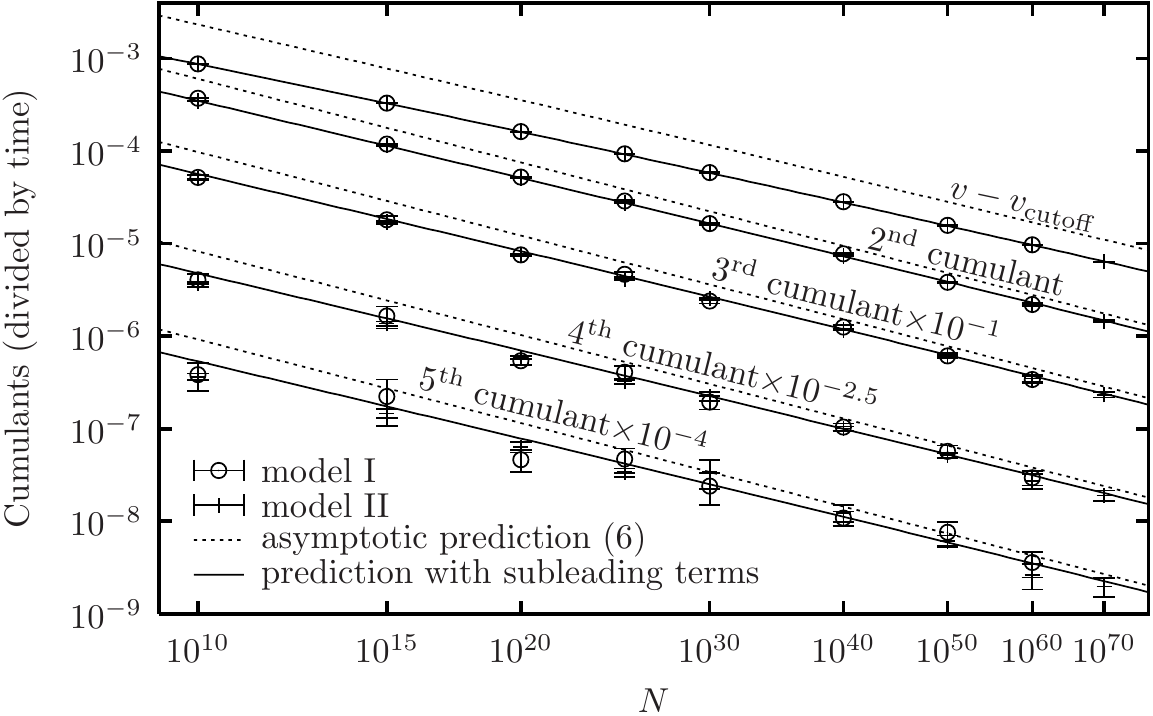}
\end{center}
\caption{\label{cumulants}
[From Ref.~\cite{Brunet:2005bz}]
From top to bottom, the correction to the
velocity given by the cutoff theory 
and the cumulants of orders 2 to 5 of the position
of the front in the stochastic model. The numerical data
are compared to our parameter-free 
analytical predictions~(\ref{eq:vcorr}),(\ref{eq:cumcorr}), 
represented by the dashed line.
}
\end{figure}

Our numerical data for the cumulants 
is shown in Fig.~\ref{cumulants} together with
the analytical
predictions obtained from~(\ref{eq:vcorr}),(\ref{eq:cumcorr})
(dotted lines in the figure).
We see that the numerical simulations get very close to the analytical 
predictions at large $N$.
However, higher-order corrections are presumably still important for
the lowest values of $N$ displayed in the figure.

We try to account for these corrections by replacing
the factor $(\ln N)/\gamma_c=L_0$ in the denominator of the expression for the
cumulants in Eqs.~(\ref{eq:vcorr}),(\ref{eq:cumcorr})
by the ansatz 
\begin{equation}
L_\text{eff}
=L_0+\frac{3 \ln(\ln N)}{\gamma_c}+c+d \frac{\ln(\ln N)}{\ln N}\ .
\label{ansatzLeff}
\end{equation}
The two first terms in the r.h.s. are suggested by our model.
We have
added two subleading terms which go beyond our theory:
a constant term, and a term that vanishes at large $N$.
The latter are naturally expected to be among the next terms in the
asymptotic expansion for large $N$. We include them
in this numerical analysis
because in the range of $N$ in which we are able to perform our numerical
simulations, they may still bring a significant contribution.

We fit (\ref{ansatzLeff})
to the numerical data obtained in the framework of
model~II, restricting ourselves to values of $N$ larger than $10^{30}$. 
In the fit, each data point is weighted by the statistical 
dispersion of its value in our sample of data.
We obtain a determination of the values of the free parameters
$c=-4.26\pm 0.01$ and $d=5.12\pm 0.27$, with a good quality of the fit 
($\chi^2/d.o.f\sim 1.15$).

Now we see that with this modification of the 
expression of the size of the front, 
the results for the cumulants shown in figure~\ref{cumulants} (full lines)
are in excellent
agreement with the numerical data over the whole range of~$N$.


\chapter{\label{sec:spatial}
Spatial correlations}

{\it
So far, all models and calculations aimed at describing QCD scattering amplitudes
assumed uniformity in impact-parameter space, or,
decoupling of the evolution between different points in
the transverse space.
Indeed, we considered one-dimensional models
while to fully describe the impact parameter, two dimensions
are necessary.
We shall address here the issue of the
correlations of the QCD evolution
between different impact parameters.
}\\

\minitoc

\section{\label{sec:relevance}
Relevance of one-dimensional models}

So far, we have argued that high-energy scattering in QCD 
at fixed coupling and fixed impact parameter
is in the universality class of the stochastic FKPP equation 
(Chap.~\ref{sec:schannel}), which is
an equation with one evolution variable (time or rapidity in QCD), 
and one spatial dimension ($x$ generically, or $\ln k^2\sim \ln(1/r^2)$ in QCD).
From the very beginning, we have simply discarded the impact parameter dependence.
It is important to understand that the spatial variable and the impact
parameter play different roles, and thus, the impact parameter may a priori
not be accounted for by a two-dimensional extension of the FKPP equation.

There are general arguments to support the assumption that the
QCD evolution is local enough
for the different impact parameters to decouple through the
rapidity evolution, which we are now going to present.

Let us start with a single dipole at rest, and bring it gradually to a higher
rapidity. As was explained in Chap.~\ref{sec:schannel},
during this process, this dipole may be replaced by
two new dipoles, which themselves may split, and so on, eventually producing
a chain of dipoles.
Figure~\ref{fig:scheme} pictures one realization of such a chain.

According to the splitting rate given in
Eq.~(\ref{splitting}), splittings to smaller-size dipoles
are favored, and thus, one expects that the sizes of the dipoles
get smaller on the average,
and that in turn, the successive splittings become more local.
The dipoles around region ``1'' and those around region ``2'' should
have an independent evolution beyond the stage pictured 
in Fig.~\ref{fig:scheme}: 
Further splittings will not mix in impact parameter space, and thus,
the traveling waves around these regions should be uncorrelated.
For a dipole in region 1 of size $r$ to migrate to region 2, it should
first split into a dipole whose size is of the order of the
distance $\Delta b$ between regions 1 and 2, up to
some multiplicative factor of order $1$. 
(We assume in this discussion 
that the dipoles in region 2 relevant to the propagation 
of the local traveling waves, that is, those
which are in the bulk of the wave front, also have sizes of order $r$).
Roughly speaking, the rate of such splittings may be estimated from
the dipole splitting probability~(\ref{splitting}): 
it is of order $\bar\alpha (r^2/(\Delta b)^2)^2$,
while the rate of splittings of the same dipole into a dipole of similar
size in region 1 is of order $\bar\alpha$. Thus the first process is strongly suppressed
as soon as regions 1 and 2 are more distant than a few units of $r$.
Note that for $\Delta b \gtrsim 1/Q_s$, saturation may further reduce
the emission of the first, large, dipole leading to an even stronger
suppression of the estimated rate.

What could also happen is that some larger dipole has, by chance, one of
its endpoints tuned to the vicinity of the coordinate
one is looking at (at a distance which is at most $|\Delta r|\ll 1/Q_s(y)$),
and easily produces a large number of dipoles there.
In this case, the position of the traveling wave at that impact
parameter would suddenly jump.
If such events were frequent enough, then they would
modify the average wave velocity and thus the one-dimensional
sFKPP picture. We may give a rough estimate of the rate at which
dipoles of size smaller than $\Delta r$ are produced.
Assuming local uniformity for the distribution $n$ of the emitting
dipoles, the rate (per unit of $\bar\alpha y$) of such events can be
written
\begin{equation}
  \int_{r_0 > \Delta r} \frac{d^2r_0}{r_0^2}
  \int_{\varepsilon < \Delta r} d^2\varepsilon\:
  n(r_0) \left(\frac{\varepsilon}{r_0}\right)^2
  \frac{1}{2\pi}\frac{r_0^2}{\varepsilon^2 (r_0-\varepsilon)^2},
\end{equation}
where we integrate over large dipoles of size $r_0 > \Delta r$
emitting smaller dipoles (of size $\varepsilon < \Delta r$) with a
probability $d^2\varepsilon\, r_0^2/(2\pi\varepsilon^2 (r_0-\varepsilon)^2)$. The factor
$(\varepsilon/r_0)^2$ accounts for the fact that one endpoint of the
dipole of size $r_0$ has to be in a given region of size $\varepsilon$ in
order to emit the dipoles at the right impact parameter.
To estimate this expression, we first use $n(r_0)=T(r_0)/\alpha_s^2$
and approximate $T$ by
\be
T(r_0) = \theta(r_0 - 1/Q_s)\, + \,(r_0^2Q_s^2)^{\gamma_c}\, \theta(1/Q_s - r_0).
\label{eq:approxt}
\ee
The front is replaced by 1 above the saturation scale
(for $r_0>1/Q_s$) and by an exponentially decaying tail
for $r_0 < 1/Q_s$.
Using $r_0-\varepsilon \approx r_0$ in the emission kernel, the
integration is then easily performed and
one finds a rate whose dominant term is
\begin{equation}
\frac{\pi}{2\alpha_s^2}\frac{((\Delta r)^2Q_s^2)^{\gamma_c}}{1-\gamma_c}.
\end{equation}
For $(\Delta r)^2\ll(\alpha_s^2)^{1/\gamma_c}/Q_s^2$, i.e.
ahead of the bulk of the front, this term is
parametrically less than 1 and is in fact of the order of the probability
to find an object in this region that contributes
to the normal evolution of the front \cite{Brunet:2005bz}.
Hence there is no extra contribution due to the fact that there
are many dipoles around at different impact parameters.

The arguments given here are based on estimates 
of average numbers of dipoles, 
on typical configurations,
and we are not able to account analytically for the
possible fluctuations. As we have seen through this review, the latter 
often play an important role. As a matter of fact, in the physics of
disordered systems, rare events sometimes dominate.
So before studying the phenomenological consequences of 
the statistical picture of high-energy QCD based on a one-dimensional equation,
one should check more precisely the locality of the evolution
in impact parameter.

A numerical check was achieved in the case of 
a toy model that has an impact-parameter dependence
in Ref.~\cite{Munier:2008cg}.
Let us briefly describe the model, before presenting the
main numerical results.

\subsection{A model incorporating an impact-parameter dependence}

In order to arrive at a model that is tractable numerically, we only keep
one transverse dimension
instead of two in 3+1-dimensional QCD. 
However, we cannot consider genuine 2+1-dimensional QCD
because we
do not wish to give up the
logarithmic collinear singularities at $x_2=x_0$ and $x_2=x_1$.
Moreover, QCD with one dimension less has very different properties at high
energies \cite{Ivanov:1998we}.
Starting from Eq.~(\ref{splitting}), 
a splitting rate which complies with our requirements is:
\begin{equation}
\frac{dP}{d(\bar\alpha y)}=\frac{1}{4}\frac{|x_{01}|}{|x_{02}| |x_{12}|}dx_2.
\label{eq6:split0}
\end{equation}
We can further simplify this probability distribution
by keeping only its collinear and infrared asymptotics 
(as in Ref.~\cite{Ciafaloni:1999yw}).
If $|x_{02}|\ll |x_{01}|$ (or the symmetrical case $|x_{12}|\ll |x_{01}|$), 
the probability reduces to $dx_2/|x_{02}|$ ($dx_2/|x_{12}|$ resp.). 
The result of the splitting is a small dipole $(x_0,x_2)$ together with 
one close in size to the parent. So for simplicity we will just add the
small dipole to the system and leave the parent unchanged.
In the infrared region, a dipole of size $|x_{02}|\gg |x_{01}|$ 
is emitted with
a rate given by the large-$|x_{02}|$ limit of the above probability.
The probability laws~(\ref{splitting}),(\ref{eq6:split0}) 
imply that a second dipole of
similar size should be produced while the parent dipole
disappears.
To retain a behavior as close as possible to that in the collinear
limit, we will instead just generate a single large dipole and
keep the parent. To do this consistently one must include a factor
of two in the infrared splitting rate, so as not to modify the average
rate of production of large dipoles.

Let us focus first on the distribution of the 
sizes of the participating dipoles.
(The simplifying assumptions made above enable one to 
choose the sizes and the impact parameters of the dipoles successively).
We call $r$ the modulus of the emitted dipole, $r_0$ the modulus of its parent
and we define $Y=\bar\alpha y$.
The splitting rate~(\ref{eq6:split0}) 
reads, in this simplified model
\begin{equation}
\frac{dP_{r_0\to r}}{dY}
 = \theta(r-r_0) \frac{r_0 dr}{r^2} + \theta(r_0-r) \frac{dr}{r},
\end{equation}
and the original parent dipole is kept.
Logarithmic variables are the relevant ones here, so we introduce
\begin{equation}
\rho = \thelog_2(1/r) \qquad \text{or} \qquad r=2^{-\rho}.
\end{equation}
We can thus rewrite the dipole creation rate as
\be
\frac{dP_{\rho_0\to \rho}}{dY}
 = \theta(\rho_0-\rho)\, 2^{\rho-\rho_0}\, \thelog 2\, d\rho
 + \theta(\rho-\rho_0)\, \thelog 2\, d\rho.
\ee
To further simplify the model,
we discretise the dipole sizes in such a way that $\rho$
is now an integer.
This amounts to restricting the dipole sizes to negative
integer powers of $2$.
The probability that a dipole at lattice site $i$ ({\em i.e.} a dipole
of size $2^{-i}$) creates a new
dipole at lattice site $j$ is
\be
\frac{dP_{i\to j}}{dY}
 =\int_{\rho_j}^{\rho_{j+1}} \frac{dP_{\rho_i\to\rho}}{dY}
 = \begin{cases}
\thelog 2 & j\ge i \\
2^{j-i} & j<i
\end{cases}.
\label{splitf}
\ee
The rates $dP_{i\pm}/dY$ for a dipole at lattice site $i$ to split to any
lattice site $j\ge i$ or $j<i$ respectively are then given by
\be
\frac{dP_{i+}}{dY}  =  \sum_{j=i}^{\imax-1} \frac{dP_{i\to j}}{dY}
 = \thelog 2 (\imax-i),\ \
\frac{dP_{i-}}{dY}  =  \sum_{j=0}^{i-1} \frac{dP_{i\to j}}{dY}
 = 1-2^{-i},
\label{lifetimef}
\ee
where we have restricted the lattice to $0\le i <\imax$, for obvious
reasons related to the numerical implementation.

Now we have to address the question of the impact parameter of the emitted dipole.
In QCD, the collinear dipoles are produced near the endpoints of the 
parent dipoles. Let us take a parent of size $r_0$ at impact parameter $b_0$.
We set
the emitted dipole (size $r$) at the impact parameter $b$ such that
\begin{equation}
b=b_0\pm \frac{r_0\pm r\times s}{2}
\label{bf}
\end{equation}
where $s$ has uniform probability between 0 and 1. It is introduced to
obtain a continuous distribution of the impact parameter unaffected by
the discretisation of $r$.
This prescription is quite arbitrary in its details, but the latter
do not influence significantly the physical observables.
Each of the two signs that appear in the above expression
is chosen to be either $+$ or $-$ with equal weights.
We apply the same prescription 
when the emitted dipole is larger than its parent.\\

Now that we have introduced a branching process similar
to QCD dipole evolution, we must define the
scattering amplitude.
We have explained above (see Sec.~\ref{sec:partonmodel}) 
that in QCD, the scattering amplitude 
of an elementary probe dipole
of size $r_i=2^{-i}$
with a dipole in an evolved Fock state
is proportional to the number of objects which have a size of the same
order of magnitude and which sit in a region of size of order $r_i$ around the
impact point of the probe dipole.
Since in our case, the sizes are discrete, the amplitude is just given,
up to a factor, by the
number of dipoles that are exactly in the same bin of
size as the probe, namely
\be
T(i,b_0)=\alpha_s^2
\times\# \{\text{dipoles of size $2^{-i}$ %
at impact parameter $b$ satisfying }|b-b_0|<r_i/2\}.
\label{Tf}
\ee

Finally, we must enforce unitarity, that is, the condition
\begin{equation}
T(i,b)\leq 1
\label{eq6:unitarity}
\end{equation}
for any $i$ and $b$.
This condition is expected to hold due to gluon saturation in QCD.
However,
saturation is not included in the
original dipole model.
The simplest choice is to veto splittings that would locally 
drive the amplitude
to values larger than 1.
In practice, for each splitting that gives birth to a new dipole
of size $i$ at impact parameter $b$, we compute
$T(i,b)$ and $T(i,b\pm r_i/2)$, and throw away the
produced dipole whenever
one of these numbers gets larger than one.

Given the definition of the amplitude $T$, this saturation rule implies
that there is a maximum number of objects in each bin of size
and at each impact parameter, which is equal to $N_\text{sat}=1/\alpha_s^2$.

\subsection{Numerical evaluation of the correlators}

We have implemented this model numerically. Let us discuss
how we operated this implementation and the results we obtained.

We take as an initial condition a number $N_\text{sat}$ 
of dipoles of size 1 ($i=0$),
uniformly distributed in impact parameter between
$-r_0/2$ and $r_0/2$.
The impact parameters $b_j$ that are considered are respectively
$0$, $10^{-6}$, $10^{-4}$, $10^{-2}$ and $10^{-1}$.
The number of events generated is typically $10^4$,
which allows one to measure the mean and variance of the position 
of the traveling waves to a sufficient accuracy.

We have checked that at each impact parameter, we get
traveling waves whose positions grow linearly with rapidity
at a velocity less than the expected mean-field velocity for
this model. $N_\text{sat}$ was varied from 10 to 200.

\begin{figure}
\begin{center}
\includegraphics[width=0.7\textwidth]{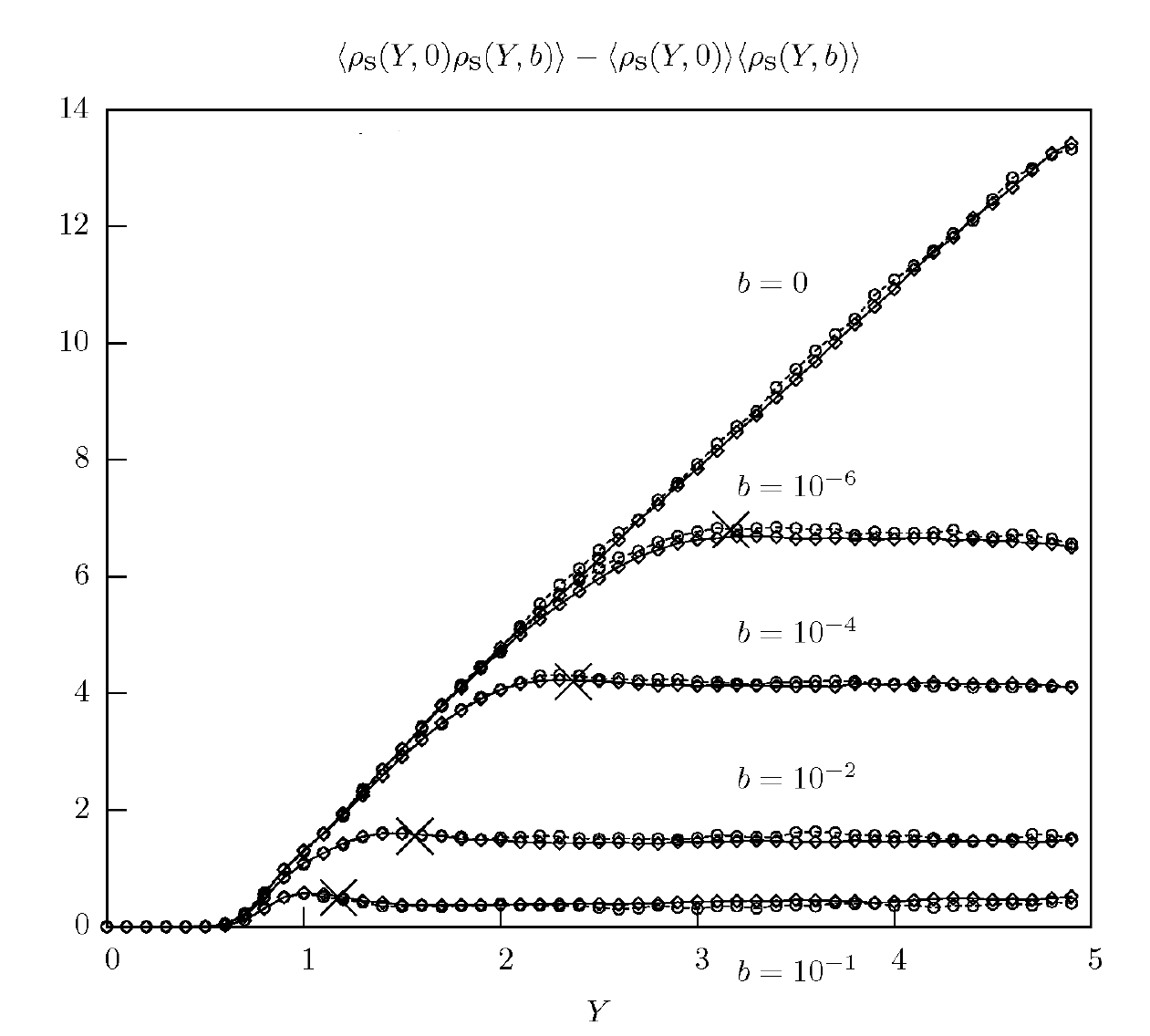}
\end{center}
\caption{\label{fig:plot_cor}
Correlations of the positions of the traveling wave fronts between
different impact parameters in the toy model of Sec.~\ref{sec:relevance}. 
The points where the correlations flatten
correspond to the decoupling of the waves in the corresponding regions
of impact parameter.
}
\end{figure}

Figure~\ref{fig:plot_cor} represents the correlations between the
positions of the wave fronts at different impact parameters, defined
as
\begin{equation}
\langle \rho_\text{s}(Y,b_1)\rho_\text{s}(Y,b_2)\rangle-
\langle \rho_\text{s}(Y,b_1)\rangle\langle\rho_\text{s}(Y,b_2)\rangle.
\end{equation}
We set $N_\text{sat}$ to 25 in that figure, 
but we also repeated the analysis for different values of $N_\text{sat}$
between 10 and 200.

We see very clearly 
the successive decouplings of the different
impact parameters in Fig.~\ref{fig:plot_cor}, from the most distant to the closest one,
as rapidity increases. Indeed, the correlation functions flatten after some
given rapidity depending on the difference in the probed impact parameters, which
means that the evolutions decouple.
This decoupling is expected as soon as the traveling wave front
reaches dipole sizes which are smaller than the distance
between the probed impact parameters, {\em i.e.} at $Y$ such that
$|b_2-b_1|\approx 1/Q_s(Y) = 2^{-\rho_\text{s}(Y)}$.
(We shall further comment on this decoupling in the next section).
From the data for $\rho_\text{s}(Y)$, we can estimate quantitatively
the values of the rapidities at which the traveling waves decouple between the
different impact parameters. (It is enough to invert the above formula for the
relevant values of $b_2-b_1$).
These rapidities are denoted by a cross in Fig.~\ref{fig:plot_cor} for the
considered impact parameter differences.
Our numerical results for the correlations are nicely consistent 
with this estimate, since the correlations start 
to saturate to a constant value precisely on the right of each such cross.

We conclude that the different impact parameters indeed decouple,
as was expected from a naive analytical estimate.
What is true for our toy model should go over
to full QCD, since we have included the main
features of QCD.
When looking at the data more carefully however,
it turns out that the model with impact parameter
does not reduce exactly to a supposedly equivalent 
one-dimensional model of the sFKPP type.
This is a point that would deserve more work.
We refer the reader to Ref.~\cite{Munier:2008cg}
for all details of our numerical investigations.

However, even if one takes the statistical
decoupling of impact parameters as soon as
$\Delta b>1/Q_s(Y)$ for granted, there may still be some
effective correlations
persisting at large rapidities since two points in impact-parameter
space share some common history. 
Indeed, fluctuations need some rapidity to affect saturation
scales, and history may be remembered way after the
rapidity at which the decoupling happened.
Such effects are negligible at small rapidities,
as was just shown numerically, but may be crucial
at large rapidities.
We shall now investigate this point.


\section{Computing the correlations}

Our method will consist in proposing 
a simple toy model which contains
the main physical features of QCD,
which may be implemented as a Monte-Carlo event generator and
for which analytical calculations will be possible.
In these respects, our approach follows the one developed
in the previous section, but while the latter work was
purely numerical, our main results will consist in
analytical expressions of the correlation of
the saturation scale between two points in impact-parameter
space, as a function of the distance between the points
and as a function of the rapidity.
The content of this section was published in 
Ref.~\cite{Mueller:2010fi,Munier:2010hv}.

The model that we will study will have the following
characteristics.
With respect to QCD, we assume
the following simplifications:
{\em (i)} Dipoles evolve by giving birth
to one dipole of half size
(the left or the right half of the parent dipole),
or to one dipole of double size (in such a way
that the parent be the left or right half of its
offspring) at some
fixed rates,
{\em (ii)} dipoles do not disappear in the evolution, 
that is to say, the parent
dipoles are not removed,
{\em (iii)} the positions and dipole sizes are discrete,
and {\em (iv)} the configuration space of the dipoles is a
line instead of the full two-dimensional space.
We thus
give up two main properties of the QCD dipole model:
The collinear singularities, which cause
the dipole endpoints to emit an arbitrary number
of dipoles of arbitrarily
small sizes, and the continuous and two-dimensional nature
of the dipole sizes and positions.
The first simplification is the diffusion approximation,
which has been studied in the context of BFKL physics
(see e.g. Ref.~\cite{CC}),
but which was not assumed in the previous model.
The second simplification was instead already assumed in there.
These model simplifications may introduce some artefacts, but
that we believe are under control, and many results
which we will obtain within such simple
models are likely
to apply to QCD since
they will not depend on the details.

\begin{figure}
\begin{center}
\includegraphics[width=0.6\textwidth]{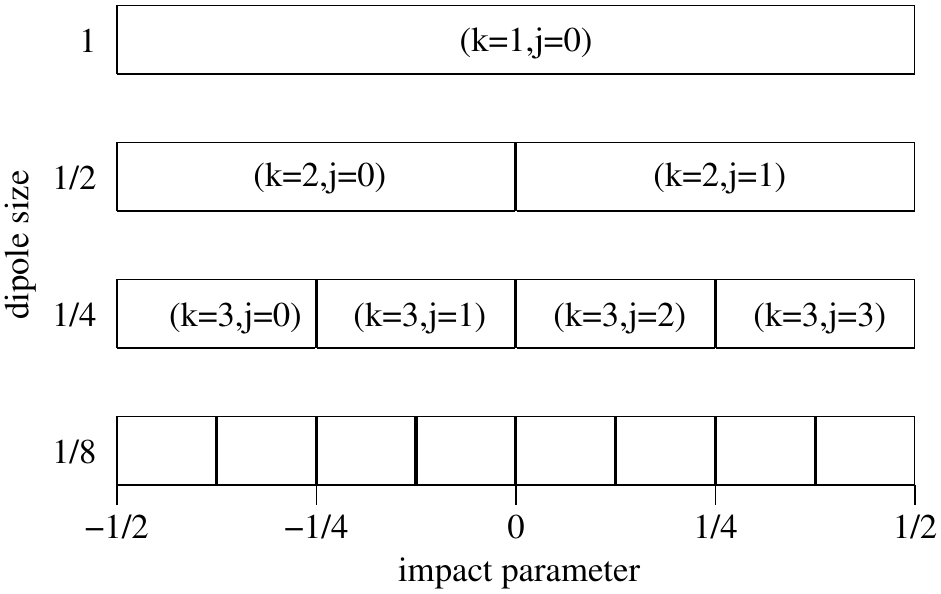}
\end{center}
\caption{\label{fig:sketch0}The hierarchical 
structure of the model.
Each box represents a bin which may contain 
up to $N$ dipoles of given sizes (vertical
axis) and positions in impact-parameter space (horizontal
axis). The conventional
numbering of the bins 
that we have chosen
is also shown for $k=1,2,3$.
}
\end{figure}

Let us now specify completely the model.
According to the evolution rules given above,
starting from a dipole of size 1,
the sizes of all dipoles present in the system
after evolution are powers of 2.
In practice, we shall only consider fractions of 1, i.e.
the sizes may be written as $2^{1-k}$, where $k\geq 1$.
For each value of $k$, there are $2^{k-1}$ possible
values of the position $b$ of the center of the dipoles: 
$b=-\frac12+2^{-k},-\frac12+3\times 2^{-k},\cdots,
\frac12-3\times 2^{-k},\frac12-2^{-k}$.
Let us number these bins
by the index $0\leq j \leq 2^{k-1}-1$ running
from the negative to the positive positions.
The model may be represented as 
a hierarchy of bins that contain
a discrete number of dipoles, see Fig.~\ref{fig:sketch0}.
Note that to any given impact parameter $b$ between 
$-\frac12$ and $\frac12$
corresponds one unique bin at each level of
size.
For example, at position $b=-\frac12$, one sees the bins
$(k=1,j=0)$, $(k=2,j=0)$, $(k=3,j=0)$ etc...
At position $-0.2$, one sees the bins
$(k=1,j=0)$, $(k=2,j=0)$, $(k=3,j=1)$ etc...
More generally, at position $-\frac12+\Delta b$, one sees 
$(k,[\Delta b\times 2^{k-1}])$, where
the square brackets represent the integer part.

During the rapidity 
(or time)
interval $dt$, a dipole
in the bin $(k,j)$
has a probability $\alpha dt$ to give
birth to a dipole in the bin $(k+1,2j)$,
$\alpha d t$ to give a dipole in the bin 
 $(k+1,2j+1)$, and $\beta dt/2$
to give a dipole
in the bin $(k-1,j/2)$ if
$j$ is even and $(k-1,(j-1)/2)$ if $j$ is odd.
Note that $dt$ may be infinitesimal (which is
generally speaking convenient
for analytical calculations), but
also finite (which is convenient
for numerical simulations).

As for the saturation mechanism, we
assume the simplest one: We veto splittings to
bins which already host the number $N$ of dipoles.

We can consider that the number density of ``gluons'' of
a given size
seen at one impact parameter
is proportional to the number of dipoles in the corresponding
bin $(k,j)$.
As rapidity is increased, the occupation of the
bins with low values of $k$ gets higher until the
number of objects they contain reaches $N$.
The subsequent filling of the bins indexed by larger
values of $k$ (smaller dipole sizes)
can be seen as the propagation of
traveling wave fronts at each impact parameter, 
with possibly complicated
relationships between them.
The (logarithm of the) saturation scale $X(b,t)$
at impact parameter $b$
is related to the position of the front
seen there at time $t$.
There are several equivalent ways to define
the position of the front.
It could be, for example,
the largest value of $k$ for which the number of
objects becomes some given
fraction of $N$. (Later, we will use a slightly different
definition).


\subsection{Basic features of the model}

Let us denote by $n_{(k,j)}(t)$ the number of dipoles
present in the bin $(k,j)$ at time $t$.
Then, according to the rules given above, we can write 
the following stochastic evolution equation:
\be
n_{(k,j)}(t+dt)=\min\bigg[N,n_{(k,j)}(t)
+\delta^\alpha_{(k-1,[j/2])}(t)
+\delta^{\beta/2}_{(k+1,2j)}(t)+\delta^{\beta/2}_{(k+1,2j+1)}(t)
\bigg],
\label{eq:defmodel1}
\ee
where the $\delta^x_{(k,j)}$ are drawn according to the binomial
distribution
\be
\mbox{Proba}\left[\delta^x_{(k,j)}(t)=l\right]=
\begin{pmatrix}{n_{(k,j)}(t)}\\{l}\end{pmatrix}
(x dt)^l(1-x dt)^{n_{(k,j)}(t)-l}.
\label{eq:defmodel2}
\ee
This is a rather complicated equation which we do not
know how to solve except numerically. 

This model does not a priori look
like a stochastic FKPP model.
We may assume uniformity in impact parameter: This would
amount to imposing the same $\delta^\alpha$ and $\delta^{\beta/2}$
respectively
for all $j$ at any given $k$.
In this case, the model
would be projected to the FKPP class, but
by definition, this would wash out the
fluctuations between the different impact parameters.
This simplified model, 
that we call ``FIP'' (for ``Fixed Impact Parameter'',
since effectively, the model is completely
defined by a single impact parameter) 
is nevertheless
useful since it provides a benchmark
to evaluate how the fluctuations
between different impact parameters
may alter the FKPP picture.
In this paper, we will rely on 
(and check again in the case of our model)
the conclusion 
reached in Ref.~\cite{Munier:2008cg}
that thanks to saturation,
locally at each impact parameter,
the full model is still well-described by
a one-dimensional FKPP equation,
and the fluctuations 
between different positions in impact-parameter space
do not qualitatively change the picture.

Let us first apply the treatment of FKPP
equations exposed in Chap.~\ref{sec:reviewtraveling} 
to the FIP case. 
We know that the large-rapidity realizations
of the model are stochastic traveling waves, whose
main features can be determined from a simple analysis
of the linear part of the evolution equation.
In this model, only the number of dipoles $n_k$
in the bins
say $(k,0)$ (i.e. at impact parameter $-\frac12$) is relevant.
The evolution equation reads\footnote{%
We could also write $2\delta^{\beta/2}_{k+1}(t)$
instead of the last term in Eq.~(\ref{eq:evolutionFIP}).
(This may even be a more literal implementation
of the FIP approximation).
But this would not make a large difference,
which anyway, we would be unable to capture analytically.
}
\be
n_k(t+dt)=\min
\bigg[
N,n_k(t)+\delta^\alpha_{k-1}(t)+\delta^\beta_{k+1}(t)
\bigg].
\label{eq:evolutionFIP}
\ee
The mean-field (or Balitsky-Kovchegov) 
approximation to the evolution leads to the equation
\be
n_k(t+dt)=\min
\bigg[
N,n_k(t)+\alpha dt\, n_{k-1}(t)
+\beta dt\, n_{k+1}(t)
\bigg],
\ee
where the $n_k$ are now real functions of $k$.
The linearized equation (equivalent to the BFKL equation) 
is simply obtained by
discarding the ``$\min$'' in the previous equation:
\be
n_k(t+dt)=
n_k(t)+\alpha dt\, n_{k-1}(t)+\beta dt\, n_{k+1}(t).
\label{eq:BFKLn}
\ee
From standard arguments, 
we know that for asymptotically large $t$ and $N$, the
velocity of the wave front, that is the time derivative
of the position $X(t)$ of the front, is given by
\be
v_0=\frac{dX}{dt}=\omega^\prime(\gamma_c),
\label{eq:v0}
\ee 
where $\omega(\gamma)$ is the
eigenvalue of the kernel of the linearized evolution 
equation~(\ref{eq:BFKLn}) corresponding to the 
eigenfunction~$e^{-\gamma k}$, namely
\be
\omega(\gamma)=\frac{1}{dt}\ln
\left(1+
\alpha dt\,e^\gamma+\beta dt\,e^{-\gamma}
\right),
\label{eq:eigenvalue}
\ee
and $\gamma_c$ minimizes $\omega(\gamma)/\gamma$.
We recall that $dt$ may be finite or infinitesimal,
in which case Eq.~(\ref{eq:eigenvalue}) is to be understood
as the derivative of $\ln(1+\cdot)$ at the origin.


Our aim 
is to study the correlations between the point
at position $b=-\frac12$ in transverse space
(left edge of the system, 
see Fig.~\ref{fig:sketch0})
and the one at position $b=-\frac12+\Delta b$ with
$0\leq \Delta b<1$.
We calculate the
average of the squared difference of the
positions of the front
between these
points, which is formally related to the
two-point correlation function
of (the logarithm of) the saturation scales,
and which we deem a good
estimator of the spatial fluctuations of the saturation
scale. In the hierarchical model,
all bins with index $k$ less than 
or equal to
$k_{\Delta b}\equiv 1+[-\log_2\Delta b]$ 
(the notation ``$[\cdots]$''
stands for the integer part) and $j=0$
overlap both impact parameters, and thus the
dipoles of size larger than $2^{-k_{\Delta b}}$
seen at these points are exactly the same.
For $k>k_{\Delta b}$ instead, the bins seen
at the two points are distinct and nonoverlaping.
So in particular,
in our model with $\beta=0$, as soon as
the position of the front
at one point or at the other is
larger than $k_{\Delta b}$, that is to say, as soon
as there are of the order of $N$ 
dipoles in the bin $(k_{\Delta b},j=0)$, then the evolutions are
completely uncorrelated
at the two points in the corresponding bins.
(We expect that for finite $\beta$ of order 1,
the discussion would not be qualitatively changed.)
This matches to the picture that we may infer for the
QCD dipole model: The dipoles at two
positions in impact-parameter space
separated by a distance larger than the typical saturation
scales in that region evolve (almost)
independently towards larger
rapidities.
Note that choosing pairs of points
around impact parameter 0, one with positive impact parameter
and another one with negative impact parameter,
 would not satisfy this property,
due to the rigidity of the sizes and positions of the dipoles.
Indeed, these two points would decorrelate very soon in 
the evolution since their common ancestors necessarily
sit in the bin $(k=1,j=0)$, 
see Fig.~\ref{fig:sketch0}.

As a consequence of these features of 
QCD reproduced in the toy model,
studying two-point
correlations between points in 
impact-parameter space
as a function of their distance $\Delta b$ and of the 
time (=rapidity) $t$
is equivalent to
studying the time dependence of the correlations of
the saturation scales of two realizations of the
model whose evolutions
are identical until the tip of the front
reaches $k_{\Delta b}$. On the average,
it takes a time $t_{\Delta b}=(k_{\Delta b}-1)/v$, $v$ being
the mean velocity of the individual
fronts, 
for the front whose tip is at $k_{\Delta b}=1$ at the beginning of the
evolution to have its tip at $k_{\Delta b}$.
Then the bins such that $k>k_{\Delta b}$
evolve independently between the two realizations
over the remaining 
time interval 
\be
\Delta t=t-t_{\Delta b},
\ \ \mbox{with}\ \ 
t_{\Delta b}=\frac{\left[-\log_2\Delta b\right]}{v}.
\label{eq:ttb}
\ee
Note that
this is very close to assuming that the realizations are identical
for $t\leq t_{\Delta b}$ and completely
uncorrelated for $t>t_{\Delta b}$.

From this discussion, we see that
the basic input of our calculation will
be the mechanism for the propagation of a FKPP front
which was explained in Chap.~\ref{sec:reviewtraveling}.
We will review it in the next subsection, then we will
proceed to the formulation of the calculation of
the correlations.


\subsection{Formulation of the calculation of the correlations}

In line with the above discussion, we wish to compute the
correlations of the position of two fronts whose evolutions are
identical for $t\leq t_{\Delta b}$ and uncorrelated for $t>t_{\Delta b}$.
Note that strictly speaking, we would need to
keep the content of all bins $k\leq k_{\Delta b}$ identical 
between the two realizations at all times, even
after time $t_{\Delta b}$.
But these two formulations give quantitatively similar
results.

Let us introduce $X(t_0,t)$ the position of the front at time $t$ in the
frame in which $X(t_0,t_0)=0$.
We focus on what happens slightly before the initial time $t_0$.
According to the mechanism of front propagation explained 
in Chap.~\ref{sec:reviewtraveling},
on one hand, $X(t_0-dt_0,t)=X(t_0,t)+v_\text{BD}dt_0$
if no fluctuation has occurred between times $t_0-dt_0$ and $t_0$,
on the other hand,
$X(t_0-dt_0,t)=X(t_0,t)+v_\text{BD}dt_0+R(t-t_0,\delta)$ if a 
fluctuation has
occurred at a position $\delta$ ahead of the front
(which happens with probability $p(\delta)d\delta\,dt_0$).
$v_{\text{BD}}$ was defined in Eq.~(\ref{eq:velocityBD}), $R(t,\delta)$
in Eq.~(\ref{eq:R}) and $p(\delta)$ in Eq.~(\ref{proba}).
It is straightforward to
write an equation for the generating function of the
cumulants of $X$:
\be
-\frac{d}{dt_0}\ln\left\langle e^{\lambda X(t_0,t)}\right\rangle
=\lambda v_\text{BD}+\int d\delta\,p(\delta)
\left(
e^{\lambda R(t-t_0,\delta)}-1
\right).
\ee
One now considers two such independent 
fronts and add up the generating functions. One gets
\be
-\frac{d}{dt_0}\ln\left(
\left\langle
e^{\lambda X_1(t_0,t)}
\right\rangle
\left\langle
e^{-\lambda X_2(t_0,t)}
\right\rangle
\right)
=\int d\delta\,p(\delta)
\left(e^{\lambda R(t-t_0,\delta)}+e^{-\lambda R(t-t_0,\delta)}-2
\right).
\ee
Expanding for $\lambda$ close to 0,
the coefficients of the second power of $\lambda$ 
obey the equation
\be
\frac{d}{dt}\left\langle (X_1-X_2)^2\right\rangle
=2\int d\delta \,p(\delta) R^2(t-t_0,\delta),
\label{eq:corr}
\ee
where we have used the fact that $X_1$ and $X_2$ are independent
random variables for $t>t_0$, and we have
traded $t_0$ for $t$ in the derivative, taking advantage
of the fact that both $X_1-X_2$ and $R$ only depend
on $t-t_0$.
In practice, $t_0$ will be equal to $t_{\Delta b}$, the time
at which the tip of the single front reaches $k_{\Delta b}$.
From Eq.~(\ref{eq:ttb}), this time is $[-\log_2\Delta b]/v$.

We see that the basic ingredient is 
the time evolution of the shift of
the front due to a forward fluctuation.
This shift was given in Eq.~(\ref{eq:R}).
It involved the expression for $\psi_\delta$ in Eq.~(\ref{eq:psi}).
In order to write a compact expression for $R$, it
is interesting to note that
$\psi_\delta$
is related to some Jacobi $\vartheta$ function \cite{AS}.
Since
\be
\vartheta_4(z|q)=1+2\sum_{n=1}^{\infty}(-1)^n\cos(2nz)q^{n^2},
\ee
we may rewrite Eq.~(\ref{eq:psi}) as
\be
\psi_\delta(y,\rho)=\frac{e^{\gamma_c\delta}}{2 L^2}
\frac{1}{q}
\bigg[
\vartheta_4\left(\frac{\pi(\bar a+\rho)}{2}\bigg|q\right)
-\vartheta_4\left(\frac{\pi(\bar a-\rho)}{2}\bigg|q\right)
\bigg].
\ee
The notation
\be
q\equiv e^{-\frac{\pi^2 y}{4}}
=e^{-\frac{\pi^2 \omega^{\prime\prime}(\gamma_c)t}{2L^2}}
\label{eq:q}
\ee
has been introduced.
Using Eq.~(\ref{eq:R}) and performing the appropriate expansion
for small $\bar a$ and $a$, we arrive at
an expression for $R(t,\delta)$ 
in terms of the $\vartheta_4$-function
which is particularly
compact:
\be
{
R(t,\delta)=\frac{1}{\gamma_c}\ln\left[
1-C_2\frac{e^{\gamma_c\delta}}{2 L^3}
\partial_q\vartheta_4(0|q)
\right],
}
\label{eq:Rtheta}
\ee
with
\be
-\partial_q\vartheta_4(0|q)=2\sum_{n=1}^{+\infty}
(-1)^{n+1}n^2 q^{n^2-1}.
\label{eq:theta4def1}
\ee
It is actually quite natural that the Jacobi theta functions appear,
since the latter are defined as solutions of the one-dimensional 
heat equation
with periodic boundary conditions.

We turn to the analysis of the obtained result.
First, for large $y$, 
only the fundamental mode contributes 
significantly
to $\psi_\delta$.
Looking back at Eq.~(\ref{eq:psi}), 
we see that higher
harmonics would give a series of exponentially
 decreasing corrections.
But at a finite time,
a large number of modes have to
be taken into account, typically all modes such that
$n\leq (L/\pi) \sqrt{2/\omega^{\prime\prime}(\gamma_c)t}$. 
A few low-lying modes are not enough
to describe the small-time behavior. Instead,
it is a saddle point
(in an appropriate integral reformulation)
that dominates the sum~(\ref{eq:psi}).
In this regime, it would be
useful to find a way to write the series of harmonics
such that at asymptotically large $y$, 
only the first term contributes instead of the whole
series. This is actually possible using the Poisson
summation formula
\be
\sum_{n=-\infty}^{+\infty}f(n)=\sum_{k=-\infty}^{+\infty}
\int dx\,f(x)e^{-2i\pi k x}.
\label{eq:poisson}
\ee
In order to get $R(t,\delta)$, we need the value 
of $\psi_\delta$ at $\rho=a$.
Hence we choose
\be
f(x)=-\frac{e^{\gamma_c\delta}}{L^2}
\sin\pi x \bar a
\sin\pi x a 
\,
q^{x^2-1}
e^{i\pi x}.
\ee
We then perform the integral
over $x$ in the r.h.s. of Eq.~(\ref{eq:poisson}).
Introducing $\gamma_+=\bar a+a$ and $\gamma_-=\bar a-a$,
we get the following expression for $\psi_\delta$:
\be
\psi_\delta(y,a)=\frac{e^{\gamma_c\delta}}{4 L^2}
\frac{1}{q}\sqrt{\frac{\pi}{-\ln q}}
\sum_{k=-\infty}^{+\infty}
\bigg(
e^{\frac{(2k-1+\gamma_+)^2\pi^2}{4\ln q}}
+e^{\frac{(2k-1-\gamma_+)^2\pi^2}{4\ln q}}
-e^{\frac{(2k-1+\gamma_-)^2\pi^2}{4\ln q}}
-e^{\frac{(2k-1-\gamma_-)^2\pi^2}{4\ln q}}
\bigg).
\ee
Since we eventually want to apply Eq.~(\ref{eq:R})
in order to get an expression of the shift of the front,
we expand the latter formula
for $\bar a,a\ll 1$.
The leading order reads
\be
\psi_\delta(y,a)=\frac{\sqrt{\pi}}{2}
\frac{e^{\gamma_c\delta}}{q(-\ln q)^{5/2}}
\frac{\pi^2 \bar a a}{L^2}
\sum_{k=1}^{+\infty}
\left[
\pi^2(2k-1)^2+2\ln q
\right]
e^{\frac{(2k-1)^2\pi^2}{4\ln q}}.
\ee
The shift of the
front due to a fluctuation
is obtained 
from $\psi_\delta$
with the help of Eq.~(\ref{eq:R}):
\be
R(t,\delta)=\frac{1}{\gamma_c}
\ln\bigg\{
1+C_2
\frac{\sqrt{2}L^2 e^{\gamma_c\delta}}{(\pi\omega^{\prime\prime}(\gamma_c)t)^{5/2}}
e^{\frac{\pi^2\omega^{\prime\prime}(\gamma_c)t}{2L^2}}
\sum_{k=1}^{+\infty}
\left[(2k-1)^2-\frac{\omega^{\prime\prime}(\gamma_c)t}{L^2}
\right]
e^{-\frac{(2k-1)^2L^2}{2\omega^{\prime\prime}(\gamma_c)t}}
\bigg\}.
\label{eq:Rcomplet}
\ee
$q$ is the function of $t$ given by Eq.~(\ref{eq:q}).
This formula is extremely useful, since the series
indexed by $k$ converges fast.
Even for moderately large values of $t$,
a few terms accurately describe the whole function.
This is actually the best formula for numerical
evaluations of $R$.

We shall now examine the limit of small $t$ ($y\ll 1$).
Then only the term $k=1$ has to be 
kept.
The expression for $R$ boils down to
\be
R(t,\delta)=\frac{1}{\gamma_c}
\ln\bigg(
1+C_2\frac{\sqrt{2}e^{\gamma_c\delta}L^2}
{(\pi\omega^{\prime\prime}(\gamma_c))^{5/2}}
\frac{e^{-\frac{L^2}
{2\omega^{\prime\prime}(\gamma_c)t}}}{t^{5/2}}
\bigg).
\ee

As a final remark, let us note that 
the Poisson summation~(\ref{eq:poisson})
 that we have used to rewrite the
series of harmonics corresponds to a Jacobi identity
for the $\vartheta$ functions \cite{AS}.
Equation~(\ref{eq:Rcomplet}) results from
Eq.~(\ref{eq:Rtheta}) with the replacement
\be
-\partial_q\vartheta_4(0|q)=\frac{\sqrt{\pi}}{2}
\frac{1}{q(-\ln q)^{5/2}}
\sum_{k=1}^{+\infty}\left(\pi^2(2k-1)^2+2\ln q\right)
e^{\frac{(2k-1)^2\pi^2}{4\ln q}}.
\label{eq:theta4def2}
\ee


\subsection{Analytical expression for the correlations}

With the elements presented in the previous sections,
we can write the expression for 
$\sigma_{12}^2\equiv\langle (X_1-X_2)^2\rangle$.
It is enough to insert the expression for the probability 
of fluctuations
(Eq.~(\ref{proba}))
and for the time-dependent shift (Eq.~(\ref{eq:Rtheta}))
into Eq.~(\ref{eq:corr}):
\be
\frac{d\sigma_{12}^2}{dt}=\frac{2C_1}{\gamma_c^2}
\int_0^{+\infty}d\delta\,e^{-\gamma_c\delta}
\ln^2\left[
1-C_2\frac{e^{\gamma_c\delta}}{2 L^3}
\partial_q\vartheta_4(0|q)
\right],
\label{eq:calsigma2}
\ee
where for $\partial_q\vartheta_4(0|q)$ we use 
either one of the equivalent
expressions~(\ref{eq:theta4def1}),~(\ref{eq:theta4def2})
according to the limit that we want to investigate.
We now have to fix the value of $L$.
In Ref.~\cite{Brunet:2005bz}, $L$ was taken to be a constant.
(The phenomenological
model predicted $L=L_0\equiv \ln N/\gamma_c$,
but empirically, we saw that it 
was better to add a subdominant correction,
namely $L=L_0+\frac{3}{\gamma_c}\ln L_0+\mbox{const}$.)
In this case, a change of variable can be made
in the integrand. All the parameters may be factored out,
leaving us with a simple numerical integral to perform:
\be
\int_0^{+\infty}\frac{dx}{x^2}\ln^2(1+x)=2\zeta(2)=\frac{\pi^2}{3}.
\ee
Thus
\be
\frac{d\sigma_{12}^2}{dt}=\frac{\pi^2C_1C_2}{3\gamma_c^3L^3}
\left[-\partial_q\vartheta_4(0|q)\right].
\ee
Replacing the product of the unknown constants by 
Eq.~(\ref{eq:C1C2}) and $q$ by Eq.~(\ref{eq:q}) and integrating
over the time variable between 0 and $\Delta t=t-t_{\Delta b}$, 
we arrive at a parameter-free expression for
$\sigma_{12}^2$ as a function of $\Delta t$, namely
\be
{
\sigma_{12}^2=\frac{2\pi^2}{3\gamma_c^3L}
\int_{e^{-\frac{\pi^2\omega^{\prime\prime}(\gamma_c)\Delta t}{2L^2}}}^1
\frac{dq}{q}\left[-\partial_q\vartheta_4(0|q)\right].
}
\label{eq:sigma122f}
\ee

We now investigate the two interesting limits,
i.e. $\Delta t\gg L^2$ and $\Delta t\ll L^2$.
For large $\Delta t$, 
the integral is dominated by the region
$q\rightarrow 0$, thus
$-\partial_q\vartheta_4(0|q)$
may be replaced by its value at $q=0$ ($-\partial_q\vartheta_4(0|0)=2$).
Performing the remaining integration, we get
\be
\sigma_{12}^2\underset{\Delta t\gg L^2}{\sim}
\frac{2\pi^4\omega^{\prime\prime}(\gamma_c)}
{3\gamma_c^3 L^3}\Delta t,
\ee
which is twice the second-order cumulant of the position
of the front in a fixed impact-parameter model, see Eq.~(\ref{eq:cumcorr}).
For small $\Delta t$ instead, say $L\ll \Delta t\ll L^2$, 
we use the expansion
of $\partial_q\vartheta_4(0|q)$
for $q\rightarrow 1$, i.e. the first term in Eq.~(\ref{eq:theta4def2}),
which reads
\be
\partial_q\vartheta_4(0|q)=
-\frac{\sqrt{\pi}}{2}\frac{\pi^2+2\ln q}{q(-\ln q)^{5/2}}
e^{\frac{\pi^2}{4\ln q}}.
\ee
Equation~(\ref{eq:sigma122f})
boils down to the following expression:
\be
\sigma_{12}^2\underset{\Delta t\ll L^2}{\sim}
\frac{4}{3\gamma_c^3}
\sqrt{
\frac{2\pi^3}{\omega^{\prime\prime}(\gamma_c)\Delta t}
}
\exp\left(-\frac{L^2}{2\omega^{\prime\prime}(\gamma_c)\Delta t}\right).
\label{eq:smalltasymptotics}
\ee

So far, we have chosen 
the size of the front
$L$ constant, of the order of $L_0$.
Another possible model for $L$
would be to promote it to a function of $\delta$
at the level of Eq.~(\ref{eq:calsigma2}), namely
\be
L=L_0+\delta+\mbox{const},
\ee
where the constant has to be determined empirically.
This choice takes maybe into account more accurately
the extension of the front by $\delta$ generated
by the fluctuations.
The $\delta$-integral cannot be performed
analytically in Eq.~(\ref{eq:calsigma2}) except in 
some limits, so
a priori,
there is no simpler expression than Eq.~(\ref{eq:calsigma2}). 
Thus
we need to know the values of $C_1$ and $C_2$
individually. 
We can consider that
$C_1=\gamma_c$ is the natural normalization
of the probability distribution $p(\delta)$. Then, we
must set $C_2=\pi^2\omega^{\prime\prime}(\gamma_c)/\gamma_c$
in order to satisfy Eq.~(\ref{eq:C1C2}).

The above-mentioned
two models,
in which $L$ is either constant or $\delta$-dependent,
differ by subleading terms in the large-$L$
limit.
Since the values of $\delta$ which dominate the $\delta$-integral
in Eq.~(\ref{eq:calsigma2})
are of order $\frac{3}{\gamma_c}\ln L_0$, 
like the first correction to $L_0$ in the case of constant $L$,
the models are not
expected to differ significantly.
We will check this statement numerically.


\subsubsection{Scaling}

Looking back at Eq.~(\ref{eq:sigma122f}),
 we see that $\sigma_{12}^2$ has a nice
scaling property.
Indeed, we may rewrite the latter equation as
\be
\sigma_{12}^2=\frac{D}{\gamma_c(v_0-v)}
\int_{e^{-\gamma_c(v_0-v)\Delta t}}^1\frac{dq}{q}
\left[-\partial_q\vartheta_4(0|q)\right]
\ee
in terms of the properties of a single front (its velocity $v$
and the diffusion constant $D$ whose analytical expressions
were given in Eq.~(\ref{eq:velocityBD}) and~(\ref{eq:cumcorr})), 
where $v_0$ can be read in Eq.~(\ref{eq:v0}).
In particular, we have the following scaling:
\be
{
\frac{\sigma_{12}^2}{D\Delta t}=\mbox{function}[(v_0-v)\Delta t].
}
\label{eq:scaling}
\ee
From Eq.~(\ref{eq:smalltasymptotics}), 
we see that the function in the right-hand side
is exponentially damped when
its argument is smaller than 1, i.e. parametrically
for $\Delta t\ll L^2$.

Once one knows the characteristics of the traveling waves
in the FIP model (i.e. $v$ and $D$), this
scaling of the correlations is a pure prediction.
Thus it will be interesting to check it in the
numerical calculations.


\subsubsection{Limits on the validity of the calculations}

Let us try and
evaluate the limits on the validity of our calculations.
The latter were essentially based on the assumption 
that the eigenvalue $\gamma=\gamma_c$
of the kernel $\omega$ dominates.
While this statement is clearly true 
at large times, when the traveling-wave
front is well formed,
it must break down at early
times right after a fluctuation has occurred:
Indeed, a fluctuation
has an initial 
shape that is far from the one of the asymptotic front, 
see Fig.~\ref{fig:psi}.

We wish to estimate the order of magnitude of the dispersion
of the relevant eigenvalues
about $\gamma_c$. To this aim, 
neglecting for the moment the boundary conditions
and the prefactors,
we write the solution of the general branching diffusion as
\be
u(\Delta t,k)\sim \int d\gamma\, e^{-\gamma k+\omega(\gamma)\Delta t}.
\ee
The interesting values of $k$ are the ones around the 
position of the
wave front,
therefore we write
$k=v_0 \Delta t+\delta k$, where $\delta k$ is of the order
of the size $L$ of the front.
Expanding
$\omega(\gamma)$ about $\gamma_c$, we write
\be
u(\Delta t,v_0\Delta t+\delta k)\sim 
e^{-\gamma_c\times\delta k}
\int d(\delta\gamma)\,
e^{-\delta\gamma\times\delta k +\frac12 \omega^{\prime\prime}(\gamma_c)
(\delta\gamma)^2\Delta t+\cdots},
\label{eq:uapprox}
\ee
where $\delta\gamma=\gamma-\gamma_c$.
It is clear from this equation that the relevant values
of $\delta\gamma$ are of the order of 
$\delta k/(\omega^{\prime\prime}(\gamma_c)\Delta t)$.
Since the order of magnitude of $\delta k$ is the size $L$ of
the front,
we would a priori conclude that the dispersion of $\gamma$ 
around $\gamma_c$ is small
and hence that the calculation is valid as soon as $\Delta t\gg L$.

However, we have also expanded $\omega(\gamma)$ to second order.
This means that for a generic kernel $\omega$, we have neglected terms
of the form
$\frac16\omega^{(3)}(\gamma_c)(\delta\gamma)^3\Delta t\sim L^3/(\Delta t)^2$ 
(which would fit in the
dots in Eq.~(\ref{eq:uapprox})).
The expansion is a good approximation if the latter 
term is small, i.e. if
\be
\Delta t\gg L^{3/2}.
\ee


\subsubsection{Back to impact-parameter space}

So far, we have been working with the minimal model,
consisting in two realizations of the FIP model
which evolve in the same way until their 
common tip reaches $k_{\Delta b}$,
and which decorrelate for $k>k_{\Delta b}$.
The only relevant parameter 
which determined the decorrelation of the positions 
of the fronts
of the realizations
was the time $\Delta t=t-t_{\Delta b}$ 
after the tip had reached $k_{\Delta b}$.
We now wish to discuss the transcription of the
obtained results
to impact-parameter space, which was our initial problem.

To this aim, we will of course 
make use of Eq.~(\ref{eq:ttb})
to express $t_{\Delta b}$ with the help of the mean front
velocity $v$. 
But we also need a length scale to which 
the distance in impact-parameter space
$\Delta b$
may be compared. The natural length
is the dipole size at the position 
of the front, namely
\be
l_s(t)=2^{-X(t)}=l_s(t_{\Delta b})2^{-v\Delta t}.
\label{eq:ls}
\ee
On the other hand, according to Eq.~(\ref{eq:ttb})
and disregarding the integer part operator,
$-\log_2\Delta b=k_{\Delta b}$ and the tip of the front
$k_{\Delta b}$ is ahead of the bulk $X(t_{\Delta b})$
by $L$: $k_{\Delta b}=X(t_{\Delta b})+L$. Using the previous equation,
we may now express $\Delta t$ as a function of $\Delta b$ and of
the length scale $l_s(t)$:
\be
\Delta t=\frac{1}{v}\left[L+\log_2\frac{\Delta b}{l_s(t)}\right].
\ee
The scaling~(\ref{eq:scaling})
reads
\be
\sigma_{12}^2\sim
\frac{L+\log_2\frac{\Delta b}{l_s(t)}}{L^3}\times
\mbox{function}
\left[\frac{L+\log_2\frac{\Delta b}{l_s(t)}}{L^2}\right].
\ee
This formula, together with the behavior of
the scaling function (see Eq.~(\ref{eq:smalltasymptotics})),
shows that there is little $b$-dependence
until $\log_2 (\Delta b/l_s(t))\sim L^2$, that is to say,
until $\Delta b\sim l_s(t) e^{\text{const}\times L^2}$.
In other terms, the size $\Delta b$ of the
domain around impact parameter $b$ in which the 
fluctuations in the
position of the fronts
are negligible is, in notations more familiar to QCD experts,
\be
{
\Delta b\sim \frac{e^{\text{const}\times \ln^2(1/\alpha_s^2)}}{Q_s(b)},
}
\ee
where $Q_s(b)$ is the usual saturation momentum at impact
parameter $b$.
Note that since the fronts are statistically independent
as soon as $\Delta b\times Q_s(b)>1$, this result may seem a bit surprising:
It says that the effective correlation length between different
points in impact-parameter space is much larger than $1/Q_s(b)$
in the parametrical limit of small $\alpha_s$.


\subsection{Numerical simulations}

In this section, we confront our analytical calculations
to numerical simulations of the toy model.
First, we consider the full model and
test the validity of the
assumption that the minimal model
is a good approximation to the full model
also for $\beta\sim 1$, i.e. when splittings
to larger-size dipoles are authorized.
Second,
we compare the minimal model to the analytical results
for the fluctuations between different positions in
impact-parameter space
(given essentially by
Eqs.~(\ref{eq:calsigma2}),(\ref{eq:sigma122f})).

\subsubsection{Full model}

The model defined by Eqs.~(\ref{eq:defmodel1}) and~(\ref{eq:defmodel2}) 
is straightforward to implement numerically
in the form of a Monte-Carlo event generator.
The simplest is to store
the number of dipoles in each bin
in an array whose index $i$ is related to $k$ and $j$
through $i=2^{k-1}+j$. 
The splitting dynamics relates bin $i$ to $2i$ (down left),
$2i+1$ (down right) and $[i/2]$ (up; the square brackets stand 
once again for
the integer part).

We have to deal
with an array whose size grows exponentially with time.
It is thus very difficult to pick large values of $t$,
and thus also large values of $N$. Indeed, the relevant time
scale grows with $N$ like $\ln^2 N$, and consequently
the minimum number of
entries in the array 
one wants to consider
grows like $e^{\ln^2 N}$.
In practice, we limit ourselves to $t\leq 4$ and $N\leq 100$.
As for the time step $dt$, the most convenient is to take it
small but finite. We set $dt=10^{-2}$.

\begin{figure*}
\begin{center}
\includegraphics[width=13cm]{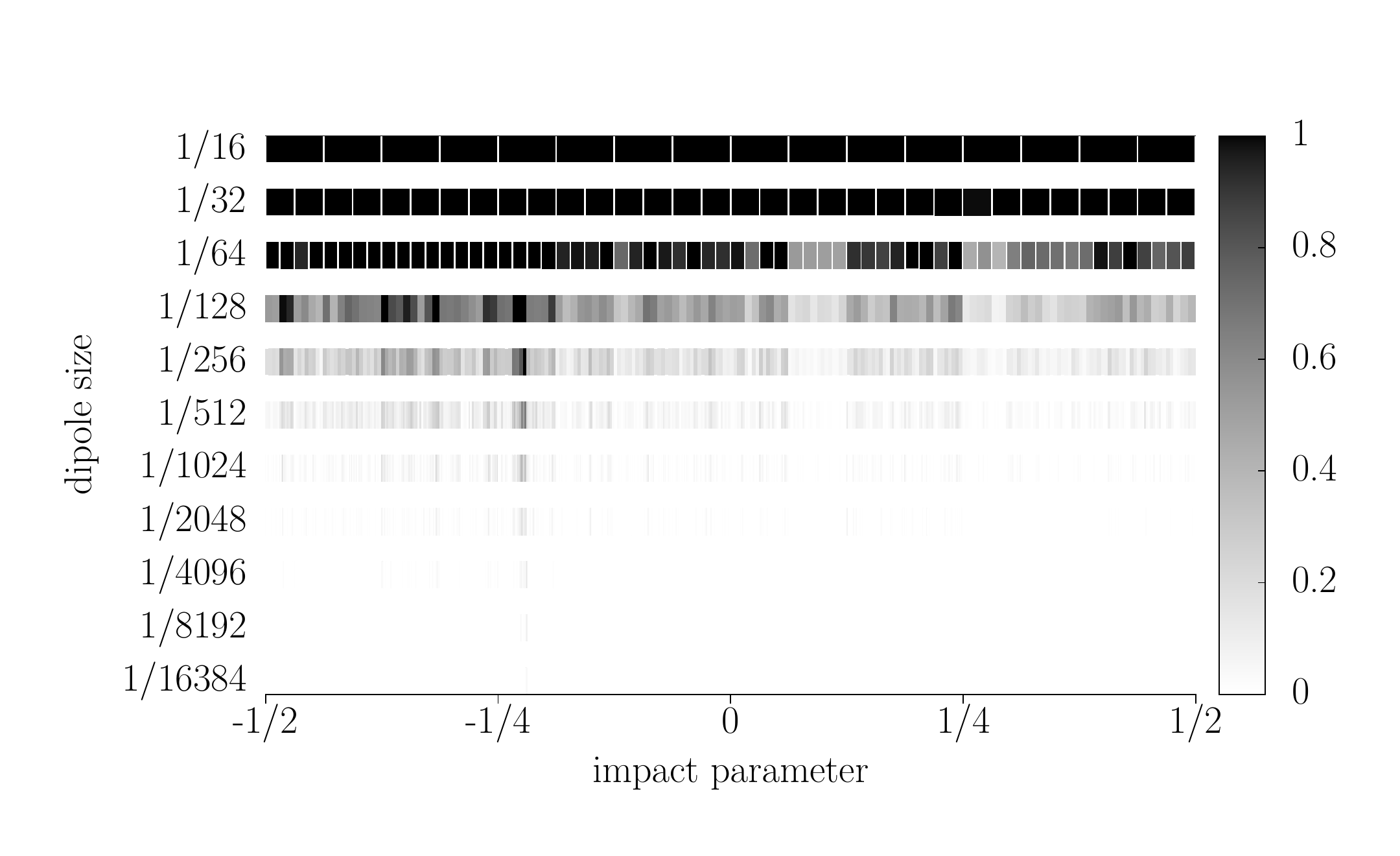}
\end{center}
\caption{\label{fig:sketch1}One event of the full model
with $\alpha=\beta=1$, $N=100$ and $t-t_{\Delta b}=4$. 
Only the bins $k\geq 5$ are
represented. (The bins for $k<5$ all contain $N$ dipoles.)
The number of dipoles in each bin is proportional to the blackness
which is displayed.
We see that in the transition region close to blackness, 
nearby bins are often of similar grey levels, which
illustrates the statement that the density of gluons varies significantly
only over scales which are larger than the relevant
length scale $l_s(t)$ (see Eq.~(\ref{eq:ls})).
}
\end{figure*}

We start with one particle and evolve it for a few hundred units of time
using the FIP version of the
model. We obtain a traveling wave
front, whose tip we eventually label $k=1$. (The complete front
sits in the bins $k\leq 1$).
From the initial condition built in this way, we evolve
all bins for which $k<1$ using the FIP model, and
all bins for
$k\geq 1$ using the full model.
One event is shown in Fig.~\ref{fig:sketch1}.
Although $N$ and $t$ are small in this calculation, 
we see that the regions 
in impact-parameter
space which have similar numbers of dipoles are larger
than the local length scale $l_s(t)$ (see Eq.~(\ref{eq:ls})).

After the evolution times $t=3$ and $t=4$ respectively, 
we measure the position of the front at various impact
parameters on a uniform tight grid
ranging from $-\frac12$ to
$+\frac12$.
We use the following definition 
of the position of the front:
\be
X(\Delta b,t)=k_0
+\sum_{k=k_0+1}^{+\infty}\frac{n_{(k,[\Delta b\times 2^{k-1}])}(t)}{N}
\ee
where $k_0$ is the largest $k$ for which $n_{(k,[\Delta b\times 2^{k-1}])}=N$.
Note that in principle, we could have chosen $X(\Delta b,t)=k_0$.
In practice however, because of the discreteness of $k$
in our model, this choice
would introduce artefacts which we do not expect in real QCD.

We compute the squared difference of the front positions
between the impact parameters $-\frac12$ and $-\frac12+\Delta b$,
and average over events.
We plot the result as a function of $t+\log_2\Delta b/v$,
where $v$ is the average front velocity measured
at impact parameter $-\frac12$.

We compare the results to the correlations
obtained in the minimal model, i.e.
when we consider two independent 
realizations of an initial front.
We do not attempt to compare to our analytical formulas
since the values of $N$ that we are able to reach
are too small for the approximations
that we had to assume to be relevant.

\begin{figure}
\begin{center}
\includegraphics[width=0.8\textwidth]{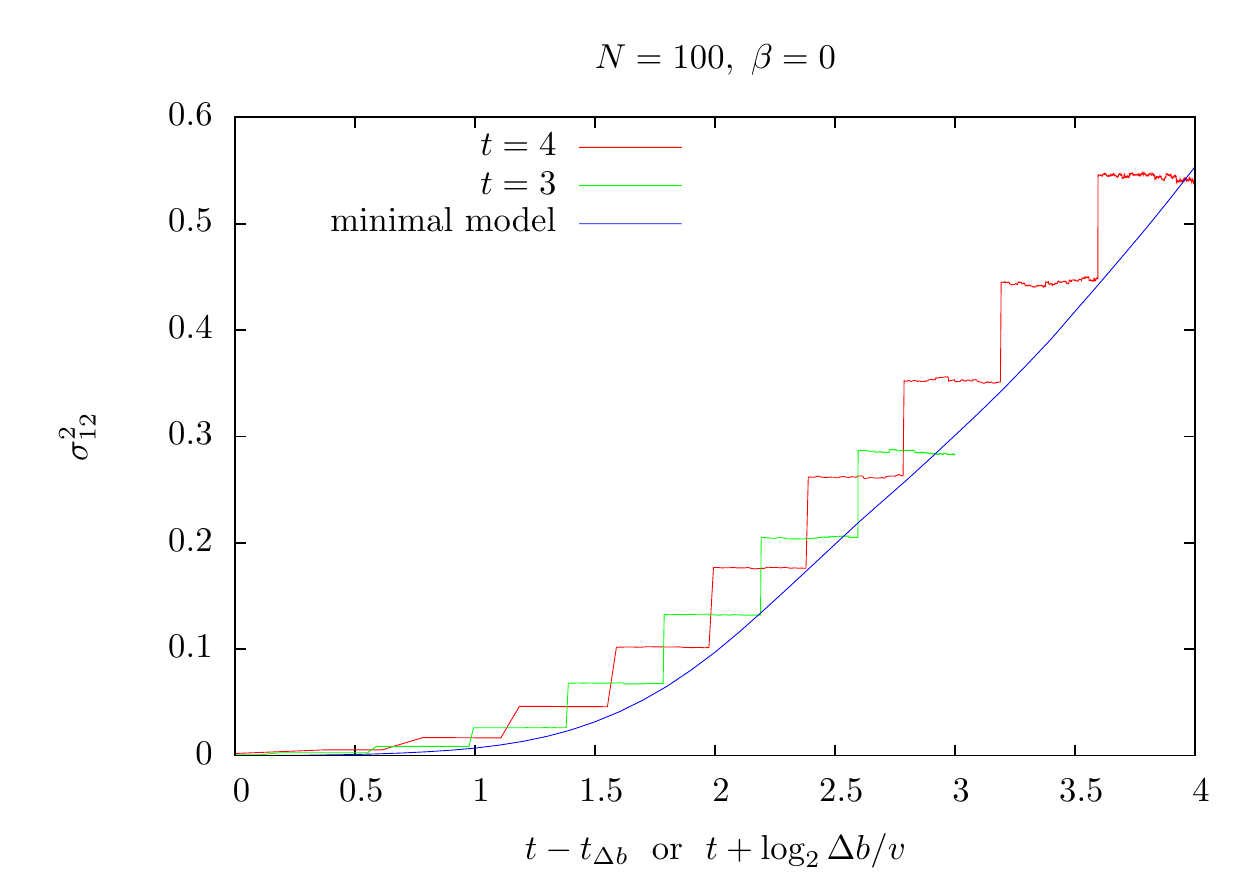}
\end{center}
\caption{\label{fig:plot_100_0}
$\sigma^2_{12}=\langle (X_1-X_2)^2\rangle$ as a function
of $\Delta t=t-t_{\Delta b}$ in the full model with $\alpha=1$ and $\beta=0$ 
(lines with steps; one corresponds to an evolution time $t=3$, the other
one to $t=4$)
and in the minimal model.
In the full model, $t_{\Delta b}=[-\log_2\Delta b]/v$, where $v$ is the measured
velocity of the front at impact parameter $-\frac12$.
In the FIP model, $t_{\Delta b}$ is a fixed time, and corresponds to
the time at which 
the tip of the front reaches $k_{\Delta b}$, the bin after which 
two uncorrelated evolutions take place.
}
\end{figure}

\begin{figure}
\begin{center}
\includegraphics[width=0.8\textwidth]{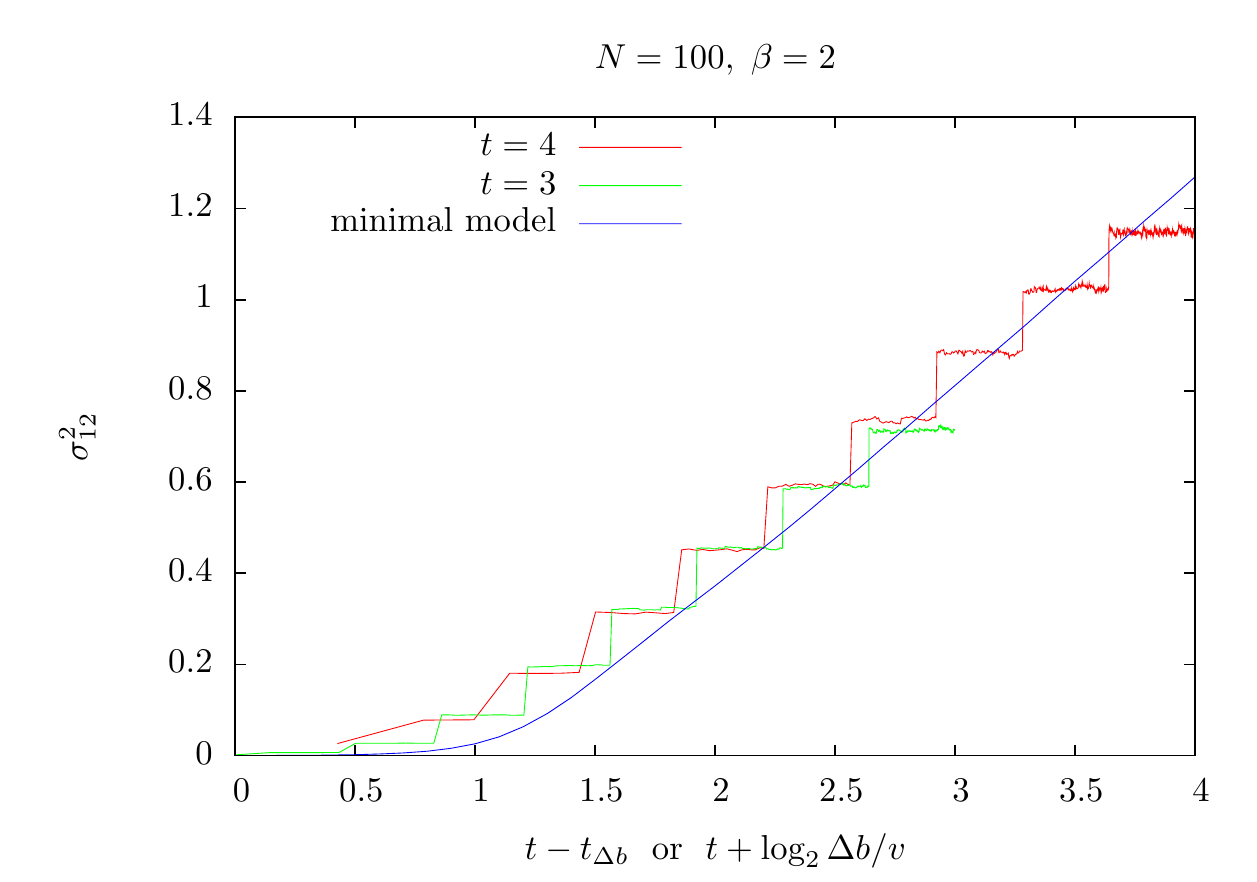}
\end{center}
\caption{\label{fig:plot_100_2}
The same as in Fig.~\ref{fig:plot_100_0} but for $\beta=2$.
}
\end{figure}

The corresponding plot is displayed in Fig.~\ref{fig:plot_100_0}
for $N=100$, $\alpha=1$, $\beta=0$, and in  Fig.~\ref{fig:plot_100_2}
with the same parameters except $\beta=2$. 
First, we see that in the full model,
the graph of $\sigma_{12}^2$ exhibits steps, i.e. 
$\sigma_{12}^2$ is constant by parts. 
This is related
to the hierarchical structure of the model:
The correlations between $b=-\frac12$ and
any of the points at $b\geq 0$ are identical;
The same is true for $-0.25\leq b<0$, $-0.375\leq b<-0.25$ etc...
The logarithmic $b$-scale on the $t$-axis makes the widths of the
steps all equal.
Next, we see that for small $t-t_{\Delta b}$ 
(i.e. impact parameters close
to $-\frac12$)
there are very little fluctuations in the front positions.

Finally, we see that for $\beta=0$, as anticipated, the full model
and the minimal ones coincide almost perfectly (Fig.~\ref{fig:plot_100_0}).
For $\beta=2$, 
i.e. when splittings towards larger dipole sizes are switched on
and therefore new correlations appear beyond
the ones taken into account in the minimal model,
there are some quantitative differences for large
$t$ (Fig.~\ref{fig:plot_100_2}). 
But we see that using the minimal model instead of the
full model that keeps all impact parameters is
a good approximation.
This corroborates the conclusions of the work in Ref.~\cite{Munier:2008cg}.

\subsubsection{Minimal model}

\begin{figure}
\begin{center}
\includegraphics[width=0.7\textwidth]{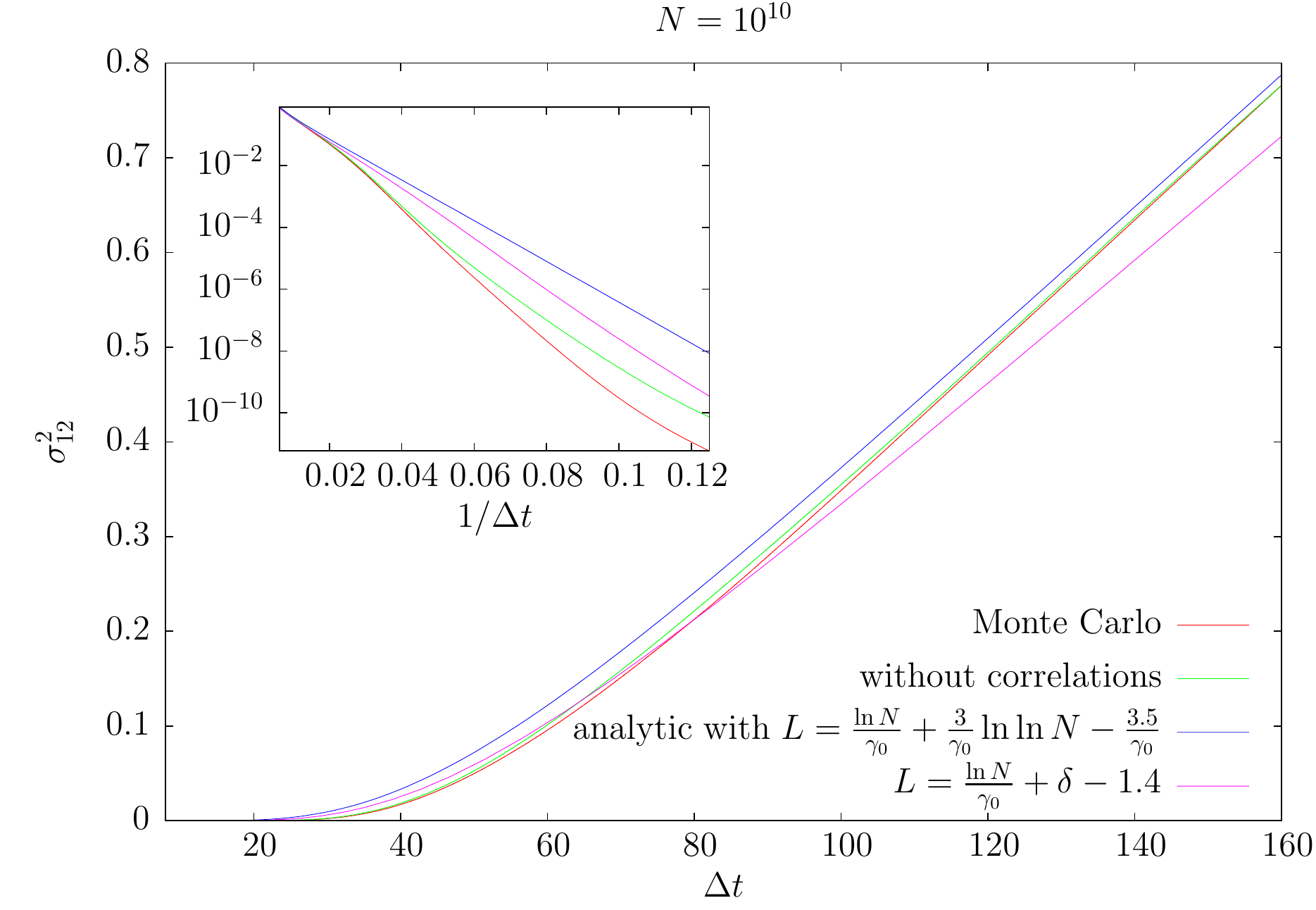}
\end{center}
\caption{\label{fig:plot10}
$\sigma_{12}^2$ as a function of $\Delta t=t-t_{\Delta b}$
in the minimal model with $\beta=0$
for $N=10^{10}$.
We display the results obtained within
the model in which the realizations
decorrelate in the bins $k>k_{\Delta b}$
(labelled ``Monte Carlo''), and within the model
in which the decorrelation is complete
after time $t_{\Delta b}$ (labelled ``without correlations'').
The theoretical curves use Eq.~(\ref{eq:sigma122f})
with the two possible choices for the front size $L$.
{\it Inset:} The same, as a function
of $1/\Delta t$ in order to highlight the
small-$\Delta t$ region where, as expected, important 
differences
appear between the models.
}
\end{figure}

\begin{figure}
\begin{center}
\includegraphics[width=0.7\textwidth]{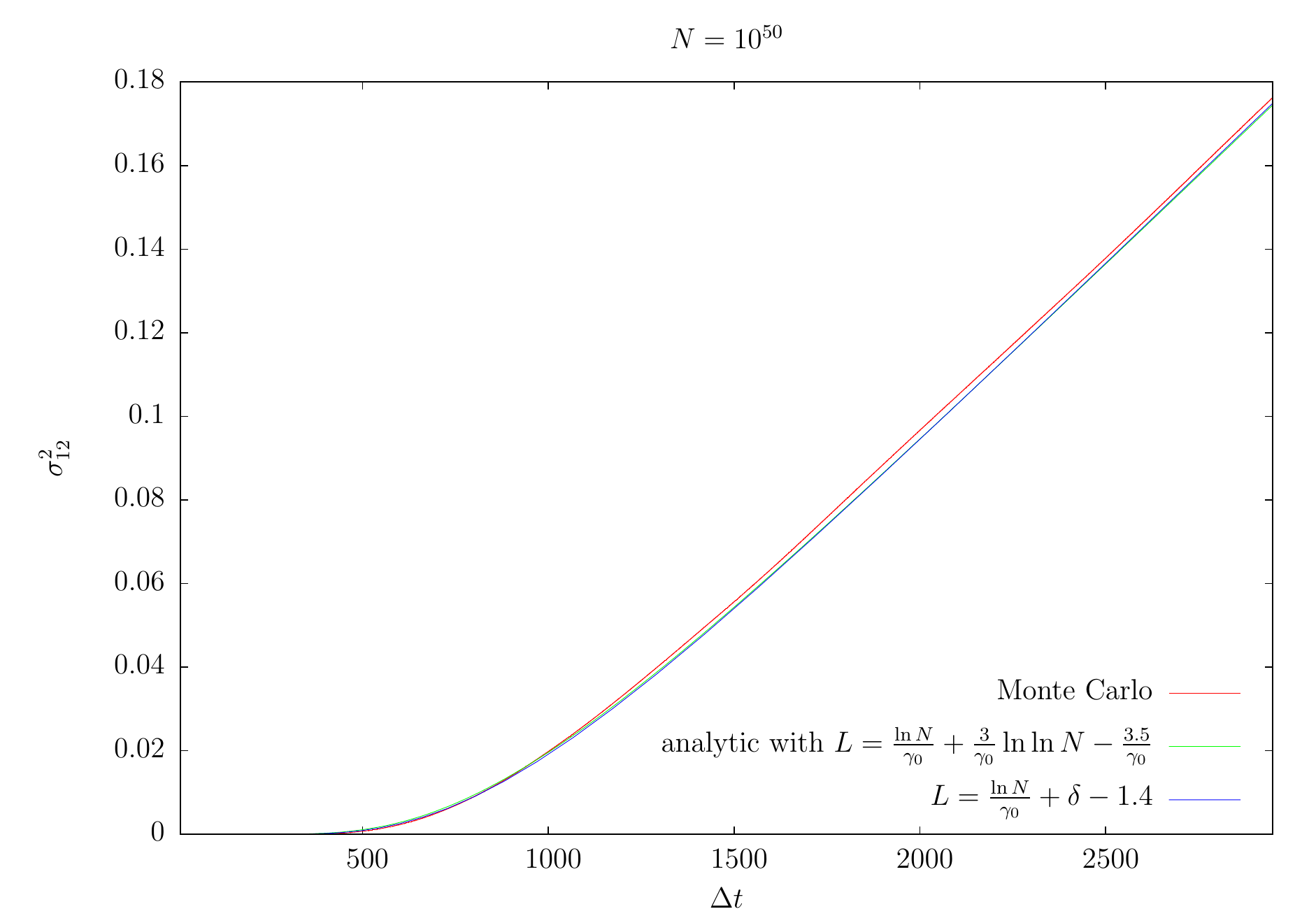}
\end{center}
\caption{\label{fig:plot50}
The same as in Fig.~\ref{fig:plot10}, for $N=10^{50}$.
All curves coincide almost perfectly.
}
\end{figure}

We now set $\beta=0$, in which case, as discussed
earlier and as checked numerically, 
the model exactly reduces to a collection
of one-dimensional FKPP-like models.
Hence, in order to compute two-point correlation functions,
it is enough to evolve two realizations of the corresponding
FIP model with the constraint
that all bins with $k\leq k_{\Delta b}$ be identical between the
two realizations, and the bins $k>k_{\Delta b}$ be completely
independent.
Alternatively, we could also generate one single realization
and evolve it for $t_{\Delta b}$ time steps, 
replicate it at time $t_{\Delta b}$,
and then evolve the two replicas completely independently of each
other. The difference between these two possible implementations
of the minimal model cannot be accounted for
in our analytical calculations,
thus the differences that we shall find 
numerically
will give an indication of 
the model uncertainty.
This time, our aim is essentially
to check our analytical formulas,
thus we will pick very large values of $N$,
even if they appear to be 
unphysical in the QCD context since
they would correspond to exponentially small values of
the strong coupling constant $\alpha_s$.

The parameters of the model are obtained from Eq.~(\ref{eq:eigenvalue})
with $\alpha=1$, $\beta=0$ and $dt=10^{-2}$:
\be
\gamma_c=1.0136\cdots\ ,\ \ v_0=2.6817\cdots\ ,\ \ 
\omega^{\prime\prime}(\gamma_c)=2.6098\cdots
\ee
These values are close to $1$, $e$ and $e$ respectively,
which would be the correct parameters if $dt$ were infinitesimal,
in which case $\omega(\gamma)=e^\gamma$ 
(see Eq.~(\ref{eq:eigenvalue})).

The numerical results are shown 
in Fig.~\ref{fig:plot10}
for $N=10^{10}$
with the two versions of the model
(we generated about $10^5$ realizations),
and compared with
the analytical predictions.
We test the two possible choices for the size $L$ of the
front: Either $L$ is a constant, which from our
previous experience with FKPP traveling waves \cite{Brunet:2005bz},
we set to
\be
L=\frac{1}{\gamma_c}\ln N+\frac{3}{\gamma_c}\ln\ln N-\frac{3.5}{\gamma_c}
\ee
(see e.g. Eq.~(\ref{ansatzLeff})), or it is $\delta$-dependent, namely
\be
L=\frac{1}{\gamma_c}{\ln N}+\delta-1.4.
\ee
The numerical
constants, which are not determined in our theory,
were chosen empirically so that they properly
describe all numerical data for $N\geq 10^{10}$.
In the first case, Eq.~(\ref{eq:sigma122f}) 
is used. In the second case,
Eq.~(\ref{eq:calsigma2}) 
is integrated numerically over
$t$ and $\delta$.
We see that the agreement between the numerical calculation
and the analytical predictions
is good, except maybe for very small values of $\Delta t$
where the calculations are not expected to be accurate.
Indeed, for the same values of $\Delta t$, we also 
see in Fig.~\ref{fig:plot10} a sizable discreapancy between
the two versions of the minimal model.
The calculations for $N=10^{50}$ are shown in Fig.~\ref{fig:plot50}.
The numerical results and the theoretical 
expectations (Eq.~(\ref{eq:sigma122f}))
coincide almost perfectly.

\begin{figure}
\begin{center}
\includegraphics[width=0.8\textwidth]{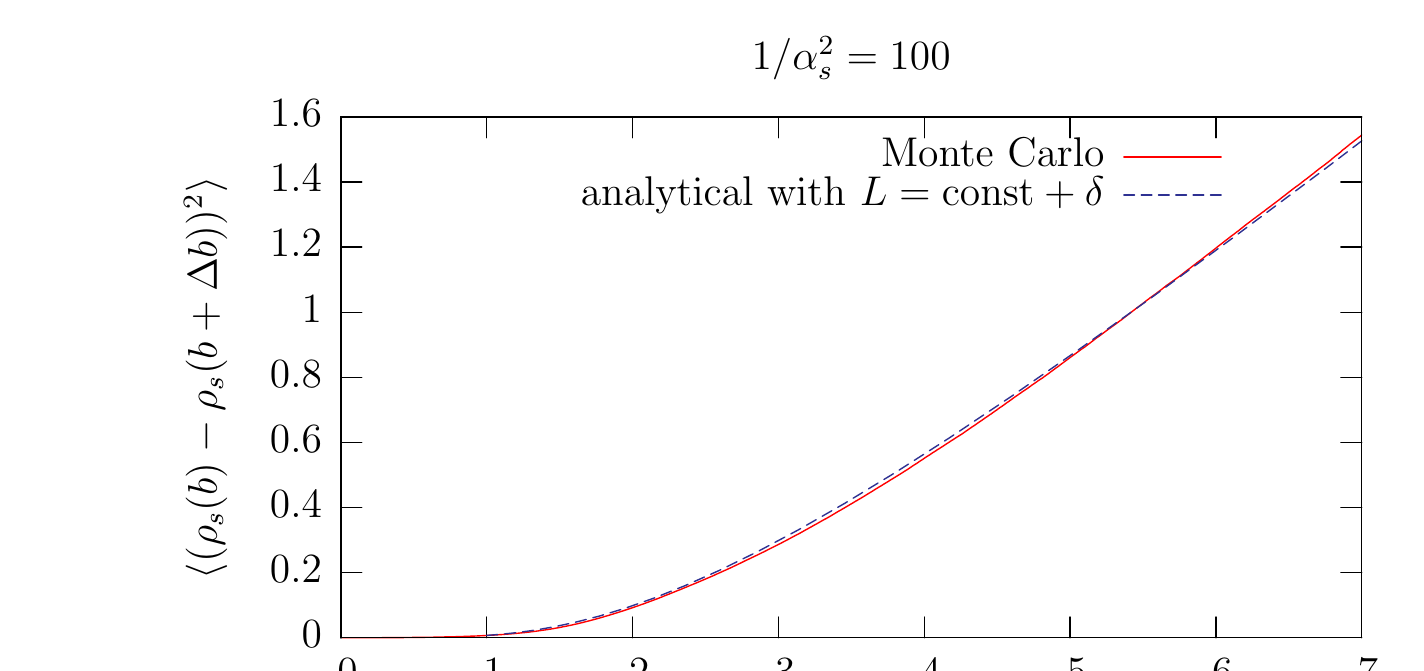}
\end{center}
\caption{\label{fig:plot100}
Comparison of a numerical Monte Carlo simulation
and our analytical formula.
The constant in the parameter $L$ (see the text)
which should be equal to
$\ln(1/\alpha_s^2)/\gamma_0$ for very small $\alpha_s$, 
has been shifted by a phenomenological
constant. Once this is done, we get 
a very good agreement
between the two calculations.
}
\end{figure}

For larger and more realistic values of $\alpha_s$, 
the persistence of the correlations
is still seen in the numerical simulations, 
but some parameters
should be modified
in the analytical expressions
and tuned to account for 
our lack of understanding of subleading corrections
important
for finite $\ln(1/\alpha_s^2)$. We show such a calculation
for $\alpha_s=0.1$ in Fig.~\ref{fig:plot100}, 
compared to a variant
of Eq.~(\ref{eq:sigma122f}).

\begin{figure}
\begin{center}
\includegraphics[width=0.8\textwidth]{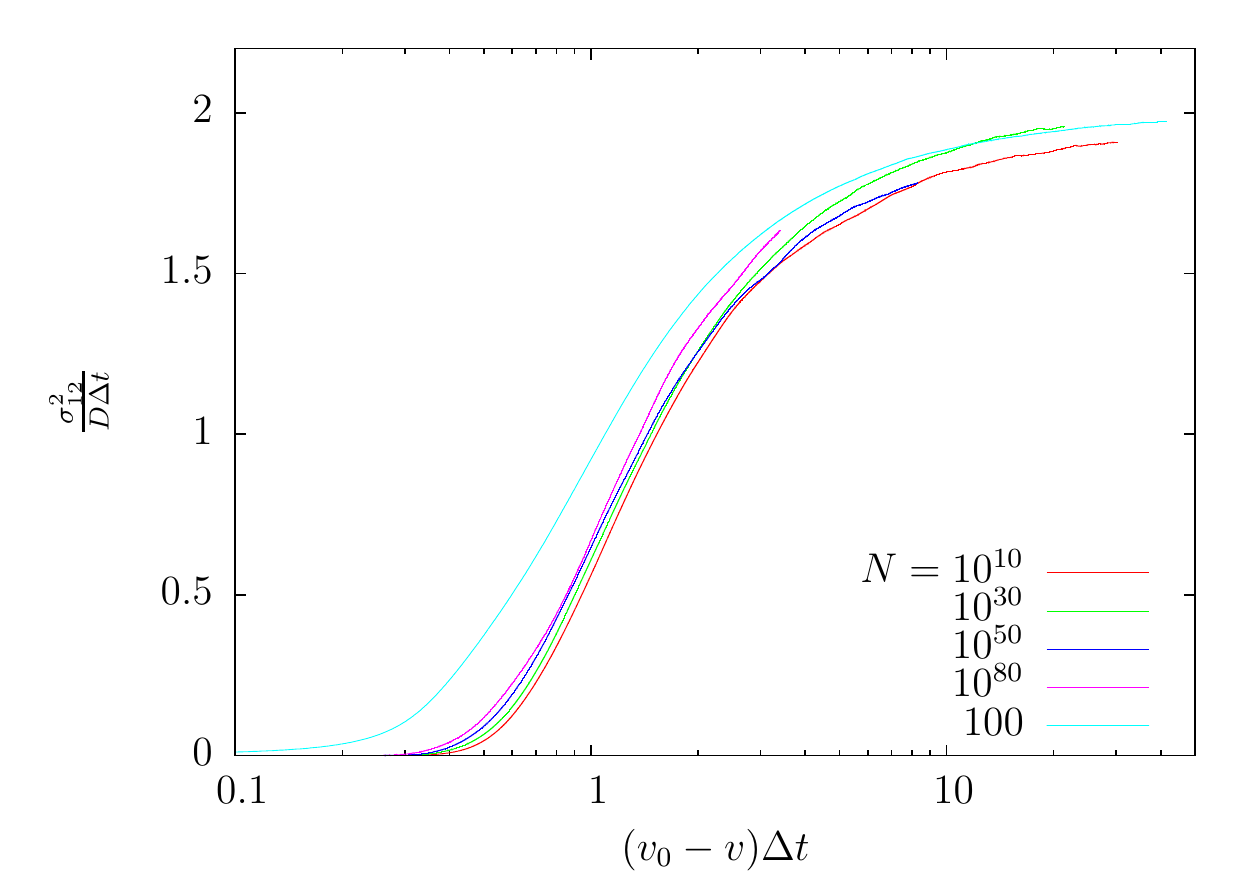}
\end{center}
\caption{\label{fig:scaling}
Numerical check of the scaling~(\ref{eq:scaling}).
The curves for the different values of $N$
are very close together for $N\geq 10^{10}$, 
but the scaling seems to break down for low values of $N$
(see the curve for $N=100$), as expected.
}
\end{figure}
Finally, we check that the scaling in Eq.~(\ref{eq:scaling})
is well reproduced by the numerical data.
The Monte-Carlo simulations are shown in Fig.~\ref{fig:scaling},
plotted in the appropriate scaling variables.
The diffusion constant of a single wave front $D$ as well as
the velocity $v$ are measured from the same data.
We see that all curves nicely superimpose for $N\geq 10^{10}$ (we
show data for values of $N$ as large as $10^{80}$),
while there are clear deviations for smaller $N$
(see the curve for $N=100$), as expected.


\chapter{\label{sec:application}
Phenomenological applications}

{\it
In this chapter, we review the phenomenological
consequences
of the results obtained from the correspondence
with statistical physics.
We derive new properties of the QCD scattering amplitudes
and discuss their impact on phenomenology.}\\

\minitoc


As was stated in the Introduction, the initially unplanned opportunity to collect
data in the high-energy regime of deep-inelastic scattering
at HERA triggered a renewed interest in small-$x$ physics among phenomenologists.
The major discoveries in this regime is the (unexpected) 
important fraction of diffractive events,
and a new scaling, {\em geometric scaling}, featured by total 
(and even semi-inclusive) cross-sections (see Fig.~\ref{fig:geometricscaling} in
the Introduction).

In order to deal theoretically with the small-$x$ regime,
one needs new factorization theorems in order to
single out the elements of the cross-sections 
that are computable in perturbation theory. High-energy, 
also called $k_\perp$-factorization
\cite{Catani:1990xk,Catani:1990eg,Collins:1991ty}, is the
appropriate tool. A practical way to implement
$k_\perp$-factorization is the color dipole model 
presented in Chap.~\ref{sec:schannel}.

\section{Dipole models and geometric scaling}

The main observable measured at HERA is the proton structure function $F_2$.
It is proportional to the sum of the virtual photon-proton cross-section
for a transversely and longitudinally polarized photon respectively.

A bare photon has no hadronic interactions, since it does not carry any color
charge. However, it may easily fluctuate into a quark-antiquark pair, 
overall color-neutral, thus forming a color dipole.
Subsequently, these dipoles interact with the target proton.
This picture is represented by the following equations:
\be\label{sigmagamma}
\begin{split}
F_2(x,Q^2)&= \frac{Q^2}{4 \pi^2 \alpha_{\rm em}}
\big(\sigma_T+\sigma_L\big),\\
\sigma_{T,L}(x,Q^2)&= \int dz d^2{r}\, |\Psi_{T,L}(z,{ r},Q^2)|^2\,
\sigma_{\rm dipole}(x,{r}).
\end{split}
\ee
Here, $\sigma_{T,L}$ are the photon-proton cross-sections for transversly and
longitudinally polarized virtual photons.
$\Psi_{T,L}$ are light-cone wavefunctions for 
$\gamma^*$, computable
within QED (see, e.g., Ref.~\cite{GolecBiernat:1998js} 
for explicit expressions to lowest order in
$\alpha_{\rm em}$).
Furthermore, $\sigma_{\rm dipole}(x,{r})$ is the cross-section 
for dipole--proton scattering (for a dipole of transverse size ${r}$),
and encodes all the information about hadronic interactions
(including unitarization effects). 
This cross-section is related to the
amplitude $A$ discussed so far by an integration over the impact
parameter. (Actually, $A$ was the forward elastic amplitude; the
optical theorem relates it to the total cross-section).

In Ref.~\cite{GolecBiernat:1998js,GolecBiernat:1999qd}, the dipole
cross-section was modeled as
\be\label{Golec}
\sigma_{\rm dipole}(x,{r})\,=\,\sigma_0\Big(1 - {\rm e}^{-{r}^2 Q_s^2(x)/4}\Big),
\ee
where $\sigma_0$ is a hadronic cross-section: It stems from
the integration over the impact parameter, when the impact parameter dependence
is supposed to be uniform over a disk of radius $\sim\sqrt{\sigma_0}$.
$Q_s(x)$ plays the role
of the saturation momentum, parametrized as 
$Q_s^2 (x)= (x_0/x)^\lambda\times 1$~GeV$^2$. 
Note that, by construction, this cross-section only depends on the combined variable
$r^2 Q_s^2(x)$ instead of $r$ and $x$ separately.
This property is transmitted to the measured photon
cross-sections $\sigma_{T,L}(x,Q^2)$, which then depend on $Q^2/Q_s^2(x)$ only
(this scaling is slightly violated by the masses of the quarks).
This is {\em geometric scaling}, predicted to be a feature of
the solutions to the BK equation at large rapidity.

Historically, geometric scaling was discovered first in the data
(see Ref.~\cite{Stasto:2000er}), after Golec-Biernat and W\"usthoff (GBW)
had written down their model: The latter happened to
feature this scaling (up to small violations
induced by the quark masses).
There was no apparent need for finite rapidity scaling violations
in the first HERA data.
However, later analysis revealed
that a significant amount of explicit scaling violations 
in the dipole cross-section,
predicted by the BK equation, were actually
required by more accurate data.

A now popular model that describes the HERA data in a way that 
takes a better account of the subasymptotics, beyond the
GBW model, was formulated in Ref.~\cite{Iancu:2003ge}.
The dipole scattering cross-section reads
$\sigma_{\rm dipole}(x,{r})=2\pi R^2 {\mathcal N}(y,rQ_s)$, with
\be\label{NFIT}
{\mathcal N}(y,rQ_s)=
\begin{cases}
{\mathcal N}_0\, \left(\frac{{r}^2 Q_s^2}{4}\right)^
{\gamma_c + \frac{\ln(2/rQ_s)}{\kappa \lambda Y}}&{\rm for}\quad rQ_s\le 2,
\\
1 - {\rm e}^{-a\ln^2(b\, rQ_s)}&{\rm for}\quad rQ_s > 2,
\end{cases}
\ee
where
$Q_s\equiv Q_s(x) = (x_0/x)^{\lambda/2}$ GeV.
The expression for the cross-section for $r$ small compared to $2/Q_s$ corresponds
to the solution of the BK equation
(compare to Eq.~(\ref{eq:frontgeneral0}) with the help of Tab.~\ref{tab:dictionary}),
in which we substituted $\omega(\gamma_c)=\omega(\gamma_c)$ 
and $\omega^{\prime\prime}(\gamma_c)=\omega^{\prime\prime}(\gamma_c)$
by the parameters $\lambda$ and $\kappa$ that we subsequently fit to the data.
The expression in the second line also has the correct functional
form for $r\gg 2/Q_s$, as obtained by solving the BK equation \cite{Levin:2000mv}.
This is strictly valid only to leading-order accuracy,
but here it is used merely as a convenient interpolation 
towards the `black disk' limit ${\mathcal N}=1$.
(The details of this interpolation 
are unimportant for the calculation of $\sigma_{\gamma^*p}$.)
The coefficients $a$ and $b$ are determined  uniquely
from the condition that ${\mathcal N}(rQ_s,Y)$ 
and its slope be continuous at $rQ_s=2$.
The overall factor ${\mathcal N}_0$ in the first line of Eq.~(\ref{NFIT})
is ambiguous, reflecting an ambiguity in the definition of $Q_s$. 
This model fits well all HERA data for structure functions, in the range 
$x\leq 10^{-2}$. All details may be found in Ref.~\cite{Iancu:2003ge}.

The model explicitely breaks geometric scaling. However,
effectively, geometric scaling remains a fairly 
good symmetry of the model,
as required by the data.
The small finite-rapidity scaling violations are needed to describe
accurately the high-precision HERA data.

The model may also accomodate less inclusive observables, such as
diffraction \cite{Forshaw:2004xd}.
It has been improved recently by including heavy quarks \cite{Soyez:2007kg}
(The crucial need for taking account of the charm quark 
was emphasized in Ref.~\cite{Thorne:2005kj}).
An impact-parameter dependence was also introduced
\cite{Kowalski:2003hm,Watt:2007nr,Sergey:2008wk} 
that was already missing
in the GBW model.

The range of validity of dipole models has been re-examined recently \cite{Ewerz:2007md}.

\section{Diffusive scaling}

At still higher energies, according to the discussion 
of Chap.~\ref{sec:reviewtraveling}, one expects
the saturation scale to acquire a dispersion from event to event
that scales with the rapidity like $\sqrt{\bar\alpha y}$ when rapidity increases.
Although this dispersion is not an observable since there is no way
to measure the saturation scale of an individual event,
it manifests itself in the total cross-section in the form
of a new scaling, different from geometric scaling.

The physical amplitude for the scattering of a dipole of size $r$ off some
target is given by the average of all realizations
of the evolution at a given $y$:
\begin{equation}
A(y,r)=\langle T(r)\rangle|_y.
\end{equation}
For large enough rapidities and small enough $\alpha_s$,
these realizations are exponentially decaying fronts 
in the variable $\rho=\ln(1/r^2)$,
fully characterized by a stochastic saturation
scale, or rather its logarithm $\rho_s=\ln Q_s^2(y)$. 
For the purpose of the present discussion,
it may be approximated in the same way as in Eq.~(\ref{eq:approxt}), namely
\begin{equation}
T(\rho)=\theta(\rho_s-\rho)+\theta(\rho-\rho_s)e^{-\gamma_c(\rho-\rho_s)}.
\label{eq:tsimple2}
\end{equation}
The statistics of $\rho_s$ is given by Eqs.~(\ref{eq:vcorr}),(\ref{eq:cumcorr})
(up to the replacements suggested in Tab.~\ref{tab:dictionary} to go from a generic
reaction-diffusion to QCD).
At ultrahigh energies (and very small $\alpha_s$), 
it is essentially a Gaussian centered at
\be
\langle\rho_s\rangle=
\left(\frac{\omega(\gamma_c)}{\gamma_c}-
\frac{\pi^2\gamma_c\omega^{\prime\prime}(\gamma_c)}
{2\left(\ln(1/\alpha_s^2)+3\ln\ln(1/\alpha_s^2)\right)^2}
\right)\bar\alpha y
\ee
and of variance
\be
\sigma^2=\langle \rho_s^2\rangle-\langle\rho_s\rangle^2=
\frac{\pi^4\omega^{\prime\prime}(\gamma_c)}{3\ln^3(1/\alpha_s^2)}\bar\alpha y.
\ee
The scattering amplitude may be expressed by the simple formula
\be
A(y,\rho)=\frac{1}{\sigma\sqrt{2\pi}}
\int d\rho_s\, T(\rho)|_y
\exp\left(\frac{(\rho_s-\langle\rho_s\rangle)^2}{2\sigma^2}\right).
\label{eq:adiffusive0}
\ee
The most remarkable feature of this amplitude 
is the scaling form for $A$ that it yields:
\be
A(y,\rho)=A\left(\frac{\rho-\langle\rho_s(y)\rangle}
{\sqrt{\bar\alpha y/\ln^3(1/\alpha_s^2)}}
\right).
\label{eq:diffusivescalingA}
\ee
This equation may be obtained by performing
the integration in Eq.~(\ref{eq:adiffusive0}) after the replacement of
$T$ by its approximation~(\ref{eq:tsimple2}).
This scaling obviously violates geometric scaling: If the
latter scaling were satisfied, then $A$ would be a function
of $\rho-\langle \rho_s(y)\rangle$ only.

In Ref.~\cite{Mueller:2004sea},
Mueller and Shoshi had already noted
that geometric scaling had to be violated beyond the BK equation.
However, the square root in the denominator of the scaling variable
in Eq.~(\ref{eq:diffusivescalingA})
was missing because their approach was relying on mean field
throughout, thus missing the stochastic nature of the evolution.

This new scaling is a firm
prediction of the correspondence with statistical physics.
However, it may not be tested at particle colliders in a simple way.
Let us work out the order of magnitude of the rapidity 
needed for the different effects (saturation, geometric scaling, diffusive
scaling) to show up.
The rapidity that is needed to reach saturation is roughly
\be
y_{\text{BFKL}}\sim
\frac{\ln({1}/{\alpha_s^2})}{\bar\alpha\omega(\frac12)}.
\ee
The BK picture is expected to be valid until the asymptotic 
exponential shape of the
front has diffused down to the point where the amplitude becomes
of the order of $\alpha_s^2$. This additional rapidity 
needed to get to the regime of geometric scaling is thus given by
Eq.~(\ref{eq:relaxtime}) once the appropriate replacements have been done
\be
y_\text{BK}\sim
\frac{1}{2\bar\alpha\omega^{\prime\prime}(\gamma_c)}
\left[\frac{\ln({1}/{\alpha_s^2})}{\gamma_c}\right]^2,
\ee
and finally, the effect of the fluctuations of the saturation scale
gets important at the rapidity
\be
y_\text{fluct}\sim\frac{3\ln^3 (1/\alpha_s^2)}
{\bar\alpha\pi^3\omega^{\prime\prime}(\gamma_c)}.
\ee
The relevant parameters in QCD are deduced from
the BFKL kernel. They read
\begin{equation}
\gamma_c=0.627549,\ \ \omega(\gamma_c)=3.0645,\ \ 
\omega^{\prime\prime}(\gamma_c)=48.5176.
\end{equation}
For some realistic strong coupling constant, $\alpha_s\sim 0.2$,
we get
\begin{equation}
y_{\text{BFKL}}\sim 6.07879
\ ,\ \
y_\text{BK}\sim 1.41965
\ ,\ \
y_\text{fluct}\sim 0.348244.
\end{equation}
Given that rapidities in the small-$x$ regime at HERA were of the order of 10,
and will be of the order of 15 at the LHC, these figures indicate
that we may observe these effects.
However, there are many criticism
to these naive estimates.

First, the values of the rapidity that delimitate the different regimes
are largely underestimated given that they rely on the 
leading-order BFKL kernel, which predicts a much too large
growth of the cross-section with the rapidity and a too fast diffusion (see the 
large value of $\omega^{\prime\prime}(\gamma_c)$).
Already the effect of the running coupling, which should
be taken into account in any detailed phenomenological
study, is expected
to still reduce the effects of the fluctuations \cite{Dumitru:2007ew}.

Next, one also has to keep in mind that the former estimates should only hold
for very small values of $\alpha_s$, such that $\ln 1/\alpha_s^2\gg 1$
which is certainly not true in real-life QCD.
Note that asymptotically, one
should have $y_\text{fluct}\gg y_\text{BK}\gg y_\text{BFKL}$. The fact that the
order is inverted means that the quantitative results obtained within the
phenomenological model for front propagation should not be trusted for values of
$\alpha_s$ as ``large'' as $0.2$.

Nevertheless,
the effect of diffusive scaling (i.e. of the event-by-event
fluctuations of the saturation scale) on observables
has already been investigated in some detail 
by several groups. Diffractive amplitudes were studied
in Ref.~\cite{Hatta:2006hs}.
The ratio of the gluon distribution
in a nucleus to the same quantity in a proton
was computed in Ref.~\cite{Kozlov:2006qw}.


\chapter{Conclusion and outlook}

We have reviewed a peculiar way of viewing high-energy scattering in QCD,
based on the physics of the parton model, and its strong similarities
with reaction-diffusion processes (Chap.~\ref{sec:schannel}).
The correspondence is best summarized in the mapping 
of Tab.~\ref{tab:dictionary}.
We have seen that the equations that describe the dynamics of
these processes are in the universality class of the stochastic FKPP
equation, and admit traveling-wave solutions whose
features are likely to be universal, in such a way
that a study of simple reaction-diffusion-like models may lead to
exact asymptotic results also for QCD scattering amplitudes.
Understanding the very mechanism of traveling wave formation
and front propagation was crucial to see how the universality may
come about (see Chap.~\ref{sec:reviewtraveling}).

In zero-dimensional stochastic models, we could perform exact calculations
and get analytical results within different formulations 
(Chap.~\ref{sec:zerodimensional}).
We understood that analyzing the structure of single events
was technically much simpler if one wants
to get leading orders at large $N$ ($=1/\alpha_s^2$),
since in individual realizations, one may factorize the fluctuating part
from the nonlinear effects.
Thanks to this observation,
in one-dimensional models which admit realizations in the form
of stochastic traveling waves,
we could also get precise analytical results on the
form and shape of the traveling waves, which are presumably exact
asymptotically (Chap.~\ref{sec:reviewtraveling}).
Universality enables one to make statements on
the form of the QCD scattering amplitudes at very high energies.
Appropriate extensions of the 
relevant statistical models which incorporate 
an additional dimension
lead to predictions for the correlations in the transverse plane 
(Chap.~\ref{sec:spatial}).
Some of these results turn into firm phenomenological predictions 
(Chap.~\ref{sec:application}), which however do not seem to be
testable at colliders in the near future.
Nevertheless, getting new analytical results for QCD in some
limit is always an interesting achievement, given the
complexity of the theory.
Furthermore, while our analytical results only apply for
exponentially small $\alpha_s$ ($\ln(1/\alpha_s^2)\gg 1$),
the picture itself should be valid in the whole perturbative range,
namely for $\alpha_s^2\ll 1$.


\paragraph{Prospects.}

There are still many open questions.
On the statistical physics side, the statistics of the front position
that we have found has not been derived rigorously, but rather
guessed, and rely on many quite ad hoc conjectures. We got confidence
on the validity of our conjectures on the basis of
numerical simulations.
Moreover, although we expect universality up to corrections of order $1/N$
(that is to say $\mathcal{O}(\alpha_s^2)$ in QCD),
we could only get analytical expressions relative to the cumulants
of the position of the front for 
the first terms in an expansion in powers of $1/\ln N$, which 
extremely large values of $N$ (small $\alpha_s$) to be valid.
But on a more general footing, the sFKPP equation seems to describe
many physical, chemical or biological problems (in particular
population evolution with selection in evolutionary biology).
We have also found recently an explicit analogy with the theory of spin
glasses \cite{Brunet:2006zn,BDMM:2007}.
This large universality is maybe the strongest incentive to
try and find more accurate solutions to that kind of equations.

On the QCD side, the correspondence with reaction-diffusion processes
strongly relies on the assumption that there is saturation
of some form of the quark and gluon densities
in the hadronic wave functions.
While this is a reasonable guess that few experts would challenge,
it is clear that we cannot consider that the problem
is solved before the saturation mechanism at work in QCD
has been exhibited.
QCD is formulated as a quantum field theory. To see the similarity
with reaction-diffusion, we basically needed to
translate it into the parton model first.
It would be better to recover the results of 
Chap.~\ref{sec:reviewtraveling}
(and hopefully get more)
directly from field theory \cite{Munier:2008rh}, 
as one could do it in the zero-dimensional model
introduced in Chap.~\ref{sec:zerodimensional}.
This requires to understand the strong field regime of field theory.
This is an exciting challenge for both particle physicists and statistical
physicists.\\

Let us finally state our personal prospects in the field.
First, we wish to go back to the simple Balitsky-Kovchegov equation
and study in more detail the properties of its solutions.
It is important for phenomenology, since it seems that in the range
of energy that may be reached at experiments, effects described by
more advanced equations (incorporating genuine saturation effects,
as discussed at length in previous chapters) are likely to be
negligible. Interestingly, since the BK equation also represents
the statistical properties of the tip of a random
walk (see Chap.~\ref{sec:reviewtraveling} and the recent
paper~\cite{BD2011}), some fine properties of scattering
amplitudes may be inferred from the study of the latter.
Work is in progress in this direction in collaboration with Al Mueller.

On the pure statistical physics side, we wish to pursue the study of
simple models like the ones presented in Chap.~\ref{sec:zerodimensional},
which could be of some interest in the 
interdisciplinary field of population evolution studies.

Last, if the formalism of the dipole model on which relies
most our work in QCD seems well-suited for electron-proton or nucleus
high-energy scattering, most of the experimental data which will become
available in the next decade are about proton and nucleus interactions
at the Large Hadron Collider (LHC).
The new challenge to phenomenologists is to formulate and compute
observables in this context. It seems that quadrupoles play an
important role for all interesting observables, see e.g. 
Ref.~\cite{Dominguez:2010xd}. Recently, we
made a first step in the direction of computing the
evolution of such objects with the energy~\cite{Dominguez:2011gc} and
we will pursue in this promising direction. 
An interesting theoretical question in
the continuation of our work would be, for example, if 
the correlations computed in
Chap.~\ref{sec:spatial} would show up 
in the evolution of these quadrupoles and hence in the corresponding
observables measured at the LHC.


\end{document}